\documentclass[phd, 10pt, a4paper, inlinechaptertoc]{psuthesis}
\usepackage[latin1]{inputenc}
\usepackage[left=3.3cm,top=4cm,right=2.1cm,bottom=3cm]{geometry}  
\usepackage{fancyhdr}
\usepackage{amsmath,booktabs}
\usepackage{amssymb}
\usepackage{amsthm}
\usepackage{amsmath}
\usepackage{mathrsfs}

\allowdisplaybreaks

\usepackage{subcaption}

\usepackage{exscale}
\usepackage{dsfont}
\usepackage{fancybox}
\usepackage[mathscr]{eucal}
\usepackage{bm}
\usepackage{eqlist} 
\usepackage[final]{graphicx}
\usepackage[dvipsnames]{color}
\usepackage{bbding}
\definecolor{maroon}         {cmyk}{0   , 0.93, 0.9   , 0.40}

\usepackage{braket}
\usepackage{tikz}

\usepackage{url}

\usepackage{soul}

\usepackage{imakeidx} 
\usepackage{xstring}
\newcommand{\important}[1]{\IfSubStr{#1}{!}{\StrBefore{#1}{!}~\StrBehind{#1}{!}\index{#1}}{#1\index{#1}}}

\usepackage{cite}

\usepackage[all]{xy}

\DeclareGraphicsExtensions{.pdf, .jpg, .png}

\usepackage[Lenny]{fncychap}
\ChTitleVar{\Huge\sffamily\scshape}

\DeclareMathOperator{\G}{\mathcal{G}}
\DeclareMathOperator{\E}{\mathcal{E}}
\DeclareMathOperator{\T}{\mathcal{T}}

\newcommand{\imap}{{\mathbb{I}}{\rm m}}
\newcommand{\realp}{{\mathbb{R}}{\rm e}}
\newcommand{\D}{{\rm d}}
\newcommand{\R}{{\mathbb{R}}}
\newcommand{\calD}{{\mathcal{D}}}
\newcommand{\calL}{{\mathcal{L}}}
\newcommand{\calV}{{\mathcal{V}}}
\newcommand{\calT}{{\mathcal{T}}}
\newcommand{\calJ}{{\mathcal{J}}}
\newcommand{\calS}{{\mathcal{S}}}
\newcommand{\calM}{{\mathcal{M}}}
\newcommand{\calH}{{\mathcal{H}}}

\newcommand{\sign}{{\rm sign}}

\title{Nonlocal Lagrangian Formalism}

\author{Carlos Heredia Pimienta}
\dept{Institute for Theoretical Physics}
\degreedate{\today}
\copyrightyear{\small{\today}}

\documenttype{Thesis}

\submittedto{Departament de F\'isica Qu\`antica i Astrof\'isica}

\makeindex[columns=2,intoc]

\usepackage[hidelinks]{hyperref}

\begin{document}

\frontmatter

%



\psutitlepageFP

\clearpage
\thispagestyle{empty}
\mbox{}
\clearpage

\psutitlepage
\clearpage
\thispagestyle{empty}
\mbox{}
\clearpage




%
\pagestyle{plain}
\chapter{Preface}
    \begin{singlespace}
~~

The more physics advances, the more mathematical tools we need. This (usually intensive) interaction between these two fields gives rise to the famous subject known as \textit{mathematical physics}. Standing on the edge between them is a highly complex task. We must always satisfy two requirements that are usually not easy to overlap. On the one hand, we must present the main results as generally and rigorously as possible. On the other hand, we must not explain them in such a way that could cause rejection by physicists, who certainly need them. The question is: did we succeed in these requirements? Only the readers can answer it. We hope they enjoy reading this manuscript as much as we have enjoyed writing it. 

We know --so far-- that matter can be fractionated to a minimum unit, quarks. A legitimate question is whether the same is true for spacetime, i.e., is there a minimal unit of it? It seems that the answer has to be so because, when we speak only about space, we certainly see that point objects cannot exist in reality. They are only a mathematical artifice to simplify our calculations. Therefore, something similar must happen with time; it should have a minimum unit as well. In our opinion, this idea suggests that the point concept should be switched to the region concept, which is \textit{nonlocality}'s physical idea.

One of the promising and challenging goals of theoretical physics for this century is the unification of gravity and quantum physics. These two seemingly wholly unrelated fields --one governing the macroscopic scale and the other the microscopic scale-- coexist in a particular region of spacetime, black holes. These exotic objects are heavy and immense but also contain quantum properties. We might say they are a clear example of the need for a \textit{quantum gravity} theory with a fair degree of certainty.
  
Nonlocality is an exotic emerging feature in quantum gravity models. Inspired by the non-presence of singularities in these models, this characteristic has also been transferred  to classical gravity as a proposal to solve \textit{singularity problems}.  However, nonlocality appears not only in classical and quantum gravity but also in other fields, such as dispersive media and Fokker-Wheeler-Feymann electrodynamics. The method commonly used to explore such models is based on the generalization of Ostrogradsky formalism. The main idea of this generalization is to treat $r^{\rm th}$-order Lagrangians with $r\to\infty$. This approach can be heuristic unless the convergence of the series involved is proved or the ``convergence'' for $r\to\infty$ is adequately defined. For this reason, we propose an alternative to this method based on functional methods, which are lighter to handle as they involve integrals instead of series.

\vspace{0.3cm}
\textbf{Abstract}
\vspace{0.3cm}

 This thesis aims to study nonlocal Lagrangians with a finite and an infinite number of degrees of freedom. We obtain an extension of Noether's theorem and Noether's identities for such Lagrangians. We then set up a Hamiltonian formalism for them. Furthermore, we show that $r^{\rm th}$-order Lagrangians can be treated as a particular case, and the expected results are recovered. Finally, the method developed is applied to different examples: nonlocal harmonic oscillator, $p$-adic particle, $p$-adic open string field, and electrodynamics of dispersive media.

\newpage

\textbf{Summary of each chapter}
\vspace{0.3cm}

For that purpose, we have divided this thesis into six chapters. In chapter \ref{Intro}, we introduce the central theme of it.  We define the term \textit{nonlocal} and its motivations in current physical models. Furthermore, we explain the importance of Noether's theorem, as well as the Lagrangian and Hamiltonian formulations. 

Chapter \ref{chap2} reviews the most relevant concepts of Lagrangian and Hamiltonian formalisms for first- and $r^{\rm th}$-order theories. We start with the concepts of configuration space, generalized coordinates, cotangent space, and differential forms. We revisit the notion of the \textit{least action principle} for regular Lagrangians --that depend explicitly on time-- and derive the \textit{Euler-Lagrange equations}. We then prove \textit{Noether's theorem} and \textit{Noether's identities}. In particular, we examine the boundary term appearing in Noether's theorem and its connection with the definition of the \textit{Legendre transformation}. In this way, we construct the \textit{Hamiltonian formalism} as it is usually done in textbooks and more modernly based on symplectic geometry. Finally, we review all the above for $r^{\rm th}$-order Lagrangians and discuss the consequences for the Euler-Lagrange equations, the initial conditions, the Legendre transform, and the Noether symmetry when $r\to\infty$. 

Chapter \ref{chap3} adapts the results of chapter \ref{chap2} to Lagrangian fields, considering all the peculiarities of field theories concerning mechanics. We start directly with $r^{\rm th}$-order Lagrangian fields. We revisit the principle of least action and derive the \textit{Euler-Ostrogradsky equations}.  Next, we prove Noether's theorems and identities. We focus on the \textit{Poincar\'e symmetry} for the first theorem and derive the \textit{canonical energy-momentum tensor} and the \textit{angular current}. Moreover, since the canonical energy-momentum tensor does not necessarily have to be symmetric because of the \textit{spin current}, we introduce the \textit{Belinfante-Rosenfeld symmetrization technique} to construct a symmetric energy-momentum tensor.  In the following, we state Noether's second theorem for gauge systems, which is not usually very familiar, and, finally, we develop the Hamiltonian formalism for such Lagrangians. 

Chapter \ref{chap4} is the cornerstone of this thesis. We first establish the principle of least action for a nonlocal Lagrangian with a finite number of degrees of freedom, which may depend explicitly on time, and derive the nonlocal Euler-Lagrange equations. As a rule, they are integrodifferential equations. Becase we do not have general theorems of existence and uniqueness for these equations, it makes no sense to take them as laws ruling the system's time evolution from a set of initial data. Instead, we take them as constraints selecting the class of \textit{dynamical trajectories} as a submanifold of the broader class of \textit{kinematical trajectories}. We especially emphasize what is meant by the \textit{time evolution} of a given trajectory, and we relate it to the standard notion of evolution in local mechanics.

It happens that any Lagrangian is the total derivative of a nonlocal function. However, this fact does not imply that the Euler-Lagrange equations vanish identically. We give a particular example. Furthermore, we find what an extra condition the nonlocal function must satisfy so that the nonlocal Euler-Lagrange equations vanish identically.

We then prove an extension of Noether's theorem and Noether's identities for nonlocal Lagrangians. This extension --inspired by the local case-- allows us to use a conserved quantity to infer a suitable definition for the Legendre transformation and thus set up a Hamiltonian formalism for them. We observe that we cannot proceed like in the standard method in which the Legendre transformation is a change of coordinates --replacing velocities with momenta-- in the initial data space. Moreover, due to the lack of theorems of existence and uniqueness for the nonlocal Euler-Lagrange equations, we neither know what the dynamic space is like nor have a specific coordinate system for it. Instead, we introduce a trivial Hamiltonian formalism on the cotangent space on the infinite-dimensional manifold of kinematic trajectories and then translate it into a Hamiltonian formalism on the space of dynamic trajectories. The instrument to do so is the pullback of the injection, which mapps the cotangent space into the dynamic space. 

There are two ways of translating the Hamiltonian formalism from a larger phase space to a submanifold. One is based on the Dirac theory of constraints. This method implies computing the Poisson brackets between constraints, identifying the second-class constraints, inverting their matrix, and finally taking the associated Dirac brackets as the effective Poisson brackets for functions defined on the reduced space. Although the whole thing is conceptually simple for a finite-dimensional phase space and a finite number of constraints, it is not so when both the number of  dimensions and the constraints are infinite. The other method is based on symplectic mechanics. The Hamiltonian and the symplectic form on the reduced space are derived by pulling them back from the larger phase space. This second procedure has the advantage of reaching further without needing to find a system of coordinates for the dynamic space, which is the most challenging part of the problem. 

Chapter \ref{chap5} presents the functional form of nonlocal Lagrangian fields, which explicitly depend on the spacetime point. We then raise the nonlocal principle of least action and derive the nonlocal Euler-Lagrange field equations. As in nonlocal mechanics, it happens that any Lagrangian field is the total four-divergence of a nonlocal current, but this fact does not imply that the nonlocal Euler-Lagrange equations vanish identically. We find what extra conditions the nonlocal current must satisfy for the nonlocal Euler-Lagrange equations to vanish identically. We prove the first and second Noether's theorem. We then concretize to Poincar\'e invariant field theories and derive both the energy-momentum and angular momentum tensors. Finally, we set up a Hamiltonian formalism for the nonlocal Lagrangian field theory and derive a precise expression for the Hamiltonian and the symplectic form.

Chapter \ref{chap6} is devoted to the applications of chapters \ref{chap4} and \ref{chap5}. We present five examples, which are divided into two groups. The first group is for the nonlocal mechanics examples. The first example shows that $r^{\rm th}$-order regular Lagrangians are a particular case of the nonlocal formulation. In the second and third examples, we obtain the Hamiltonian and the symplectic form of the nonlocal harmonic oscillator and the $p$-adic particle. However, the latter example only in a perturbative way. The second group is for the examples of nonlocal fields. The first example is the $p$-adic open string, in which the hamiltonian and symplectic form are obtained perturbatively. Furthermore, we derive the components of the energy-momentum tensor in a closed form. The second example is the electromagnetic field in dispersive media. We obtain the energy-momentum tensor of a wave packet solution.

\vspace{0.3cm}
\textbf{Acknowledgements}
\vspace{0.3cm}

I want to thank everyone I have met during this fabulous journey. Thanks to all of them, achieving this would not have been possible. In particular, I would like to thank my colleagues at the University of Barcelona, Javier, Cristian, I\~{n}igo, Ivan, Paula, Aniol, and Diego, for our great discussions. They have been amazing! Also, I would like to thank my workmates at EY: Oscar, Alex, Toni, Jesus, Pol, Maria, Sara, Feyza, David Hita, David Ballano, Jordi, Julien, and Abraham. Thank you for teaching and training me as a consultant and for all your excellent advice over those three years. Besides, I want to thank also Prof. Anupam Mazumdar and Dr. Ivan Kol\'a\v{r} for the intense, profound, magnificent discussions on nonlocality. Thank you very much for teaching me and allowing me to have been with you in Groningen. Without forgetting Eliska and Felix, thank you very much for your hospitality. The experience was unique! 

En cuanto a mi c\'irculo m\'as cercano, me gustar\'ia agradecer a mis amig@s de toda la vida Roberto, Macarena, Rub\'en, Daniel Guirado y Daniel D\'iaz por la cantidad de veces que me hab\'eis hecho re\'ir. Cuando m\'as lo necesitaba, sin daros cuenta, me ayudasteis a desconectar y mirar los ``problemas" con otra perspectiva. No sab\'eis las veces que he presumido de vosotros y lo orgulloso que estoy de ello. Por cierto, enhorabuena Macarena y Roberto por traer a este mundo a vuestra preciosa hija Nora! Tambi\'en me gustar\'ia agradecer a mis compa\~{n}er@s que hicieron la carrera de F\'isica conmigo y que ahora forman parte de mi vida. Gracias a Carlos, Anna, Marta, Tablero, Javier, Berga, C\'ambara, Cantero, Eric, Fede, Miguel por las interminables discusiones sobre f\'utbol, y por las palabras de apoyo que siempre he recibido por vuestra parte. Por cierto, enhorabuena Marta y Alberto por vuestro maravilloso hijo Guillermo! Para acabar, me gustar\'ia agradecer todo esto a una de las personas m\'as importantes de mi vida en estos \'ultimos a\~{n}os. Gracias Eva por todo el apoyo, las horas (y horas y horas) que me has escuchado, la estabilidad, y por ense\~{n}arme y aceptarme tal y como soy. Siempre lo he dicho, y nunca lo dejar\'e de decir, fuiste un \'angel ca\'ido del cielo. !`Te quiero mucho! 

\vspace{0.3cm}
\textbf{A mi madre, a mi padre, a mi hermano y a mis abuelos}
\vspace{0.3cm}

Hay ciertas personas que quiero mostrar mi agradecimiento a parte. Estas personas son: a mi madre \'Angeles, a mi padre Manuel, a mi hermano Marcos, a mi abuelo Francisco y a mi abuela Ces\'area. No pod\'eis ni imaginar lo orgulloso que estoy de mi familia. Gracias por todo el apoyo que me hab\'eis dado durante todos estos a\~{n}os y, sobre todo, por haber confiado en m\'i. Rechazo tras rechazo, nunca me dejasteis de animar, tanto moralmente como econ\'omicamente. Sab\'iais que era mi sue\~{n}o, y, ahora s\'i, puedo deciros que lo consegu\'i. Esta sociedad est\'a acostumbra a seguir a la multitud, sin embargo, vosotros me ense\~{n}asteis a seguir lo que a MI me apasionaba; gracias, y mil gracias, por ense\~{n}arme a ser libre.

\vspace{0.3cm}
\textbf{To Prof. Josep Llosa}
\vspace{0.3cm}

Josep, let me write my thanks in English. I want everyone to understand how proud I am to have done my doctoral thesis with you. I thank you with all my heart for believing in me. Thank you for all the hours you have spent listening to and teaching me. For all the patience you had when I did not understand something, and for all the excellent advice you gave me during all these years. I want you to know that I have always been bragging about how lucky I have come across you. I have fulfilled a dream thanks to you. Gr\`acies per tot Josep, i que s\`apigues que, encara que et retiris, no et lliurar\`as tan f\`acilment de mi!

\vspace{0.3cm}
\textbf{Articles}
\vspace{0.3cm}

This manuscript has been the result of the following publications:
\begin{enumerate}
\item C. Heredia and J. Llosa, ``Energy-momentum tensor for the electromagnetic field in a dispersive medium," Journal of Physics Communications 5 no. 5, (May, 2021) 055003.
\item C. Heredia and J. Llosa, ``Non-local Lagrangian mechanics: Noether's theorem and Hamiltonian formalism," Journal of Physics A: Mathematical and Theoretical 54 no. 42, (Sep, 2021) 425202.
\item C. Heredia, I. Kol\'a\v{r}, J. Llosa, F. J. M. Torralba, and A. Mazumdar, ``Infinite-derivative linearized gravity in convolutional form," Classical and Quantum Gravity 39 no. 8, (Mar, 2022) 085001.
\item C. Heredia and J. Llosa, ``Nonlocal Lagrangian fields: Noether's theorem and hamiltonian formalism," Physical Review D 105 no. 12, (Jun, 2022).
\end{enumerate}
and the following preprints:
\begin{enumerate}\setcounter{enumi}{4}
\item C. Heredia and J. Llosa, ``Infinite Derivatives vs Integral Operators. The Moeller-Zwiebach paradox,"  arXiv: 2302.00684 [hep-th] (2023) .
\item C. Heredia and J. Llosa, ``Non-local Lagrangian fields and Noether theorem. Non-commutative U(1) gauge theory," (under construction).
\end{enumerate}

\vspace{0.3cm}
\noindent
\textbf{Financial support:} This thesis was partially supported by the Spanish MINCIU, ERDF and MCIN  (projects ref.: RTI2018-098117-B-C22 and PID2021-123879OB-C22).

\vspace{0.3cm}
\noindent
\textbf{Thesis committee:} I would like to express my sincere gratitude to the committee of my Ph.D. defense, Prof. Anupam Mazumdar, Prof. Jorge Russo, and Prof. Carles Batlle. I am deeply grateful for the time and effort you have dedicated to reviewing my thesis. Once again, thank you for having been a part of this important milestone in my academic journey.

\vspace{0.3cm}
\noindent
\textbf{To remember:} \textit{Working in a company for more than ten hours a day and studying in the evenings during the first three years. Sacrificing my personal stability and savings for the last two years. And I could make it happen because I had an incredible professor, a family that supported me in this decision, and an incalculable passion and love for physics and mathematics.}

    \end{singlespace}
\newpage







\thesistableofcontents

\thesismainmatter

\allowdisplaybreaks{
%

\pagestyle{fancy}
\fancyhead{}  
\fancyfoot{}  
\fancyhead[LE,RO]{\thepage}

\chapter{Introduction}\label{Intro}

One of the most fundamental characteristics of human beings is the desire to know and understand what surrounds us. Over the years, fundamental questions such as \textit{How could we describe the motion of this object?} or \textit{What would the trajectory of this object be?} have been a significant focus of interest. \textit{Joseph-Louis Lagrange}, \textit{Leonhard Paul Euler}, \textit{William Rowan Hamilton} --eminent physicists and mathematicians of the 18th-19th century, among others-- were involved in answering these questions. Thanks to their hard work and dedication, we have reached a point of great understanding of this topic known as \textit{Mechanics}. 

Due to the emergence of one of the most interesting, fascinating, mesmerizing theories of the modern era of physics, \textit{Quantum Physics}, the terms \textit{Classical} and \textit{Quantum} were added to distinguish those theories in which the quantum world is involved. Throughout this thesis, we shall concentrate on the branch of \textit{Classical Mechanics} and stand on the shoulders of these great giants to do our bit on this topic. 

The presentation of theories involving classical mechanics is based on one fundamental feature: \textit{locality}. In our context, we shall define a \textit{\important{local theory}} as a theory in which a finite number of derivatives describes the dynamics. There is no doubt that such theories have helped us to describe (with outstanding accuracy) the reality that surrounds us. Without them, we would never have advanced to the point of compression where we are right now. However, the modern physics of this century suggests that this attribute  should be relaxed. 

Gravitational theories are the hot spots where this feature is questioned \cite{Deser2007, Mashhoon2017, Capozziello2021, Calcagni2022}. The scientific community agrees that understanding gravity is one of the significant challenges of this century, and many physicists and mathematicians are trying to explain a phenomenon as typical in our lives as gravitational attraction. Einstein's theory of gravity (also known as General Relativity) pioneered giving answers and generating unexpected results, such as black holes and gravitational waves. Even though it is the most successful theory of gravity, unsolved problems still make us think it is only an effective theory, which works exceptionally well at low energies but breaks down at high energies (or UV regime). This theory contains singularities --places where the curvature of spacetime is infinite, i.e., in the Big Bang and black holes-- that make it incomplete to describe gravity correctly, and it is believed that locality may be the cause.

Gravitational theories are not the only place where this feature is questioned. A brief reading of the literature already shows the extensive diversity where it is relaxed: string theory \cite{ELIEZER1989}, and non-commutative theories \cite{WITTEN1986}, among others. It seems that modern physics needs the concept of \textit{nonlocality}. 

\section{What is a nonlocal theory?}
As far as this thesis is concerned, we shall define a \textit{\important{nonlocal!theory}} as a theory involving integrodifferential operators or an infinite number of derivatives. 

An example of a nonlocal theory is String Theory. Its nonlocality is displayed in its interactions, characterized by its infinite derivative structure. From a more physical point of view, the interactions are not point-wise but are given in a specific finite region \cite{Calcagni2014}. A similar idea occurs in the case of effective models of string theory, such as \textit{$p$-adic strings} \cite{Volovich1987, BREKKE1988}.

Other examples are nonlocal gravity models \cite{Biswas2006}. They are inspired by string theory's UV finiteness and are being proposed to solve both cosmological and black hole singularities \cite{Biswas2010, Buoninfante2022}. An essential improvement in the UV regime was to add infinite derivatives (through the D'Alembert-Beltrami operator $\Box$) to the Lagrangian without introducing new degrees of freedom \cite{Biswas2012}. These nonlocal gravity models are called Infinite Derivative Theories of Gravity, and their results are pretty promising; for instance, they can show the regularisation of the gravitational potential $1/r$ of pointlike sources at the linearised level \cite{Edholm2016}, as well as other sorts of sources \cite{Boos2021,Kol2020,Kol2021,Buoninfante2018_2,FrolovValeri2015,Dengiz2020,Kumar2021,Vinckers2022}. Likewise, other nonlocal gravity models are also being used to explain the cosmic expansion of the Universe \cite{Capozziello2021}. It was shown that the $1/\Box$ operator applied on the $R$-curvature scalar results in an accelerated expansion of the Universe without relying on a contribution from dark energy \cite{Deser2007}. 

Another example of the nonlocal theory could be the one describing \textit{dispersive media}. For an isotropic non-dispersive linear medium, the constitutive equations are $\mathbf{D} = \varepsilon \mathbf{E}\,$, and $\mathbf{H}=\mu^{-1}\,\mathbf{B}\,$, where $\varepsilon$ and $\mu$ are the dielectric and magnetic constants, respectively. Furthermore, $\mathbf{E}$ and $\mathbf{H}$ are the electric and magnetic fields, $\mathbf{D}$ is the electric displacement, and $\mathbf{B}$ is the magnetic induction. In natural (dispersive) media, $\varepsilon$ and $\mu $ are not constant and generally depend on frequency and wavelength; it is, then, when we speak of dispersive media \cite{Jackson1999, Landau1984, schwinger1998}.  This fact causes the above-mentioned constituent equations to be transformed into integral equations, $\mathbf{D} =(2\pi)^{-2} (\varepsilon\ast \mathbf{E})$ and $\mathbf{H} =(2\pi)^{-2} (\mu^{-1}\ast\mathbf{B})\,$, suggesting that the theory describing this type of medium must be nonlocal \cite{Heredia1}. 

\section{Searching for symmetries}
A reader who does not come from a physics background may wonder:  \textit{Why do physicists search for symmetries?} To answer this question, we must first clarify what we mean by symmetry. We shall define \textit{\important{symmetry}} as an operation on a (physical or mathematical) object that leaves such an object unchanged. An intuitive example would be, for instance, the rotation of a circle. The circle remains invariant, whatever the rotation angle is applied to it.

There are two types of symmetries: \textit{discrete} and \textit{continuous}. An example of discrete symmetry is the rotation of a square. The square is invariant (only) under ninety-degree rotations. On the other hand, an example of continuous symmetry is the example given above, the rotation of a circle. As we have explained, the circle remains unchanged under any rotation angle. Continuous symmetries can be subdivided into: \textit{global} and \textit{local}. We shall define a global symmetry as one in which finite sets of parameters can parametrize the transformation. An example of this type of symmetry is the \textit{Poincar\'e symmetry}, where the transformation parameters are the rotation angles, boosts, and spacetime translations. On the other hand, we shall define a local symmetry as one in which arbitrary functions of the coordinates parametrize the transformation. An example of this type of symmetry is \textit{gauge transformations}. 

The famous and brilliant mathematician \textit{Emily Noether} undoubtedly comes to mind when we think of symmetries. She showed (formally) that every global symmetry has a conserved quantity and that every local symmetry constrains the dynamic evolution of the system \cite{Noether1918}. Her theorems are currently known as \textit{Noether's first} and \textit{second theorem} and are considered some of the most beautiful, simple, and powerful theorems in mathematical physics. Therefore, one of the main reasons for searching for symmetries is to understand (through Noether's theorems) more fundamentally nature's laws.

Consider now that nonlocal theories might describe the phenomena of nature. The next question quickly arises: \textit{Could we use Noether's theorems for such theories?} Unfortunately, we cannot because these two theorems were developed for local $r^{\rm th}$-order theories but not for nonlocal ones. For this reason, we undoubtedly see the necessity of extending these theorems due to the requirements of modern physics. 

\section{Why Lagrangian and Hamiltonian formulations?}
In classical mechanics, there are three formalisms to describe dynamics: \textit{Newtonian formalism}, \textit{Lagrangian formalism}, and \textit{Hamiltonian formalism}. 

In the Newtonian formalism, the law that describes the dynamics of objects (and constitutes the basis of all classical mechanics) is \textit{Newton's second law}, $\mathbf{F} = m \ddot{\mathbf{x}}$. This equation tells us that the object's acceleration $\ddot{\mathbf{x}}$ depends on a constant $m$ (known as the object's mass) and the net force $\mathbf{F}$  resulting from the forces exerted on it. This formalism is a potent tool to describe the dynamics of simple systems; however, it becomes tough to handle when working with complicated ones. 

The Lagrangian formalism was developed with the aim of succinctly and effectively representing highly complex systems, while uncovering crucial properties that may have otherwise remained overlooked. Although both formalisms are physically equivalent, the Newtonian formulation is based on Cartesian coordinates, whereas the Lagrangian formulation is independent of the coordinate system. Furthermore, it  focuses on energies rather than forces and assumes that there is a fundamental principe of nature: \textit{the principle of least action}. This principle tells us (in simple words) that, over all the possible trajectories that an object can take from point A to point B, the trajectory chosen will be the most optimal one. This fact is (somehow) mesmerizing as nature is telling us that it works in the most efficient way. Furthemore, the equations of motion can be derived from this principle and are known as the \textit{Euler-Lagrange equations}, which are second-order differential equations in the simplest case. 

The Hamiltonian formulation allows us to reduce their order to two pairs of first-order equations (again, in the simplest case). It describes the same physics as Newtonian mechanics but has a particular advantage: it recognizes conserved quantities more optimally. If a quantity is conserved, i.e., it does not change over time, it is no longer necessary to see how it evolves, simplifying our analysis. Finally, the Hamiltonian formulation has become necessary in physics over the years because of quantum mechanics. This formalism is the first step toward quantization; therefore, it is essential to have it well-posed to transition from classical to quantum systems \cite{Berndt2001}.

As far as this manuscript is concerned, we shall only focus on the Lagrangian and Hamiltonian formalisms. Both formalisms were derived using operators that incorporate solely a finite number of derivatives, rather than nonlocal ones. However, in the 1990s and 2000s, a Hamiltonian formalism for nonlocal Lagrangians was proposed \cite{Jaen1987,Llosa1994} and was known as the $(1+1)$-dimensional Hamiltonian formalism \cite{Gomis2001}. The main idea of this formalism is to rewrite the nonlocal Lagrangian into a local-in-time one by using an extra dimension and thus be able to formulate the Hamiltonian formalism in this equivalent theory. Later, this formalism was applied in \cite{Gomis2001_2,SUYKENS2009,Kolar2020}, among other cases. Unfortunately, as Ferialdi et al. \cite{Ferialdi2012} correctly pointed out, this approach is lacking in considering nonlocal Lagrangians that explicitly depend on time.  Hence, this thesis improves the previously proposed nonlocal (Lagrangian and Hamiltonian) formalisms.


\chapter{Local Lagrangian Mechanics}\label{chap2}

In this chapter, we shall review and summarise the most relevant results of the Lagrangian formalism, Noether's theorem, and the Hamiltonian formalism for systems with a finite number of degrees of freedom. For this purpose, we shall rely on \cite{Landau1976,Gelfand2000,Goldstein2002,Berndt2001,Arnold1989,Lee2012}.

\section{The configuration space and generalized coordinates}\label{chap21}
We shall refer to a \textit{\important{dynamic!system}} as an object --or ensemble of objects-- whose shape can be neglected when describing its motion. Additionally, the \textit{\important{degrees of freedom}} will be defined as the independent number of coordinates that uniquely determine the dynamic system's position. These independent coordinates $(q^1,q^2,\ldots, q^s)$ are known as \textit{\important{generalized!coordinates}}, and the set of possible positions constitutes the so-called \textit{\important{configuration space}} $Q$. For the sake of simplicity of notation, we shall refer to $q$ as the $s$-tuple of all generalized coordinates $(q^1, q^2, \ldots,q^s)$.

 We shall consider ``the system's motion" as the \textit{\important{curve}} (or \textit{\important{trajectory}}) 
\begin{equation}\label{chap2-c}
q: \;\;t \mapsto q(t)\,, \qquad \mathrm{with} \qquad q(t_0) := q_0\,,
\end{equation}
that the system follows through the configuration space and the \textit{time} variable as its curve parameter. Given a function $f(q)$ on $Q$, the derivative following the curve $q(t)$
\begin{equation}
\left.\frac{\D f(q(t))}{\D t}\right|_{q_0}:= \mathbf{v}_{q_0} f
\end{equation}
defines the tangent vector $\mathbf{v}_{q_0}$ at point $q_0$ --figure (\ref{fig:chp21-TQ})--. From the chain rule, it quickly follows that
\begin{equation}
\mathbf{v}_{q_0} = \sum^s_{i=1} \dot q^i_0 \left.\frac{\partial}{\partial q^i}\right|_{q_0}\,,\quad\mathrm{with} \quad \dot q^i_0 := \left.\frac{\D q^i(t)}{\D t}\right|_{t_0}\,,
\end{equation}
where $\dot q^i_0$ are the well-known \textit{\important{generalized!velocities}} and compose the components of the tangent vector $\mathbf{v}_{q_0}$. Note that by changing the generalized coordinates to $q^{\prime i} = q^{\prime i}(q^j)$, the corresponding  generalized velocities transform with the Jacobian, just as a vector does
\begin{equation}
\dot q^{i\prime} = \frac{\partial q^{i\prime}}{\partial q^j} \dot q^j\,.
\end{equation}
As above, we shall refer to $\dot q_0$ as the $s$-tuple of all generalized velocities $(\dot q^1_0, \ldots,\dot q^s_0)$. 

\begin{figure}[t]
\centering
\includegraphics[width=0.4\textwidth]{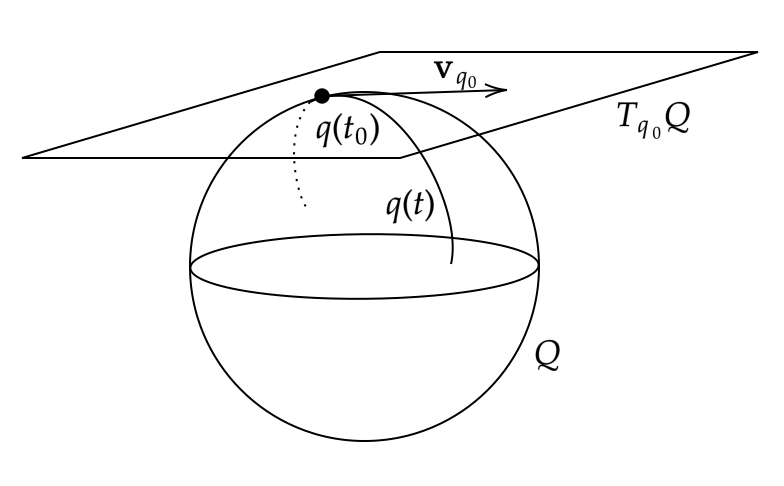}
\caption{Visual representation of tangent space $T_{q_0}Q$.}
\label{fig:chp21-TQ}
\end{figure}

All the tangent vectors at $q_0$ constitute the so-called \textit{\important{tangent!space}} $T_{q_0}Q$ and their disjoint union, $\,\dot{\bigcup}_{q_j\in Q} T_{q_j}Q\,$, the \textit{\important{tangent!bundle}} of $Q$, denoted by $TQ$. The $2s$ components $(q,\dot q)$ give rise to the local coordinate system on $TQ$\footnote{We suggest reading chapter 3 of \cite{Lee2012} for those readers interested in a more formal introduction to smooth manifolds.}. 

In addition, we shall define a \textit{\important{vector field}} as an application $\mathbf{X}\,:\, Q \rightarrow TQ$ where each point $q_j\in Q$ is assigned a tangent vector $\mathbf{X}_{q_j}\in T_{q_j}Q$. The components of $\mathbf{X}$ (in some coordinates) are smooth functions that transform by employing the Jacobian matrix of the change of coordinates. For each coordinate system, the vector field $\frac{\partial}{\partial q^i}$ is associated.

\section{The cotangent space and differential forms}\label{chapCTDF}

A \textit{\important{linear form}} --also known as $1$-form-- at $q_0$ is a linear application $\;\alpha_{q_0}: \,T_{q_0}Q \rightarrow \R\,$ in which, for each vector $\mathbf{v}_{q_0}$, a real value $(\alpha_{q_0},\mathbf{v}_{q_0})\in\R$ is assigned. The set of $1$-forms is the dual of the tangent space $T_{q_0} Q$, known as the \textit{\important{cotangent!space}} and denoted by $T^*_{q_0} Q$. The dual basis of the coordinate basis $\left.\frac{\partial}{\partial q^i}\right|_{q_0}\,, \, i=1,\ldots, s\,,$ is constituted by the $1$-forms $\D_{q_0} q^j\,,\,j=1,\ldots, s\,,$ such that
\begin{equation}
\left( \D_{q_0} q^j, \mathbf{v}_{q_0}\right) = v^j \,, \qquad {\rm with} \qquad \mathbf{v}_{q_0} = \sum_{i=1}^s v^i\,\left. \frac{\partial \;}{\partial q^i}\right|_{q_0} \,,
\end{equation}
where the relation 
\begin{equation}
\displaystyle{\left( \D_{q_0} q^j, \left. \frac{\partial \;}{\partial q^i}\right|_{q_0}\right) = \delta^j_i }
\end{equation}
is understood. In the same way as above, the disjoint union of the cotangent spaces, $\dot{\bigcup}_{q_j\in Q} T^*_{q_j}Q\,$, is the so-called \textit{\important{cotangent!bundle}} and is denoted by $T^*Q$. 

A \textit{differential $1$-form} is a linear $1$-form field $\alpha:\,Q \rightarrow T^\ast Q\,$, in which each $q_0\in Q$ is assigned a cotangent $1$-form $\alpha_{q_0} \in T^*_{q_0} Q$. At given coordinates, it has the expression
 \begin{equation}
 \alpha =\sum^s_{i=1} p_i(q)\,\D q^i \,,
 \end{equation}
 where $p_i(q)$ are smooth functions and transform under a change of coordinates through the inverse Jacobian matrix 
 \begin{equation}
 p_i(q)\, \frac{\partial q^i}{\partial q^{\prime j}} = p_j(q^\prime)\,.
 \end{equation}
 Furthermore, for each coordinate system, the differential 1-forms $\D q^i\,, i=1,\ldots,s\,,$ are characterized by $\left(\D q^i, \frac{\partial}{\partial q^i}\right)=\delta^j_i$, and the $2s$ components $(q^1,\ldots, q^s, p_1,\ldots,p_s)$ constitute a collection of local coordinates for $T^*Q$.
 
For what lies ahead\footnote{Section \ref{ch241-SHF}: Symplectic Hamiltonian formalism.}, it is necessary to introduce some specific mathematical tools \cite{Choquet1982,Berndt2001}. We shall define the $k$-\textit{\important{differential form}} $\omega\in\Lambda^kQ$ as 
\begin{equation}
\omega = \sum_{i_1<\ldots<i_k} \sum^s_{i_1,\ldots,i_k=1}\omega_{i_1 i_2 \ldots i_k}(q^j)\, \D q^{i_1} \wedge \, \D q^{i_2}\wedge \ldots \wedge \, \D q^{i_k}\,,
\end{equation}
where $\wedge$ denotes the \textit{\important{exterior!product},} and the components $\omega_{i_1 i_2 \ldots i_k}(q^j)$ are fully antisymmetric and transform as 
\begin{equation}
\omega_{i_1 i_2 \ldots  i_k}(q^j)\,\frac{\partial q^{i_1}}{\partial q^{\prime l_1}}\, \ldots \,\frac{\partial q^{i_k}}{\partial q^{\prime l_k}} = \omega^\prime_{l_1 l_2 \ldots  l_k}(q^{\prime j})\,.
\end{equation}

We shall introduce the \textit{\important{exterior!derivative}} as a mapping from $k$-forms to $(k+1)$-forms, namely, $\rm d:\, \Lambda^k Q \rightarrow \Lambda^{k+1} Q$. In coordinates, if $\omega\in\Lambda^kQ$, $\D \omega$ is
\begin{equation}
 \D \omega = \sum_{i_1<\ldots<i_k}\sum^s_{i_1,\ldots,i_k,l=1} \partial_l \omega_{i_1 i_2 \ldots  i_k} \,\D q^{l} \wedge \,\D q^{i_1} \wedge \, \D q^{i_2}\wedge \ldots \wedge \, \D q^{i_k} \in \Lambda^{k+1}Q\,.
\end{equation}
Notice that the exterior derivative satisfies the property $\rm d(\rm d \omega) = \rm d^2 \omega = 0$ for any $k$-form $\omega$. 

To conclude this section, given a vector field $\mathbf{X} = X^l\partial/\partial q^l$ and a differential $k$-form $\omega\in\Lambda^kQ$, the \textit{\important{inner product}} is defined as
\begin{equation}
i_\mathbf{X} \omega= \sum_{i_2<\ldots<i_k}\sum^s_{i_2,\ldots,i_k,l=1} X^l\,\omega_{l i_2 \ldots i_k}(q^j)\, \D q^{i_2} \wedge \ldots \wedge \, \D q^{i_k}  \in \Lambda^{k-1}Q\,.
\end{equation} 

 \section{The principle of least action}\label{chap22}
The most general way to present the motion's law of particle systems is by employing \textit{the \important{principle of least action}} (also known as the Hamilton principle). It states that, among all possible (kinematically admissible) trajectories connecting two points $q(t_0):=q_0$ and $q(t_1) = q_1$ in $Q$, the dynamic trajectory $q(t)$ is the one that makes the curvilinear integral 
\begin{equation}\label{ch22-S}
S([q])= \int^{t_1}_{t_0}\D t\,L(q,\dot q,t) 
\end{equation}
stationary. The function $L$ is the \textit{\important{Lagrangian}} and is defined in an extended\footnote{The word ``extended" refers to the fact that we are considering explicitly time-dependent Lagrangians.} space $TQ\times\R$. Moreover, it is of class $\mathcal{C}^2(TQ\times\R; \R)$, i.e., its derivatives are continuous up to the second order. On the other hand, $S([q])$ is a real functional $\mathscr{DQ}\rightarrow \R$ defined on the space of curves
\begin{equation}
\mathscr{DQ} = \left\{ q \in \mathcal{C}^2([t_0,t_1], Q)\,, \quad q(t_0) = q_0 \;\; {\rm and} \;\;  q(t_1) = q_1 \right\}\,,
\end{equation}
which is an infinite-dimensional Banach space \cite{Schuller2015} with the norm
\begin{equation}
\|q\|_\infty = \underset{ t_0 \leq t\leq t_1}{\rm sup}\left\{ |q^i(t)|,\, |\dot q^j(t)|\right\}\,,
\end{equation}
where $sup$ denotes the supremum.  This real functional is the so-called \textit{\important{action integral}}. The elements of $\mathscr{DQ}$ that make action integral stationary are those for which its variation $\delta S$ vanishes.

The principle of least action equivalently reads  
\begin{equation}\label{ch22-dS}
\delta S([q]) = \int^{t_1}_{t_0}\D t\,\left(\frac{\partial L}{\partial q} \delta q + \frac{\partial L}{\partial \dot q} \delta \dot q\right) = 0
\end{equation}
for all variations $\delta q$  that vanish at the extremes. The variations $\delta q$ and $\delta \dot q$ are not independent; for that reason, we integrate the second term by parts. Consequently, we get
\begin{equation}\label{ch22-dSb}
\delta S([q]) = \left.\frac{\partial L}{\partial \dot q} \delta q\right|^{t_1}_{t_0} + \int^{t_1}_{t_0}\D t\, \delta q\,\left(\frac{\partial L}{\partial q} - \frac{\D}{\D t}\frac{\partial L}{\partial \dot q}\right)= 0\,.
\end{equation}
As $\delta q(t_0) = \delta q(t_1) = 0$, the first term disappears, and only the integral remains. Because it must cancel for all $\delta q$, we find the following necessary condition for the action integral to be stationary  
\begin{equation}\label{ch22-EL1}
\begin{split}
\mathcal{E}(q,\dot q,\ddot q,t)&:= \frac{\partial L}{\partial q} - \frac{\D}{\D t}\frac{\partial L}{\partial \dot q} = 0\,.
\end{split}
\end{equation}
Indeed, the time derivative operator $\D/\D t$ is referred to as the \textit{\important{time!evolution generator}} $\mathbf{D}$ on $T(TQ\times\R)$
\begin{equation}\label{chap23-D}
\mathbf{D}:= \frac{\D}{\D t} = \frac{\partial}{\partial t} + \dot q \frac{\partial}{\partial q} + \ddot q \frac{\partial}{\partial \dot q} \,. 
\end{equation}  
Considering all the $s$ degrees of freedom explicitly and expanding the time derivative, we get $s$ second-order differential equations
 \begin{equation}\label{ch22-ELs}
\sum^s_{j=1} \frac{\partial^2L}{\partial \dot q^i \partial \dot q^j} \ddot q^j = \frac{\partial L}{\partial q^i} - \frac{\partial^2L}{\partial t \partial \dot q^i} - \sum^s_{j=1}\frac{\partial^2 L}{\partial \dot q^i \partial q^j }\dot q^j,\qquad  i=1,2,\ldots,s\,,
\end{equation}
known as the \textit{\important{Euler-Lagrange!equations}}. \\
\indent A Lagrangian is \textit{regular} if the determinant of the Hessian matrix is
\begin{equation}\label{ch22-HMD}
\left|\frac{\partial^2L}{\partial \dot q^i \partial \dot q^j}\right| \neq 0\,.
\end{equation}
Under this condition, we can always isolate the acceleration $\ddot q$ as a function of the generalized coordinates, velocity, and time, 
\begin{equation}\label{ch22-ddq}
\ddot q^j = f^j(q, \dot q, t)\,,
\end{equation}
constituting a system whose solutions are determined by $2s+1$ arbitrary constants. The theorems of existence and uniqueness \cite{barbu2016,Arnold1992} establish that if $f^j(q, \dot q, t)$ satisfies certain continuity and differentiability conditions, for every point $(q_0, \dot q_0, t_0)$ of the domain of $L$, there is a single solution of the Euler-Lagrange equations 
\begin{equation}
q^j = \varphi^j(t,q_0,\dot q_0, t_0) \qquad \mathrm{such}\,\,\,\mathrm{that}\qquad q^j_0 = \varphi^j(t_0;q_0,\dot q_0,t_0), \quad \dot q^j_0 =\frac{\D \varphi^j(t_0;q_0,\dot q_0,t_0)}{\D t}\,.
\end{equation}
They allow us to establish a one-to-one correspondence between the space constituted by the curves satisfying the Euler-Lagrange equations, called the\footnote{We add the word ``extended" to distinguish the case where the submanifold is constituted by all curves satisfying the non-explicit-time-dependent Euler-Lagrange equations.} \textit{\important{extended!dynamic space}} $\calD^\prime$, 
\begin{equation}
\calD^\prime := \left\{\,(q, t)\in\mathcal{C}^{\infty}(\R,\R^s)\times\R \quad | \quad \mathcal{E}(q,\dot q, \ddot q,t) = 0\,\right\}\,,
\end{equation}
 and the \textit{\important{initial data space}} $TQ\times\R$. This space $\calD^\prime$ is a manifold of dimensions $2s+1$.\\
 \indent Nevertheless, the principle of least action is not well suited for initial points. Bear in mind that it is suited for the contour data where the generalized coordinates $q^j$ are fixed at the extremes without specifying the generalized velocities $\dot q^j$ anywhere. Therefore, we must remember that the initial conditions $(q_0^j,\dot q_0^j,t_0)\,$, which are provided to solve the ordinary differential equation (\ref{ch22-ELs}), must be compatible with the chosen boundary conditions $(q_0^j, t_0, q_1 ^j,t_1)$. As far as this manuscript is concerned, we shall assume this is the case since it is the most common for regular Lagrangians or second-differential equations such as (\ref{ch22-ELs}). See \cite{Dyer2009} for an example where the choice of the contour problem is not equivalent to the initial value formulation.  \\
 \indent Those Lagrangians for which the Hessian determinant (\ref{ch22-HMD}) vanishes are called \textit{singular}. Through equation (\ref{ch22-ELs}), it might be seen that the accelerations $\ddot q^j$ will not be uniquely determined by the positions and velocities, leading to the fact that the existence and uniqueness theorems of ordinary differential equations cannot be used in this context. The study of this issue gives rise to an extensive Lagrangian formalism branch known as the so-called  \textit{Dirac-Bergmann constrained systems} \cite{PONS2005,Henneaux1992,Dirac1964}. However, we shall not discuss this subject in this manuscript and only focus on regular ones. \\
 \indent It is essential to emphasize that the Lagrangian $L(q,\dot q,t)$ is not the only one that leads to the equations of motion (\ref{ch22-ELs}); the following Lagrangian does too \vspace{0.1cm}
\begin{equation}\label{ch22-TD}
L^\prime(q, \dot q,t ) =  L(q, \dot q,t ) + \frac{\D F(q,t)}{\D t},
\end{equation}
where $F(q,t)$ is a derivable function from any of its variables. This transformation is a symmetry in the principle of least action called the \textit{\important{Noether!symmetry}}. It will play a significant role in the next section. 

\section{The Noether theorem}\label{chap23}
The \important{Noether!theorem} is a formal proof that every symmetry has a conserved quantity. It was proved for higher-order Lagrangians \cite{Noether1918}; however, in this section, we shall only illustrate it for first-order Lagrangians for simplicity.

The theorem begins with an assumption: The time transformation is infinitesimal
\begin{equation}\label{ch23-IT}
t^\prime = t + \delta t (t)
\end{equation}
and leads to the following generalized coordinate transformation
\begin{equation}\label{ch23-ITQ}
q^\prime(t) = q(t) + \delta q(t)\,.
\end{equation}
The Lagrangian shall transform in such a way that the value of the action integral (\ref{ch22-S}) is invariant under such infinitesimal transformations, namely,
\begin{equation}\label{ch23-LpL}
L^\prime(q^\prime,\dot q^\prime, t^\prime) = \left|\frac{\partial t}{\partial t^\prime}\right|  L(q,\dot q, t) \approx \left(1 - \dot{\delta t}\right)L(q,\dot q,t)\,. 
\end{equation}
Therefore, if the time interval $[t_0^\prime,t_1^\prime]$ is the transformed interval of $[t_0,t_1]$ according to (\ref{ch23-IT}), we have that
\begin{equation}\label{ch23-SSp}
 \int^{t^\prime_1}_{t^\prime_0}\D t^\prime\, L^\prime(q^\prime(t^\prime),\dot q^\prime(t^\prime), t^\prime) =  \int^{t_1}_{t_0}\D t\, L(q(t),\dot q(t), t).
\end{equation}
From equation (\ref{ch23-SSp}), it follows that 
\begin{equation}\label{ch23-dS0}
 \int^{t^\prime_1}_{t^\prime_0}\D t\, L^\prime(q^\prime(t),\dot q^\prime(t), t) -  \int^{t_1}_{t_0}\D t\, L(q(t),\dot q(t), t)  =0,
\end{equation}
where the dummy variable $t^\prime$ has been replaced with $t$ in the first integral. Given the time infinitesimal transformation (\ref{ch23-IT}) and equation (\ref{ch23-LpL}), the first integral can be approximated to the leading order as 
\begin{equation}
 \int^{t_1+ \delta t_1}_{t_0+\delta t_0}\D t\, L^\prime(q^\prime,\dot q^\prime, t) = \int^{t_1}_{t_0}\D t\,\left\{L^\prime(q^\prime,\dot q^\prime, t) + \frac{\D}{\D t}\left[L(q,\dot q, t)\delta t\right]\right\}\,, 
\end{equation}
where $\delta t(t_0)= \delta t_0$ and $\delta t(t_1)= \delta t_1$ are understood. Consequently, the difference between the integrals of equation (\ref{ch23-dS0}) is \newpage
\begin{equation}\label{ch23-LpLdT}
\int^{t_1}_{t_0}\D t \left\{\left[L^\prime(q^\prime,\dot q^\prime, t) - L(q,\dot q,t)\right] + \frac{\D}{\D t}\left[L(q,\dot q, t)\delta t\right]\right\} = 0\,. 
\end{equation}
Since we are working with infinitesimal approximations, we shall naturally assume that the Lagrangian $L^\prime$ transforms as 
\begin{equation}
L^\prime(q^\prime, \dot q^\prime, t) = L(q^\prime, \dot q^\prime, t) + \delta L(q^\prime, \dot q^\prime,t)\,.
\end{equation}
By considering the generalized coordinates transformation (\ref{ch23-ITQ}), the last equation becomes  
\begin{equation}\label{ch23-relNTdL}
\begin{split}
L^\prime(q^\prime, \dot q^\prime, t) - L(q,\dot q,t)  &= \frac{\partial L(q,\dot q,t)}{\partial q}\delta q + \frac{\partial L(q, \dot q,t)}{\partial \dot q} \delta \dot q + \delta L(q, \dot q,t)
\end{split}
\end{equation}
up to the leading order. Thus, integrating by parts, we get
\begin{align}
L^\prime(q^\prime,\dot q^\prime, t) - L(q,\dot q,t) =&  \frac{\partial L(q,\dot q,t)}{\partial q}\delta q + \frac{\D}{\D t}\left[\frac{\partial L(q,\dot q,t)}{\partial \dot q} \delta q\right]-  \frac{\D}{\D t}\left[\frac{\partial L(q,\dot q,t)}{\partial \dot q}\right] \delta q + \delta L(q,\dot q,t)\,.
\end{align}
Regarding the last expression, equation (\ref{ch23-LpLdT}) is 
\begin{equation}\label{ch23-LpLdTNS}
\int^{t_1}_{t_0}\D t \left\{\delta L +\left[\frac{\partial L}{\partial q}- \frac{\D}{\D t}\frac{\partial L}{\partial \dot q} \right]\delta q +  \frac{\D}{\D t}\left[L\,\delta t + \frac{\partial L}{\partial \dot q}\delta q\right]\right\} = 0\,.
\end{equation}
Note that the second term is equation (\ref{ch22-EL1}), $\mathcal{E}(q,\dot q, \ddot q,t)$. Furthermore, assuming that our Lagrangian presents a Noether symmetry, we have that $\delta L(q, \dot q,t) := \D F(q,t)/\D t$. Hence, equation (\ref{ch23-LpLdTNS}) finally becomes
\begin{equation}
\int^{t_1}_{t_0}\D t \left\{\mathcal{E}\,\delta q +  \frac{\D}{\D t}\left[L\,\delta t + F + \frac{\partial L}{\partial \dot q}\delta q\right]\right\} = 0\,.
\end{equation}
Since the time interval choice $[t_0,t_1]$ is entirely arbitrary, the integrand must be identically 0 for every interval we take, therefore, 
\begin{equation}\label{ch23-IN_V2}
\mathcal{E}\,\delta q + \frac{\D J}{\D t} \equiv 0\,,
\end{equation}
 where the function $J$ is defined as
 \begin{equation}\label{ch23-NC}
 J := L\,\delta t + F + \frac{\partial L}{\partial \dot q}\delta q\,
 \end{equation}
 and is known as the Noether constant of motion (conserved quantity or first integral) \cite{Landau1976}. As seen from equation (\ref{ch23-IN_V2}), if the trajectories satisfy the Euler-Lagrange equations (namely, $(q,t)\in\calD^\prime$), the function $J$ is a conserved quantity
\begin{equation}\label{ch23-JT}
\frac{\D J}{\D t} = 0\,.
\end{equation}

The last equation gives rise to the theorem's conclusion. It states that, given a transformation (symmetry) that leaves the value of the integral action (\ref{ch23-SSp}) unchanged, a conserved quantity is associated with it.

\newpage

\subsection{The energy function}\label{chap221}
\textit{\important{Energy conservation}} is one of the most fundamental concepts in physics. As shown below, Noether's theorem indicates that energy conservation occurs due to \textit{\important{time!translation invariance}}. 

Assume that the Lagrangian of our system does not explicitly depend on time, $L(q,\dot q)$. Time translation is the transformation
\begin{equation}\label{ch231-tt}
\delta t = t^\prime - t = \epsilon\,, \qquad \mathrm{and} \qquad q^\prime(t^\prime) = q(t)\,,
\end{equation}
where $\epsilon$ is a real-valued constant. Using equations (\ref{ch231-tt}) and (\ref{ch23-LpL}), the relation between the transformed Lagrangian $L^\prime$ and $L$ becomes
\begin{equation}
L^\prime(q^\prime,\dot q^\prime) = L(q,\dot q)\,. 
\end{equation}
This fact immediately yields $F=0$. Furthermore, the infinitesimal variation of the generalized coordinates becomes
\begin{equation}
q^\prime(t^\prime) = q(t) = q(t^\prime-\epsilon) \approx q(t^\prime) - \epsilon \dot q(t^\prime)\,
\end{equation}
up to higher-order terms, and renaming the dummy variable $t^\prime$ to $t$,
\begin{equation}\label{ch231-tq}
\delta q(t) = q^\prime(t) - q (t) = - \epsilon \dot q(t) \,.
\end{equation}
Plugging now the infinitesimal transformations (\ref{ch231-tt}) and (\ref{ch231-tq}) into (\ref{ch23-NC}),  we get the so-called \textit{\important{energy function}} $E:=-J/\epsilon$
\begin{equation}\label{ch231-JE}
E(q,\dot q) =  \frac{\partial L(q,\dot q)}{\partial \dot q} \dot q - L(q,\dot q)\,,
\end{equation}
which is preserved due to (\ref{ch23-JT}).
 
\section{Hamiltonian formalism}\label{ch24-SH}

So far, we have assumed that Euler-Lagrange's equations (\ref{ch22-ELs})  rule the evolution of the mechanical system. As discussed, being a system of second-order differential equations, we need to specify either $2s+1$ initial conditions to determine the system's state fully.

\textit{\important{Hamiltonian!formalism}} is an alternative approach. It is (fundamentally) based on describing the particle's dynamics through a system of first-order differential equations. As the $2s$ conditions to determine the system's state must be preserved, we need $2s$ independent first-order equations described by $2s$ independent variables. As is well known, these independent variables are the generalized coordinates $q^i$ and the \textit{\important{generalized!momenta}} $p_i$, which are introduced by the \textit{\important{Legendre transformation}}
\begin{equation}\label{ch24-p}
p_i(q,\dot q,t):=\frac{\partial L(q,\dot q, t)}{\partial \dot q^i}, \qquad i= 1,\ldots,s\,. 
\end{equation}
The new variables, called \textit{\important{canonical!variables}}, are $(q,p)$ and coordinate the well-known \textit{\important{phase space}} $\Gamma$. The extension for the case where the system is non-autonomous and time appears explicitly $(q,p,t)$ will be named the \textit{\important{extended!phase space}} $\Gamma^\prime$. \\
\indent Let us see how the Lagrangian formulation transforms due to (\ref{ch24-p}). The total differential of the Lagrangian $L(q,\dot q,t)$ is
\begin{equation}\label{ch24-dL}
\D L= \sum^N_{i=0} \frac{\partial L}{\partial q^i}\D q^i + \sum^N_{i=0} \frac{\partial L}{\partial \dot q^i}\D \dot q^i + \frac{\partial L}{\partial t}\D t\,.
\end{equation}
Employing the Euler-Lagrange equations (\ref{ch22-EL1}), we can observe that $\dot p_i = \partial L/\partial q^i$. Therefore, equation (\ref{ch24-dL}) can be written as 
\begin{equation}
\D L= \sum^N_{i=0} \dot p_i \D q^i + \sum^N_{i=0}p_i \D \dot q^i + \frac{\partial L}{\partial t}\D t \,.
\end{equation}
Integrating by parts the second term on the right-hand side, we arrive at 
\begin{equation}\label{ch24-dH}
\D H = - \sum^N_{i=0} \dot p_i \D q^i + \sum^N_{i=0}\dot q^i \D p_i - \frac{\partial L}{\partial t}\D t\,, 
\end{equation}
where we have introduced the \textit{\important{Hamiltonian!function}}
\begin{equation}\label{ch24-H}
H(q,p,t):=\sum^N_{i=0} p_i \dot q^i - L(q,\dot q,t)\,,
\end{equation}
and $\dot q^i = \dot q^i(q,p,t)$ due to (\ref{ch24-p}). We can quickly infer from equation (\ref{ch24-dH}) that
\begin{equation}\label{ch24-HE}
\dot p_i = - \frac{\partial H}{\partial q^i},\qquad  \dot q^i = \frac{\partial H}{\partial p_i}\,.
\end{equation}
These are known as the \textit{\important{Hamilton equations}}, which describe the system's dynamics on the phase space, and they constitute a system of $2s$ first-order differential equations, as anticipated above. Suppose now we calculate the total time derivative of the Hamiltonian $H(q,p,t)$
\begin{equation}
\frac{\D H}{\D t} =\sum^N_{i=0} \frac{\partial H}{\partial q^i}\dot q^i + \sum^N_{i=0}  \frac{\partial H}{\partial p_i}\dot p_i + \frac{\partial H}{\partial t}
\end{equation}
 and  use Hamilton's equations (\ref{ch24-HE}) and (\ref{ch24-dH}), then 
\begin{equation}
\frac{\D H}{\D t} = \frac{\partial H}{\partial t} = - \frac{\partial L}{\partial t}\,.
\end{equation} 
In particular, if the Lagrangian does not depend explicitly on time, we conclude that the Hamiltonian is preserved over time; namely, it is a constant of motion. Now, observe that equation (\ref{ch24-H}) is simply equation (\ref{ch231-JE}) combined with the Legendre transformation (\ref{ch24-p}). Therefore, if our system does not explicitly depend on time, the Hamiltonian is nothing more than the energy function expressed as a function of the generalized coordinates and momenta. \\
\indent Another significant observation is that the Noether theorem (more precisely, the boundary terms) suggests the definition of the Legendre transformation, namely, the third term on the right-hand side of (\ref{ch23-NC}). This point will be crucial when defining the Legendre transformation for nonlocal systems. 

\subsection{Symplectic Hamiltonian formalism}\label{ch241-SHF}
Hamiltonian formalism can be set in a more geometric language, namely, \textit{\important{symplectic!mechanics}}. A \textit{\important{symplectic!form}} on a manifold $\calM$ is a differential $2$-form $\omega= \sum^s_{i,j=1} \omega_{ij}\D x^i \wedge \D x^j \in\Lambda^2\calM$ such that it is closed 
\begin{equation}
\D\omega=0 \quad \mathrm{or,\,\, in\,\, coordinates,} \quad \partial_l \omega_{ij} + \partial_i \omega_{jl} + \partial_j \omega_{li} = 0
\end{equation}
and non-degenerate
\begin{equation}
i_\mathbf{X} \omega = 0 \quad \Rightarrow \quad \mathbf{X} =0 \quad \mathrm{or,\,\, in\,\, coordinates,} \quad \mathrm{det}\,\omega_{ij} \neq 0 \,.
\end{equation}
Under the change of coordinates, the $\omega_{ij}$ components behave as a $2$-covariant antisymmetric tensor. Likewise, from the second condition, it is evident that the number of dimensions of $\calM$ must be even \cite{Castellet2000}. Thus, we can pair the coordinates as $(x^1,y_1,x^2,y_2,\ldots,x^s,y_s)$. Indeed, it can be shown --Darboux's theorem \cite{Choquet1982,Berndt2001}-- that, locally, one can find coordinate systems in which the $2$-form is
\begin{equation}
\omega = \sum^s_{i=1} \D y_i \wedge \D x^i\,.
\end{equation}
Such coordinates are called \textit{\important{canonical!coordinates}}. 

We shall say that a vector field $\mathbf{X}$ is \textit{\important{Hamiltonian}} (concerning $\omega$) when there exists a function $f\in\Lambda^0\calM$ such that
\begin{equation}\label{chap251-iWx}
i_\mathbf{X} \omega = - \D f\,.
\end{equation}
In canonical coordinates, this means that
\begin{equation}
-\D f =\sum^s_{j=1}\left\{\mathbf{X}\left(y_j\right) \D x^j - \mathbf{X}\left(x^j\right)\D y_j\right\}\,, 
\end{equation}
where 
\begin{equation}
\mathbf{X}\left(y_j\right) := - \frac{\partial f}{\partial x^j},\qquad \mathbf{X}\left(x^j\right) := \frac{\partial f}{\partial y_j}\,. 
\end{equation}
Note the clear parallelism with equation (\ref{ch24-HE}). Indeed, a field $\mathbf{X}$ is \textit{locally Hamiltonian} when its flow leaves $\omega$ invariant, namely,
\begin{equation}
\mathscr{L}_\mathbf{X} \omega = 0\,,
\end{equation}
where $\mathscr{L}$ denotes the Lie derivative. Using Cartan's formula \cite{Choquet1982}, we have that 
\begin{equation}
0 = \mathscr{L}_{\mathbf{X}} \omega = i_{\mathbf{X}} \circ \D \omega + \D \circ i_{\mathbf{X}}\omega = \D(i_\mathbf{X}\omega)\,,
\end{equation}
which implies that $i_{\mathbf{X}}\omega\in\Lambda^1\calM$ is closed. By Poincar\'e's lemma \cite{Choquet1982}, it is also locally exact and, at any point, there exists a function $f$ such that $i_\mathbf{X}\omega = - \D f$.

The relation (\ref{chap251-iWx}) allows us to establish a biunivocal correspondence between functions $f$ on $\calM$ and the \textit{\important{Hamiltonian!vector field}}, i.e.,
\begin{equation}\label{chap251-XfDf}
f \rightarrow \mathbf{X}_f, \qquad i_{\mathbf{X}_f}\omega = - \D f\,.
\end{equation}
Note that this last equality --equation (\ref{chap251-XfDf})-- is a differential system that can be solved in $\mathbf{X}_f$ because $\omega$ is non-degenerate. In other words, in (non-canonical) coordinates, the above relation is written 
\begin{equation}
\sum^s_{i=1}X^i_f \omega_{ij} = \partial_j f \qquad \mathrm{and\,\, hence\,\,} \qquad X^l_f = \sum^s_{j=1}\omega^{lj}\partial_j f\,,
\end{equation}
where $\omega^{lj}$ is the inverse matrix of $\omega_{jk}$. In the same way, it allows us to define a dual --contravariant-- structure of the symplectic form, i.e., the \textit{\important{Poisson bracket}} of two functions, $f$ and $g$, on $\calM$
\begin{align}\label{chap251-PBG}
\left\{f,g\right\} &:= -\omega\left(\mathbf{X}_f,\mathbf{X}_g\right) = -\sum^s_{i,j=1} \omega_{ij} X^i_f X^j_g \nonumber \\
&= - \sum^s_{i,j,l,k=1}\omega_{ij}\,\omega^{il}\,\partial_l f\,\omega^{jk}\,\partial_k g = \sum^s_{l,k = 1}\omega^{lk}\partial_l f \partial_k g\,.
\end{align}

All these ideas can be transferred to the phase space $T^\ast Q$ of dimension $2s$. It has a naturally (canonical) associated symplectic form which, in the coordinates $(q^j,p_j)$ presented in Section \ref{chapCTDF}, is expressed by
\begin{equation}
\omega = \sum^s_{j= 1} \D p_j \wedge \D q^j \in \Lambda^2(T^*Q)\,.
\end{equation}
 Notice that it is closed and non-degenerate. 

Let $H(q,p)$ be a Hamiltonian function independent of $t$, and $\mathbf{X}_H$ be the tangent field to the Hamiltonian flow, i.e., defined by the Hamilton equations (\ref{ch24-HE}), 
\begin{equation}
\mathbf{X}_H =  \sum^s_{i=1} \left(\frac{\partial H}{\partial p_i} \frac{\partial}{q^i} - \frac{\partial H}{\partial q^i}\frac{\partial}{\partial p_i}\right)\,,
\end{equation}
then
\begin{equation}
i_{\mathbf{X}_H} \omega = -\sum^s_{i=1}\left( \frac{\partial H}{\partial p_i} \D p_i+ \frac{\partial H}{\partial q^j} \D q^j\right) =  -\D H\,.
\end{equation}

On the other hand, if a Hamiltonian function depends on $t$, the relevant manifold is the extended phase space $\Gamma^\prime = \Gamma\times \R$ and the generating field of the time evolution is 
\begin{equation}\label{chap251-HVF}
\mathbf{X}_H := \frac{\partial}{\partial t} + \sum^s_{i=1}\left(\frac{\partial H}{\partial p_i} \frac{\partial}{q^i} - \frac{\partial H}{\partial q^i}\frac{\partial}{\partial p_i}\right)\,. 
\end{equation}
Consequently, instead of the 2-form $\omega = \sum^s_{i=1} \D p_j \wedge \D q^j$, which is symplectic over each manifold $T^*Q\times \{t\}$ at constant $t$, we shall use the contact form 
\begin{equation}
\Omega = \omega - \D H \wedge \D t\,.
\end{equation}
Furthermore, the Hamiltonian equations (\ref{ch24-HE}) become
\begin{equation}\label{chap251-HEQ}
i_{\mathbf{X}_H} \Omega = 0\,.
\end{equation}  
Particularising the Poisson brackets (\ref{chap251-PBG}) in canonical coordinates, we have that 
\begin{equation}\label{ch241-PB}
\left\{f,g\right\} = \sum^s_{i=1}\left(\frac{\partial f}{\partial q^i}\frac{\partial g}{\partial p_i} - \frac{\partial f}{\partial p_i}\frac{\partial g}{\partial q^i}\right)\
\end{equation}
and, for the particular case of $f:=q^i$ and $g:=p_i$, the so-called \textit{\important{canonical!Poisson brackets}} 
\begin{equation}
\left\{q^i, p_j\right\} = \delta^i_j, \qquad \mathrm{and} \qquad \left\{q^i,q^j\right\} = \left\{p_i,p_j\right\} =0\,.
\end{equation}
It is worth noting that the Hamiltonian vector field (\ref{chap251-HVF}) can be written employing the Poisson bracket 
\begin{equation}
\mathbf{X}_H = \frac{\partial}{\partial t} + \left\{\cdot,H\right\}\,,
\end{equation}
giving rise to an equivalent form of writing the Hamiltonian equations
\begin{equation}\label{ch241-PH}
\dot q^j =\mathbf{X}_H\,q^j  = \left\{q^j,H\right\}\,, \qquad \mathrm{and} \qquad  \dot p_j = \mathbf{X}_H\,p_j =  \left\{p_j,H\right\}\,.
\end{equation}
From the last expression, we can understand  that any element that commutes with the Hamiltonian (i.e., the Poisson bracket with the Hamiltonian is equal to 0) is a constant of the motion. So, we intuitively notice that the Poisson bracket allows us to measure how the elements vary along the particle's motion.\\
\indent The Legendre transformation establishes a biunivocal correspondence (for the non-singular case) between the extended dynamical space $\calD^\prime = TQ\times\R$ and the extended phase space $\Gamma^\prime=T^*Q \times\R$:
\begin{equation}
 \left.  \begin{array}{lccc}
          j: &   \mathcal{D}^\prime & \longrightarrow & \Gamma^\prime \\
						 &  (q, \dot q,t)     & \longmapsto &  (q,p,t) 
					\end{array}   \right\}  \,, \qquad {\rm with} \qquad   p_j(q,\dot q,t):=\frac{\partial L(q,\dot q,t)}{\partial \dot q^j} \,.
\end{equation}
\indent The vector field $\mathbf{D}$ (\ref{chap23-D}) generates the time evolution tangent to $\calD^\prime$. As for the Jacobian application $j^T$, 
\begin{equation}\label{chap251-JT}
j^T(\mathbf{D}) := \mathbf{D} \circ j
\end{equation}
transforms the time evolution generator $\mathbf{D}$ into the Hamiltonian vector field $\mathbf{X}_H$  tangent to the submanifold of $\Gamma^\prime$, that is,
\begin{equation}\label{ch241-JTD}
j^T\left(\mathbf{D}\right) = \mathbf{X}_H\,.
\end{equation}
Indeed, $j^T\left(\mathbf{D}\right)$ is a tangent vector field with local components
\begin{align}
&j^T\left(\mathbf{D}\right) q_i = \mathbf{D}\,q_i = \dot q_i\,,\nonumber\\
&j^T\left(\mathbf{D}\right) p_i = \mathbf{D}\left[\frac{\partial L(q,\dot q, t)}{\partial \dot q^i}\right] = \frac{\partial L}{\partial q^i}\,.
\end{align}
Then, using (\ref{ch241-PH}), we get
\begin{equation}
j^T\left(\mathbf{D}\right) q_i = \mathbf{X}_H\,q_i, \quad \mathrm{and} \quad  j^T\left(\mathbf{D}\right) p_i = \mathbf{X}_H\,p_i\,,
\end{equation}
which shows (\ref{ch241-JTD}). 

\section{Ostrogradsky formalism}

The previous sections have been developed assuming a first-order Lagrangian for simplicity. However, they can be extended by considering a Lagrangian that depends on higher-order derivatives. This extension is known as the \textit{\important{Ostrogradsky!formalism}} \cite{Ostrogradski,Leon1985,Pons1989,Woodard2015}. In order to simplify the notation and not make it cumbersome to follow, we shall only write explicitly down the degrees of freedom when we deem it necessary. 

\subsection{The Euler-Ostrogradsky equations}

Consider an $r^{\rm th}$-order Lagrangian on an $(r\times s)$-dimensional tangent bundle $T^rQ\times\R$. The action integral between $t_0$ and $t_1$ is
\begin{equation}\label{ch25-SHL}
S([q]) := \int^{t_1}_{t_0}\D t\, L(q,\dot q,\ldots,q^{(r)},t)\,.
\end{equation}
Applying the \textit{\important{principle of least action}} (\ref{ch22-dS}) with the boundary conditions $\delta q^{(n)}(t_0) = \delta q^{(n)}(t_1) = 0,\,\,\,n = \{0,1,\ldots,r-1\}$ and integrating by parts successively, we get
\begin{equation}\label{ch25-dSOF}
\begin{split}
\delta S([q]) = \int^{t_1}_{t_0}\D t\, \left\{\left[\sum^{r}_{n=0} \left(-\frac{\D}{\D t} \right)^n \frac{\partial L}{\partial q^{(n)}}\right]\delta q + \frac{\D}{\D t}\left[\sum^{r-1}_{n=0} p_n \delta q^{(n)}\right]\right\} = 0\,,
\end{split}
\end{equation} 
where
\begin{equation}\label{ch25-OM}
p_n := \sum^{r}_{k=n+1} \left(-\frac{\D}{\D t}\right)^{k-n-1}\left[\frac{\partial L}{\partial q^{(k)}}\right]\,.
\end{equation}
For this case, the time derivative operator $\D/\D t$ is referred to as the $r^{\rm th}$-order \textit{\important{time!evolution generator}} $\mathbf{D}$ on $T(T^rQ\times\R)$
\begin{equation}\label{chap261-D}
\mathbf{D} := \frac{\D}{\D t} =\frac{\partial}{\partial t} + \sum^r_{m=0}q^{(m+1)}\frac{\partial}{\partial q^{(m)}}\,.
\end{equation}
The second term in integral (\ref{ch25-dSOF}) is a total derivative, which vanishes at $t_0$ and $t_1$ due to the chosen boundary conditions. Then, since $\delta q(t)$ is completely arbitrary except for the boundary conditions, the integrand must be zero $0$, giving rise to the so-called \textit{\important{Euler-Ostrogradsky equations}}\vspace{0.2cm}
\begin{equation}\label{ch25-LOE}
\mathcal{E}(q,\ldots,q^{(2r)},t):=\sum^{r}_{n=0} \left(-\frac{\D}{\D t}\right)^n\left[\frac{\partial L}{\partial q^{(n)}}\right] = 0\,.
\end{equation} 
By considering all degrees of freedom explicitly, these differential equations constitute a system of $2\,r^{\rm th}$-order $s$-differential equations such that, provided that the Hessian is regular,
\begin{equation}
\left|\frac{\partial^2 L}{\partial q^{i,(r)} \partial q^{j,(r)}}\right| \neq 0, \qquad i,j=\{1,\ldots,s\}\,,
\end{equation}
they can be solved in the generalized coordinate derivatives $q^{j,(2r)}$ and the remaining derivatives up to the order $2r-1$
 \begin{equation}
 q^{j,(2r)} = f^j(q ,\dot q,\ldots,q^{(2r-1)}, t)\,.
 \end{equation}
As long as we provide $2r\times s+1$ initial conditions $(q_0,\dot q_0, \ldots,q_0^{(2r-1)}, t_0)$, the solution of this differential equation always exists and is unique by the theorems of existence and uniqueness of ordinary differential systems. The class of these solutions can be coordinated by the $2r \times s +1$ initial data $(q_0,\dot q_0, \ldots, q_0^{(2r-1)}, t_0)$.

\subsection{Total derivative for higher-order Lagrangians}
For the first-order Lagrangian case, we can always add a \textit{\important{total derivative}} of the form (\ref{ch22-TD}) without modifying the equations of motion. Assuming that an $r^{\rm th}$-order Lagrangian can be written as a \textit{\important{total derivative}}
\begin{equation}\label{ch25-relTDL}
L(q,\dot q,\ldots,q^{(r)},t) = \frac{\D W(q,\dot q,\ldots, q^{(r-1)},t)}{\D t }\,,
\end{equation}
the Euler-Ostrogradski equations (\ref{ch25-LOE}) vanish identically
\begin{equation}\label{ch25-EOTD}
\frac{\partial L}{\partial q} - \frac{\D}{\D t}\left(\frac{\partial L}{\partial \dot q}\right)+ \ldots + (-1)^r\frac{\D^r}{\D t^r}\left(\frac{\partial L}{\partial q^{(r)}}\right) \equiv 0\,.
\end{equation}
Indeed, from definition (\ref{chap261-D}) and the fact that
 \begin{equation}
 \left[\frac{\partial}{\partial q^{(n)}}, \frac{\D}{\D t}\right]W = W_{(n-1)}, \qquad \mathrm{with}\qquad W_{(n)}:= \frac{\partial W}{\partial q^{(n)}}\,,
 \end{equation}
 where $[A,B]= AB-BA$ is the anticommutator, we observe that the following relation is satisfied
 \begin{equation}\label{ch25-relTD}
 \frac{\partial L}{\partial q^{(n)}} = \left[\frac{\partial}{\partial q^{(n)}},\frac{\D}{\D t}\right] W + \frac{\D}{\D t}\left(\frac{\partial W}{\partial q^{(n)}}\right) = W_{(n-1)} + \frac{\D W_{(n)}}{\D t}\,.
 \end{equation}
 Plugging (\ref{ch25-relTD}) into (\ref{ch25-EOTD}), we arrive at 
 \begin{equation}
\frac{\D W_{(0)}}{\D t} - \left[\frac{\D W_{(0)}}{\D t} + \frac{\D^2 W_{(1)}}{\D t^2}\right] + \left[\frac{\D^2 W_{(1)}}{\D t^2} + \frac{\D^3 W_{(2)}}{\D t^3}\right]+\ldots + (-1)^r\left[\frac{\D^r W_{(r-1)}}{\D t^r} + \frac{\D^{r+1} W_{(r)}}{\D t^{r+1}}\right]\,.
 \end{equation}\vspace{0.2cm}
Notice that all the terms cancel in pairs except the last one, $\D^{r+1} W_{(r)}/\D t^{r+1}$, which also vanishes because of $W_{(r)} = 0$. Therefore, the equations of motion are identically zero. 

An alternative proof is based on the principle of least action for higher-order Lagrangians. Introducing (\ref{ch25-relTDL}) into (\ref{ch25-SHL}), we find
 \begin{equation}
 S= \left[W(q,\dot q,\ldots, q^{(r-1)},t)\right]^{t_1}_{t_0}\,.
 \end{equation}  
For this action, the principle of least action with $\delta q^{(n)}(t_0) = \delta q^{(n)}(t_1)=0$ (for $n=0,\ldots,r-1$) reads
\begin{equation}
\delta S = \left[\delta W(q,\dot q,\ldots, q^{(r-1)},t)\right]^{t_1}_{t_0} \equiv 0,  
\end{equation}
which is trivial for any trajectory fulfilling the boundary conditions. 

Therefore, since we are free to add a total derivative without modifying the equations of motion, we also conclude that the definition of \important{Noether!symmetry} can still be applied to higher-order Lagrangians.

\subsection{The Noether theorem for higher-order Lagrangians}

With all that has been developed so far, the extension of the \textit{\important{Noether!theorem}} is practically straightforward. All the steps prior to equation (\ref{ch23-relNTdL}) are not affected by considering higher-order Lagrangians; therefore, we can reuse them by replacing the first-order Lagrangian with the higher-order one. Then, equation (\ref{ch23-relNTdL}) becomes
\begin{equation}\label{chap263-relNTdL}
L^\prime(q^\prime,\ldots, q^{(r)\prime}, t) - L(q,\ldots, q^{(r)},t) = \frac{\partial L}{\partial q} \delta q + \ldots + \frac{\partial L }{\partial q^{(r)}} \delta q^{(r)} + \delta L\,.
\end{equation}
Next, by computing the difference of the square brackets of equation (\ref{ch23-LpLdT})  and integrating by parts successively, we get, up to the leading order,
\begin{equation}
\begin{split}
\left[L^\prime(q^\prime,\ldots, q^{(r)\prime}, t)-L(q,\ldots, q^{(r)},t)\right] =\delta L+  \mathcal{E}\,\delta q+ \frac{\D}{\D t}\left[\sum^{r-1}_{n=0} p_n \delta q^{(n)}\right]\,,
\end{split} 
\end{equation}
where $p_n$ is defined in (\ref{ch25-OM}) and $\mathcal{E}(q,\ldots,q^{(2r)},t)$ are the Euler-Ostrogradsky equations (\ref{ch25-LOE}). Regarding the last expression, equation (\ref{ch23-LpLdT}) becomes 
\begin{equation}
\int^{t_1}_{t_0}\D t \left\{\delta L +\mathcal{E}\,\delta q +  \frac{\D}{\D t}\left[L\,\delta t +\sum^{r-1}_{n=0} p_n \delta q^{(n)}\right]\right\} = 0\,.
\end{equation}
If the infinitesimal transformation is a Noether symmetry, it carries an associated conserved quantity $\delta L = \D W /\D t$.  Hence, equation (\ref{ch23-LpLdTNS}) is finally 
\begin{equation}
\int^{t_1}_{t_0}\D t \left\{\mathcal{E}\,\delta q +  \frac{\D}{\D t}\left[L\,\delta t + W +\sum^{r-1}_{n=0} p_n \delta q^{(n)}\right]\right\} = 0\,.
\end{equation}

Since the choice of the time interval $[t_1,t_2]$ is entirely arbitrary, the integrant must be identical to $0$ for every interval we take; therefore, we get the extension of identity (\ref{ch23-IN_V2}) for higher-order Lagrangians 
\begin{equation}\label{ch23-IN}
\mathcal{E} \,\delta q + \frac{\D J }{\D t} \equiv 0\,,
\end{equation}
 where 
 \begin{equation}\label{ch253-NCO}
 J:= L\,\delta t + W + \sum^{r-1}_{n=0} p_n \delta q^{(n)}
 \end{equation}
 is the Noether constant of motion for them.
  
As seen in Section \ref{chap221}, invariance under time translations rises energy conservation. In the same way, we shall assume that the $r^{\rm th}$-order Lagrangian does not depend explicitly on time, $L = L(q,\ldots,q^{(r)})$. Therefore, it is also invariant under such a transformation, so $W=0$. By taking (\ref{ch231-tt}), calculating the $n^{\rm th}$-order derivative of (\ref{ch231-tq}), $\delta q^{(n)} = - \epsilon\, q^{(n+1)}$, and plugging them into (\ref{ch253-NCO}), the \important{energy function} $E:= -J/\epsilon$ becomes
\begin{equation}\label{chap263-EF}
E = \sum^{r-1}_{n=0} p_n\,q^{(n+1)} - L\,,
\end{equation} 
which is conserved (once the equations of motion are applied) due to (\ref{ch23-IN}).

\subsection{Hamiltonian formalism for higher-order Lagrangians}\label{chap264-HF}

As discussed in Section \ref{ch24-SH}, Noether's theorem (more precisely, the boundary terms) provided us with the definition of the Legendre transformation to build the Hamiltonian formalism. We shall use this analogy and define it for higher-Lagrangian employing those elements which are proportional to $\delta q^{(n)}$. Thus, considering equation (\ref{ch253-NCO}), we define the \textit{\important{Legendre-Ostrogradsky transformation}} as
\begin{equation}\label{ch254-OMD}
p_n := \sum^{r}_{k=n+1} \left(-\frac{\D}{\D t}\right)^{k-n-1} \frac{\partial L}{\partial q^{(k)}}\,, \qquad n=\{0,\ldots,r-1\}\,.
\end{equation}
These $p_n$ are known as \textit{\important{Ostrogradsky!momenta}}.  Observe that, by using the definition (\ref{ch254-OMD}) for the case $n=r-1$, we get
\begin{equation}
p_{r-1}= \frac{\partial L(q,\ldots,q^{(r)},t)}{\partial q^{(r)}}\,.
\end{equation}
We can isolate $q^{(r)}$ (as long as the Hessian is regular) as a function of the remaining derivatives
\begin{equation}\label{ch254-qr}
q^{(r)}=q^{(r)}(q,\dot q,\ldots, q^{(r-1)},p_{r-1},t)\,.
\end{equation}
Proceeding in the same way for the following order, $n=r-2$, we find that 
 \begin{equation}
p_{r-2} = -\frac{\partial^2 L}{\partial q^{(r)} \partial q^{(r)}}\,q^{(r+1)} + f(q,\ldots, q^{(r)},t)\,,
\end{equation}
which depends on derivatives up to order $r+1$ and allows us to isolate\vspace{0.1cm} 
\begin{equation}
q^{(r+1)}= q^{(r+1)}(q,\dot q,\ldots,q^{(r-1)},p_{r-1},p_{r-2},t)\,.
\end{equation}
For the following orders, proceed in the same way. 

The Legendre transformation uniquely connects  $\calD^\prime=T^rQ\times\R$ with the extended phase space $\Gamma^\prime = T^*(T^{r-1}Q)\times\R$ by
\begin{equation}  \label{ch245-cHL} 
j:\,\,(q,\dot q,\ldots,q^{(2r-1)}, t)\in \mathcal{D}^\prime\quad \longrightarrow\quad (q,\dot q,\ldots,q^{(r-1)},p_0,\ldots, p_{r-1},t)\in\Gamma^\prime\,,
\end{equation}
where $p_n$ with $n=0,\ldots,r-1$ are defined by (\ref{ch254-OMD}). 

Employing the definition of the Legendre transformation (\ref{ch254-OMD}), we can naturally extend\footnote{If the reader is intrigued by a more formal introduction to these geometric structures for higher-order cotangent bundles, we suggest \cite{Pons1991} and references therein.} the Hamiltonian function and the geometric structures, such as the symplectic form or the Hamiltonian vector field, simply ``swapping" the momentum $p$ for the Ostrogradsky momenta $p_n$.  Thus, using the energy function (\ref{chap263-EF}) as a reference, we shall define the $r^{\rm th}$-order \textit{\important{Ostrogradsky!Hamiltonian}} on $\Gamma^\prime$ as
\begin{align}\label{ch254-HO}
&H(q,\dot q,\ldots,q^{(r-1)},p_0,\ldots, p_{r-1},t) := \sum^{r-1}_{n=0} p_n q^{(n+1)} - L\nonumber\\
&\hspace{3cm} =p_0 \dot q + p_1 \ddot q + \ldots + p_{r-1} q^{(r)}(q,\dot q,\ldots, q^{(r-1)},p_{r-1},t)\nonumber\\
&\hspace{5cm}- L(q,\dot q,\ldots, q^{(r-1)},q^{(r)}(q,\dot q,\ldots, q^{(r-1)},p_{r-1},t),t)\,,
\end{align}
where we have replaced the term $q^{(r)}$ with equation (\ref{ch254-qr}). Furthermore, the $r^{\rm th}$-order \textit{\important{canonical!Poisson brackets}} become
\begin{equation}
\left\{q^{(n)}, p_k\right\} = \delta^n_k, \quad \left\{q^{(n)}, q^{(k)}\right\} =0, \quad \left\{p_n, p_k\right\} =0
\end{equation}
characterized by
\begin{equation}\label{chap264-PB}
\left\{f,g\right\} = \sum^{r-1}_{n=0}\left(\frac{\partial f}{\partial q^{(n)}}\frac{\partial g}{\partial p_n} - \frac{\partial f}{\partial p_n}\frac{\partial g}{\partial q^{(n)}}\right)\,.
\end{equation}
Finally, the \textit{\important{Hamilton equations}} are
\begin{align}\label{cha264-EH}
&\frac{\D q^{(n)}}{\D t}= \mathbf{X}_H\,q^{(n)}  = q^{(n+1)}\,,\nonumber\\
&\frac{\D p_n}{\D t}= \mathbf{X}_H\,p_n = - p_{n-1} + \frac{\partial L}{\partial q^{(n)}}\,,
\end{align}
where $p_{-1} \equiv 0$ is understood, and $\mathbf{X}_H$ is the $r^{\rm th}$-order \textit{\important{Hamiltonian!vector field}}
\begin{equation}\label{chap264-HV}
\mathbf{X}_H = \partial_t + \sum^{r-1}_{n=0}\left(\frac{\partial H}{\partial p_n}  \frac{\partial }{\partial q^{(n)}}-\frac{\partial H}{\partial q^{(n)}}\frac{\partial }{\partial p_n}\right)\,.
\end{equation}
As done above, they could also be expressed by the Poisson bracket (\ref{chap264-PB}), $q^{(n+1)} = \{q^{(n)},H\}$ and $\dot p_n = \{p_n,H\}$.

The Ostrogradsky Hamiltonian (\ref{ch254-HO}) depends linearly on the momenta $p_0,\ldots,p_{r-2}$. Since they may be negative, the Hamiltonian has not a definite sign. Therefore, it could be the case that the system's energy might be negative since, as noticed in Section \ref{ch24-SH}, the Hamiltonian might be related to the energy function as long as it does not depend on time. In literature \cite{Chen2013,Woodard2015,Ganz2021}, this fact is interpreted as if the solutions (or trajectories) were not stable and is referred to as \textit{\important{Ostrogradsky!instability}}. However, in \cite{Llosa2003} is argued (and even exemplified) that the fact that the energy is not bounded from below does not imply that the system is unstable. The only claim we can make is that the energy (or, equivalently, for this case, the Hamiltonian) cannot be considered a \important{Lyapunov!function} \cite{Boyce2008,Gant2003,gantmacher1975}. Positive energy is a sufficient condition for stability; however, it does not imply that the system is unstable if it is not fulfilled\footnote{In the Lagrange theorem on the stability criterion of the equilibrium position, the positive energy function is used to ensure that the system is stable. However, the \important{Lyapunov!theorem} (a generalization of Lagrange's theorem) shows that there could be a function (that is not necessarily the energy function) that indicates the stability of our equilibrium point. Indeed, these functions used in the theorem are known as Lyapunov's functions. The significant problem (and from a mathematical point of view, as far as we know, is still open) is that there is no systematic way to find these functions, which makes it challenging to analyze the stability of our system \cite{gantmacher1975}.}. We shall illustrate this fact for nonlocal theories in one of our examples --Section \ref{chap421}--.

As seen above, the application  $j$ (\ref{ch245-cHL}) puts in biunivocal correspondence the solutions of the Ostrogradsky equations and the solutions of the Hamilton equations; strictly speaking, the time evolution generators $\mathbf{D}$ on $\calD^\prime$ can be connected with the Hamiltonian vector field $\mathbf{X}_H$ through the Jacobian map $j^T$, i.e., $j^T\left(\mathbf{D}\right) = \mathbf{X}_H$. Indeed, proceeding in the same way as (\ref{ch241-JTD}), we find that the components of the tangent vector field  $j^T\left(\mathbf{D}\right)$ are
\begin{align}
j^T\left(\mathbf{D}\right)q^{(n)}= \mathbf{D}\,q^{(n)} = q^{(n+1)}\,
\end{align}
and
\begin{align}
j^T\left(\mathbf{D}\right)\,p_n&= \mathbf{D}\left[\sum^{r}_{k=n+1} \left(-\mathbf{D}\right)^{k-n-1} \frac{\partial L}{\partial q^{(k)}}\right] = - \sum^{r}_{k=n+1} \left(-\mathbf{D}\right)^{k-n}\left[\frac{\partial L}{\partial q^{(k)}}\right] + \frac{\partial L}{\partial q^{(n)}} - \frac{\partial L}{\partial q^{(n)}}\nonumber\\
& = - \sum^{r}_{k=n} \left(-\mathbf{D}\right)^{k-n}\left[\frac{\partial L}{\partial q^{(k)}}\right] + \frac{\partial L}{\partial q^{(n)}} = - p_{n-1} +  \frac{\partial L}{\partial q^{(n)}}\,.
\end{align}
Then, using (\ref{cha264-EH}), we get 
\begin{equation}
j^T\left(\mathbf{D}\right)\,q^{(n)} = \mathbf{X}_H\,q^{(n)} \qquad \mathrm{and} \qquad j^T\left(\mathbf{D}\right)\,p_n = \mathbf{X}_H\,p_n\,.
\end{equation}

As in Section \ref{ch241-SHF}, the Hamiltonian formalism can be put into a geometrical form using the \important{contact form} 
\begin{equation}\label{ch254-w}
\Omega^\prime  = \Omega - \D H\wedge \D t, \qquad \mathrm{where} \qquad \Omega:= \sum^{r-1}_{n=0} \D p_n \wedge \D q^{(n)}\,.
\end{equation}
Keeping in mind that the time evolution generator is (\ref{chap264-HV}), we get that the Hamilton equation can be written as
\begin{equation}
i_{\mathbf{X}_H} \Omega = 0\,.
\end{equation}
 
The infinite-order Lagrangian framework is the generalization of Ostrogradsky's formalism,  based on merely putting  $\infty$ instead of $r$. This way of proceeding can lead to particular crossroads. One of them is the \textit{\important{Euler-Ostrogradsky equations}} (\ref{ch25-LOE}). Note that they are no longer differential equations, properly speaking. Since there is no higher term, we cannot solve the derivative of a particular order, $q^{(n)}$, as a function of the lower-order ones, $q^{(k)}$ for $k<n,$ and $n,k\in\mathbb{N}$. Therefore, they are no longer the dynamic equations that give us information about how our system evolves.  A similar case happens with the \textit{\important{Legendre-Ostrogradsky transformation}} (\ref{ch254-OMD}). It is unclear how to invert such a transformation since it is unknown which derivative of $q$ should be substituted by the momenta. 

Another problem with this generalization is identifying the elements (initial conditions) that coordinatize the extended dynamic space $\calD^\prime$. Extrapolating what we know about the initial data, we would conclude (by analogy) that we need an infinite number of initial data to determine the solution.  Since we do not have existence and uniqueness theorems for such systems, the framework of ``initial data evolving over time" is lost. 

Another implication of this generalization is the \textit{\important{Noether!symmetry}}. By adding an infinite-order \textit{\important{total derivative}}, one might think the equations of motion --analogous to case (\ref{ch25-relTDL})-- will remain invariant. However, it might be not true. For instance, consider case (\ref{ch25-relTDL}) but with an infinite-order boundary term $W=W(q,\ldots, q^{(r-1)},\ldots,t)$. Therefore, the Euler-Ostrogradsky equations (\ref{ch25-EOTD}) are
 \begin{align}
 &\frac{\D W_{(0)}}{\D t} - \left[\frac{\D W_{(0)}}{\D t} + \frac{\D^2 W_{(1)}}{\D t^2}\right] + \left[\frac{\D^2 W_{(1)}}{\D^2 t} + \frac{\D^3 W_{(2)}}{\D t^3}\right]\nonumber\\
 & \qquad \qquad + \ldots + (-1)^r\left[\frac{\D^r W_{(r-1)}}{\D t^r} + \frac{\D^{r+1} W_{(r)}}{\D t^{r+1}}\right]+ \ldots\,.
\end{align}
Although the terms still cancel in pairs, the succession of partial sums is
\begin{equation}
\frac{\D W_{(0)}}{\D t}, \,0,\, \frac{\D^2 W_{(1)}}{\D t^2},\, 0,\,\frac{\D^3W_{(2)}}{\D t^3},\,0, \ldots, 0,\frac{\D^n W_{(n-1)}}{\D t^n},\,0,\ldots
\end{equation}
and the vanishing of the addition depends on whether the ``infinite" number of terms added is even or odd. Hence, as the sum depends on the order of addition, the series is not summable. This fact leads us to conclude that the equations of motion might be modified by adding an infinite-order total derivative. Consequently, we are forced to look for additional conditions to hold Noether's symmetry. Perhaps, this casuistic may shed some light on the problem of defining the functional space since it will restrict the class of functions we can consider. 

Finally, the use of infinite series without specifying a prescription of how to sum them is another unpleasant feature, as just seen.  This generalization might only be heuristic unless the convergence of the series is proved. However, as this proof can be remarkably complex, we propose following the line of research of \cite{Marnelius1973,Llosa1994} and dealing with nonlocal Lagrangians using functional methods without going through Ostrogradsky's formalism.


\chapter{Local Lagrangian Fields}\label{chap3}

In the previous chapter, we focused on systems with a finite number of degrees of freedom. For this one, we shall extend the previous results for systems with an infinite number of degrees of freedom, also called \textit{\important{fields}}. Likewise, we shall work directly with the  \textit{\important{Ostrogradsky!formalism}} to make it as general as possible. We shall rely on \cite{Leon1985, Henneaux1992, Peskin1995, Goldstein2002,Pons2011}.  Before starting, it is indispensable to mention that we shall use the Einstein summation convention throughout this chapter. Summation over repeated indices is always understood unless the contrary is indicated.

\section{The principle of least action}\label{chap31}
In a local field theory, the $r^{\rm th}$-order Lagrangian $L$ on $T^rQ \times \R$   is defined as a spatial integral of a \textit{\important{Lagrangian!density}}, denoted by $\calL$, which is a function of one or more fields $\phi$ (or field components $\phi^A$, $A\in\mathbb{N}$) and their partial derivatives $\phi_{|_{b_1\ldots b_r}}$,
\begin{equation}\label{chap31-L}
L(\phi^A,\ldots,\phi^A_{|_{4 \ldots (r) \ldots 4}} ,t):= \int_{\R^d} \D \mathbf{x}\,\calL(\phi^A,\ldots,\phi^A_{|_{b_1\ldots b_r}},\mathbf{x},t),
\end{equation}
where $d$ denotes the spatial dimensions, and the ``stroke" means ``partial derivative" with respect to the manifold coordinates. 

The \important{action integral} is defined as
\begin{equation}\label{chap31-S}
S([\phi]) = \int_\calV\D x\,\calL(\phi^A,\ldots,\phi^A_{|_{b_1\ldots b_r}},x)
\end{equation}
for all $(d+1)$-dimensional volume $\calV$, where we have denoted $x^a =(\mathbf{x},t)$.  We shall take $d=3$ for concreteness; however, the following also holds for any number of spatial dimensions. 

The principle of least action reads $\delta S = 0$, where
\begin{equation}
\delta S([\phi])  = \int_\calV \D x \left(\frac{\partial \calL}{\partial \phi^A} \delta \phi^A + \ldots + \frac{\partial \calL}{\partial \phi^A_{|_{b_1\ldots b_r}}}\delta \phi^A_{|_{b_1\ldots b_r}}\right)\,,
\end{equation}
for all variations $\delta \phi^A_{|_{b_1\ldots b_n}}$ with $n=\{0,1,\ldots,r-1\}$ that vanish at the boundary $\partial \calV$. Since $\delta \phi^A$ and the derivatives are not independent, we integrate successively by parts and get 
\begin{equation}\label{chap31-dS1}
\delta S([\phi]) =  \int_\calV \D x\left\{\left(\sum^r_{n=0} (-1)^n \partial_{b_1\ldots b_n}\left[\frac{\partial \calL}{\partial \phi^A_{|_{b_1\ldots b_n}}}\right]\right)\delta \phi^A + \partial_b\left[\sum^{r-1}_{n=0} \Pi^{c_1\ldots c_n b}_A \delta \phi^A_{|_{c_1\ldots c_n}}\right]\right\} 
\end{equation}
with
\begin{equation}\label{chap31-P}
\Pi^{c_1\ldots c_n b}_A:= \sum^r_{k=n+1} (-1)^{k-n-1} \partial_{a_1\ldots a_{k-n-1}}\left[\frac{\partial \calL}{\partial \phi^A_{|_{a_1\ldots a_{k-n-1} c_1\ldots c_n b}}}\right]\,,
\end{equation}
where $\partial_{a_1\ldots a_{k-n-1}}$ denotes $\partial_{a_1} \partial_{a_2}\ldots\partial_{a_{k-n-1}}$ . Note that the second term in (\ref{chap31-dS1}) is a four-divergence. Using Gauss' theorem, it becomes a surface integral on $\partial\calV$, which vanishes because $\left.\delta \phi^A_{|_{c_1\ldots c_n}}\right|_{\partial\calV} = 0$. As the volume $\calV$ has been arbitrarily chosen, the remaining term must vanish identically. Consequently, we get the \textit{\important{Euler-Ostrogradsky equations}} for fields
\begin{equation}\label{chap31-EL}
\mathcal{E}_A(\phi^B,\ldots, \phi^B_{|_{b_1\ldots b_{2r}}},x):=\sum^r_{n=0} (-1)^n \partial_{b_1\ldots b_n}\left[\frac{\partial \calL}{\partial \phi^A_{|_{b_1\ldots b_n}}}\right] = 0\,.
\end{equation}
In particular, for $r=1$, we have the \textit{\important{Euler-Lagrange!field equations}} as expected
\begin{equation}
\mathcal{E}_A(\phi^B,\phi^B_{|_{b_1}},\phi^B_{|_{b_1b_2}},x):=\frac{\partial \calL}{\partial \phi^A} - \partial_b\left[\frac{\partial \calL}{\partial \phi^A_{|_b}}\right] = 0\,.
\end{equation}
\indent Equation (\ref{chap31-EL}) is a partial differential equation (PDE) system of order $2 r$. According to the Cauchy-Kowalewski theorem \cite{john1991}, given a non-characteristic hypersurface $\Sigma$ in $\R^4$ with normal vector $\,n^b\,$ and $\,2 r\,$ functions, $\; u_j\,,\; j= 0,\ldots,2r-1\,$, on $\Sigma$, there exists a solution $\,\phi^A(x)\,$ of the PDE (\ref{chap31-EL}) such that
\begin{equation}
n^{b_1} \ldots n^{b_j} \phi_{|_{b_1 \ldots b_j}}(x) = u_j(x)\,, \qquad j= 0,\ldots,2r-1\,,\qquad \forall x\in\Sigma \,.
\end{equation} 
In case that $\Sigma$ is the hyperplane $\,t=x^4 = 0\,$, then $n^b= (0,0,0,1)\,$, and the \textit{\important{Cauchy-Kowalevski theorem}} is the basis for interpreting the \textit{\important{Cauchy data}} $\; u_j(\mathbf{x})\,,\; j= 0,\ldots,2r-1\,$ as ``the state of the field'' at $t=0$, which evolves in time steered by the field equations (\ref{chap31-EL}). Furthermore, and similarly to what the theorems of existence and uniqueness do for systems with a finite number of degrees of freedom, the Cauchy-Kowalevski theorem allows parametrizing each solution $(\phi,x) \in \mathcal{D}^\prime$ with a well-defined --although an infinite-- set of ``parameters'', namely, the Cauchy data.\\
\indent The infinitesimal generators of spacetime translations are the vector fields $\mathbf{D}_a\,$, $\;a = 1,\ldots, 4$, which are tangent to the curves $(\phi^A,\ldots,\phi^A_{|_{b_1\ldots b_r}}, x)\,.$ Therefore, for a function $F$, we have that
\begin{equation}  \label{chp31-L4} 
\mathbf{D}_a F(\phi^A,\ldots,\phi^A_{|_{b_1\ldots b_r}}, x) := \left[\frac{\partial F\left(\phi^A(x+\varepsilon), \ldots, \phi^A(x+\varepsilon)_{|_{b_1\ldots b_r}}, x + \varepsilon\right)}{\partial \varepsilon^a}\right]_{\varepsilon^a=0}\,,
\end{equation}
which, including the chain rule, can be written as\vspace{0.1cm}
\begin{equation}  \label{L5} 
\mathbf{D}_a =  \partial_{a} + \sum^{r}_{m=0} \phi^A_{|_{a\,b_1\ldots b_m}} \frac{\partial}{\partial \phi^A_{|_{b_1\ldots b_{m}}}}\,.
\end{equation}
In particular, $\mathbf{D}_4$ is the \textit{\important{time!evolution generator}}. 

\section{Four-divergence for fields}

Let us focus on transformations that leave the equations of motion invariant, i.e., \textit{\important{Noether!symmetry}} transformations. Provided that the Lagrangian density is a \textit{\important{four-divergence}}
\begin{equation}\label{chap31-LW}
\calL(\phi^A,\ldots,\phi^A_{|_{b_1\ldots b_r}},x) = \partial_b W^b(\phi^A,\ldots,\phi^A_{|_{b_1\ldots b_{r-1}}},x)\,,
\end{equation}
the Euler-Ostrogradsky field equations $\mathcal{E}_A(\phi^B,\ldots, \phi^B_{|_{b_1\ldots b_{2r}}},x) \equiv 0$. Indeed, by plugging (\ref{chap31-LW}) into (\ref{chap31-S}),  the action integral becomes
\begin{equation}
S([\phi^A]) = \int_{\partial\calV} \D \Sigma_b\,W^b(\phi^A,\ldots,\phi^A_{|_{b_1\ldots b_{r-1}}},x)\,.
\end{equation}
Consequently, the principle of least action reads 
\begin{equation}
\delta S([\phi^A]) = 0, \qquad \mathrm{for\,\,all} \qquad \left.\delta\phi^A_{|_{b_1\ldots b_j}}\right|_{\partial\calV} = 0\,, \quad j=0,\ldots,r-1\,, 
\end{equation}
or, equivalently, 
\begin{equation}
 \int_{\partial\calV} \D \Sigma_b\, \delta W^b(\phi^A,\ldots,\phi^A_{|_{b_1\ldots b_{r-1}}},x)= 0 \quad \mathrm{for\,\,all} \quad \left.\delta\phi^A_{|_{b_1\ldots b_j}}\right|_{\partial\calV} = 0\,, \quad j=0,\ldots,r-1\,,
\end{equation} 
which is satisfied trivially for any $(\phi^A,\ldots,\phi^A_{|_{b_1\ldots b_r}}, x)$. Therefore, Noether's symmetry holds also for fields.

\section{The Noether theorem}\label{chap32}
Consider an infinitesimal transformation of the coordinates
\begin{equation}\label{chap32-T}
x^{\prime a} = x^a + \delta x^a\,,
\end{equation}
which induces the following transformation of the fields
\begin{equation}\label{chap32-TP}
\phi^{\prime A} = \phi^{A} + \delta \phi^A\,.
\end{equation}
The Lagrangian density shall be transformed so that the action integral (\ref{chap31-S}) remains unchanged, i.e.,
\begin{equation}\label{chap32-LLp}
\calL^\prime(\phi^{\prime A}(x^\prime),\ldots, \phi^{\prime A}_{|_{b_1\ldots b_r}}(x^\prime), x^\prime) = \left|\frac{\partial x}{\partial x^\prime}\right| \calL(\phi^A(x),\ldots, \phi^A_{|_{b_1\ldots b_r}}(x), x)\,.
\end{equation}
Thus, if $\calV^\prime$ is the transformed four-volume element of $\calV$ according to (\ref{chap32-T}), we have that
\begin{equation}\label{chap32-SN}
\int_{\calV} \D x\,\calL(\phi^A(x),\ldots, \phi^A_{|_{b_1\ldots b_r}}(x), x) = \int_{\calV^\prime} \D x^\prime\,\calL^\prime(\phi^{\prime A}(x^\prime),\ldots, \phi^{\prime A}_{|_{b_1\ldots b_r}}(x^\prime), x^\prime)\,.
\end{equation}
By subtracting both sides of (\ref{chap32-SN}) and replacing the dummy variable $x^\prime$ with $x$, we get  
\begin{equation}\label{chap32-SN2}
\int_{\calV} \D x\,\calL(\phi^A(x),\ldots, \phi^A_{|_{b_1\ldots b_r}}(x), x) - \int_{\calV^\prime} \D x\,\calL^\prime(\phi^{\prime A}(x),\ldots, \phi^{\prime A}_{|_{b_1\ldots b_r}}(x), x) = 0 \,.
\end{equation}
  \begin{figure}[h]
    \centering
    \includegraphics[width=0.7\textwidth]{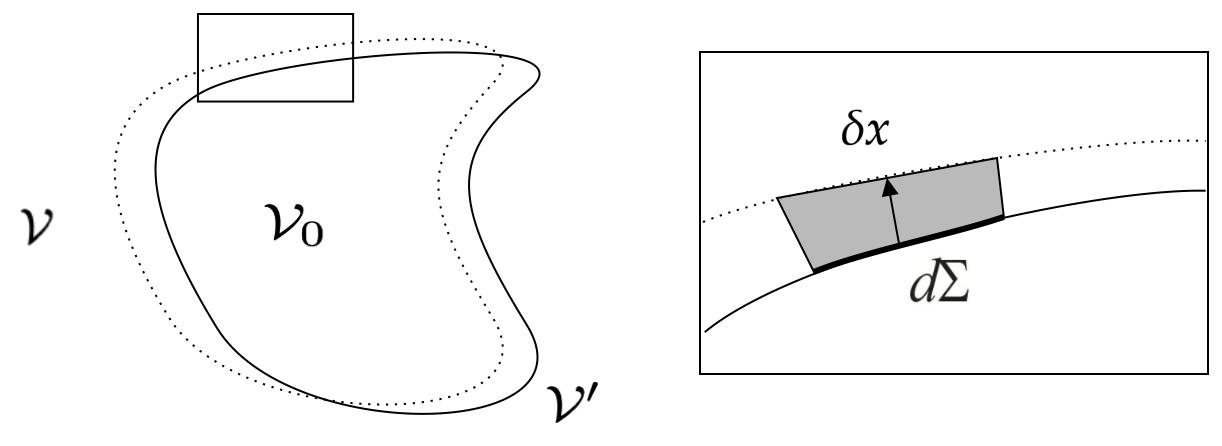}
    \caption{The spacetime domain $\calV$ variation.}
    \label{fig:chp32-V}
\end{figure}
As shown in figure (\ref{fig:chp32-V}), the volumes $\calV$ and $\calV^\prime$ share a common part $\calV_0$ and only differ by an infinitesimal amount at the boundary $\partial\calV_0$. The volume element on the boundary is $\D x = \D\Sigma_a \delta x^a$, so equation (\ref{chap32-SN2}) can be expressed as
\begin{equation}\label{chap32-DS}
\begin{split}
&\int_{\calV} \D x\left[ \calL(\phi^A(x),\ldots, \phi^A_{|_{b_1\ldots b_r}}(x), x) -\calL^\prime(\phi^{\prime A}(x),\ldots, \phi^{\prime A}_{|_{b_1\ldots b_r}}(x), x)\right]\\
&\hspace{5cm}  + \int_{\partial\calV} \D\Sigma_a\,\calL (\phi^A(x),\ldots, \phi^A_{|_{b_1\ldots b_r}}(x), x)\,\delta x^a = 0 \,,
\end{split}
\end{equation}
where equation (\ref{chap32-LLp}) has been used, and the second-order infinitesimals were neglected. 

We shall say that the transformation (\ref{chap32-T}-\ref{chap32-TP}) is a Noether symmetry transformation if
\begin{equation}
\calL^\prime = \calL + \partial_b W^b \qquad \mathrm{where} \qquad  W^b(x) = W^b(\phi^A(x),\ldots, \phi^A_{|_{b_1\ldots b_{r-1}}}(x),x)\,.
\end{equation} 
In such a case, equation (\ref{chap32-DS}) becomes 
\begin{equation}
\begin{split}
&\int_{\calV} \D x\left\{ \calL(\phi^A,\ldots, \phi^A_{|_{b_1\ldots b_r}}, x) -\calL(\phi^{\prime A},\ldots, \phi^{\prime A}_{|_{b_1\ldots b_r}}, x)\right.\\
&\left. \qquad \qquad + \partial_b\left[W^b(\phi^A,\ldots, \phi^A_{|_{b_1\ldots b_{r-1}}},x) + \calL (\phi^A(x),\ldots, \phi^A_{|_{b_1\ldots b_r}}(x), x)\,\delta x^b \right] \right\} = 0\,,
\end{split}
\end{equation}
where Gauss'  theorem has been used. Computing the difference, we arrive --after successive integrations by parts-- at 
\begin{equation}\label{chap3.2-InNC}
\int_{\calV} \D x\left\{ \mathcal{E}_A\,\delta \phi^A  + \partial_b \left[ \sum^{r-1}_{n=0} \Pi^{c_1\ldots c_n b}_A \delta \phi^A_{|_{c_1\ldots c_n}} + W^b + \calL\,\delta x^b \right] \right\}\, = 0\,,
\end{equation}
where $\Pi^{c_1\ldots c_n b}$ are given by (\ref{chap31-P}) and $\mathcal{E}_A$ are the Euler-Ostrogradsky equations (\ref{chap31-EL}). Finally, since the choice of volume $\calV$ has been completely arbitrary, the expression under the integral sign must be identically $0$, which leads to the following identity
\begin{equation}\label{chap32-IDN}
\mathcal{E}_A\,\delta \phi^A  + \partial_b J^b \equiv 0\,,
\end{equation}
where
\begin{equation}\label{chap321-J}
J^b:= \sum^{r-1}_{n=0} \Pi^{c_1\ldots c_n b}_A \delta \phi^A_{|_{c_1\ldots c_n}} + W^b + \calL\,\delta x^b\,. 
\end{equation}
The first term of (\ref{chap32-IDN}) is identically zero if the field equations are satisfied, which leads to the fact that the four-divergence is locally conserved,
\begin{equation}
\partial_b J^b = 0\,.
\end{equation}

\subsection{The energy- and angular momentum currents}\label{chap331}
Infinitesimal \textit{\important{Poincar\'e!transformations}} act on coordinates (\ref{chap32-T}) as 
\begin{equation}  \label{chap331-P1}
\delta x^a = \varepsilon^a + \omega^a_{\;b} x^b\,, \qquad \omega_{ab}+\omega_{ba}=0 \,,
\end{equation}
where $\,\varepsilon^a\,$ and $\,\omega^a_{\;b}\,$ are constants, $\; \omega_{ab} = \eta_{ac}\omega^c_{\;b} \,$ and $\eta_{ac}=\mathrm{diag}(1,1,1,-1)$ is the Minkowski matrix to raise and lower indices. Likewise, the field $\phi^A$ transforms as a tensor object
\begin{equation}  \label{P2}
\phi^{\prime\,A}(x^\prime) = \phi^A(x) + \tilde{M}^A_{\; B} \phi^B(x)\,, \qquad 
\end{equation}
where the constant matrix $\; \tilde{M}^A_{\; B} = \omega^{ab}\,M^A_{\; B[ab]} \;$ depends on the tensor type of the field. Thus,
\begin{equation}  \label{chap331-P3}
\delta \phi^A(x) = \omega^{ab}\,M^A_{\; B[ab]} \phi^B(x) - \phi^A_{|c}(x) \,\left(\varepsilon^c + \omega^c_{\;b} x^b\right)\,, 
\end{equation}
where  (\ref{chap331-P1}) and (\ref{chap32-T}) have been included. For derivatives of any order, we have 
\begin{equation} \label{chap331-P4}
\delta \phi^A_{|_{c_1\ldots c_n}} = \omega^{ab}\left[M^A_{B[ab]} \phi^B_{|_{c_1\ldots c_n}} - \eta_{b c_1} \phi^A_{|_{a c_2\ldots c_n}} - \ldots - \eta_{bc_n}\phi^A_{|_{a c_1\ldots c_{n-1}}}\right] - \phi^A_{|_{b c_1\ldots c_n}} \delta x^b\,.
\end{equation}
Then, plugging (\ref{chap331-P4}) and (\ref{chap331-P1}) into (\ref{chap321-J}), and assuming that the Lagrangian density is \textit{\important{Poincar\'e!invariant}} --so, $W^b= 0$--, we find that the conserved current (\ref{chap321-J}) can be written as
\begin{equation}
J^b = -\varepsilon^a \calT^{\,\,\,b}_a - \frac{1}{2}\omega^{ac}\calJ_{ac}^{\,\,\,\,\,b}\,,  
\end{equation}
where 
\begin{equation}
\calT^{\,\,\,b}_a:= \sum^{r-1}_{n=0}\Pi^{c_1\ldots c_n b}_A \phi^A_{|_{a c_1\ldots c_n}} - \calL\, \delta^b_a
\end{equation}
is the canonical \textit{\important{energy-momentum tensor}},\vspace{0.2cm}
\begin{equation}
\calJ_{ac}^{\,\,\,\,\,b}:= 2\,\calT^{\,\,\,b}_{[a} x_{c]} + \calS^{\,\,\,\,\,b}_{ac}
\end{equation}
is the \textit{\important{angular momentum tensor}}, 
\begin{equation}
\frac{1}{2} \calS^{\,\,\,\,\,b}_{ac} := \sum^{r-1}_{n=0} \Pi^{c_1\ldots c_n b} \left[- M^A_{B[ac]} \phi^B_{|_{c_1\ldots c_n}} +\eta_{[c c_1} \phi^A_{|_{a] c_2\ldots c_n}} + \ldots + \eta_{[c c_n}\phi^A_{|_{a] c_1\ldots c_{n-1}}}\right]\,
\end{equation}
is the \textit{\important{spin current}}, and finally $2\,\calT^{\,\,\,b}_{[a} x_{c]}$ is the \textit{\important{orbital angular momentum tensor}}. Using the symmetry of the cross-derivatives and $\Pi^{c_1\ldots c_n b}$ concerning all indices, we have that the latter expression can be simplified as
\begin{equation}
\frac{1}{2} \calS^{acb} = -  M^{A[ac]}_B \sum^{r-1}_{n=0} \Pi^{c_1\ldots c_n b}_A \phi^B_{|_{c_1\ldots c_n}} + \sum^{r-1}_{n=1} n\,\Pi^{b c_1\ldots c_{n-1}[c}_A \phi^{A|a]}_{_{\,\,\,\,\,\,\,\,\,\,\,c_1\ldots c_{n-1}}}\,.
\end{equation}

As the ten parameters $\varepsilon^a$ and $\omega^{ac}$ are independent, the local conservation of the current $\,J^b\,$ implies that the currents $\mathcal{T}^{\; b}_a$ and $\,\mathcal{J}^{\; \,\;b}_{ac}\,$ are separately conserved, that is,
\begin{equation}  \label{P7_LF}
 \partial_b \mathcal{T}^{\; b}_a = 0 \qquad {\rm and}\qquad \partial_b\mathcal{S}^{\; \,\;b}_{ac} + 2 \,\mathcal{T}_{[ac]} = 0\,.
\end{equation}
 As a rule, the canonical energy-momentum tensor $\calT^{\,\,\,b}_a$ is not symmetric. As a consequence of the second equation (\ref{P7_LF}), it is symmetric if, and only if, the divergence of the spin current vanishes. Indeed, it happens for scalar fields ruled by first-order Lagrangians. However, there is a spin current even for them in the case of higher-order ones.

In all cases, an energy-momentum tensor $\Theta^{ab}$, the \textit{\important{Belinfante-Rosenfeld energy-momentum tensor}}\cite{BELINFANTE1940,rosenfeld1940,dixon1978,Llosa2004}, 
\begin{equation}\label{chap331-BT}
 \Theta^{a b} = \mathcal{T}^{a b} + \partial_c \mathcal{W}^{cba}
\end{equation}
with
\begin{equation}\label{chap331-W}
\mathcal{W}^{cba} :=\frac12\,\left(\mathcal{S}^{cba} + \mathcal{S}^{cab} - \mathcal{S}^{bac} \right)\;,
\end{equation}
can be found so that it is symmetric and, in some sense, is equivalent to the canonical energy-momentum tensor. We say ``equivalent" because: \textbf{(a)} the total energy-momentum contained in a hyperplane $t$ (constant) is the same for both tensors
\begin{equation}
\int_{\R^3} \D\mathbf{x}\, \Theta^{\;4}_a (\mathbf{x},t)=   \int_{\R^3} \D\mathbf{x}\, \mathcal{T}^{\;4}_a (\mathbf{x},t)\,,
\end{equation}
\textbf{(b)} the four-divergences are equal, $\partial_b \Theta^{\;b}_a = \partial_b\mathcal{T}^{\;b}_a = 0\,,$ which implies that the current $\Theta_a^{\;b}$ is also conserved, and \textbf{(c)} the new orbital angular momentum current $2\,x_{[c} \Theta_{a]}^{\;b}$ and the new spin current $\Sigma^{\;\, \;b}_{ac}:= \mathcal{J}^{\;\, \;b}_{ac} - 2\,x_{[c} \Theta_{a]}^{\;\,b} \, $ are also separately conserved.

Let us examine the expression of the \textit{\important{energy function}}. The $\mathcal{T}^{\,\,\,4}_4(\phi^A,\ldots, \phi^A_{|_{b_1\ldots b_r}},x)$  component of the canonical energy-momentum tensor\footnote{The Belinfante-Rosenfeld tensor can also be used. In one of the examples ($p$-adic strings), we will explicitly calculate both and show that the symmetrization process does not modify the total energy.} is the energy density, so the total energy of the system is \vspace{0.2cm}
\begin{equation}\label{chap321-EF}
E(t) := \int_{\R^3}\D \mathbf{x}\left\{\sum^{r-1}_{n=0}\Pi^{c_1\ldots c_n 4}_A \dot \phi^A_{|_{c_1\ldots c_n}} - \calL\right\}\,,
\end{equation}
where $\dot \phi^A = \phi^A_{|_4}$, and it is conserved, provided that the fields decay fast enough at $|\mathbf{x}|\rightarrow\infty$.

Noether's theorem is extended for infinite-order Lagrangians by simply taking $r\rightarrow\infty$. This procedure gives rise to non-compact results (infinite series), making it challenging to handle. However, these infinite series corresponding to the energy-momentum and spin tensors can be surprisingly summed, as detailed in \cite{Marnelius1973,Heredia2020v1}. 

\subsection{The second Noether theorem: gauge symmetries}\label{chap332}
Consider now that the $r^{\rm th}$-order Lagrangian density has symmetries of the type \cite{Henneaux1992,Pons2011}
\begin{equation}\label{chap322-CS}
\delta \phi^A = R^A_a \epsilon^a + R^{Ab}_a \epsilon^a_{|_b}\,,
\end{equation}
where $\epsilon^a$ are arbitrary smooth functions of compact support corresponding to infinitesimal parameters of the symmetry, and $R^A_a$ and $R^{Ab}_a$ are functions of the fields and their derivatives. These symmetries are called \textit{\important{gauge transformation}s}. 

For the sake of simplicity, we consider $\delta x^b=0$ and assume that the $r^{\rm th}$-order Lagrangian is invariant under such a transformation. Therefore, $ W^b=0$. Because $\epsilon^a$ are compact support functions, the integral of $J^b$ on the boundary $\partial\mathcal{V}$ shall vanish because, according to (3.27), $\delta\phi^A|_{\partial\calV}
$ is 0. Consequently, equation (\ref{chap3.2-InNC}) becomes
\begin{equation}\label{chap322-InNC}
\int_{\calV} \D x\,\mathcal{E}_A\,\delta \phi^A  = \int_{\calV} \D x\left\{ \mathcal{E}_AR^A_a - \partial_b\left[\mathcal{E}_AR^{Ab}_a\right] \right\} \epsilon^a \, = 0\,.
\end{equation}
Due to the arbitrariness of $\epsilon^a$, we have what is known as the \textit{\important{second Noether theorem}}\footnote{This result can be easily generalised to a more general gauge symmetry; for instance, $\delta \phi^A = R^A_a \epsilon^a + R^{Ab}_a \epsilon^a_{|_b} +R^{Abc}_a \epsilon^a_{|_{bc}} + \ldots$.}
\begin{equation}
\mathcal{E}_AR^A_a - \partial_b\left[\mathcal{E}_AR^{Ab}_a\right] \equiv 0\,,
\end{equation}
which are relations between field equations. 

\section{Hamiltonian formalism}\label{chap34}
In the following, we shall build the \textit{\important{Hamiltonian!formalism}} for the $r^{\rm th}$-order Lagrangian fields. To do so, we shall rely on the table (\ref{chap34-TableMF})
  \begin{table}[h]
    \centering
    \includegraphics[width=0.32\textwidth]{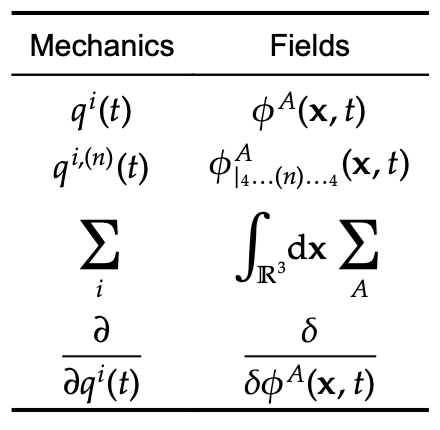}
    \caption{This table shows the changes from mechanics to fields. }
    \label{chap34-TableMF}
\end{table}
that summarises all the transformations we have to make to translate from mechanics to fields; everything described in Section \ref{chap264-HF} also applies to fields\footnote{One could define the Legendre-Ostrogradsky transformation by inferring the energy function (\ref{chap321-EF}). However, this way of proceeding and the one we are considering differ in boundary terms. }. Furthermore, note that: \textbf{(a)} the finite number $i=1,\ldots,s$ of degrees of freedom of  mechanics is substituted by continuous infinity degrees of freedom, indexed by $\mathbf{x} \in \R^3$ and the field component $A$, and \textbf{(b)} the partial derivative for $q^j$ is replaced by the functional derivative concerning $\phi^A(\mathbf{x})$.

Based on Sections  \ref{chap264-HF} and \ref{ch24-SH}, we shall define the \textit{\important{Legendre-Ostrogradsky transformation}} as
\begin{equation}
 p_{A/n}(\mathbf{x},t):=\sum^{r}_{k=n+1} \left(-\frac{\D}{\D t}\right)^{k-n-1}\left[\frac{\delta L(t)}{\delta \phi^A_{|_{4\ldots(k)\ldots4}}(\mathbf{x},t)}\right]\,, \qquad n=\{0,\ldots,r-1\}\,,
\end{equation}
where $p_{A/n}(\mathbf{x},t)$ are known as the \textit{\important{Ostrogradsky!momenta}} for fields \cite{Leon1985}. As the Lagrangian $L$ is derived from a Lagrangian density $\calL$, for $n=r-1$, we have 
\begin{equation}
p_{A/r-1}(\mathbf{x},t) = \frac{\delta L(t)}{\delta \phi^A_{|_{4\ldots(r)\ldots4}}(\mathbf{x},t)} =\left[\frac{\partial \calL}{\partial \phi^A_{|_{4\ldots(r)\ldots4}}}\right]_{(\mathbf{x},t)}\,.
\end{equation}
By assuming that the Lagrangian density is regular, we can isolate $\phi^A_{|_{4\ldots(r)\ldots4}}$ as a function of the remaining derivatives and the momentum $p_{A/r-1}$, i.e.,
\begin{equation}\label{chap33-phir1}
\phi^A_{|_{4\ldots(r)\ldots4}} = \phi^A_{|_{4\ldots(r)\ldots4}}(\phi^A,\ldots,\phi^A_{|_{b_1\ldots b_{r-1}}},\tilde \phi^A_{|_{b_1\ldots b_r}},p_{A/r-1},x)\,,
\end{equation}
where $\tilde \phi^A_{|_{b_1\ldots b_r}}$ denotes all the $r^{\rm th}$-order partial derivatives\footnote{Indeed, if we consider the set $X=\{4,3,2,1\}$, which represents each of the partial derivative, we can show that the number of elements $\tilde \phi^A_{|_{b_1\ldots b_r}}$ is $\binom{4+r-1}{r} -1$.} except $\phi^A_{|_{4\ldots(r)\ldots4}}$. Technically, the Legendre-Ostrogradsky transformation relates, in a biunivocal way (for the regular case), the extended dynamical space $\calD^\prime$ and the extended phase space $\Gamma^\prime$ by
\begin{align}  
&j:\,\,(\phi^A,\ldots,\phi^A_{|_{b_1\ldots b_{2r-1}}}, x) \quad\longrightarrow\quad (\phi^A,\ldots,\phi^A_{|_{b_1\ldots b_{r-1}}},\tilde \phi^A_{|_{b_1\ldots b_r}}, p_{A/0},\ldots, p_{A/r-1},x)\,.
\end{align}

We shall define the \textit{\important{Ostrogradsky!Hamiltonian} function} $H$ on $\Gamma^\prime$ as
\begin{align}
&H:= \int_{\R^3}\D\mathbf{x}\,\calH(\phi^A,\ldots,\phi^A_{|_{b_1\ldots b_{r-1}}},\tilde \phi^A_{|_{b_1\ldots b_r}}, p_{A/0},\ldots, p_{A/r-1},x)\,,
\end{align}
 where $\calH$ is the \textit{\important{Ostrogradsky!Hamiltonian} density}
 \begin{align}
\calH:= \sum^{r-1}_{n=0} p_{A/n} \phi^A_{|_{4\ldots(n+1)\ldots4}} - \calL(\phi^A,\ldots,\phi^A_{|_{b_1\ldots b_r}},x)\,,
 \end{align}
and the element $\phi^A_{|_{4\ldots(r)\ldots4}}$ is given by equation (\ref{chap33-phir1}).\\
\indent Likewise, given two functionals $f$ and $g$ of the fields $\phi^A$ and the conjugate momenta $p_{A}$ at time $t$, the \textit{\important{Poisson bracket}} is defined as
\begin{equation}\label{chap33-H}
\left\{f,g\right\}_{(t)} =\int_{\R^3}\D\mathbf{x}  \sum^{r-1}_{n=0}  \left(\frac{\delta f(t)}{\delta \phi^A_{|_{4\ldots(n)\ldots4}}(\mathbf{x},t)}\frac{\delta g(t)}{\delta p_{A/n}(\mathbf{x},t) } - \frac{\delta f(t)}{\delta p_{A/n}(\mathbf{x},t) } \frac{\delta g(t)}{\delta\phi^A_{|_{4\ldots(n)\ldots4}}(\mathbf{x},t)}\right)\,,
\end{equation}
with the corresponding elementary Poisson brackets
\begin{align}
&\left\{\phi^A_{|_{4\ldots(s)\ldots4}} (\mathbf{z},t),p_{B/k}(\mathbf{y},t)\right\} = \delta^A_B\, \delta_{s\,k}\,\delta(\mathbf{z}-\mathbf{y}),\\
&\left\{\phi^A_{|_{4\ldots(s)\ldots4}} (\mathbf{z},t),\phi^A_{|_{4\ldots(k)\ldots4}}(\mathbf{y},t)\right\} = 0,\\
 &\left\{p_{A/s} (\mathbf{z},t),p_{B/k}(\mathbf{y},t)\right\} = 0\,.
\end{align}
\indent Finally, the \textit{\important{Hamilton equations}} are 
\begin{align}\label{chap33-HEQ}
\mathbf{X}_H\,\phi^A_{|_{4\ldots(s)\ldots4}}(\mathbf{x},t) &= \phi^A_{|_{4\ldots(s+1)\ldots4}}(\mathbf{x},t)\,, \nonumber \\
\mathbf{X}_H\,p_{A/s}(\mathbf{x},t) &= - p_{A/s-1}(\mathbf{x},t)+ \int_{\R^3}\D\mathbf{y} \frac{\delta \calL(\phi^A,\ldots,\phi^A_{|_{b_1\ldots b_r}},\mathbf{y},t)}{\delta \phi^A_{|_{4\ldots(s)\ldots4}}(\mathbf{x},t)}\,,
\end{align}
where $\mathbf{X}_H$ is the \textit{\important{Hamiltonian!vector field}}, $\mathbf{X}_H = \partial_t + \left\{ \cdot, H\right\}$, with the Poisson bracket (\ref{chap33-H}). As seen in the previous chapter, they can be written in a more compact form through the contact 2-form 
\begin{equation}
\Omega^\prime = \Omega - \D H \wedge \D t\,,
\end{equation}
where $\Omega$ is the symplectic form defined as
\begin{equation}
\Omega(t) = \sum^{r-1}_{n=0} \int_{\R^3} \D \mathbf{x}\,\,\D p_{A/n}(\mathbf{x},t) \wedge \D \phi^A_{|_{4\ldots(n)\ldots4}}(\mathbf{x},t)\,.
\end{equation}
Therefore, the Hamilton equations (\ref{chap33-HEQ}) become
\begin{equation}
i_{\mathbf{X}_H} \Omega^\prime = 0\,.
\end{equation}
\indent To conclude this section, as studied in mechanics --Section \ref{chap264-HF}--, the time evolution generator $\mathbf{D}_4$ --equation (\ref{L5})-- in $\calD^\prime$ is connected with the Hamiltonian vector field $\mathbf{X}_H$. For the case of fields, it is also true, i.e., $j^T(\mathbf{D}_4) = \mathbf{X}_H$. Indeed, the components of the tangent vector field $j^T(\mathbf{D}_4)$ are\vspace{0.2cm}
\begin{equation}
j^T(\mathbf{D}_4)\,\phi^A_{|_{4\ldots(s)\ldots4}}(\mathbf{x},t) = \mathbf{D}_4\,\phi^A_{|_{4\ldots(s)\ldots4}}(\mathbf{x},t)  = \phi^A_{|_{4\ldots(s+1)\ldots4}}(\mathbf{x},t)\,, 
\end{equation}
and
\begin{align}
j^T(\mathbf{D}_4)\,p_{A/s}(\mathbf{x},t) &= \mathbf{D}_4\left[\sum^{r}_{k=s+1} \left(-\frac{\D}{\D t}\right)^{k-s-1}\left[\frac{\delta L(t)}{\delta \phi^A_{|_{4\ldots(k)\ldots4}}(\mathbf{x},t)}\right]\right]\nonumber\\
& = - \sum^{r}_{k=s} \left(-\frac{\D}{\D t}\right)^{k-s}\left[\frac{\delta L(t)}{\delta \phi^A_{|_{4\ldots(k)\ldots4}}(\mathbf{x},t)}\right] + \frac{\delta L(t)}{\delta \phi^A_{|_{4\ldots(s)\ldots4}}(\mathbf{x},t)}\nonumber\\
& =  - p_{A/s-1}(\mathbf{x},t)+ \int_{\R^3}\D\mathbf{y} \frac{\delta \calL(\phi^A,\ldots,\phi^A_{|_{b_1\ldots b_r}},\mathbf{y},t)}{\delta \phi^A_{|_{4\ldots(s)\ldots4}}(\mathbf{x},t)}\,.
\end{align}
Thus, using (\ref{chap33-HEQ}), we get 
\begin{equation}
j^T(\mathbf{D}_4)\,\phi^A_{|_{4\ldots(s)\ldots4}} = \mathbf{X}_H\,\phi^A_{|_{4\ldots(s)\ldots4}} \qquad \mathrm{and} \qquad j^T(\mathbf{D}_4)\,p_{A/s} = \mathbf{X}_H\,p_{A/s}.
\end{equation}


\chapter{Nonlocal Lagrangian Mechanics}\label{chap4}

The previous chapters focused on the Lagrangian and Hamiltonian formalisms for classical mechanics and fields. Both have been expected to fulfill a fundamental property: locality. The trajectory or field and its corresponding derivatives depend on a single point. However, in the following sections, we shall relax this property and consider that both exhibit \textit{\important{nonlocality}}. 

There are two types of nonlocality: the one described by operators with infinite derivatives and the one described by integrodifferential operators. As far as this manuscript is concerned, we shall only focus on the second type, i.e., by those described, for example, as 
\begin{equation}
\label{chap3-I}
I[q]_{(t)} = \int_{\R}\D \sigma \,G(\sigma) q(\sigma+t)\,,
\end{equation}
where $G(\sigma)$ is the kernel of the integrodifferential operator $I$. Note that by assuming the analyticity of the curve $q$, we can expand it into a formal Taylor series, $q(\sigma+t) = \sum^\infty_{n=0} \frac{\sigma^n}{n!} q^{(n)}(t)$, giving rise to 
\begin{equation}
I[q]_{(t)} = \sum^\infty_{n=0} c_n\, q^{(n)}(t)\,, \qquad \mathrm{where} \qquad c_n :=\int_{\R}\D\sigma \,G(\sigma) \frac{\sigma^n}{n!}\,.
\end{equation}
This fact allows us to establish a one-to-one correspondence between the ``infinite-order" Ostrogradsky formalism and the type of nonlocality we are considering. In contrast to Ostrogradsky's formalism, our way of proceeding is based on functional methods and, as it involves integrals instead of series, is much lighter to handle.

This particular chapter is based on \cite{Heredia2}; however, we shall incorporate some improvements made later in \cite{Heredia2022}. Furthermore, throughout the chapter, we shall highlight the differences between local Lagrangian and  Hamiltonian formalisms --Chapter \ref{chap2}-- and nonlocal ones.

\section{The principle of least action}\label{chap31}
Consider a dynamic system ruled by the following nonlocal action integral
\begin{equation}\label{chap31-IA}
S([q^i]) = \int_{\R}\D t\,L\left([q^i],t\right)\,.
\end{equation}
We shall say that it is nonlocal because the Lagrangian\footnote{An example of such a Lagrangian would be one containing an integrodifferential operator as (\ref{chap3-I}).} $L$ may depend on all the values $q^i(\sigma)\,,\,i=1,\ldots,s\,,$ at times $\sigma$ other than $t$. The class of all (possible) kinematic trajectories is the function space $\mathcal{K}= \mathcal{C}^\infty(\R,\R^s)$, which we shall call \textit{\important{kinematic space}}. We must resort to the \textit{\important{extended!kinematic space}}, $\mathcal{K}^\prime= \mathcal{K} \times\R$, for time-dependent nonlocal Lagrangians. 

The \textit{\important{nonlocal!Lagrangian}} is a real-valued functional defined on $\mathcal{K}^\prime$:
\begin{equation}
 ([q^i],t)\in\mathcal{K}^\prime \longrightarrow L([q^i],t) \in \R \,.
\end{equation}
It is worth stressing that $[q]$ means the functional dependence on the whole $q^i(\sigma)\,, \sigma\in\R\,$, and not only on $q^i(t)$ and a finite number of derivatives at $t$, as in the local case. Therefore, the curve $q(\sigma)$ contains all the necessary information about time evolution in $\mathcal{K}^\prime$. 

It is possible to establish a one-to-one correspondence between the ``infinite-order" Ostrogradsky formalism and the nonlocal one based on formal Taylor's series (FTS) 
\begin{equation}\label{chap31-FTS}
\left(\left\{q^{i,(r)}(t)\right\}_{r\in\mathbb{N}}, t\right) \longleftrightarrow ([q^i],t)\qquad \mathrm{with} \qquad q^i(\sigma+t) = \sum^\infty_{k=0}\frac{\sigma^k}{k!}q^{i,(k)}(t)\,,
\end{equation}
which have only a heuristic value. We label it as ``formal" because we will proceed without caring about its summability or the range of its convergence domain.

We shall define --see figure (\ref{fig:chp31-DQS})-- the \textit{\important{time!evolution operator}} $T_t$ according to 
\begin{equation}\label{chap31-Tt}
([q^i],0) \stackrel{T_t}{\longrightarrow} ([T_t q^i], t) \,,\qquad {\rm where} \qquad  T_t q^i(\sigma) = q^i(\sigma+t)\,.
\end{equation}
Notice that the additive property $T_{\tau_1}\circ T_{\tau_2} = T_{\tau_1+\tau_2}$ is satisfied. The correspondence (\ref{chap31-FTS}) is preserved by the time evolution operator $T_t$ or, in mathematical terms, the following diagram is commutative 
\begin{equation}\label{chap31-Diag}
\xymatrix{
\left(\left\{q^{i,(r)}_0\right\}_{r\in\mathbb{N}}, 0\right) \ar @{<->}[d]^{FTS} \ar @{<->}[r]^{T_t} &\left(\left\{q^{i,(r)}(t)\right\}_{r\in\mathbb{N}}, t\right) \ar @{<->}[d]^{FTS}\\
([q^i],0) \ar @{<->}[r]^{T_t}&([T_tq^i],t)\,.}
\end{equation}

\begin{figure}[t]
    \centering
    \includegraphics[width=0.65\textwidth]{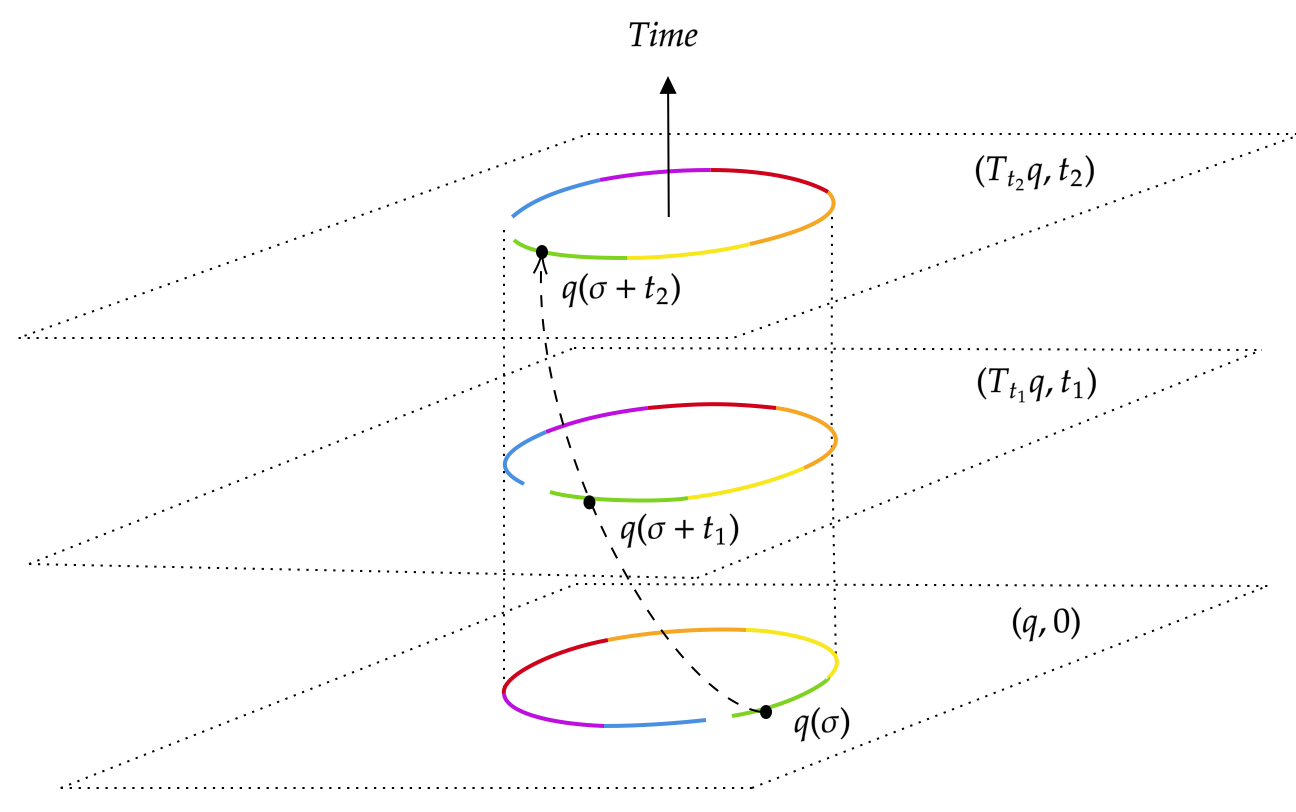}
    \caption{Time evolution of the $T_t q(\sigma)=q(\sigma+t)$ curve. It is important to recognize the distinction from the local scenario. In the local case, the initial conditions (points) are what undergo changes. However, in nonlocal theories, the entire curve evolves rather than just individual points.}
    \label{fig:chp31-DQS}
\end{figure}

The \textit{nonlocal \important{action integral}} (\ref{chap31-IA}) should be better understood as 
\begin{equation}
S(q):=\int_{\R}\D t\,L\left(T_t q,t\right)\,,
\end{equation}
where we have written $L(T_t q^i,t)$ instead of $L([T_t q^i],t)$ and $q$ instead of $q^i$ to make the notation lighter without misunderstanding risk. Because the Lagrangian $L$ depends on all the values $q(\sigma)$, we need an unbounded integration domain since the variation should be set at the interval boundaries and outside. This fact may cause it to diverge. For this reason, we propose an alternative and more consistent formulation based on introducing the one-parameter family of finite nonlocal action integrals 
\begin{equation} \label{chap31-SR}
 S(q,R) = \int_{|t|\leq R} \D t\,L(T_t q,t)  \,, \qquad \forall R\in \R^+ \,.
\end{equation}
Therefore, the \textit{\important{principle of least action}} reads 
\begin{align} \label{A2}
\lim_{R\rightarrow\infty} \delta S(q,R) &\equiv \lim_{R\rightarrow\infty}  \int_{|t|\leq R} \D t\,\int_{\R} \D \sigma\,\frac{\delta L(T_t q,t)}{\delta q(\sigma)} \,{\delta q(\sigma)} \nonumber\\
&=  \int_{\R} \D\sigma\,\delta q(\sigma) \int_\R \D t\, \frac{\delta L(T_t q,t)}{\delta q(\sigma)} = 0\,,
\end{align} 
which is well-defined as long as the limit exists $\forall\,\delta q(\sigma)$ with compact support. Furthermore, we have assumed that the variation behaves correctly so that the integrals and the limit can be permuted. Then, the \textit{\important{nonlocal!Euler-Lagrange equations}} are
\begin{equation}  \label{chap31-EOMS} 
\psi(q,\sigma) = 0 \,, \quad {\rm where} \quad \psi(q,\sigma) := \int_\R \D t\,\lambda(q,t,\sigma) 
\end{equation}
with
\begin{equation}\label{chap31-Lambda}
\lambda(q,t,\sigma) := \frac{\delta L\left(T_t q, t \right)}{\delta q(\sigma)}\,. 
\end{equation}

\newpage

\subsection{Local case embedded in nonlocal one}\label{TEST}
Let us see how a standard first-order Lagrangian $L_L(q(t),\dot{q}(t),t)$ fits in this formalism developed so far. The local action integral is $\int^{t_1}_{t_0} \D t\,L_L(q(t),\dot{q}(t),t)$, which has the form (\ref{chap31-SR}) provided that we take
\begin{equation}
L(T_t q, t) := L_L(q(t),\dot{q}(t),t) \,.
\end{equation}
Therefore, it follows that
\begin{equation}\label{chap411-LcNL}
\lambda(q,t,\sigma) = \left(\frac{\partial L_L}{\partial q}\right)_{(q,\dot q, t)}\,\delta(\sigma-t) - \left(\frac{\partial L_L}{\partial \dot{q}}\right)_{(q,\dot q, t)}\,\dot\delta(\sigma-t) \,,
\end{equation}
where $q(t) = \int_\R \D\sigma\,q(\sigma)\,\delta(\sigma-t)$ and $\dot{q}(t) = -\int_\R \D\sigma\,q(\sigma)\,\dot\delta(\sigma-t)$ were used and 
\begin{equation}
\left(\frac{\partial L_L}{\partial q}\right)_{(q,\dot q, t)} := \frac{\partial L_L(q(t),\dot{q}(t),t)}{\partial q}
\end{equation}
is understood, and so on. Substituting (\ref{chap411-LcNL}) in (\ref{chap31-EOMS}), we finally arrive at
\begin{equation}  \label{L2b} 
 \psi(q,\sigma) \equiv \frac{\partial L_L\left(q(\sigma),\dot{q}(\sigma),\sigma\right)}{\partial q} -\frac{\D\;}{\D \sigma} \left(\frac{\partial L_L\left(q(\sigma),\dot{q}(\sigma),\sigma\right)}{\partial \dot{q}}\right)\,,
\end{equation}
which are the Euler-Lagrange equations for a local (first-order) Lagrangian.

\subsection{Two ways of coordinating the extended kinematic space}\label{chap-CEK}
We might coordinate a point $z\in\mathcal{K}^\prime$ in --see figure (\ref{fig:EspaiK})-- two different ways:
\begin{equation}
\begin{array}{cccccccc}
\mathbf{(a)} \quad & \mathcal{K}^\prime & \longrightarrow & \mathcal{K} \times \R & \hspace*{4em} \mathbf{(b)} \quad & \mathcal{K}^\prime & \longrightarrow & \mathcal{K} \times \R   \\
  & z & \longmapsto & (\tilde{q},t) &  & z & \longmapsto & (q,t)
\end{array}
\end{equation}
where $q(\sigma) = T_t \tilde{q}(\sigma)\,$, namely, $q(\sigma) = \tilde{q}(t+\sigma)$.

The time evolution expression in each of these coordinate systems is
\begin{equation}
\mathbf{(a)} \quad (\tilde{q},t) \stackrel{T_\tau}{\longrightarrow} (\tilde{q}, t+\tau) \qquad  \qquad {\rm and} \qquad  \qquad \mathbf{(b)} \quad (q,t) \stackrel{T_\tau}{\longrightarrow}  (T_\tau q, t+\tau) \,.
\end{equation}
Notice that $\mathbf{(a)}$ embodies a sort of ``Heisenberg view", whereas $\mathbf{(b)}$ is more like a ``Schr\"odinger view". Hereon, we shall refer to $\mathbf{(a)}$ as \textit{\important{static coordinates}} and $\mathbf{(b)}$ as \textit{\important{moving coordinates}}.

\begin{figure}[h!] 
\begin{center}
\includegraphics[width=0.55\textwidth]{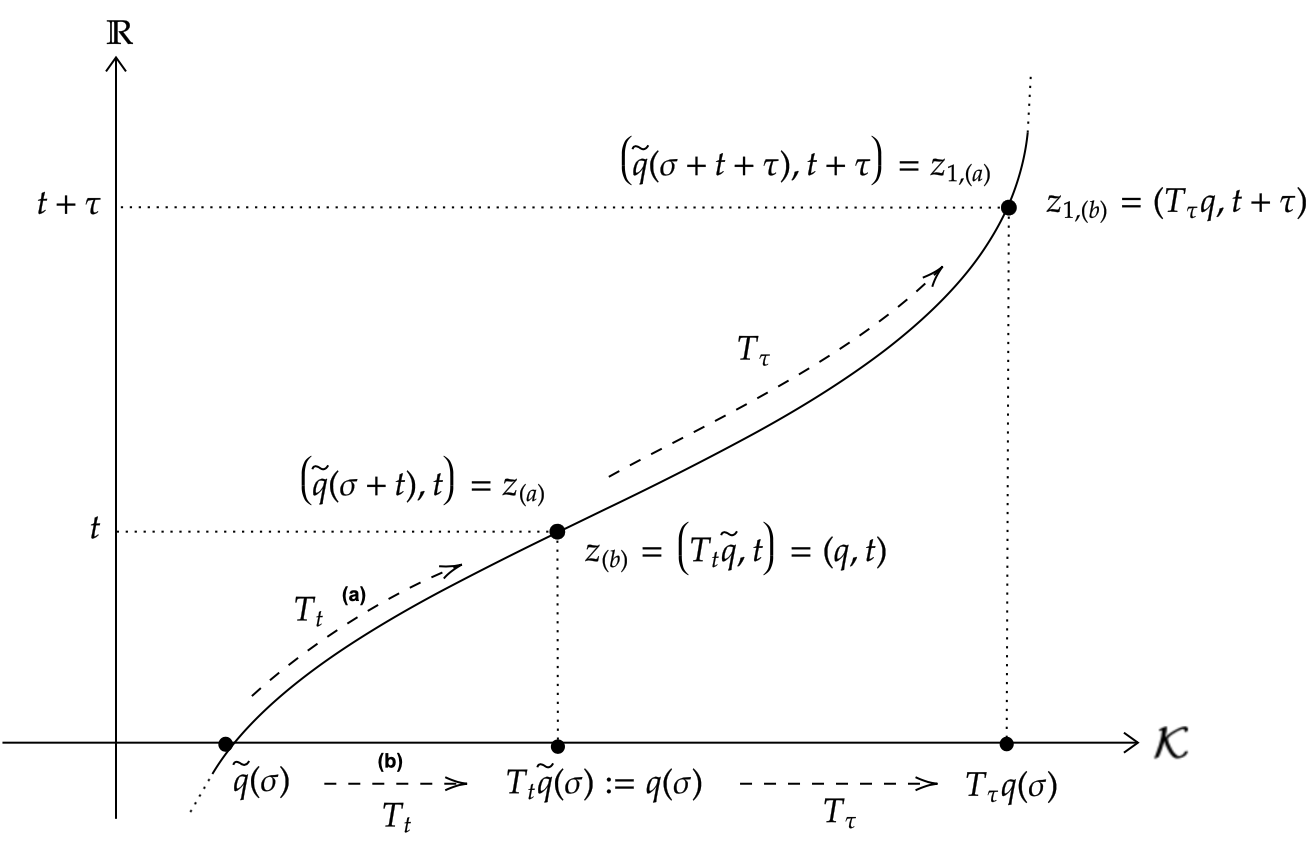}
\end{center}
\caption{This figure showcases the two distinct methods for coordinating the extended kinematic space $\mathcal{K}^\prime$. Option $\mathbf{(a)}$ employs static coordinates, whereas option $\mathbf{(b)}$ uses moving coordinates. It is worth bearing in mind that points like $z_{(a)}$ and $z_{(b)}$ correspond to the same position within the extended kinematic space, but are denoted in different coordinate systems. To transition between these coordinate systems, one must utilize $q(\sigma) = \tilde{q}(\sigma+t)$.}
\label{fig:EspaiK}
\end{figure}

A close examination of the derivation of the Euler-Lagrange equations (\ref{chap31-EOMS}) reveals that they were obtained in the context of the static coordinates; namely, they are limited to trajectories $(T_t \tilde{q}, t)\in \mathcal{K}^\prime$ starting at $(\tilde q,0)$. Therefore, they should instead read
\begin{equation}\label{chap-EOMsc}
\psi(\tilde{q},\sigma) = 0 \,, \quad {\rm with} \quad \psi(\tilde{q},\sigma) := \int_\R \D t\,\lambda(\tilde{q},t,\sigma) \quad {\rm and} \quad
\lambda(\tilde{q},t,\sigma) := \frac{\delta L\left(T_t \tilde{q}, t\right)}{\delta \tilde{q}(\sigma)}\,. 
\end{equation}
In order to have an expression more suitable for moving coordinates, we use that $(T_\tau q,t+\tau) = (T_{t+\tau} \tilde{q},t+\tau)$. Therefore, equation (\ref{chap-EOMsc}) becomes\footnote{It is worth highlighting that moving coordinates can be particularly advantageous for Lagrangians that explicitly rely on time since the corresponding trajectories are not limited to start at $t=0$.} 
\begin{equation}  \label{chap-EOMmc} 
\Psi(q,t,\sigma) = 0 \,, \qquad {\rm with} \quad  \Psi(q,t,\sigma):= \int_\R \D \tau^\prime\, \frac{\delta L\left(T_{\tau^\prime} q, t+\tau^\prime \right)}{\delta q(\sigma)} \equiv \psi(\tilde{q},t+\sigma)  \,,
\end{equation}
where $\tau = t+ \tau^\prime\,$. It is evident that
\begin{equation}
\psi(\tilde{q},\sigma) = \Psi(q,0,\sigma)  \,.
\end{equation}

The action of the time evolution generator $\mathbf{D}$ on a function $F$ on $\mathcal{K}^\prime$ has a different expression, depending on the coordinates; namely, if $F(q,t)$ and $\tilde F(\tilde q,t)$ denote the expressions of a function in each coordinate system, then
\begin{equation}
\mathbf{D} \tilde F(\tilde{q},t) = \partial_t \tilde F(\tilde{q},t) \qquad {\rm and} \qquad \mathbf{D} F(q,t) = \left[\frac{\partial F(T_\varepsilon q,t+\varepsilon)}{\partial \varepsilon} \right]_{\varepsilon=0}\,.
\end{equation}
Particularly, if $F(q,t)$ is a smooth function belonging to $\mathcal{C}^\infty(\R)\,$, and we write $F(q,t)_{(\sigma)}:=F(q,t,\sigma)$, then
\begin{equation}
\mathbf{D} \tilde F(\tilde{q},t,\sigma) = \partial_t \tilde F(\tilde{q},t,\sigma) \qquad {\rm and} \qquad \mathbf{D} F(q,t,\sigma) = \left[\frac{\partial F(T_\varepsilon q,t+\varepsilon,\sigma)}{\partial \varepsilon} \right]_{\varepsilon=0}\,.
\end{equation}
Thus, the expressions for $ \mathbf{D} $ in static and moving coordinates are
\begin{equation}  \label{L4} 
\mathbf{D} = \partial_t \qquad \quad {\rm and} \qquad \quad \mathbf{D} := \partial_t  + \int_{\R} \D\sigma\,\dot{q}(\sigma)\,\frac{\delta \quad }{\delta q(\sigma)}\,,
\end{equation}
 respectively.

\subsection{The Euler-Lagrange equations vs constraints}\label{chap31}
All kinematic trajectories $(q,t)\in\mathcal{K}^\prime$ do not fulfill equations (\ref{chap-EOMmc})  but only those that belong to the \important{extended!dynamic space} $\mathcal{D}^\prime$, i.e., the class of all dynamic trajectories. Therefore, $\calD^\prime$ defines a submanifold of $\mathcal{K}^\prime$.

As discussed in Section \ref{chap22}, in the local first-order case, the Euler-Lagrange equations turn out to be a second-order ordinary differential system that can be solved in the accelerations $\ddot{q} $ as functions of coordinates, velocities, and time\footnote{Similarly, the Euler-Lagrange equations for a regular local $r^{\mathrm{th}}$-order Lagrangian are an ordinary differential system of order $2 r$. }. The theorems of existence and uniqueness imply that, given the coordinates and velocities at an initial time $(q_0,\dot{q}_0, t_0)$, there is a unique solution
\begin{equation} \label{A20}
 q(\sigma) = \varphi(q_0,\dot{q}_0,t_0;\sigma) \quad\mbox{such that} \quad 
q_0 = \varphi(q_0,\dot{q}_0,t_0;0)\,,\quad \dot{q}_0 = \partial_\sigma\varphi(q_0,\dot{q}_0,t_0;0) \, .
\end{equation}
This fact is usually read as though the Euler-Lagrange equations govern the system's evolution. Every trajectory in the extended dynamic space $\calD^\prime$ may be labeled with those $2s+1$ coordinates and identified with the \textit{\important{initial data space}}.

The case of a nonlocal Lagrangian is not as simple \cite{Moeller2002,Barnaby2008,Calcagni2018} because equations (\ref{chap31-EOMS}) are usually integrodifferential equations and, as a rule, there are no general theorems of existence and uniqueness supporting the above interpretation in terms of evolution from an initial data set. Furthermore, the extended dynamic space $\mathcal{D}^\prime$ may have an infinite number of dimensions.

This fact leads us to propose an alternative view and take the nonlocal Euler-Lagrange equations (\ref{chap31-EOMS}) as the constraints that define $\mathcal{D}^\prime$ as a submanifold of $\mathcal{K}^\prime$ in implicit form; that is, $\psi(\tilde q,\sigma)$ acts as an infinite number of constraints, one for each $\sigma\in \R$. 

Indeed, this picture also holds for the standard local case, but the theorems of existence and uniqueness imply that the shape of the dynamic trajectories is (\ref{A20}), namely, the explicit parametric equations of the submanifold $\mathcal{D}^\prime$. Therefore, these theorems determine the number of essential parameters to individualize a dynamic solution, i.e., they provide the dynamic space coordinates with the initial positions and velocities.

We shall write the constraints in moving coordinates as    
\begin{equation}  \label{chap413-PsiEOM} 
 \Psi\left(q, t\right)_{(\sigma)} = 0 \,, \qquad \qquad \Psi\left( q, t\right)_{(\sigma)} := \Psi\left( q, t, \sigma\right)
\end{equation}
The notation is meant to suggest that $\Psi$ maps the extended kinematic space onto a space of smooth functions of the real variable $\sigma$, and the trajectories in $\mathcal{D}^\prime$ are those $( q,t)$ such that they make $\Psi$ null.

By its very construction, the constraints (\ref{chap413-PsiEOM}) are stable under time evolution, or, what is equivalent, the vector $\mathbf{D}$ is tangent to the dynamic space $\mathcal{D}^\prime$. Indeed, from (\ref{chap-EOMmc}), we have that 
\begin{align}
\Psi\left(T_\varepsilon  q, t+\varepsilon,\sigma \right) &= \int_\R \D\tau\,\frac{\delta  L\left(T_{\tau+\varepsilon}  q, t+\varepsilon+\tau \right)}{\delta T_\varepsilon  q(\sigma)} \nonumber\\
&= \int_\R \D\tau^\prime\,\frac{\delta  L\left(T_{\tau^\prime}  q, t+\tau^\prime \right)}{\delta  q(\sigma+\varepsilon)} = \Psi\left( q, t,\sigma+\varepsilon \right) \,,
\end{align}
where  $\tau^\prime:=\tau+\varepsilon\,$. Thus, if $\Psi( q,t,\sigma)=0\,\,$$\,\,\forall\sigma$, then $\Psi\left(T_\varepsilon  q, t+\varepsilon,\sigma \right) =0$ as well; therefore,
\begin{equation}
 \mathbf{D} \Psi( q,t,\sigma) = \left[\frac{\partial \Psi\left(T_\varepsilon  q, t+\varepsilon,\sigma  \right)}{\partial \varepsilon} \right]_{\varepsilon=0} = 0\,. 
 \end{equation}
 Notice that, in static coordinates, the constraints $\psi(\tilde{q},\sigma)$ satisfy trivially 
 \begin{equation}
 \mathbf{D}\psi(\tilde{q},\sigma) = \partial_t\psi(\tilde{q},\sigma) = 0\,.
 \end{equation}  
 
\subsection{Nonlocal total derivative}\label{chap32}
In the previous chapters, we have seen that the \textit{\important{Noether!symmetry}} plays an important role in Noether's theorem. Therefore, let us see what happens when we add a \textit{nonlocal \important{total derivative}} to a nonlocal Lagrangian. Recall --Section \ref{chap264-HF}-- that, for the case of infinite derivatives, we observed that the series is not summable and could lead to changes in the equations of motion. 

First, let us show that, for any nonlocal Lagrangian, one can find a functional $W(T_t \tilde{q}, t)$ such as
\begin{equation}\label{chap414-LDW}
L(T_t \tilde{q}, t) = \frac{\D W(T_t \tilde{q}, t)}{\D t}\,.
\end{equation}
Indeed, a particular solution of this equation is 
\begin{equation}\label{chap32-W}
W(T_t \tilde{q}, t) = \int_{\R}\D\sigma\left[\theta(\sigma)- \theta(\sigma-t)\right]L(T_\sigma\tilde{q},\sigma)\,,
\end{equation}
where $\theta(x)$ denotes the Heaviside step function. Notice that it is nonlocal and always exits. 

Now, consider the following nonlocal quadratic Lagrangian
\begin{equation}\label{chap32-L}
L(T_t \tilde{q}, t) = \dot{\tilde{q}}(t)\left(G\ast \tilde{q}\right)_{(t)}\,,
\end{equation}
where $G(x)$ is a given function --more precisely, the kernel of the integral operator-- that vanishes for $|\tau|\rightarrow\infty$, and $(G\ast-)_{(t)}$ denotes the convolutional operator\footnote{Note that this nonlocal Lagrangian is consistent with our notation: $L(T_t \tilde{q}, t) = L(T_t \tilde{q}) = \dot{\tilde{q}}(t) \int_{\R}\D \sigma\,G(\sigma)\,\tilde{q}(t-\sigma) = T_t \dot{\tilde{q}}(0) \int_{\R} \D\sigma\,G(\sigma)\,T_t \tilde{q}(-\sigma)\,.$}. The nonlocal action integral is 
\begin{equation}
S(\tilde{q},R)= \int_{|t|\leq R} \D t \int_\R\D\tau\, \dot{\tilde{q}}(t)\, \tilde{q}(\tau)\,G(t-\tau)\,,
\end{equation}
whose nonlocal Euler-Lagrange equations (\ref{chap31-EOMS}) are 
\begin{equation}\label{chap32-G+}
\psi(\tilde q,\sigma) \equiv (\dot G_{|_+}\ast \tilde{q})_{(\sigma)} =0\,, 
\end{equation}
with $\dot{G}(x)_{|_+}:= \dot{G}(x)+\dot{G}(-x)$.  Because of (\ref{chap32-G+}), only the odd part of the kernel matters as
\begin{equation}
\dot G(t-\tau) = \dot G(\tau-t) \qquad \mathrm{implies} \qquad G(t-\tau) = - G(\tau-t)\,.
\end{equation}

Next, combining (\ref{chap32-L}) and (\ref{chap32-W}), we find that
\begin{equation}
W(T_t \tilde{q}, t) = \tilde{q}(t)\left(G\ast \tilde{q}\right)_{(t)} - \int_{\R^2}\D\rho\,\D\sigma\left[\theta(\rho)-\theta(\rho-t)\right] \dot G(\rho-\sigma)\,\tilde{q}(\rho)\,\tilde{q}(\sigma)\,.
\end{equation}
To check whether the nonlocal Euler-Lagrange equations for the nonlocal Lagrangian (\ref{chap32-L}) vanish identically, we first derive
\begin{equation}
\frac{\delta L(T_t \tilde{q},t)}{\delta \tilde{q}(\sigma)} = \dot \delta(t-\sigma)\left(G\ast \tilde{q}\right)_{(t)} + \dot{\tilde{q}}(t) G(t-\sigma)\,,
\end{equation}
where we have taken into account equation (\ref{chap414-LDW}). Then,
\begin{equation}\label{chap32-EL}
\psi(\tilde{q},\sigma) = - (\dot G\ast \tilde{q})_{(\sigma)} + \int_{\R} \D t\,\dot{\tilde{q}}(t)\,G(t-\sigma) = \int_{\R}\D t\,\dot{\tilde{q}}(t)\left[G(t-\sigma)-G(\sigma-t)\right]\,
\end{equation}
that only vanishes identically if $G(x)$ is even, which is not the case. Thus, unless the nonlocal action vanishes identically, the nonlocal Euler-Lagrange equations (\ref{chap32-EL}) are not identically zero.

Let us find a sufficient condition so that the nonlocal Euler-Lagrange equations vanish identically. As stated above, the nonlocal Lagrangian can always be written in the form (\ref{chap32-L}), and the nonlocal action integral and its variation, respectively, become
\begin{equation}
S(\tilde{q},R) = \int_{|t|\leq R}\D t \frac{\D W(T_t \tilde{q},t)}{\D t} = W(T_R \tilde{q}, R) - W(T_{-R}\tilde{q},-R),
\end{equation}
and
\begin{equation}
\delta S(\tilde{q},R) = \delta W(T_R \tilde{q}, R) - \delta W(T_{-R}\tilde{q},-R)\,
\end{equation}
so that the nonlocal Euler-Lagrange equations are
\begin{equation}
\psi(\tilde{q},\sigma) := \lim_{R\rightarrow\infty}\left[\frac{\delta W(T_R \tilde{q}, R)}{\delta \tilde{q}(\sigma)} - \frac{\delta W(T_{-R} \tilde{q},-R)}{\delta \tilde{q}(\sigma)}\right] \equiv 0\,.
\end{equation}
Therefore,
\begin{equation}\label{chap32-ACW}
\lim_{R\rightarrow\pm\infty}\frac{\delta W(T_R \tilde{q}, R)}{\delta \tilde{q}(\sigma)}=0
\end{equation}
is a sufficient condition for the nonlocal Euler-Lagrange equations to vanish identically. This condition is met in the standard case of local $W(T_t \tilde{q},  t)$, as discussed in Section \ref{chap22}.

\section{The Noether theorem}\label{chap33}
The proof of the \textit{\important{Noether!theorem}} for a local Lagrangian, which might contain derivatives up to $r^{\rm th}$  order, involves some integrations by parts to remove the derivatives of $\delta q(t)$ of orders higher than $r-1$. As a consequence, the variation of the action contains some boundary terms that, in the end, give rise to a conserved quantity. In a nonlocal Lagrangian, there is no such highest-order derivative, and the latter scheme does not make sense. Therefore, we shall use a trick that will bring out the equivalent of these boundary terms without resorting to the integration by parts.\\
\indent Consider the infinitesimal transformations
\begin{equation}  \label{chap33-IT}
t^\prime(t) = t + \delta t(t) \,, \qquad \mathrm{and} \qquad \quad \tilde{q}^{\prime}(t) = \tilde{q}(t) + \delta \tilde{q}(t)\,. 
\end{equation}
The nonlocal Lagrangian shall transform so that it leaves the nonlocal action integral (\ref{chap31-SR}) invariant, that is, 
 \begin{equation}\label{chap33-RL}
L^\prime(T_{t^\prime} \tilde{q}^\prime,t^\prime) =\left|\frac{\D t}{\D t^\prime}\right| L(T_t \tilde{q}, t) \approx (1- \delta \dot t)L(T_t \tilde{q}, t)\,.
 \end{equation}
Thus, if $[t_0,t_1]$ is a time interval and $[t^\prime_0,t^\prime_1]$ is the transformed one according to (\ref{chap33-IT}), we get that
 \begin{equation}
\int^{t^\prime_1}_{t^\prime_0}\D t^\prime\,L^\prime(T_{t^\prime} \tilde{q}^\prime,t^\prime) =  \int^{t_1}_{t_0}\D t\,L(T_t \tilde{q}, t)\,.
 \end{equation}
As $S^\prime(\tilde{q}^\prime, t^\prime_0, t^\prime_1) = S(\tilde{q},t_0,t_1)$, it follows
 \begin{equation}\label{chap33-DS}
 \int^{t^\prime_1}_{t^\prime_0}\D t\,L^\prime(T_{t} \tilde{q}^\prime,t) - \int^{t_1}_{t_0}\D t\,L(T_t \tilde{q}, t) = 0\,,
 \end{equation}
 where we have replaced the dummy variable $t^\prime$ with $t$ in the first integral. Given the infinitesimal transformations (\ref{chap33-IT}) and taking into account equation (\ref{chap33-RL}), the first integral can be approximated to the leading order as
 \begin{equation}
  \int^{t^\prime_1}_{t^\prime_0}\D t\,L^\prime(T_{t} \tilde{q}^\prime,t) = \int^{t_1}_{t_0}\D t\left\{L^\prime(T_{t} \tilde{q}^\prime,t) + \frac{\D}{\D t}\left[L(T_t \tilde{q}, t)\delta t \right]\right\}\,.
 \end{equation}
 Hence, equation (\ref{chap33-DS}) becomes 
\begin{equation}\label{chap33-ST2}
\int^{t_1}_{t_0}\D t\left\{L^\prime(T_{t} \tilde{q}^\prime,t) - L(T_t \tilde{q},t) + \frac{\D}{\D t}\left[L(T_t \tilde{q}, t) \delta t \right]\right\} = 0\,.
 \end{equation}
As in the local case, we say that transformation (\ref{chap33-IT}) is a Noether symmetry if
\begin{equation} \label{chapNTNL-Lp}
L^\prime(T_t \tilde{q}^\prime, t)= L(T_t \tilde{q}^\prime, t) + \frac{\D}{\D t}W(T_t \tilde{q}^\prime, t)\,,
\end{equation}
where the functional $W$ satisfies the asymptotic condition (\ref{chap32-ACW}). This way, we can ensure that the nonlocal Euler-Lagrange equations remain invariant under such a transformation. Consequently, using the infinitesimal transformations (\ref{chap33-IT}) and equation (\ref{chapNTNL-Lp}), we find that equation (\ref{chap33-ST2}) up to the leading order is
\begin{equation}
\int^{t_1}_{t_0}\D t\left\{ \int_\R \D\sigma \lambda(\tilde{q},t,\sigma)\,\delta \tilde{q}(\sigma) + \frac{\D}{\D t}\left[L(T_t \tilde{q}, t) \delta t + W(T_t \tilde{q}, t)\right]\right\} = 0\,,
\end{equation}
 which, adding and subtracting (\ref{chap31-EOMS}), becomes\vspace{0.2cm}
 \begin{align}
&\int^{t_1}_{t_0}\D t\left\{ \psi(\tilde{q},t) \delta \tilde{q}(t) + \frac{\D}{\D t}\left[L(T_t \tilde{q}, t) \delta t + W(T_t \tilde{q}, t)\right]\right.\nonumber\\
&\left.\qquad \qquad \qquad \qquad+  \int_\R \D\sigma \lambda(\tilde{q},t,\sigma) \delta \tilde{q}(\sigma) - \int_\R \D\sigma \lambda(\tilde{q},\sigma,t)\delta \tilde{q}(t) \right\} = 0\,.
\end{align}
 Now, the trick. After suitable changes of the variable $\sigma$ in the last two integrals, we arrive at
 \begin{align}\label{chap31-DSL}
&\int^{t_1}_{t_0}\D t\left\{ \psi(\tilde{q},t) \delta \tilde{q}(t) + \frac{\D}{\D t}\left[L(T_t \tilde{q}, t) \delta t + W(T_t \tilde{q}, t)\right]\right.\nonumber\\
&\left.\qquad \qquad \qquad \qquad+  \int_\R \D\xi\,\left[\lambda(\tilde{q},t,t+\xi)\,\delta \tilde{q}(t+\xi) - \lambda(\tilde{q},t-\xi,t)\,\delta \tilde{q}(t) \right] \right\} = 0\,.
\end{align}
We can write the integrand in the last term on the right-hand side as
\begin{eqnarray}
\lefteqn{\lambda(\tilde{q},t,t+\xi)\,\delta \tilde{q}(t+\xi) -\lambda(\tilde{q},t-\xi,t)\,\delta \tilde{q}(t) =} \nonumber \\[1ex]
 \qquad &  \qquad \quad= &  \int_0^1\D\eta \,\frac{\partial}{\partial\eta}\left[\lambda(\tilde{q},t+(\eta-1)\xi,t+\eta\xi)\,\delta \tilde{q}(t+\eta\xi) \right] \nonumber \\ \label{A4a}
 &  \qquad \quad = & \xi\, \int_0^1\,\D\eta\,\frac{\partial}{\partial t} \left[\lambda(\tilde{q},t+(\eta-1)\xi,t+\eta\xi)\,\delta \tilde{q}(t+\eta\xi)\right]\,.
\end{eqnarray}
Provided that the nonlocal Lagrangian is a well-behaved functional since $q(t)$ is a smooth function, the theorems of differentiation under the integral sign are applied \cite{Apostol}. Thus, the integral and the partial derivative commute. Substituting the latter expression into (\ref{chap31-DSL}), we obtain
\begin{equation}
\int^{t_1}_{t_0}\D t\left\{ \psi(\tilde{q},t) \delta \tilde{q}(t) + \frac{\D}{\D t}\left[L(T_t \tilde{q}, t) \delta t + W(T_t \tilde{q}, t) + U(T_t\tilde{q},t)\right]\right\}=0\,,
\end{equation}
with
\begin{equation}   \label{A6}
U(T_t\tilde{q},t) := \int_\R\D\xi\,\xi\,\int_0^1\,\D\eta \,\lambda(\tilde{q},t+(\eta-1)\xi,t+\eta\xi)\,\delta \tilde{q}(t+\eta\xi) \,.
\end{equation}
By replacing now $\eta\xi=\rho$,  the latter can be written as
\begin{equation}
U(T_t\tilde{q},t) = \int_\R \D\xi\,\int_0^{\xi}\lambda(\tilde{q},t+\rho-\xi,t+\rho)\,\delta \tilde{q}(t+\rho)\,\D\rho\,, 
\end{equation}
which, after inverting the order of the integrals, leads to 
\begin{equation}
 U(T_t\tilde{q},t) = \int_\R \D\rho\,\delta \tilde{q}(t+\rho) \int_\R\D\xi\,\left[\theta(\rho)\,\theta(\xi-\rho)- \theta(-\rho)\,\theta(\rho-\xi)\right]\,\lambda(\tilde{q},t+\rho-\xi,t+\rho)\,,
 \end{equation}
or
\begin{equation}
U(T_t\tilde{q},t) = \int_\R \D\rho\,\delta \tilde{q}(t+\rho)\, \int_\R\D\xi\,\left[\theta(\xi-\rho)-\theta(-\rho)\right]\,\lambda(\tilde{q},t+\rho-\xi,t+\rho)\,,
\end{equation}
where $\theta(x)$ is the Heaviside step function. Replacing $\zeta:=\rho-\xi$, it becomes
\begin{equation}   \label{A7}
U(T_t\tilde{q},t) = \int_\R \D\rho\,\delta \tilde{q}(t+\rho)\, P(\tilde{q},t,\rho) \,,
\end{equation}
with 
\begin{equation}\label{chap42-P}
 P(\tilde{q},t,\rho):= \int_\R\D\zeta\,\left[\theta(\rho)- \theta(\zeta)\right]\,\frac{\delta L(T_{t+\zeta} \tilde q, t+\zeta)}{\delta \tilde q(t+\rho)}  \,.
\end{equation}
Notice that the resemblance of $U(T_t\tilde{q},t)$ with the ``boundary terms'' one encounters in Noether's theorem for local Lagrangians is obvious.

Finally, since the choice of the interval $[t_0,t_1]$ has been completely arbitrary, we have that the integrant must be identically zero, giving rise to \vspace{0.2cm}
\begin{equation}  \label{A9}
 N(\tilde{q},t):=\psi(\tilde{q},t)\,\delta \tilde{q}(t) +\frac{\D}{\D t} \left[L\left(T_{t}\tilde{q},t\right)\delta t(t)+ W(T_t \tilde{q},t) + U(T_t\tilde{q},t) \right] \equiv 0\,,
\end{equation}
which is an extension of the identity (\ref{ch23-IN}) and Noether's constant of motion (\ref{ch253-NCO}) to nonlocal Lagrangians that may explicitly depend on $t$. For dynamic trajectories, i.e., solutions of the nonlocal Euler-Lagrange equations, this identity implies that
\begin{equation}\label{chap42-DDTJ}
 \frac{\D \;}{\D t} J(T_t\tilde{q},t) = 0 \,,
\end{equation}
namely, $J(T_t\tilde{q},t)$ is a locally conserved quantity, where 
\begin{equation}\label{chap332-J}
J(T_t\tilde{q},t) := L\left(T_{t}\tilde{q},t\right)\delta t(t)+ W(T_t \tilde{q},t) + U(T_t\tilde{q},t)\,.
\end{equation}

These last results can be expressed in moving coordinates making the change $q=T_t\tilde{q}$, that is,
\begin{equation}\label{chap42-DDTJMC}
 \frac{\D \;}{\D t} J(q,t) = 0\,, \qquad \mathrm{with} \qquad J(q,t) = L\left(q,t\right)\delta t(t)+ W(q,t) + U(q,t),
\end{equation}
and $U(q,t) = \int_\R \D\rho\,\delta q(\rho)\,P(q,t,\rho)$ with 
\begin{equation}\label{chap42-PMC}
P(q,t,\rho) = \int_\R\D\zeta\,\left[\theta(\rho)- \theta(\zeta)\right]\,\frac{\delta L(T_\zeta q, t+\zeta)}{\delta q(\rho)}  \,.
\end{equation}

\subsection{The energy function}\label{chap331}
If the nonlocal Lagrangian does not explicitly depend on $t$, it is invariant under time translations; therefore, $W(T_t\tilde{q})=0$, and the above results can be applied to the infinitesimal transformations (\ref{ch231-tt}) and (\ref{ch231-tq}). Hence, the \textit{\important{energy function}} --in moving coordinates-- becomes
\begin{equation} \label{A12z}
E(q):=-\epsilon^{-1}J(q) = - L\left(q\right) + \int_\R \D\rho\,\dot q(\rho)\, P(q,\rho) 
\end{equation}
with 
\begin{equation}
P(q,\rho) := \int_\R\D\zeta\,\left[\theta(\rho)- \theta(\zeta)\right]\,\frac{\delta L(T_\zeta q)}{\delta q(\rho)}
\end{equation}
that is preserved due to (\ref{chap42-DDTJMC}).

Let us now particularize the latter for a first-order Lagrangian $L_L(q,\dot{q})$. Substituting (\ref{chap411-LcNL}) into definition (\ref{chap42-P}), we have that\vspace{0.5cm}
\begin{align}\label{chap331-PFL}
  P(q,\rho) &= \int_\R \D\zeta \left[\theta(\rho)-\theta(\zeta)\right]\left[\left(\frac{\partial L_L}{\partial q}\right)_{(q,\dot q,\zeta)} \delta (\zeta-\rho)+\left(\frac{\partial L_L}{\partial \dot q}\right)_{(q,\dot q,\zeta)} \dot \delta(\zeta-\rho)\right]\nonumber\\
  & = \delta (\rho) \frac{\partial L_L(q_0,\dot q_0)}{\partial \dot q_0}=\delta(\rho)\frac{\partial L_L(\tilde{q},\dot{\tilde{q}})}{\partial \dot{\tilde{q}}}\,,
\end{align}
where we have included that $q_0 := q(0)=\tilde{q}(t)$ and $\dot q_0 := \dot q(0)=\dot{\tilde{q}}(t)$. The functional dependence of $P(q,\rho)$ on $q(\sigma)$ is reduced to the initial coordinate and velocity, which are the parameters of $q(\sigma)$ in $\calD$:
\begin{equation}
P(q,\rho) = p(q_0,\dot q_0)\,\delta (\rho) \qquad \mathrm{with} \qquad p(q_0,\dot q_0):= \frac{\partial L_L(q_0,\dot q_0)}{\partial \dot q_0}\,.
\end{equation}
Finally, substituting (\ref{chap331-PFL}) into (\ref{chap332-J}) yields
\begin{equation}  \label{A11b}
J(q,t) :=  L_L(q_0,\dot{q}_0)\,\delta t(t) + \frac{\partial L_L(q_0,\dot q_0)}{\partial \dot q_0}\delta q_0 \,,
\end{equation}
where we have used that $L(q) = L_L(q_0,\dot q_0)$. The functional $J(q,t)$ is a constant of motion (once the nonlocal Euler-Lagrange equations are applied) and only depends on the trajectory $q(\sigma)$ through its initial values $(q_0,\dot q_0)$. Now, plugging the infinitesimal transformations (\ref{ch231-tt}) and (\ref{ch231-tq}) into (\ref{A11b}), we reach the desired energy function for a first-order Lagrangian
\begin{equation}
E(q_0,\dot q_0) := \frac{\partial L_L(q_0,\dot q_0)}{\partial \dot q_0}\dot q_0  -  L_L(q_0,\dot{q}_0)\,.
\end{equation}

\section{Hamiltonian formalism}\label{chap34}
Our next aim is to set up a \textit{\important{nonlocal!Hamiltonian formalism}} for the nonlocal Euler-Lagrange equations (\ref{chap31-EOMS}). The standard procedure for  first-order local Lagrangians consists of introducing the canonical momenta and, by inverting the Legendre transformation, replacing one-half of the variables, namely the velocities, with the momenta as coordinates in $\mathcal{D}^\prime$. With these new coordinates, the extended dynamic space becomes the exended phase space.

This procedure is not feasible in the nonlocal case because: (a) we still have no coordinates for $\mathcal{D}^\prime$ --it might depend on the integrodifferential equations--, and (b) the fact that $\mathcal{D}^\prime$ likely has an infinite number of dimensions. However, it is worth noticing that, in the local first-order case, the canonical momentum for the Legendre transformation is the prefactor of $\delta q(t)$ in the conserved quantity (\ref{ch231-JE}).  Therefore, we shall use this fact to make an ``educated guess'' and, thus, define the Legendre transformation for nonlocal Lagrangians.

\subsection{The Legendre transformation}\label{chap431-LT}

We first introduce the \textit{\important{nonlocal!Hamiltonian}} on the \textit{\important{extended!phase space}} $\,\Gamma^\prime =\mathcal{K}^2\times \mathbb{R}\,$ made of points $(q, \pi,t)$ in moving coordinates, where $q, \,\pi \in\mathcal{K}\,$ are smooth functions, as follows
\begin{equation}  \label{L6} 
H(q,\pi,t) = \int_\R \D\sigma\,\pi(\sigma)\,\dot{q}(\sigma) -  L(q,t)\,,
\end{equation}
which is equipped with the following \textit{\important{Poisson bracket}}
\begin{equation}
 \left\{ F, G \right\} = \int_\R \D\sigma\,\left(\frac{\delta F}{\delta q(\sigma)}\,\frac{\delta G}{\delta \pi(\sigma)} - \frac{\delta F}{\delta \pi(\sigma)}\,\frac{\delta G}{\delta q(\sigma)} \right) \,.
 \end{equation}
The \textit{\important{Hamilton equations}} are
\begin{eqnarray}  \label{L7a}
\mathbf{X}_H {q}(\sigma) & = & \frac{\delta H}{\delta \pi(\sigma)} = \dot{q}(\sigma)  \\[1ex]  \label{L7b}
\mathbf{X}_H {\pi}(\sigma) & = & -\frac{\delta H}{\delta q(\sigma)} = \dot{\pi}(\sigma) + \frac{\delta L(q,t)}{\delta q(\sigma)} \,,
\end{eqnarray}
where $\mathbf{X}_H$ is the \textit{\important{Hamiltonian!vector field}}
\begin{equation}
 \mathbf{X}_H = \partial_t + \int_\R \D\sigma\,\left( \dot{q}(\sigma)  \,\frac{\delta \quad}{\delta q(\sigma)} + \left[\dot{\pi}(\sigma) + \frac{\delta L(q,t)}{\delta q(\sigma)} \right] \,\frac{\delta \quad}{\delta \pi(\sigma)} \right) \,.
 \end{equation}
 
As seen previously in Section \ref{ch241-SHF},  Hamilton's equations can be written in a more compact form using the contact differential 2-form 
\begin{equation}  \label{L7c} 
 \Omega^\prime = \Omega - \delta H \wedge \delta t\,, \qquad {\rm where} \qquad \Omega = \int_\R \D\sigma\;\delta \pi(\sigma) \wedge \delta q(\sigma) \,,
\end{equation}
is the symplectic form (and we have written``$\delta$'' to distinguish between the differential on the manifold $\Gamma^\prime$ and the ``$\D$'' occurring in the notation for integrals that we have adopted here). The Hamilton equations (\ref{L7a}-\ref{L7b}) are then equivalent to
\begin{equation}  \label{L7d} 
i_{\mathbf{X}_H} \Omega^\prime = 0  \,.
\end{equation}
\indent So far, this Hamiltonian system in the extended phase space $\Gamma^\prime$ has almost nothing to do with the nonlocal Euler-Lagrange equations (\ref{chap31-EOMS}) nor the time evolution generator  $\mathbf{D}$ in the space $\mathcal{D}^\prime\,$. However, we can connect both through the injection 
\begin{equation}  \label{L8} 
 (q, t)\in \mathcal{D}^\prime \stackrel{j}{\lhook\joinrel\longrightarrow} (q,\pi,t)\in\Gamma^\prime \,, \qquad  \mbox{where } \qquad \pi(\sigma) := P(q,t,\sigma)   \,,
\end{equation}
and $P(q,t,\sigma)$ is the prefactor of $\delta q(\sigma)$ in the Noether conserved quantity (\ref{chap42-PMC}), that is,
\begin{equation}  \label{L8a} 
P(q,t,\sigma) = \int_\R\D\zeta\,\left[\theta(\sigma)- \theta(\zeta)\right]\,\frac{\delta L(T_\zeta q, t+\zeta)}{\delta q(\sigma)}\,.
\end{equation}
\indent $j$ defines a one-to-one map from $\mathcal{D}^\prime$ into its range, $j(\mathcal{D}^\prime)\subset \Gamma^\prime$, i.e., the submanifold implicitly defined by the constraints
\begin{equation}  \label{L9} 
\Psi\left(q,t,\sigma \right)= 0  \qquad {\rm and} \qquad 
\Upsilon\left(q, \pi, t,\sigma\right):= \pi(\sigma) - P(q,t,\sigma) = 0  \,\qquad \forall\sigma\in \R\,,
\end{equation}
and the Jacobian map $\,j^T\,$ maps the  infinitesimal time evolution generator $\mathbf{D}$ in $\mathcal{D}^\prime$ into $\mathbf{X}_H$, the Hamiltonian flow generator in $\,\Gamma^\prime\,$. Indeed, we have that
\begin{equation}
j^T\left(\mathbf{D}\right) q(\sigma) = \mathbf{D}\,q(\sigma) =\left[\frac{\partial T_\varepsilon q(\sigma)}{\partial \varepsilon}\right]_{\varepsilon=0} = \dot q(\sigma) = \mathbf{X}_H\,q(\sigma)\,.
\end{equation}
Furthermore,
\begin{align}
 j^T\left(\mathbf{D}\right) \pi(\sigma) &= \mathbf{D} P(q,t,\sigma) = \left[\frac{\partial P(T_\varepsilon q,t+\varepsilon,\sigma)}{\partial \varepsilon}\right]_{\varepsilon=0}\nonumber\\
 & = \frac{\partial}{\partial \varepsilon}\left[\int_\R \D\zeta \left[\theta(\sigma)-\theta(\zeta)\right] \frac{\delta L(T_{\varepsilon+\zeta} q,t+\varepsilon+\zeta)}{\delta q(\sigma+\varepsilon)}\right]_{\varepsilon=0}\,.
 \end{align}
Substituting $\zeta^\prime=\zeta+\varepsilon$, we obtain
 \begin{align}
\mathbf{D} P(q,t,\sigma) & = \frac{\partial}{\partial\varepsilon}\left[\int_\R \D\zeta^\prime \left[\theta(\sigma)-\theta(\zeta^\prime-\varepsilon)\right]\frac{\delta L(T_{\zeta^\prime} q,t+\zeta^\prime)}{\delta q(\sigma+\varepsilon)}\right]_{\varepsilon=0}\nonumber\\
 & = \frac{\delta L(q,t)}{\delta q(\sigma)} + \int_{\R}\D\zeta^\prime \left[\theta(\sigma)-\theta(\zeta^\prime)\right] \partial_\sigma\left[\frac{\delta L(T_{\zeta^\prime}q,t+\zeta^\prime)}{\delta q(\sigma)}\right]\nonumber\\
 & =  \frac{\delta L(q,t)}{\delta q(\sigma)} +\partial_\sigma\left[ \int_{\R}\D\zeta^\prime \left[\theta(\sigma)-\theta(\zeta^\prime)\right] \frac{\delta L(T_{\zeta^\prime}q,t+\zeta^\prime)}{\delta q(\sigma)}\right] - \delta(\sigma) \Psi(q,t,\sigma)\,,
 \end{align}
where we have used that $\dot\theta(\zeta)=\delta(\zeta)$ and equation (\ref{chap-EOMmc}). Therefore, we find  that
 \begin{equation}
 \mathbf{D} P(q,t,\sigma) = \frac{\delta L(q,t)}{\delta q(\sigma)} + \partial_\sigma P(q,t,\sigma) - \,\delta(\sigma) \,\Psi(q,t,\sigma) \,,
 \end{equation}
where equation (\ref{chap42-PMC}) was used. As the point $(q,t)\in\mathcal{D}^\prime$, the last term on the right vanishes by the constraints (\ref{L9}), and we finally obtain
\begin{equation}
 j^T\left(\mathbf{D}\right) \pi(\sigma) = \frac{\delta L(q,t)}{\delta q(\sigma)} + \partial_\sigma P(q,t,\sigma) = \mathbf{X}_{H} \pi(\sigma)\,. 
 \end{equation}
 
As a corollary, $\mathbf{X}_H= j^T\left(\mathbf{D}\right)$ is tangent to the submanifold $j(\mathcal{D}^\prime)$, and therefore, the constraints (\ref{L9}) are stable by the Hamiltonian flow.

 To translate the Hamiltonian formalism in $\Gamma^\prime$ into a Hamiltonian formalism in the extended dynamic space $\mathcal{D}^\prime$, we use the fact that the pullback $j^\ast$ maps the contact form (\ref{L7c}) onto the differential 2-form 
\begin{equation} \label{L10}
\omega^\prime  = j^\ast \Omega^\prime = \int_\R \D\sigma \,\delta P(q,t,\sigma)\wedge \delta q(\sigma) - \delta h \wedge \delta t \,, \qquad \quad \omega^\prime\in\Lambda^2(\mathcal{D}^\prime)\,,
\end{equation} 
where $\,h = j^* H = H\circ j\,$. As $\,j^T\left(\mathbf{D}\right) = \mathbf{X}_H\,$, the pullback of equation (\ref{L7d}) then reads\footnote{This result has been obtained by using the definition of the pullback of a $k$-form and the property $j^*\mathrm{d} H = \mathrm{d}(j^*H)$ \cite{Choquet1982}. } 
\begin{equation} \label{L11}
i_{\mathbf{D}} \omega^\prime = 0 \,.
\end{equation}
\indent The reduced Hamiltonian $h(q,t)$ and the contact form $\omega^\prime$ on $\mathcal{D}^\prime\,$ can be derived from equations (\ref{L6}), (\ref{L7c}), and (\ref{L8a}), and they are respectively
\begin{equation}  \label{L12} 
h(q,t) = \int_{\R^2} \D\zeta\,\D\sigma\,\left[\theta(\sigma)-\theta(\zeta)\right]\,\dot q(\sigma)\,\frac{\delta L(T_\zeta q, t+\zeta)}{\delta q(\sigma)} - L(q,t)\,, 
\end{equation}
and $\omega^\prime(q,t) = \omega(q,t) - \delta h(q,t) \wedge \delta t$, where
\begin{align}  \label{L13} 
\omega(q,t) =& \frac12\,\int_{\R^3} \D\zeta\,\D\sigma\,\D\rho\,\left[\theta(\sigma)-\theta(\rho)\right]\,\frac{\delta^2 L(T_\zeta q, t+\zeta)}{\delta q(\sigma) \delta q(\rho)}\,\delta q(\rho) \wedge \delta q(\sigma) \nonumber \\
&\quad + \int_{\R^2} \mathrm{d}\zeta\,\mathrm{d}\sigma\left[\theta(\sigma)-\theta(\zeta)\right]\,\frac{\delta^2 L(T_\zeta q, t+\zeta)}{\delta q(\sigma)\,\delta t}\,\delta t \wedge \delta q(\sigma),
\end{align}
is the (pre)symplectic form\footnote{We have included the skew-symmetry of $\delta q(\rho) \wedge \delta q(\sigma)$ to derive the latter expression.}, that is, it is a closed differential 2-form, but it is not clear whether it is non-degenerate at constant $t$.

\begin{figure}[h!] 
\begin{center}
\includegraphics[width=0.7\textwidth]{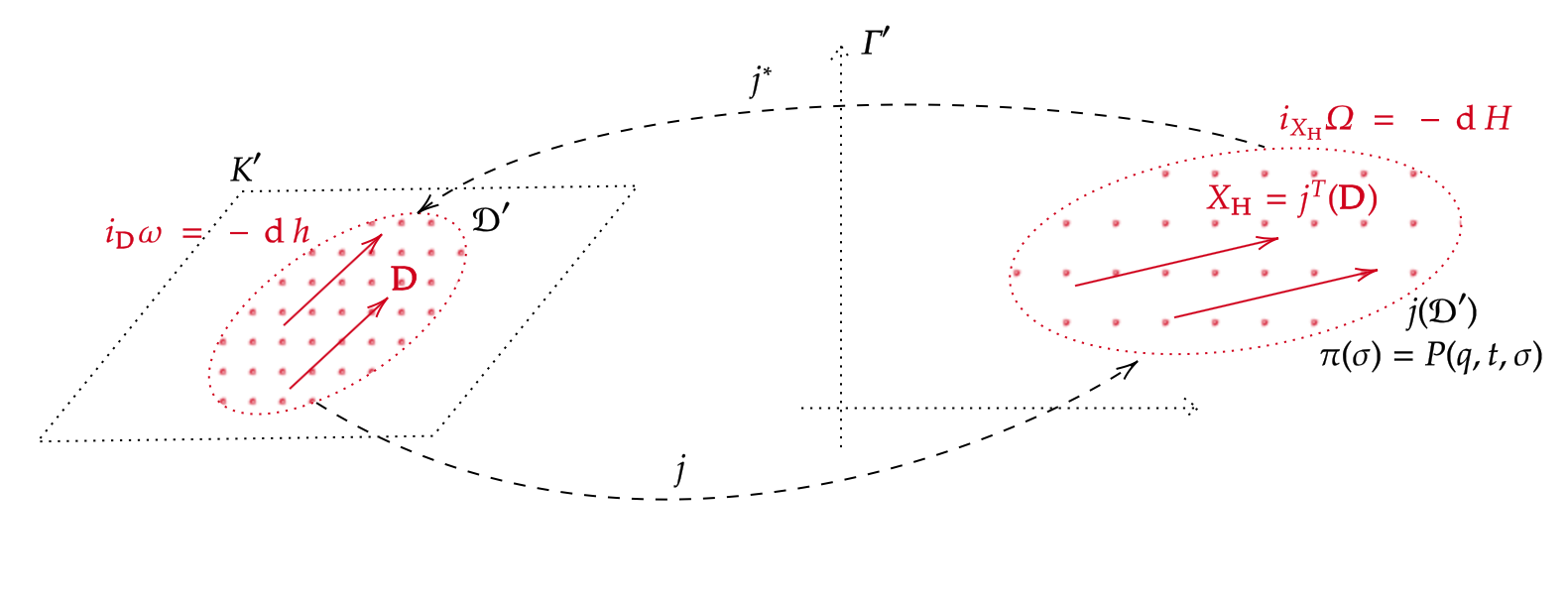}
\end{center}
\caption{This figure illustrates the step-by-step process we have employed to construct the Hamiltonian formalism for nonlocal Lagrangians. Our approach involves injecting the space $\mathcal{D}^\prime$ into the extended phase space to establish a connection with the Hamiltonian formalism we have (trivially) built. Due to this injection, the submanifold $j(\mathcal{D}^\prime)$ is then defined by the curves that satisfy the nonlocal Euler-Lagrange equations and by the momenta defined by $\pi(\sigma)= P(q,t,\sigma)$. Because $\mathbf{X}_H= j^T\left(\mathbf{D}\right)$ is satisfied on $j(\mathcal{D}^\prime)$, we are able to perform a pullback, which enables us to translate the Hamiltonian system on $\mathcal{D}^\prime$. Note that ensuring the presence of a symplectic form on $\mathcal{D}^\prime$ requires coordinating this space, which is the most complex aspect of nonlocal theories. }
\label{fig:HF}
\end{figure}

We have not reached our goal yet. Due to the constraints $\Psi(q,t,\sigma)=0$ that characterize the extended dynamic space as a submanifold of the extended kinematic space $\mathcal{K}^\prime$, $q$ and $t$ are not  independent coordinates in $\mathcal{D}^\prime$\,. Consequently, the final step consists of providing coordinates to $\mathcal{D}^\prime$ and showing that $\omega(q,t) \in \Lambda^2(\mathcal{D}^\prime)$ is non-degenerate. Therefore, we need to obtain the explicit parametric form of the submanifold $\mathcal{D}^\prime$ instead of the implicit form provided by the nonlocal Euler-Lagrange equations. This fact is easy for regular local Lagrangians (that depend on derivatives up to the $r^{\rm th}$ order) because the Euler-Lagrange equations are an ordinary differential system of order $2r$, and the theorems of existence and uniqueness provide the required parametric form. However, in the general case, deriving the explicit equations of $\mathcal{D}^\prime$ from the implicit equations is a complex task that, as far as we consider, depends on each case.


\chapter{Nonlocal Lagrangian Fields}\label{chap5}

This chapter focuses on studying the nonlocal Lagrangian and Hamiltonian formalisms with an infinite number of degrees of freedom. We shall adapt the latter results to nonlocal Lagrangian fields, considering all the peculiarities of field theories concerning mechanics. This chapter is based on \cite{Heredia2022,Heredia2023}.

\section{The principle of least action}\label{chap51-PLS}
Consider the \textit{nonlocal \important{action integral}}
\begin{equation} \label{A0}
  S = \int_{\R^4} \D x\,\mathcal{L}([\phi^A],x)  \,, 
\end{equation}
where the \textit{\important{nonlocal!Lagrangian} density} $\mathcal{L}$ depends on all the values $\phi^A(z)$, $\,A=1,\ldots,m\,$, of the field variables at points $z$ other than $x$.  We shall take  $x\in \R^4$  for concreteness; however, the following also holds for any number of dimensions.

As in mechanics, the class of all possible fields constitutes the \textit{\important{kinematic space}} $\mathcal{K}$, whether or not they meet the field equations. This space is the subspace of all smooth functions $\mathcal{C}^\infty(\R^4;\R^m) $ such that $\mathcal{L}([\phi^A],x)\,$ is locally summable. For nonlocal Lagrangians depending explicitly on the point $x^b$, we have to resort to the {\em \important{extended!kinematic space}} $\mathcal{K}^\prime= \mathcal{K} \times\R^4$. 

The nonlocal Lagrangian density is a real-valued functional  
\begin{equation}
([\phi^A],x^b) \in \mathcal{K}^\prime \longrightarrow \mathcal{L}([\phi^A],x^b) \in \R \,,
\end{equation}
and the function $\phi^A(z)$ contains all information about the evolution in $\mathcal{K}^\prime$. 

There is a one-to-one correspondence between the ``infinite-order" Ostrogradsky formalism and the nonlocal one based on the formal Taylor series (FTS) 
\begin{equation}\label{chap51-1-1}
\left(\left\{\phi^A_{|_{b_1,\ldots,b_r}}(x)\right\}, x\right) \longleftrightarrow \left([\phi^A],x\right) \qquad \mathrm{with} \qquad \phi^A(z+x) = \sum^\infty_{|\alpha| =0} \frac{z^\alpha}{\alpha!} \phi^A_{|_\alpha}(x)\,,
\end{equation}
where the multi-index notation has been included, namely, $(\alpha_1,\ldots,\alpha_4)\in\mathbb{N}^4_0$, 
\begin{equation}
\phi^A_{|_\alpha} := \frac{\partial^{|\alpha|} \phi^A}{\partial (x^1)^{\alpha_1} \ldots \partial (x^4)^{\alpha_4}}\,, \qquad z^\alpha := (z^1)^{\alpha_1} \ldots (z^4)^{\alpha_4}, \qquad \alpha! := \alpha_1! \ldots \alpha_4!
\end{equation}
and $|\alpha| := \alpha_1 + \ldots + \alpha_4$.
 
Given $x\in \R^4$, we shall define the \textit{\important{spacetime translation} operator} $T_x$ as
\begin{equation}  \label{L1} 
  ([\phi^A],0) \stackrel{T_x}{\longrightarrow} ([T_x \phi^A], x) \,,\qquad {\rm with} \qquad T_x \phi^A(z) = \phi^A(z+x)\;,
\end{equation}
which has the obvious additive property $T_{ x_1}\circ T_{x_2} = T_{x_1+ x_2}\,$ and preserves the correspondence (\ref{chap51-1-1}); namely, the following diagram
\begin{equation}\label{chap51-Diag}
\xymatrix{
\left(\left\{\phi^A_{|_{b_1,\ldots,b_r}}(0)\right\}, 0\right)\ \ar @{<->}[d]^{FTS} \ar @{<->}[r]^{T_x} &\left(\left\{\phi^A_{|_{b_1,\ldots,b_r}}(x)\right\}, x\right) \ar @{<->}[d]^{FTS}\\
([\phi^A],0) \ar @{<->}[r]^{T_x}&([T_x \phi^A],x)}
\end{equation}
is commutative. Therefore, we shall refer to the subset $\{([T_x \phi^A], x), \, x\in \R^4 \}\subset \mathcal{K}^\prime$ as the field trajectory starting at $([\phi^A],0)\,$.

The nonlocal action integral (\ref{A0}) should be better understood as the functional on $\mathcal{K}^\prime$ 
\begin{equation}   \label{A1o}
 S(\phi) := \int_{\R^4} \D  x\, \mathcal{L}\left(T_x\phi,  x \right) \,,
\end{equation}
where the functional dependence is understood --$\mathcal{L}\left(T_x\phi^A,  x \right)$ instead of $\mathcal{L}\left([T_x\phi^A],  x \right)$--, although the square bracket does not emphasize it. Moreover, we also omit the superindices in the field variables and the point coordinates unless the context makes it necessary for simplicity. Notice that it may be divergent because we need an unbounded integration domain because the nonlocal Lagrangian density $\mathcal{L}$ depends on all the values $\phi(z)\,$. For this reason, we shall introduce the alternative and more consistent formulation based on the one-parameter family of finite nonlocal action integrals 
\begin{equation} \label{A1}
 S(\phi,R) = \int_{|x|\leq R} \D x\,\mathcal{L}(T_x\phi,x)  \,, \qquad \forall R\in \R^+ \,,
\end{equation}
where $|x|= \sqrt{\sum_{j=1}^4 (x^j)^2 }$ is the Euclidean length.  Then, the \textit{\important{principle of least action}} reads
\begin{equation} \label{A2}
\lim_{R\rightarrow\infty} \delta S(\phi,R) \equiv \lim_{R\rightarrow\infty}  \int_{|x|\leq R} \D x\,\int_{\R^4} \D z\,\frac{\delta \mathcal{L}\left(T_x \phi,  x\right)}{\delta \phi(z)} \,{\delta \phi(z)} = 0\,,  
\end{equation}
for all variations $\delta \phi(z)$ with compact support. Consequently, the \textit{\important{nonlocal!Euler-Lagrange field equations}} are
\begin{equation}  \label{L2o} 
\psi(\phi,z) = 0 \,, \quad {\rm with} \quad \psi(\phi,z) := \int_{\R^4} \D  x\,\lambda(\phi,x,z) \quad {\rm and} \quad
\lambda(\phi,x,z) := \frac{\delta \mathcal{L}\left(T_x\phi, x\right)}{\delta \phi(z)} \:.
\end{equation}
The \textit{\important{dynamic!fields}} are those $\phi$ fulfilling these equations.

\subsection{Local case embedded in nonlocal one}
Let us see how a local Lagrangian density $\calL_L(\phi, \ldots, \phi_{|_{b_1 \ldots b_r}},x)$, which depends on the field derivatives up to the $r^{\rm th}$ order, fits in this formalism developed so far. The action integral 
\begin{equation}
S(\phi) = \int_{\mathcal{V}} \D x\,\calL_L(\phi,\ldots,\phi_{|_{b_1\ldots b_r}},x)
\end{equation}
has the form (\ref{A1}) provided that we take
\begin{equation}  \label{A2a}
\mathcal{L}(T_x\phi, x) := \calL_L(\phi(x),\ldots,\phi_{|_{b_1\ldots b_r}}(x), x)\:.
\end{equation}
Whence, it follows from (\ref{L2o}) that
\begin{equation}  \label{L2a} 
\lambda(\phi, x, z) = \frac{\delta \mathcal{L}\left(T_x\phi,  x\right)}{\delta \phi( z)} = \sum_{j=0}^r \left(\frac{\partial \calL_L}{\partial \phi_{|_{c_1\ldots c_j}}}\right)_{(\phi(x), \ldots, \phi_{|_{b_1\ldots b_r}}(x), x)}\,(-1)^j\delta_{|_{c_1\ldots c_j}}(z-x)  \,,
\end{equation}
where we have included that 
\begin{equation}
\phi_{|_{c_1\ldots c_j}}(x) = (-1)^j\,\int_{\R^4} \D z\,\phi( z)\,\delta_{|_{c_1\ldots c_j}}(z-x)\,.
\end{equation}
Substituting (\ref{L2a}) in (\ref{L2o}), we finally arrive at
\begin{equation}  \label{L2b} 
 \psi(\phi,z) \equiv \sum_{j=0}^r (-1)^j\,\frac{\partial^j}{\partial  z^{c_1} \ldots\partial z^{c_j}} \left(\frac{\partial \calL_L}{\partial \phi_{|_{c_1\ldots c_j}}}\right)_{(\phi( z), \ldots ,\phi_{|_{b_1\ldots b_r}}( z),z)} = 0\;,
\end{equation}
which are the \textit{\important{Euler-Ostrogradsky equations}} (\ref{chap31-EL}).

\subsection{Two ways of coordinating the extended kinematic space for fields}

Following a similar procedure as in Section \ref{chap-CEK}, we might coordinate the point $a\in\mathcal{K}^\prime$ in two ways: either through $\mathbf{(a)}$ the \textit{\important{static coordinates}} or through $\mathbf{(b)}$ the \textit{\important{moving coordinates}},
\begin{equation}
\begin{array}{cccccccc}
\mathbf{(a)} \quad & \mathcal{K}^\prime & \longrightarrow & \mathcal{K} \times \R^4 & \hspace*{4em} \mathbf{(b)} \quad & \mathcal{K}^\prime & \longrightarrow & \mathcal{K} \times \R^4   \\
  & a & \longmapsto & (\tilde{\phi},x) &  & a & \longmapsto & (\phi,x) \,,
\end{array}
\end{equation}
where $\phi(z)=T_x\tilde{\phi}(z)$, namely, $\phi(z) = \tilde{\phi}(z+x)$. The spacetime translation operator acts in these coordinate systems as
\begin{equation}
\mathbf{(a)} \quad (\tilde{\phi},x) \stackrel{T_y}{\longrightarrow} (\tilde{\phi}, x+y) \qquad  \qquad {\rm and} \qquad  \qquad \mathbf{(b)} \quad (\phi,x) \stackrel{T_y}{\longrightarrow}  (T_y \phi, x+y) \,.
\end{equation}

The nonlocal Euler-Lagrange equations (\ref{L2o}) derived in the previous section were obtained in the context of the static coordinates; namely, they are limited to trajectories $(T_x\tilde{\phi},x)\in\mathcal{K}^\prime$ starting at $(\tilde{\phi},0)$. Thus, they should read as 
\begin{equation}\label{chap512-LSC} 
\psi(\tilde{\phi},z) = 0 \,, \quad {\rm with} \quad \psi(\tilde{\phi},z) := \int_{\R^4} \D  y\,\lambda(\tilde{\phi},y,z) \quad {\rm and} \quad
\lambda(\tilde{\phi},y,z) := \frac{\delta \mathcal{L}(T_y\tilde{\phi}, y)}{\delta \tilde{\phi}(z)} \:.
\end{equation}

To transform equation (\ref{chap512-LSC}) into moving coordinates, we use that $(T_y\phi,x+y)=(T_{x+y}\tilde{\phi},x+y)$, and hence it becomes 
\begin{equation}\label{chap512-PsiMC}
\Psi(\phi,x,z) = 0\,, \qquad \mathrm{with} \qquad \Psi(\phi,x,z):= \int_{\R^4}\D y^\prime \frac{\delta L(T_{y^\prime} \phi, x+y^\prime)}{\delta \phi(z)} = \psi(\tilde{\phi},x+z)\,,
\end{equation}
where $y^\prime = y+x$ was used. Obviously, the following property is satisfied
\begin{equation}
\psi(\tilde{\phi},z) = \Psi(\phi,0,z)\,.
\end{equation}

The \textit{infinitesimal generators of \important{spacetime translation}s} $\mathbf{D}_a$, $a= 1,\ldots,4$, act differently depending on the coordinates used, i.e., if $F(\phi,x)$ and $\tilde{F}(\tilde{\phi},x)$ denote functions expressed in each of the coordinate systems, then 
\begin{equation}
\mathbf{D}_a \tilde{F}(\tilde{\phi},x) = \partial_a \tilde{F}(\tilde{\phi},x) \qquad \mathrm{and} \qquad \mathbf{D}_a F(\phi,x) =  \left[\frac{\partial F\left(T_{x+\varepsilon} \phi, x+ \varepsilon\right)}{\partial \varepsilon^a} \right]_{\varepsilon=0} \,.
\end{equation}
Particularly, if $F(\phi,x)$ is a smooth function in $\mathcal{C}^\infty(\R^4)$, and we write $F(\phi,x)_{(z)} := F(\phi,x,z)$, then
\begin{equation}\label{chap512D}
\mathbf{D}_a \tilde{F}(\tilde{\phi},x,z) = \partial_a \tilde{F}(\tilde{\phi},x,z) \qquad \mathrm{and} \qquad \mathbf{D}_a F(\phi,x,z) =  \left[\frac{\partial F\left(T_{x+\varepsilon} \phi, x+ \varepsilon, z\right)}{\partial \varepsilon^a} \right]_{\varepsilon=0} \,.
\end{equation}
They are vector fields on $\mathcal{K}^\prime$ that, including the chain rule, can be written in static and moving coordinates, respectively, as
\begin{equation}
\mathbf{D}_a = \partial_a \qquad \mathrm{and} \qquad \mathbf{D}_a = \partial_a +  \int_{\R^4} \D\sigma\,\phi_{|a}(\sigma)\,\frac{\delta \quad}{\delta \phi(\sigma)}\,, 
\end{equation}
where $\mathbf{D}_4$ is the \textit{\important{time!evolution generator}}. 

\subsection{Euler-Lagrange equations vs constraints}
Equation (\ref{chap512-PsiMC}) is not met by any $(\phi, x) \in\mathcal{K}^\prime\,$. Therefore, the nonlocal Euler-Lagrange equations act as implicit equations defining the dynamic space $\mathcal{D}^\prime$, i.e., the class of all dynamic fields, as a submanifold of $\mathcal{K}^\prime\,$.

Similarly, as discussed in the case of nonlocal mechanics --Section \ref{chap31}--, the nonlocal field equation (\ref{chap512-PsiMC}) is a partial integrodifferential system rather than a partial differential system, and, as a rule, we do not have an equivalent to the \textit{\important{Cauchy-Kowalevski theorem}} to turn to.  For this reason, we  take (\ref{chap512-PsiMC}) as implicit equations or constraints defining $\mathcal{D}^\prime$ as a submanifold of $\mathcal{K}^\prime\,$. 

\noindent We write them as
\begin{equation}  \label{L3} 
 \Psi(\phi, x,z):=\Psi(\phi,x)_{(z)}= 0 \qquad \forall z\in\R^4 \,.
\end{equation}
This notation indicates that $\Psi$ maps $\mathcal{K}^\prime\,$ on the space of smooth functions of $z\in \R^4$. The dynamic fields are those $(\phi,x)$ that make $\Psi$ null.

The constraints (\ref{L3}) are stable under spacetime translations, and therefore the generators $\,\mathbf{D}_a\,$ are tangent to the submanifold $\mathcal{D}^\prime \subset \mathcal{K}^\prime\,$. Indeed, including (\ref{L3}) and (\ref{chap512D}), we have that 
\begin{align}
\mathbf{D}_a \Psi(\phi,x,z) &= \mathbf{D}_a\left[\int_{\R^4}\D y \frac{\delta \calL(T_y \phi, x+y)}{\delta \phi(z)}\right]=\frac{\partial}{\partial \varepsilon^a}\left[\int_{\R^4} \D y \frac{\delta \calL(T_{y+\varepsilon}\phi,x+y+\varepsilon)}{\delta \phi(z+\varepsilon)}\right]_{\varepsilon=0}  \nonumber\\
&=\frac{\partial}{\partial \varepsilon^a}\left[\int_{\R^4} \D y^\prime \frac{\delta \calL(T_{y^\prime}\phi,x+y^\prime)}{\delta \phi(z+\varepsilon)}\right]_{\varepsilon=0} = \left.\frac{\partial \Psi(\phi,x,z+\varepsilon)}{\partial \varepsilon^a}\right|_{\varepsilon=0}\,,
\end{align}
where the replacement $y^\prime=y+ \varepsilon$ has been made. Hence, if $\Psi(\phi,x,z) = 0\,\,\,\forall z$, then $\Psi(\phi,x,z+\varepsilon) =0$ as well; therefore
\begin{equation}
\mathbf{D}_a\Psi(\phi,x,z) = 0\,.
\end{equation}
In static coordinates, 
\begin{equation}
\mathbf{D}_a\psi(\tilde{\phi},z)= 0
\end{equation}
is trivially satisfied.

\subsection{Nonlocal four-divergence for fields}

As discussed in Section \ref{chap31}, a well-known feature of local theories is that, when the Lagrangian density is a \textit{\important{four-divergence}},
\begin{equation}  \label{TD1}
 \mathcal{L}_L(x) = \partial_b W^b(x) \,,
\end{equation}
the Euler-Lagrange equations vanish identically --\textit{\important{Noether!symmetry}}--. The nonlocal case is more nuanced than the local one since equation (\ref{TD1}) always has a solution (in fact, infinitely many). Indeed, the general solution is
\begin{equation} 
 W^b(x) = \delta_4^b\,\int_\R \D \tau\,\left[\theta(\tau)-\theta(\tau-t)\right]\,\mathcal{L}_L(\mathbf{x},\tau) + \partial_c\Omega^{bc}(x) \,,
 \end{equation}
where $x=(\mathbf{x},t)$ and $\Omega^{bc}+\Omega^{cb} = 0\,$. 
However, as the solution $W^b(x)$ is not necessarily local, it does not imply that the Euler-Lagrange equations for any nonlocal Lagrangian density are identically null.

Let us now search for a sufficient condition on $W^b(T_x\tilde\phi,x) $ for the Lagrangian $\partial_b W^b(T_x\tilde\phi,x)$ to produce null nonlocal Euler-Lagrange equations. The family of nonlocal actions (\ref{A1}) for such a Lagrangian density is
\begin{equation}
S(\tilde{\phi},R) = \int_{|x|\leq R} \D y\,\partial_b W^b(T_x\tilde{\phi},x)  = \int_{|x|=R} \D \Sigma_b(x)\,W^b (T_x \tilde{\phi},x)\,,
\end{equation}
where Gauss' theorem has been applied, and $\D \Sigma_b(x)$ is the volume element on the hypersphere $|x|=R$. Consequently, the principle of least action (\ref{A2}) yields the nonlocal Euler-Lagrange equations\vspace{0.2cm}
\begin{equation}
\psi(\tilde{\phi},z) :=\lim_{R\rightarrow\infty} \frac{\delta S(\tilde{\phi},R)}{\delta \tilde{\phi}(z)} \equiv 
\lim_{R\rightarrow\infty} \int_{|x|=R} \D \Sigma_b(x)\,\frac{\delta W^b (T_x \tilde{\phi},x)}{\delta\tilde{\phi}(z)}\,, 
\end{equation}
and, as $\D \Sigma_b(x)$ scales as $|x|^3$, they are identically null provided that 
\begin{equation} \label{TD3}
 \lim_{|x|\rightarrow\infty} \left\{|x|^3\,\frac{\delta W^b (T_x \tilde{\phi},x)}{\delta \tilde{\phi}(z)} \right\} \equiv 0\,,  
\end{equation}
where the symbol $\equiv$ means that the equalities hold for any $\tilde{\phi}$. This condition is met if $W^b (T_x \tilde{\phi},x)$ is local, i.e., it depends only on a finite number of derivatives of $\phi$ at $x$.  

\section{The Noether theorem}
To include non-scalar fields, we shall restore the superindex $A$ in the field variable. Consider the infinitesimal transformations
\begin{equation}  \label{A3}
x^{\prime a}(x) = x^a + \delta x^a(x)  \qquad\mathrm{and}\qquad \tilde{\phi}^{\prime A}(x) = \tilde{\phi}^A(x) + \delta \tilde{\phi}^A(x) \,.
\end{equation}

The nonlocal Lagrangian density transforms so that the action integral over any four-volume is preserved $S^\prime (\mathcal{V}^\prime)=S(\mathcal{V})$, namely,
\begin{equation}
\mathcal{L}^\prime(T_{ x^\prime}\tilde{\phi}^{\prime A}, x^\prime) = \mathcal{L}(T_x\tilde{\phi}^A, x) \,\left|\frac{\partial x}{\partial x^\prime}\right|\,. 
\end{equation}
Therefore, if $\mathcal{V}^\prime$ is the transformation of the spacetime volume $\mathcal{V}$ according to (\ref{A3}), we get
\begin{equation}\label{chap52-SSp}
\int_{\mathcal{V}^\prime} \D  x^\prime\, \mathcal{L}^\prime(T_{ x^\prime}\tilde{\phi}^{\prime A}, x^\prime) = \int_{\mathcal{V}} \D x \,\mathcal{L}(T_x\tilde{\phi}^A, x)\,,
\end{equation}
and, consequently,
\begin{equation}  \label{A4}
 \int_{\mathcal{V}^\prime} \D  x\,\mathcal{L}^\prime(T_x\tilde{\phi}^{\prime A}, x) -\int_{\mathcal{V}}\D  x\,\mathcal{L}(T_x\tilde{\phi}^A, x)  = 0\,,
\end{equation}
where we have replaced the dummy variable $ x^\prime$ with $ x\,$.  As considered in the local case --see fig.(\ref{fig:chp32-V})--, let us assume that the volumes $\mathcal{V}^\prime$ and $\mathcal{V}$ share a large region and only differ in an infinitesimal layer close to the boundary $\partial\mathcal{V}$. If $\D \Sigma_a$ is the hypersurface element on the boundary, then the volume element close to the boundary is $\D  x = \D \Sigma_b\,\delta  x^b$. Hence, by neglecting second-order infinitesimals, equation (\ref{A4}) becomes
\begin{equation}  \label{A5}
 \int_{\mathcal{V}} \D x\,\left[ \mathcal{L}^\prime(T_x\tilde{\phi}^{\prime A}, x) - \mathcal{L}(T_x\tilde{\phi}^A, x)\right]\, + \int_{\partial\mathcal{V}} \mathcal{L}(T_x\tilde{\phi}^A, x)\,\delta  x^b \,\D \Sigma_b = 0\,.
\end{equation}

For a Noether symmetry, we have that 
\begin{equation} \mathcal{L}^\prime(T_x\tilde{\phi}^{\prime A}, x) = \mathcal{L}(T_x\tilde{\phi}^{\prime A}, x) +  \partial_b W^b(T_x\tilde{\phi}^{\prime A}, x)\,,
\end{equation}
where $W^b(T_{x}\tilde{\phi}^{\prime A},x)$ is a first-order infinitesimal fulfilling the asymptotic condition (\ref{TD3}); therefore,
\begin{align}\label{chap52-NTNS}
 \mathcal{L}^\prime(T_x\tilde{\phi}^{\prime A}, x) - \mathcal{L}(T_x\tilde{\phi}^A, x) =  \partial_b W^b(T_{ x}\tilde{\phi}^A, x) + \int_{\mathbb{R}^4} \D  y\,\lambda_A(\tilde{\phi}, x, y)\,\delta \tilde{\phi}^A( y)  \,,
\end{align}
where $\lambda_A(\tilde{\phi}, x, y)$ is defined in (\ref{L2o}), $\partial_b$ is the partial derivative for $x^b$, and second-order infinitesimals have been neglected. Introducing the variable $z= y - x$ in (\ref{chap52-NTNS}), substituting it in (\ref{A5}), and applying Gauss' theorem, we obtain that
\begin{align}
&\int_{\mathcal{V}}\D  x\, \left\{ \partial_b\left[\mathcal{L}(T_x\tilde{\phi}^A, x)\,\delta x^b + 
 W^b(T_x\tilde{\phi}^A, x) \right]+ \int_{\R^4} \D z\,\lambda_A(\tilde{\phi}, x,z+ x)\,\delta\tilde{\phi}^A(z+ x) \right\} = 0\,.
 \end{align}
Furthermore, including (\ref{L2o}), we can write
\begin{align}  \label{A6}
&\lefteqn{ - \int_{\mathcal{V}}\D  x\,\psi_A(\tilde{\phi}, x) \,\delta\tilde{\phi}^A( x) = \int_{\mathcal{V}}\D  x\, \left\{\partial_b\left[\mathcal{L}(T_x\tilde{\phi}^A, x)\,\delta  x^b +  W^b(T_x\tilde{\phi}^A, x)\right]  \right.} \nonumber \\ 
 &\left. \hspace*{4em}+\int_{\mathbb{R}^4} \D z \,\left[ \lambda_A(\tilde{\phi}, x,z+ x)\,\delta\tilde{\phi}^A(z+ x) - \lambda_A(\tilde{\phi}, x-z, x)\,\delta\tilde{\phi}^A( x)\right]\right\}\;.
\end{align}
\indent Now, the trick for fields. We use the identity
\begin{align}
\lefteqn{\lambda_A(\tilde{\phi}, x,z+ x)\,\delta\tilde{\phi}^A(z+ x) - \lambda_A(\tilde{\phi}, x-z, x)\,\delta\tilde{\phi}^A( x) =} \nonumber\\
 &\qquad \qquad= \int_0^1 \D s \,\frac{\D\;}{\D s}\left\{\lambda_A(\tilde{\phi}, x+[s-1] z, x+s z)\,\delta\tilde{\phi}^A( x+s z) \right\}
\nonumber\\
 &\qquad \qquad = \int_0^1 \D s \, z^b \,\frac{\partial}{\partial  x^b} \left\{\lambda_A(\tilde{\phi}, x+[s-1]z, x+s z)\,\delta\tilde{\phi}^A( x+s z) \right\} 
\end{align}
that, combined with (\ref{A6}), leads to
\begin{equation}  \label{A7}
  \int_{\mathcal{V}}\D  x\,\left\{\psi_A(\tilde{\phi}, x) \,\delta\tilde{\phi}^A( x) +\frac{\partial}{\partial  x^b} \left[\mathcal{L}(T_x\tilde{\phi}, x)\,\delta x^b + W^b(T_x\tilde{\phi}, x) + \Pi^b(T_x\tilde{\phi}, x) \right] \right\} = 0\,,
\end{equation}
where  $\Pi^b(T_x\tilde{\phi},x)$ is
\begin{equation}  \label{A8}
\Pi^b(T_x\tilde{\phi}, x) :=  \int_{\R^4} \D z\, z^b \int_0^1 \D s\,\lambda_A(\tilde{\phi}, x+[s-1] z, x+s z)\,\delta\tilde{\phi}^A( x+s  z)\;.
\end{equation}
As equation (\ref{A7}) holds  for any spacetime volume $\mathcal{V}$, it follows that
\begin{equation}  \label{A9}
N(\tilde{\phi}, x) := \partial_b J^b(T_x\tilde{\phi}, x) +\psi_A(\tilde{\phi}, x) \,\delta\tilde{\phi}^A(x) \equiv 0  \,,
\end{equation}
where
\begin{align}  \label{A10}
\quad J^b(T_x\tilde{\phi}, x) :&= \mathcal{L}(T_x\tilde{\phi},x)\,\delta x^b + W^b(T_x\tilde{\phi},x)\nonumber\\ 
&\qquad \qquad +\int_{\R^4} \D z\,z^b \int_0^1 \D s\,\lambda_A(\tilde{\phi},x+[s-1] z, x+s z)\,\delta\tilde{\phi}^A(x+s z)\;. 
\end{align}

Equation (\ref{A9}) is an identity and holds for any kinematic field $\tilde{\phi}$. For dynamic fields, this identity implies that the current $J^b(T_x\tilde{\phi}, x)$ is locally conserved
\begin{equation}  \label{A10a}
  \partial_b J^b = 0 \,.
\end{equation}

\subsection{The energy- and angular momentum currents}\label{chap521-TS}
Let us particularise the conserved current (\ref{A10}) for a \textit{\important{Poincar\'e!symmetry}}. By substituting (\ref{chap331-P1}) and (\ref{chap331-P3}) into (\ref{A10}) and assuming that the nonlocal Lagrangian density is Poincar\'e invariant --- therefore, $W^b = 0$---, we find that the conserved current  can be written as
\begin{equation}  \label{P4}
 J^b(T_x\tilde{\phi},x) = -\varepsilon^a\,\mathcal{T}^{\; b}_a(T_x\tilde{\phi},x) - \frac12\, \omega^{ac} \mathcal{J}^{\; \,\;b}_{ac}(T_x\tilde{\phi},x) \,,
\end{equation}
where
\begin{equation}  \label{P5}
 \mathcal{T}^{\; b}_a := -\mathcal{L}(T_x\tilde{\phi},x)\,\delta^b_a +  \int_{\R^4} \D z\,z^b \int_0^1 \D s\,\lambda_A(\tilde{\phi},x+[s-1]z, x+s z)\,\tilde{\phi}^A_{|a}(x+s z)\,,
\end{equation}
and
\begin{equation}\label{P6}
  \mathcal{J}^{\; \,\;b}_{ac} :=  2\, x_{[c}\mathcal{T}^{\; b}_{a]} + \mathcal{S}^{\;\,\;b}_{ac}
  \end{equation}
with 
\begin{align} \label{P6a}  
\mathcal{S}^{\;\,\;b}_{ac}(T_x\tilde{\phi},x) :=2 \int_{\R^4} \D z\, z^b \int_0^1 \D s\,&\lambda_A(\tilde{\phi}, x+[s-1]z, x+s z)\nonumber\\
&\qquad \times \left[ s\,z_{[c} \tilde{\phi}^A_{|a]}(x+s z) - M^A_{\; B[ac]}\tilde{\phi}^B(x+s z) \right]
\end{align}
are the \textit{canonical \important{energy-momentum tensor}}, the \textit{\important{angular momentum tensor}}, the \textit{\important{orbital angular momentum tensor}}, and the \textit{\important{spin current}}, respectively.

Since the ten parameters $\varepsilon^a$ and $\omega^{ac}$ are independent, the local conservation of the current $J^b(T_x\tilde{\phi},x)$ implies that the currents $\mathcal{T}^{\; b}_a(T_x\tilde{\phi},x)$ and $\,\mathcal{J}^{\; \,\;b}_{ac}(T_x\tilde{\phi},x)\,$ are separately conserved, that is,
\begin{equation}
\partial_b\mathcal{T}^{\; b}_a(T_x\tilde{\phi},x) = 0 \qquad {\rm and}\qquad \partial_b\mathcal{J}^{\; \,\;b}_{ac}(T_x\tilde{\phi},x) = 0\,,
\end{equation}
or 
\begin{equation}  \label{P7}
 \partial_b \mathcal{T}^{\; b}_a = 0 \qquad {\rm and}\qquad \partial_b\mathcal{S}^{\; \,\;b}_{ac} + 2 \,\mathcal{T}_{[ac]} = 0\,.
\end{equation}

Note that in the same way as in the local case --Section \ref{chap331}--, employing the \textit{\important{Belinfante symmetrization technique}}, we can build the \textit{\important{Belinfante-Rosenfeld energy-momentum tensor}} that is somehow equivalent to the canonical one.

\subsubsection{The energy density}

The component $\,\mathcal{T}^{\; 4}_4(T_x\tilde{\phi},x)\,$ of the canonical energy-momentum tensor is the energy density\footnote{It also applies to Belinfante-Rosenfeld's energy-momentum tensor. }. Therefore, using $x^a = (\mathbf{x},t)\,$, the \textit{\important{energy function}} for a dynamic field is\vspace{0.2cm}
\begin{equation}  \label{P9}
E(T_t\tilde{\phi},t) := \int_{\R^3} \D\mathbf{x}\,\mathcal{T}^{\; 4}_4(T_x\tilde{\phi},x)\,.
\end{equation}
\indent It is well-known \cite{Landau1975} that if the field decays fast enough at spatial infinity, the continuity equation (\ref{A10a}) implies that the total energy and momentum do not depend on $t$. In the particular case of the energy function, this fact implies that
\begin{equation}  \label{P9a}
 E(T_t\tilde{\phi},t) =  E(\tilde{\phi},0) =:E(\tilde{\phi})\;,
\end{equation}
namely, it is conserved. In a general case, we have
\begin{align} \label{P10}
 E(T_t\tilde{\phi},t) :=&  - L(T_t\tilde{\phi},t) + \int_{\R^6} \D\mathbf{x}\, \D\mathbf{z}\int_\R \D\zeta \int_0^1\D s\,\zeta\,\dot{\tilde{\phi}}^A( \mathbf{x}+s\,\mathbf{z},t+s\,\zeta)\, 
\nonumber \\
 &  \hspace*{6em}\times \lambda_A(\tilde{\phi},\mathbf{x}+(s-1)\,\mathbf{z},t+(s-1)\zeta, \mathbf{x}+s\,\mathbf{z}, t+ s\,\zeta)\,, 
\end{align}
where $L(T_t\tilde{\phi},t) :=\int_{\R^3}\D\mathbf{x}\,\mathcal{L}(T_x\tilde{\phi},x)$, $z^a =(\mathbf{z},\zeta)$ and $\dot{\tilde{\phi}}^A := \tilde{\phi}^A_{|4}\,$. After transforming the variables $\mathbf{u} = \mathbf{x} + s\,\mathbf{z}$ and $\rho =t+s\,\zeta$, the integral on the right-hand side becomes
\begin{align}
& \int_{\R^6} \D\mathbf{u}\, \D\mathbf{z}\int_\R \D\zeta \int_t^{t+\zeta} \D\rho\,\lambda_A(\tilde{\phi},\mathbf{u}-\mathbf{z},\rho-\zeta, \mathbf{u},\rho)\,\dot{\tilde{\phi}}^A( \mathbf{u},\rho) \nonumber\\
&\qquad= \int_{\R^6} \D\mathbf{u}\, \D\mathbf{z}\int_\R \D\zeta \int_t^{t+\zeta} \D\rho\,\lambda_A(\tilde{\phi},\mathbf{u}-\mathbf{z},\rho-\zeta, \mathbf{u},\rho)\,\dot{\tilde{\phi}}^A( \mathbf{u},\rho) \nonumber\\
&\qquad = \int_{\R^6} \D\mathbf{u}\, \D\mathbf{z}\int_{\R^2} \D\zeta\D\rho\,\left[\theta(t+\zeta-\rho) - \theta(t-\rho)\right]\lambda_A(\tilde{\phi},\mathbf{u}-\mathbf{z},\rho-\zeta, \mathbf{u},\rho)\,\dot{\tilde{\phi}}^A( \mathbf{u},\rho)\nonumber\\
&\qquad = \int_{\R^4}\D u\,\dot{\tilde{\phi}}(u) \int_{\R^4}\D y\left[\theta(t-y^4)-\theta(t-u^4)\right]\lambda_A(\tilde{\phi}, y, u)\,,
\end{align}
where we have taken $u^a=(\mathbf{u},\rho)$ and $y^a=(\mathbf{y},\sigma)$,  with $\mathbf{y}:=\mathbf{u}-\mathbf{z}$ and $\sigma:=\rho-\zeta$. On replacing $t-u^4 = - v^4$ and $t-y^4=-w^4$, the last expression becomes
\begin{align}
&\int_{\R^4}\D u\,\dot{\tilde{\phi}}(u) \int_{\R^4}\D y\left[\theta(t-y^4)-\theta(t-u^4)\right]\lambda_A(\tilde{\phi}, y, u)\nonumber\\
&\qquad= \int_{\R^4}\D \mathbf{u}\,\D v^4 \,\dot{\tilde{\phi}}(\mathbf{u},t+v^4) \int_{\R^4}\D \mathbf{y}\,\D w^4\left[\theta(v^4)-\theta(w^4)\right]\lambda_A(\tilde{\phi}, \mathbf{y},t+w^4, \mathbf{u}, t+ v^4)\,.
\end{align}
Then, going back to (\ref{P10}), we arrive at
\begin{equation}  \label{P11}
  E(T_t\tilde{\phi}, t)  = - L(T_t\tilde{\phi},t) + \int_{\R^4} \D u\, \dot{\tilde{\phi}}^A(\mathbf{u},t+u^4)\,P_A(\tilde{\phi},t,u)\,, 
\end{equation}
where $P_A(\phi,t,u)$ is 
\begin{equation}  \label{P11a}
P_A(\tilde{\phi},t,u) := \int_{\R^4} \D y \left[\theta(u^4)-\theta(y^4)\right]\,\lambda_A(\tilde{\phi},\mathbf{y}, t+y^4,\mathbf{u}, t+ u^4)\,. 
\end{equation}
\indent Note that the result (\ref{P11}) can be expressed in moving coordinates --$\phi^A=T_t\tilde{\phi}^A$--, i.e.,
\begin{equation}
E(\phi,t) := - L(\phi,t) + \int_{\R^4} \D u\,\dot{\phi}^A(\mathbf{u},u^4) \,P_A(T_{-t}\phi,t,u)\,,
\end{equation}
 where\footnote{It is worth mentioning that this result can be found more directly by using the transformations defined in Table (\ref{chap34-TableMF}), which makes us verify the consistency of the results.} 
 \begin{align}
 P_A(T_{-t}\phi,t,u)&:=\int_{\R^4} \D y \left[\theta(u^4)-\theta(y^4)\right]\frac{\delta \calL(T_y\phi, \mathbf{y},t+y^4)}{\delta \phi^A(u)}\,.
 \end{align}
Therefore, the total energy in moving coordinates can be written as $E(\phi, t) =\int_{\R^3} \D\mathbf{x}\, \mathcal{E}(T_{\mathbf{x}}\phi,x)$, where the energy density is
\begin{equation}  \label{P12}
\mathcal{E}(T_{\mathbf{x}}\phi,x) := - \mathcal{L}(T_{\mathbf{x}}\phi,x) + \int_\R \D u^4\,\dot{\phi}^A(\mathbf{x},u^4) \,P_A(T_{-t}\phi,t,\mathbf{x},u^4)\;.
\end{equation}

\subsection{The second Noether theorem: gauge transformations}\label{chap522-GS}
In the \textit{\important{gauge transformation}} case, the infinitesimal transformations (\ref{A3}) may depend on some arbitrary functions $\varepsilon^\alpha(x)$, $\alpha = 1,\ldots,N$, in a nonlocal manner\footnote{For the sake of simplicity, we shall only consider the case $\delta x^b = 0$.},
\begin{equation}  \label{GG1}
\delta\tilde{\phi}^A (x) = \int_{\R^4} \D y\, R^A_\alpha(x,y)\,\varepsilon^\alpha(y) \,.
\end{equation} 
Therefore, the term $\psi_A(\tilde{\phi}, x) \,\delta\tilde{\phi}^A(x)$ in (\ref{A9}) is
\begin{equation}  \label{GG1a}
 \psi_A(x) \,\delta\tilde{\phi}^A(x) =  \int_{\R^4} \D z \, \psi_A(x)\, R^A_\alpha(x,x+z )\,\varepsilon^\alpha(x+z) \,,
\end{equation} 
where we have taken $y=x+z$ and written $\psi_A(x) $ instead of $\psi_A(\tilde{\phi},x)$ to avoid an overloaded notation. Then, using the following identity
\begin{align}
\lefteqn{\psi_A(x)\, R^A_\alpha(x,x+z )\,\varepsilon^\alpha(x+z) - \psi_A(x-z )\, R^A_\alpha(x-z ,x)\,\varepsilon^\alpha(x) =} \nonumber\\
&\qquad \qquad \qquad  \int_0^1\D s \,z^b \partial_b\left[ \psi_A(x+[ s -1]z )\, R^A_\alpha(x+[ s -1]z ,x+ s  z )\,\varepsilon^\alpha(x+ s  z )\right] \,,
\end{align}
equation (\ref{GG1a}) becomes
\begin{equation}  \label{GG2a}
\psi_A(x) \,\delta\tilde{\phi}^A(x) = \varepsilon^\alpha(x)\,N(x) + \partial_b K^b(x)  \,, 
\end{equation}
with 
\begin{equation}  \label{GG2b}
N_\alpha(x):= \int_{\R^4} \D y\,\psi_A(y)\, R^A_\alpha(y,x)
\end{equation}
and
\begin{equation}
K^b(x) := \int_{\R^4} \D  z \,\int_0^1\D s \, z^b\, \psi_A(x+[ s -1] z )\, R^A_\alpha(x+[ s -1] z ,x+ s z)\,\varepsilon^\alpha(x+ s z)\,,
\end{equation}
which, on introducing the variable $y=x+s z \,,$ becomes
\begin{equation}  \label{GG3}
K^b(x) := \int_{\R^4} \D y \,\varepsilon^\alpha(y)\,\int_0^1 \frac{\D s}{s^{n+1}}\, (y^b - x^b) \, \psi_A(\xi )\, R^A_\alpha(\xi,y)\,,
\end{equation}
with $\displaystyle{\xi = \frac{s-1}s\,y +\frac1s\,x}$. Now, substituting (\ref{GG2a}) into the identity (\ref{A9}), we obtain the \textit{\important{second Noether theorem}} for nonlocal Lagrangian fields\footnote{This point is of interest in topics like the non-commutative U(1) theory \cite{Gomis2001_2}. In this particular case, the gauge transformation does not leave the field equations invariant, as would correspond to a Noether symmetry, but transforms them into equivalent ones, i.e., leaves the space $\mathcal{D}^\prime$ invariant \cite{Heredia2023}.}
\begin{equation}  \label{GG3a}
 \varepsilon^\alpha(x)\, N_\alpha(x) + \partial_b[K^b(x) + J^b(x)] + \delta \mathcal{L} (x)=0  \,.
\end{equation}
Aiming to encompass more general transformations than only Noether symmetries, we have used the identity (\ref{A9}) with $\delta \mathcal{L}(x)$ instead of the total divergence $\partial_b W^b(x)$.  Thus, the variation of the nonlocal Lagrangian also depends on the parameters $\varepsilon^\alpha(y)$ as 
\begin{equation}
\delta\mathcal{L}(x) := \int_{\R^4}\D y \,\varepsilon^\alpha(y)\,l_\alpha(x,y)\,.
\end{equation}
Furthermore, using (\ref{GG1}) and including that $\delta x^b = 0\,$, the current (\ref{A10}) can be written in the following manner
\begin{equation}  \label{GG3z}
J^b(x) := \int_{\R^4} \D y\,\varepsilon^\alpha(y)\,  \int_0^1 \frac{\D s}{s^{n+1}}\,\int_{\R^4} \D u\, (u^b - x^b) \, \lambda_A(\tilde\xi,u )\, R^A_\alpha(u,y)\,,
\end{equation}
where the variable $u=x+s z$ has been used and $\;\displaystyle{\tilde\xi = \frac{s-1}s\,u +\frac1s\,x}\,$.

\section{Hamiltonian formalism}

We shall now set up a \textit{\important{nonlocal!Hamiltonian formalism}} for the nonlocal Euler-Lagrange equations (\ref{L2o}). The procedure is similar to the one designed in Section \ref{chap34} for nonlocal mechanics. For this reason, we shall rely on table (\ref{chap34-TableMF}) to extend it from mechanics to fields.

\subsection{The Legendre transformation}
We introduce the \textit{\important{extended!phase space}} $\Gamma^\prime =\mathcal{K}^2\times\R$, which consists of the elements $(\phi, \pi, t)$.  The \textit{\important{nonlocal!Hamiltonian}} function in moving coordinates is defined as
\begin{equation}  \label{H1} 
 H(\phi, \pi,t)= \int_{\R^3}\D \mathbf{x}\,\calH(T_\mathbf{x}\phi,T_\mathbf{x}\pi,\mathbf{x},t)  \,,
\end{equation}
where $\calH(T_\mathbf{x}\phi,T_\mathbf{x}\pi,\mathbf{x},t)$ is the \textit{\important{nonlocal!Hamiltonian density}} 
\begin{equation}  
\calH(T_\mathbf{x}\phi,T_\mathbf{x}\pi,\mathbf{x},t)= \int_\R \D \sigma\,\pi_A(\mathbf{x},\sigma)\,\dot\phi^A(\mathbf{x},\sigma) -  \calL(T_\mathbf{x}\phi,\mathbf{x},t)  \,,
\end{equation}
$\calL(T_\mathbf{x}\phi, \mathbf{x},t)$ is a Lagrangian density, $\phi=(\phi^1 \ldots\phi^m),\,\pi=(\pi_1 \ldots\pi_m) \in\mathcal{K}$ are smooth functions. \\
\indent Furthermore, the \textit{\important{Poisson bracket}} is
\begin{equation}
\left\{ F, G \right\}_{(t)} = \int_{\R^4} \D \mathbf{x}\,\D\sigma\left(\frac{\delta F(t)}{\delta\phi^A(\mathbf{x},\sigma)}\,\frac{\delta G(t)}{\delta\pi_A(\mathbf{x},\sigma)} - \frac{\delta F(t)}{\delta \pi_A(\mathbf{x},\sigma)}\,\frac{\delta G(t)}{\delta\phi^A(\mathbf{x},\sigma)} \right) \,,
\end{equation} 
and the \textit{\important{Hamilton equations}} are 
\begin{eqnarray}  \label{H2a}
\mathbf{X}_H {\phi^A}(\mathbf{x},\sigma) & = & \frac{\delta H}{\delta \pi_A(\mathbf{x},\sigma)} = \dot\phi^A(\mathbf{x},\sigma)  \\ \label{H2b}
\mathbf{X}_H {\pi_A}(\mathbf{x},\sigma) & = & -\frac{\delta H}{\delta\phi^A(\mathbf{x},\sigma)} = \dot\pi_A(\mathbf{x},\sigma) + \int_{\R^3} \D\mathbf{z}\,\frac{\delta \calL(T_\mathbf{z}\phi,\mathbf{z},t)}{\delta\phi^A(\mathbf{x},\sigma)} \,,
\end{eqnarray}
where $\mathbf{X}_H$ is the \textit{\important{Hamiltonian!vector field}}, $\mathbf{X}_H = \partial_t + \{-,H\}$, that acts as follows:
\begin{equation}
(\phi^A, \pi_B, t)\longrightarrow \left(T_\tau\phi^A,T_\tau\pi_B, t+\tau \right)\,, \qquad \quad \tau^b = \tau\,\delta_4^b \,.
\end{equation}
Hamilton's equations can be written in a compact form through the contact differential 2-form
\begin{equation}  \label{H4} 
 \Omega^\prime = \Omega - \delta H \wedge \delta t\,, \qquad {\rm where} \qquad \Omega = \int_{\R^4} \D \mathbf{x}\,\D\sigma\,\delta \pi_A(\mathbf{x},\sigma) \wedge \delta \phi^A(\mathbf{x},\sigma) 
\end{equation}
is the symplectic form. Note (again) that we have written ``$\delta$'' for the differential on the manifold $\Gamma^\prime$ to distinguish it from the ``$\D$'' used in the notation for integrals we have adopted here. Then, Hamilton's equations (\ref{H2a}-\ref{H2b}) become
\begin{equation}  \label{H5} 
i_{\mathbf{X}_H} \Omega^\prime = 0\;.
\end{equation}
\indent So far, this Hamiltonian system in the extended phase space $\Gamma^\prime$ has little to do with the Lagrangian system (\ref{L3}) or the time evolution generator $\mathbf{D}_4$ in $\mathcal{D}^\prime\,$. However, they can be connected by the injection map
\begin{equation}  \label{H6} 
 (\phi, t)\in \mathcal{D}^\prime \stackrel{j}{\lhook\joinrel\longrightarrow} (\phi,\pi,t)\in\Gamma^\prime \,, \qquad  \mbox{where} \qquad \pi_A(\mathbf{x},\sigma) := P_A(\phi,t,\mathbf{x},\sigma)  \,,
\end{equation}
and $P_A(\phi,t,\mathbf{x},\sigma)$ is
\begin{equation}  \label{H7} 
P_A(\phi,t,\mathbf{x},\sigma) = \int_{\R^4} \D \mathbf{z}\,\D\zeta \left[\theta(\sigma)-\theta(\zeta)\right]\,\frac{\delta \calL(T_{\mathbf{z}}T_\zeta \phi,\mathbf{z},t+\zeta)}{\delta \phi^A(\mathbf{x},\sigma)}.
\end{equation}
\indent $j$ defines a one-to-one map from $\mathcal{D}^\prime$ into its range, $j(\mathcal{D}^\prime)\subset \Gamma^\prime$, i.e., the submanifold implicitly defined by the constraints
\begin{equation}  \label{H8} 
\Psi_A\left(\phi,x, z\right)= 0  \qquad {\rm and} \qquad 
\Upsilon_A\left(\phi, \pi, t,\mathbf{x},\sigma\right):= \pi_A(\mathbf{x},\sigma) - P_A(\phi,t,\mathbf{x},\sigma) = 0\;.
\end{equation}
\indent The Jacobian map $j^T$ maps the infinitesimal time evolution generator $\mathbf{D}_4$ in $\mathcal{D}^\prime$ with $\mathbf{X}_H$, i.e., with the Hamiltonian flow generator in $\Gamma^\prime$. Indeed, we have that\vspace{0.2cm}
\begin{equation}
j^T\left(\mathbf{D}_4\right) \phi^A (\mathbf{x},\sigma)= \mathbf{D}_4 \phi^A(\mathbf{x},\sigma) = \dot\phi^A(\mathbf{x},\sigma) =\mathbf{X}_H\phi^A(\mathbf{x},\sigma)\,.
 \end{equation}
Furthermore,
\begin{equation}
 j^T\left(\mathbf{D}_4\right) \pi_A(\mathbf{x},\sigma) = \mathbf{D}_4 P_A(\phi,t,\mathbf{x},\sigma) = \left[\partial_\varepsilon P_A(T_\varepsilon\phi,t+\varepsilon,\mathbf{x},\sigma)\right]_{\varepsilon=0}
\end{equation}
and, using (\ref{H7}) and (\ref{H2b}), we obtain 
\begin{align}
& j^T\left(\mathbf{D}_4\right) \pi_A(\mathbf{x},\sigma) = \partial_\varepsilon\left[\int_{\R^3} \D\mathbf{z}\int_\R \D \zeta\,\left[\theta(\sigma)-\theta(\zeta)\right]\,\frac{\delta\calL(T_{\mathbf{z}} T_{\zeta+\varepsilon} \phi,\mathbf{z},t+\zeta+ \varepsilon)}{\delta\phi^A(\mathbf{x},\sigma+\varepsilon)}\right]_{\varepsilon=0} \nonumber\\
 &= \left[\partial_\varepsilon \int_{\R^3} \D\mathbf{z}\int_\R \D \zeta^{\prime} \,\left[\theta(\sigma)-\theta(\zeta^{\prime} -\varepsilon)\right]\,\frac{\delta\calL(T_\mathbf{z}T_{\zeta^\prime} \phi,\mathbf{z},\zeta^{\prime}+t)}{\delta\phi^A(\mathbf{x},\sigma+\varepsilon)}\right]_{\varepsilon=0} \nonumber\\
 &=  \int_{\R^3} \D\mathbf{z}\int_\R \D \zeta^{\prime}\left\{\delta(\zeta^{\prime}) \frac{\delta\calL(T_\mathbf{z}T_{\zeta^\prime}\phi,\mathbf{z},t+\zeta^{\prime})}{\delta \phi(\mathbf{x},\sigma)} + \left[\theta(\sigma)-\theta(\zeta^{\prime})\right]\partial_{\sigma}\left[\frac{\delta\calL(T_\mathbf{z}T_{\zeta^\prime}\phi,\mathbf{z},t+\zeta^{\prime})}{\delta \phi(\mathbf{x},\sigma)} \right]\right\}\nonumber\\
&=  \partial_{\sigma} P_A(\phi,t,\mathbf{x},\sigma) +\int_{\R^3} \D\mathbf{z}\,\frac{\delta \calL(T_\mathbf{z}\phi, \mathbf{z},t)}{\delta\phi^A(\mathbf{x},\sigma)} = \mathbf{X}_H \pi_A(\mathbf{x},\sigma)\,,
\end{align}
where we have taken $\zeta^{\prime}=\zeta+\varepsilon$ and have used the second term on the right-hand side of the last but one line vanishes because of the nonlocal Euler-Lagrange equations (\ref{L3}).\\
\indent Consequently, $\mathbf{X}_H= j^T\left(\mathbf{D}_4\right)$ is tangent to the submanifold $j(\mathcal{D}^\prime)$, and therefore the constraints (\ref{H8}) are stable by the Hamiltonian flow.\\
\indent To translate the Hamiltonian formalism in $\Gamma^\prime$ into a Hamiltonian formalism in the extended dynamic space $\mathcal{D}^\prime$, we use that the pullback $j^\ast$ maps the contact form (\ref{H4}) onto the differential 2-form 
\begin{equation} \label{H9}
\omega^\prime  = j^\ast \Omega^\prime = \int_{\R^4} \D \mathbf{x}\,\D\sigma\,\delta P_A(\phi,t,\mathbf{x},\sigma)\wedge \delta \phi^A(\mathbf{x},\sigma) - \delta h \wedge \delta t \,, \qquad \quad \omega^\prime\in\Lambda^2(\mathcal{D}^\prime)\,,
\end{equation} 
where $h = H\circ j$. Then, since $j^T\left(\mathbf{D}_4\right) = \mathbf{X}_H$, the pullback of equation (\ref{H5}) implies that  
\begin{equation} \label{H11}
i_{\mathbf{D}_4} \omega^\prime = 0\;. 
\end{equation}
\indent The reduced Hamiltonian $h(\phi,t)$ and the contact form $\omega^\prime$ on $\mathcal{D}^\prime$ are derived using equations (\ref{H1}) and (\ref{H7}), and they are 
\begin{equation}  \label{H12} 
h(\phi,t) = \int_{\R^8} \D \mathbf{x}\,\D\sigma\,\D\zeta\,\D\mathbf{z}\left[\theta(\sigma)-\theta(\zeta)\right]\dot\phi^A(\mathbf{x},\sigma)\,\frac{\delta \calL(T_\mathbf{z}T_\zeta \phi,\mathbf{z},t+\zeta)}{\delta \phi^A(\mathbf{x},\sigma)}- L(\phi,t) \,, 
\end{equation}
where $L(\phi,t) = \int_{\R^3}\D\mathbf{x}\,\calL(T_\mathbf{x}\phi,\mathbf{x},t)$, and $\omega^\prime(\phi,t) = - \delta h(\phi,t) \wedge \delta t + \omega(\phi,t)$, and
\begin{align}  \label{H13} 
\omega(\phi,t) =& \frac{1}{2}\int_{\R^{12}} \D \mathbf{x}\,\D\sigma\,\D\mathbf{z}\,\D\zeta\,\D\mathbf{w}\,\D\rho \left[\theta(\sigma)-\theta(\rho)\right]\frac{\delta^2\calL(T_\mathbf{z}T_\zeta \phi,\mathbf{z},t+\zeta)}{\delta \phi^A(\mathbf{x},\sigma) \delta\phi^B(\mathbf{w},\rho)} \delta \phi^B(\mathbf{w},\rho) \wedge \delta\phi^A(\mathbf{x},\sigma)\nonumber\\
&\quad +\int_{\R^{8}} \D \mathbf{x}\,\D\sigma\,\D\mathbf{z}\,\D\zeta\, \left[\theta(\sigma)-\theta(\zeta)\right]\,\frac{\delta^2\calL(T_\mathbf{z}T_\zeta \phi,\mathbf{z},t+\zeta)}{\delta \phi^A(\mathbf{x},\sigma)\,\delta t}\, \delta t\wedge \delta \phi^A(\mathbf{x},\sigma)
\end{align}
is the (pre)symplectic form.\\
\indent As in the previous chapter,  we have not reached our goal yet because the constraints that characterize the extended dynamic space as a submanifold of the extended kinematic space $\mathcal{K}^\prime$ are $\Psi_A(\phi,x,z)=0$, and $\phi^A$ and $x^a$ are not independent coordinates in $\mathcal{D}^\prime$\,. Therefore, the remaining final step consists of coordinatizing $\mathcal{D}^\prime$. As in the case of nonlocal mechanics, we need to obtain the explicit parametric form of the submanifold $\mathcal{D}^\prime$ instead of the implicit form provided by the nonlocal Euler-Lagrange field equations. This fact is easy for regular local Lagrangian fields that depend on derivatives up to the $r^{\rm th}$ order since Euler-Lagrange's field equations are a partial differential system of order $2r$, and the Cauchy-Kowalewski theorem provides the sought parametric form. However, as a rule, deriving the explicit equations of $\mathcal{D}^\prime$ from the implicit equations in nonlocal case is a complex task that depends, as far as we consider, on each specific case.



\chapter{Applications}\label{chap6}
\section{Regular local Lagrangians}\label{chap41}
Let us consider a regular local Lagrangian $L_L(\tilde{q},\dot{\tilde{q}},t)=L_L(T_t\tilde{q}_0,T_t\dot{\tilde{q}}_0, t) = L_L(q_0,\dot q_0,t)$. We shall apply the previous section's results to determine the canonical momenta and the Hamiltonian in the dynamic space $\mathcal{D}^\prime\,$. 

As seen in Section \ref{chap22},  $\calD^\prime$ is coordinated by $(q,\dot q,t)$.  In order to calculate the momenta (\ref{L8a}) in moving coordinates, we first compute the functional derivative $\delta L(T_\zeta q,t+\zeta)/\delta q(\sigma)$, where we take $L(T_\zeta q,t+\zeta):=L_L(T_\zeta q_0,T_\zeta\dot q_0,t+\zeta)$. Indeed, it yields
\begin{equation}
\frac{\delta L(T_\zeta q,t+\zeta)}{\delta q(\sigma)}= \delta(\zeta-\sigma)\left(\frac{\partial L_L}{\partial q}\right)_{(q(\zeta),\dot q(\zeta),t+\zeta)} + \dot \delta(\zeta-\sigma)\left(\frac{\partial L_L}{\partial \dot q}\right)_{(q(\zeta),\dot q(\zeta),t+\zeta)}\,.
\end{equation}
Next, by plugging the latter into (\ref{L8a}), we get that the momenta are 
\begin{equation}\label{chap41-p}
P(q_0,\dot q_0, t,\sigma ) = \delta(\sigma)\,p(q_0,\dot q_0, t) \qquad \mathrm{with} \qquad p(q_0,\dot q_0, t):=\frac{\partial L_L(q_0,\dot q_0,t)}{\partial \dot q_0}\,.
\end{equation}
Therefore, the Hamiltonian (\ref{L12}) in the dynamic space $\calD^\prime$ is 
\begin{equation}
h(q_0,\dot q_0,t) = p(q_0,\dot q_0,t)\,\dot q_0 - L_L(q_0,\dot q_0,t) \,,
\end{equation}
and the contact form (\ref{L10}) is
\begin{equation}
\omega^\prime = \delta p \wedge \delta q_0 - \delta h\wedge \delta t \in \Lambda^2(\mathcal{D}^\prime)\,.
\end{equation}
The Hamilton equations $i_\mathbf{D} \omega^\prime = 0$ are then analogous to
\begin{equation}
\mathbf{D} q_0 = \dot q_0  \qquad {\rm and} \qquad  \mathbf{D} p = \frac{\partial L_L(q_0,\dot q_0,t)}{\partial q_0} \,,
\end{equation}
where $\mathbf{D}$ is the time evolution generator (\ref{chap23-D}). Notice that they are equivalent to the Euler-Lagrange equations in moving coordinates
 \begin{equation}
 \frac{\partial L_L(q_0,\dot q_0,t)}{\partial q_0} - \mathbf{D}\left(\frac{\partial L_L(q_0,\dot q_0,t)}{\partial \dot q_0}\right) = 0 \,.
\end{equation}

It is important to remark that, if one wishes, one could translate the above results into static coordinates $q_0=T_t \tilde q_0 =\tilde q(t)$, leading to the usual way of presenting these results in mechanics textbooks \cite{Goldstein2002}. 
 
 We could proceed similarly with a regular $r^{\rm th}$-order Lagrangian $L_L(\tilde{q},\dot{\tilde{q}},\ldots,\tilde{q}^{(r)},t)$ to obtain the Ostrogradsky momenta (\ref{ch254-OMD}) in moving coordinates. Indeed, for this case, we get that the momenta are
 \begin{equation}
 P(q,t,\sigma) =\int_\R \D\zeta\, g(\sigma,\zeta)\,  \sum^r_{i=1} \delta^{(i)}(\zeta-\sigma)\left(\frac{\partial L_L}{\partial q^{(i)}}\right)_{(q(\zeta),\ldots,q^{(r)}(\zeta),t+\zeta)}\,,
 \end{equation}
 where $g(\sigma,\zeta):= \left[\theta(\sigma)-\theta(\zeta)\right]$. Now, using the properties of the Dirac delta distribution and the $r^{\rm th}$-derivative Leibniz General Rule formula, we arrive at
 \begin{equation}
  P(q,t,\sigma) = \sum^r_{i=1} \int_\R \D \zeta (-1)^i \sum^{i}_{k=1} \binom{i}{k} g^{(k)}(\sigma,\zeta) \frac{\D^{i-k}}{\D \zeta^{i-k}}\left[\frac{\partial L_L}{\partial q^{(i)}}\right] \delta(\zeta-\sigma)\,,
 \end{equation}
where $g^{(k)}(\sigma,\zeta) := \frac{\D^k}{\D\zeta^k} g(\sigma,\zeta)$. Redefining the summations with $i= j+1$ and $k= l+1$ and using $\dot \theta(x) = \delta (x)$, we get 
 \begin{equation}
  P(q,t,\sigma) = \sum^{r-1}_{j=0} \int_\R \D \zeta (-1)^j \sum^{j}_{l=0} \binom{j+1}{l+1} \delta^{(l)}(\zeta) \frac{\D^{j-l}}{\D \zeta^{j-l}}\left[\frac{\partial L_L}{\partial q^{(j+1)}}\right] \delta(\zeta-\sigma)\,.
 \end{equation}
 Employing the ``Hockey-stick identity," we can show that the following relation for the binomial coefficient holds  
 \begin{equation}
 \binom{j+1}{l+1} = \sum^{j-l}_{k=0} \binom{j-k}{l}\,.
 \end{equation}
 Therefore, we can write the second summation as
 \begin{align}
 &\sum^{j}_{l=0} \binom{j+1}{l+1} \delta^{(l)}(\zeta) \frac{\D^{j-l}}{\D \zeta^{j-l}}\left[\frac{\partial L_L}{\partial q^{(j+1)}}\right] = \sum^j_{l=0} \sum^{j-l}_{k=0} \binom{j-k}{l} \delta^{(l)}(\zeta) \frac{\D^{j-l}}{\D\zeta^{j-l}}\left[\frac{\partial L_L}{\partial q^{(j+1)}}\right] \nonumber\\
 &= \sum^j_{k=0} \sum^{j-k}_{l=0} \binom{j-k}{l} \delta^{(l)}(\zeta) \frac{\D^{j-l}}{\D\zeta^{j-l}}\left[\frac{\partial L_L}{\partial q^{(j+1)}}\right]=\sum^j_{k=0} \frac{\D^{j-k}}{\D\zeta^{j-k}} \left[\delta(\zeta) \frac{\D^k}{\D\zeta^{k}}\left[\frac{\partial L_L}{\partial q^{(j+1)}}\right]\right]\,.
 \end{align}
 All together,
 \begin{equation}
 P(q,t,\sigma) = \sum^{r-1}_{j=0} \sum^{j}_{k=0}(-1)^{j-k} \left.\left[-\frac{\D}{\D \zeta}\right]^k\left(\frac{\partial L_L}{\partial q^{(j+1)}}\right)\right|_{\zeta=0} \delta^{(j-k)}(\sigma)\,,
 \end{equation}
where we have used (again) the Dirac delta distribution properties. Redefining the summations with $s= j+1$ and $m=s-k-1$, we get 
 \begin{align}
P(q,t,\sigma) = \sum_{m=0}^{r-1}  p_m\,(-1)^m\,\delta^{(m)}(\sigma)\,,
\end{align} 
where $p_m$ are the Ostrogradsky momenta (\ref{ch254-OMD}) in moving coordinates.

\section{Nonlocal Lagrangian mechanics}\label{chap42}

\subsection{Nonlocal harmonic oscillator}\label{chap421}
Consider the nonlocal action integral 
\begin{equation}  \label{OS1}
S(\tilde q, R) = \int_{|t|\leq R} \D t\,\left[\frac12\,\dot{\tilde{q}}^2(t) -\frac{\omega^2}2\, \tilde{q}^2(t) + \frac{g} 4 \, \tilde{q}(t)\,\int_\R \D \zeta \,K(\zeta)\, \tilde{q}(t-\zeta) \right]
\end{equation}
with $ K(\zeta) = e^{-|\zeta|}\,$ and $g\in\R$. Comparing it with expression (\ref{chap31-SR}), we have that
\begin{align}
L\left(T_t\tilde{q}\right) &= \frac12\,\dot{\tilde{q}}^2(t) -\frac{\omega^2}2\, \tilde{q}^2(t) + \frac{g} 4\,\tilde{q}(t)\,\int_\R \D \zeta \,K(\zeta)\, \tilde{q}(t-\zeta)\nonumber\\
&=  \frac12\, T_t\dot{\tilde{q}}^2_0 -\frac{\omega^2}2\, T_t\tilde{q}^2_0 + \frac{g} 4\,T_t\tilde{q}_0\,\int_\R \D \zeta \,K(\zeta)\, T_t\tilde{q}(-\zeta)\,.
\end{align}
Hence, definition (\ref{chap31-Lambda}) yields
\begin{equation}  \label{OS2}
\lambda(\tilde{q},t,\sigma) = \dot{\tilde{q}}(t)\,\dot\delta(t-\sigma) +\left[\frac{g}4\,(K\ast \tilde{q})_{(t)} - \omega^2\,\tilde{q}(t)\right]\,\delta(t-\sigma)+ \frac{g}4\,\tilde{q}(t)\,K(t-\sigma)  \,,
\end{equation}
where $(K\ast \tilde{q})_{(t)}$ is the convolution. Consequently, the Euler-Lagrange equations (\ref{chap31-EOMS}) are
\begin{equation}  \label{OS3}
\psi(\tilde{q},\sigma) = -\ddot{\tilde{q}}(\sigma) - \omega^2 \tilde{q}(\sigma) +\frac{g}2\,(K\ast\tilde{q})_{(\sigma)}=0\,,
\end{equation}
which are nonlocal due to the convolution product. 

We must now set up the dynamic submanifold $\mathcal{D}^\prime$ given by the constraints (\ref{OS3}) in the parametric form:
\begin{equation}  \label{OS5}
 \ddot{\tilde{q}}(\sigma) + \omega^2 \tilde{q}(\sigma) -\frac{g}2\,(K\ast \tilde{q})_{(\sigma)}=0 \,.
\end{equation}
On differentiating twice with respect to $\sigma$ and including that 
\begin{equation}
\frac{\D^2 e^{-|\sigma|}}{\D\sigma^2} = e^{-|\sigma|}- 2 \,\delta(\sigma) \,,
\end{equation}
the constraints become
\begin{equation}  \label{OS6}
\tilde{q}^{(iv)} + (\omega^2-1)\,\ddot{\tilde{q}} +(g -\omega^2)\tilde{q} =0  \,.
\end{equation}
Notice that any solution of the nonlocal equation (\ref{OS5}) is also a solution of the local differential equation (\ref{OS6}), while the converse is not necessarily true.  
The general solution of (\ref{OS6}) is 
\begin{equation}  \label{OS7}
\tilde{q}(\sigma) = \sum_{j=1}^4 A^j \,e^{\sigma \,r_j} \,,
\end{equation}
where $r_j$ are the roots of the characteristic equation  $r^4 +(\omega^2 - 1) r^2 + g-\omega^2 = 0\,$; that is, 
\begin{equation}  \label{OS8}
r_{\pm\pm} = \pm\, r_\pm\,, \qquad \quad r_\pm = \sqrt{\frac{1-\omega^2}2 \pm \sqrt{\Delta}} \,,\qquad {\rm with} \qquad \Delta = \frac{(\omega^2+1)^2}4 - g \,.
\end{equation}
For such a function $\tilde{q}(\sigma)$ to be a solution of (\ref{OS5}), the convolution $(K\ast \tilde{q})$  must exist, which implies that $\int_{-\infty}^\infty \D\tau\, e^{-|\tau| + \tau\,r_j} < +\infty\,$, that is, $|\R e (r_j) | < 1\,$. Therefore, the general solution of (\ref{OS5}) is then
\begin{equation}  \label{OS9}
\tilde{q}(\sigma) = \sum^4_{j=1} A^j \,e^{\sigma\, r_j} \,, \quad \mbox{for those $r_j$ such that} \quad |\R e (r_j) | < 1  \,.
\end{equation}
This equation is the parametric equation of the dynamic submanifold $\mathcal{D}^\prime$, and the coordinates are $A^j\,$. 

Since the nonlocal Lagrangian does not explicitly depend on time, we note that the nonlocal Euler-Lagrange equations in moving coordinates are equivalent to the nonlocal Euler-Lagrange equations in static coordinates: 
\begin{equation}
\Psi(q,\sigma) = -\ddot{q}(\sigma) - \omega^2 q(\sigma) +\frac{g}2\,(K\ast q)_{(\sigma)} = 0\,.
\end{equation}
Therefore, the solution structure (\ref{OS9}) is the same for moving coordinates, namely,
\begin{equation}\label{chap721-qs}
q(\sigma) = \sum^4_{j=1} B^j \,e^{\sigma\, r_j} \,, \quad \mbox{for those $r_j$ such that} \quad |\R e (r_j) | < 1  \,.
\end{equation}
It is worth noting that $q(0) = \sum_j B^j = \tilde q(t) = \sum_j A^j e^{r_j t}$; therefore, it implies that the relation between $A^j$ and $B^j$ is:  $B^j =A^j e^{r_j t}$. 

\begin{figure}[t]
\begin{center}
\includegraphics[width=0.5\textwidth]{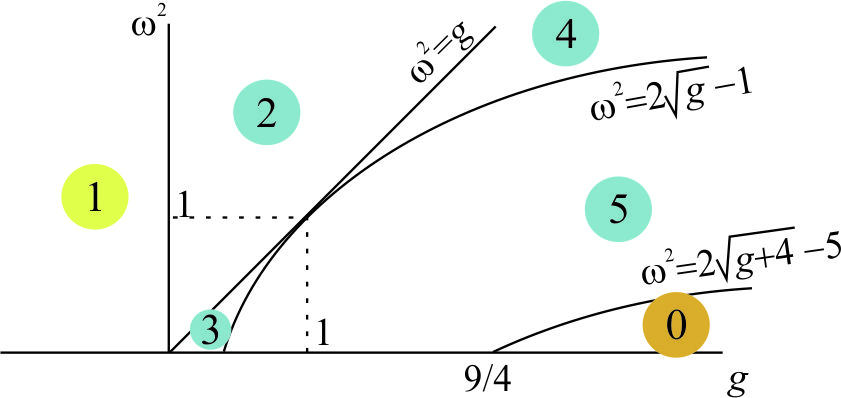}
\end{center}
\caption{Regions in the parameter space $(g,\omega^2)$}  \label{F1}
\end{figure}

In Figure \ref{F1}, the parameter space is divided into several regions according to the number and roots $r_j$, which are listed in the following table:
\begin{table}[t]
\begin{center}
\renewcommand{\arraystretch}{1.2}
\begin{tabular}{|c|c||c|c||c|}
\hline
\multicolumn{2}{|c|}{Region} & Real $r_j$ & Imaginary $r_j$ & $|\R e (r_j) | < 1$ \\
\hline
0 & $\omega^2 \leq 2\,\sqrt{g+4} - 5 $ & 0 & 0 & 0 \\
\hline
1 & $g \leq 0$ & 2 & 2 & 2 \\
\hline
2 & $\omega^2 \geq g > 0$ & 2 & 2 & 4 \\
\hline
3 & $ 2 \sqrt{g}-1 \leq \omega^2 < \min\{g,\,1\}$ & 4 & 0 & 4 \\
\hline
4 & $ \max\{1,\,2 \sqrt{g}-1\} < \omega^2 < g $ & 0 & 4 & 4\\
\hline
5 & $ 2\sqrt{g+4}-5 < \omega^2 <  2 \sqrt{g}-1$ & 0 & 0 & 4 \\
\hline
\end{tabular}
\caption{This table shows the number of roots in each region. The column ``Real $r_j$" represents the number of purely real roots while the column ``Imaginary $r_j$" refers to the number of purely imaginary roots. Finally, the last column ``$|\R e (r_j) | < 1$" indicates the number of roots that satisfy this condition.}
\end{center}
\label{T:Tparm}
\end{table}
\newpage
Discussing the stability of the solutions of (\ref{OS6}) is a relatively easy task because it is a linear equation with constant coefficients. The solutions are stable if the characteristic roots are simple and their real part is non-positive \cite{pontriaguine1969, Dettman1967}. As the characteristic equation only contains even powers of $r$, its roots come in pairs with different signs. Hence, the solutions are stable if all characteristic roots are simple and imaginary, namely, whenever the parameters $(g,\omega^2)$ lie in regions $1$ or $4$ in Figure \ref{F1}. Notice that, in region $1$, we have two real characteristic roots with $|\R e (r_j) | > 1$, so they do not contribute to the solution of (\ref{OS5}). They do not affect stability. 

The latter contradicts the widespread belief that nonlocal Lagrangians, and also local higher-order Lagrangians, suffer from \textit{\important{Ostrogradsky!instability}} \cite{Chen2013,Woodard2015}: as energy (the Hamiltonian) is not bounded from below, the system is unstable. In our view, this inference results from taking a sufficient condition of stability, namely, ``energy is bounded from below'', as a necessary condition. We shall illustrate this fact later in this Section.

On the other hand, the canonical momenta (\ref{L8a}) are
\begin{equation}  \label{OS4}
P(q,\sigma) = \delta(\sigma)\,\dot q(\sigma) + \frac{g}{4}\theta(\sigma)(K\ast q)_{(\sigma)} - \frac{g}{4}\int^\infty_0 \D\zeta\,K(\zeta-\sigma)\,q(\zeta)\,,
\end{equation}
which, by substituting (\ref{chap721-qs}) into (\ref{OS4}), we then obtain that, when $\sigma<0$,  
\begin{equation} 
P(q,\sigma) = \sum^4_{j=1} B^j \left(r_j \,\delta(\sigma) + \frac{g}4\,\frac{e^{\sigma}}{r_j -1} \right)  \,,
\end{equation}
and, when $\sigma>0$,
\begin{equation}  
P(q,\sigma) = \sum^4_{j=1} B^j \left(r_j \,\delta(\sigma) + \frac{g}4\,\frac{e^{-\sigma}}{r_j +1} \right)  \,.
\end{equation}
Thus, the canonical momentum is then
\begin{equation}\label{OS10}
P(q,\sigma) = \sum^4_{j=1} B^j \left(r_j \,\delta(\sigma) + \frac{g}4\,\frac{e^{-|\sigma|}}{r_j +\sign(\sigma)} \right)  \,.
\end{equation}

\subsubsection{The reduced Hamiltonian formalism when ${\rm dim}\,\mathcal{D}^\prime=5$}\label{chap7211}
Since the nonlocal Lagrangian does not explicitly depend on time, let us note that $\delta L(T_\zeta q, t+\zeta)/\delta q(\sigma) = \delta L(T_\zeta q)/\delta q(\sigma) = \lambda(q,\zeta,\sigma)$, and $\omega^\prime=\omega$. By directly replacing equation (\ref{OS2}) into equation (\ref{L13}), we arrive at the expression for the contact form
\begin{equation}
\omega^\prime = \sum^4_{j,k=1}\left(r_j + \frac{g(r_k-r_j)}{2 (r_k^2-1)(r_j^2-1)} \right)\,\delta B^j \wedge \delta B^k\,, 
\end{equation}
which can be easily factored as
\begin{equation}  \label{OS10a}
\omega^\prime = \left(\sum^4_{j=1}r_j\,\delta B^j\right) \wedge \left(\sum^4_{k=1} \,\delta B^k \right) + g\,
\left(\sum^4_{j=1} \frac{\delta B^j}{1-r_j^2}\right) \wedge \left(\sum^4_{k=1} \,\frac{ r_k\,\delta B^k}{1-r_k^2} \right)\,.
\end{equation}
Similarly, the reduced Hamiltonian is derived by combining (\ref{OS10}) and (\ref{L12}), and it is
\begin{equation}  \label{OS10b}
h = \frac12 \dot q^2_0 + \frac{\omega^2}2\,q^2_0 -\frac{g}2 \,\sum^4_{j,k=1} B^j B^k\,\frac{1+r_j r_k - r_j^2-r_k^2}{(1- r_j^2)(1-r_k^2)}  \,.
\end{equation}

The coordinates $B^j$ are related to $q_0\,,\,\dot q_0\,,\ldots,\,\dddot{q}_0$ through equation (\ref{OS7}), which implies that
\begin{equation}
 \sum_j r_j^n \,B^j= q^{(n)}_0 \,, \qquad \quad n= 0,\ldots,3 \,.
 \end{equation}
Solving the latter for $B^j$ and substituting the result into (\ref{OS10a}), we obtain that the (pre)symplectic form is
\begin{equation}
\omega = \delta\dot q_0 \wedge \delta q_0 + \frac1g\,\delta\left[\omega^2 q_0 +\ddot q_0\right] \wedge \delta\left[\omega^2\dot q_0 +\dddot{q}_0\right] \,, 
\end{equation}
which is non-degenerate and, therefore, symplectic. Furthermore, two pairs of canonical coordinates are apparent  
\begin{equation}  \label{OS11}
 q_0  \,,  \quad p_0 := \dot q_0 \,, \quad \pi_0 := \frac1{\sqrt{g}}\,(\omega^2\, q_0 +\ddot q_0), \quad \xi_0 := \frac1{\sqrt{g}}\,(\omega^2\,\dot q_0 + \dddot{q}_0) \,.
\end{equation}
In terms of them, the symplectic form becomes
\begin{equation}  \label{OS12}
 \omega = \delta p_0 \wedge \delta q_0 + \delta \pi_0 \wedge \delta \xi_0 \,.
\end{equation}
The symplectic form has an associated Poisson bracket, and the non-vanishing elementary Poisson brackets are
\begin{equation}
\{q_0,p_0\}=\{\xi_0,\pi_0\} = 1\,.
\end{equation}

Similarly, the reduced Hamiltonian (\ref{OS10b}) in term of (\ref{OS11}) is 
\begin{equation}  \label{OS13}
 h(p_0,q_0,\xi_0,\pi_0) = \frac12\, p^2_0 + \frac{\omega^2}2\,q^2_0 - \sqrt{g}\,q_0\,\pi_0 + \frac12\,\pi^2_0 - \frac12\,\xi^2_0\,,
\end{equation}
\newpage 
\noindent and it can be easily checked that the Hamilton equations are equivalent to the fourth-order equation (\ref{OS6}).  It can also be checked that the coordinate change
\begin{equation}
 q_0= \hat{q}_0\,, \qquad p_0 = \hat{p}_0 + \frac{g}2\,\hat{q}_1 \,, \qquad \sqrt{g}\,\pi_0 = \hat{p}_1 + \frac{g}2\,\hat{q}_0
\,, \qquad \xi_0 = \sqrt{g}\,\hat{q}_1
\end{equation}
transforms the Hamiltonian (\ref{OS13}) and the symplectic form (\ref{OS12}) into the Hamiltonian system derived in \cite{Gomis2004} for the same nonlocal oscillator. 

As commented above, if the parameters $(\omega^2,g)$ lie in subregion $4$ in Figure \ref{F1}, then the system is stable: a small perturbation in the initial data may cause the perturbed solution to oscillate slightly around the non-perturbed solution. At the same time, the oscillation neither blows up nor fades away asymptotically. However, the energy --an integral of motion-- is not bounded from below. As commented above, this does not imply that the system is unstable. Indeed, the property that ``energy is bounded from below'' is a sufficient, but not necessary, condition for stability.

\subsubsection{The reduced Hamiltonian formalism when ${\rm dim}\,\mathcal{D}^\prime=3$\label{S5.2.2} } 
When the parameters fall in region 1 in Figure \ref{F1}, ${\rm dim}\,\mathcal{D}^\prime=2+1$, only two characteristic roots $\pm\,r_-\,$ contribute to the solution (\ref{OS9}) of equation (\ref{OS6}). By a similar procedure as in Section \ref{chap7211}, we can derive the contact form
\begin{equation}  \label{OS14}
 \omega^\prime = \delta p_0 \wedge \delta q_0 \,, \qquad {\rm with} \qquad p_0 := M\,\dot q_0\,, 
\end{equation}
and the Hamiltonian
\begin{equation}  \label{OS15}
h(p_0,q_0) = \frac{p^2_0}{2 M} + \frac{K}2\, q^2_0\, \qquad {\rm with} \qquad M := 1 - \frac{g}{x_-^2}  \qquad {\rm and} \qquad K := \omega^2 - g\, \frac{2x_- - 1}{x_-^2}  \,, 
\end{equation}
i.e., an oscillator with frequency 
\begin{equation}  \label{OS16}
 \tilde\omega^2 = \frac{\omega^2 x_-^2 - g (2 x_- - 1)}{x_-^2 - g} \,, \qquad \mathrm{and}\qquad x_- = \frac{1+\omega^2}2 + \sqrt{\frac{(1+\omega^2)^2}4 - g} \,.
\end{equation}
The latter corresponds to the perturbative sector's Hamiltonian  (43-45) in \cite{Gomis2004} \footnote{If the reader is interested in another example similar to the nonlocal harmonic oscillator, we recommend \cite{Heredia2} where the nonlocal delayed harmonic oscillator is presented using this nonlocal formalism. }.

\subsection{$p$-adic particle}\label{chap423}
We now consider the $p$-adic open string Lagrangian \cite{GHOSHAL2000,Moeller2002,Gomis2004} neglecting the space dependence
\begin{equation}
 L= -\frac12\,\tilde{q}(t)\,e^{-r \partial_t^2}\,\tilde{q}(t) + \frac1{p+1}\,\tilde{q}^{p+1}(t)\,,  \qquad r=\frac12\,\log p \,,
 \end{equation} 
where $p$ is a prime integer. The linear operator $e^{-\partial_t^2}\,\tilde{q}(t)$ can be treated either as an infinite formal Taylor series that includes the coordinate derivatives at any order or as the following convolution \cite{Vladimirov2006}\newpage
\begin{equation}  \label{pS1}
e^{-\partial_t^2}\,\tilde{q}(t) \equiv (G\ast \tilde{q})_{(t)} = \int_{\R} \D t^\prime\,G(t^\prime)\,\tilde{q}(t-t^\prime) \,,\qquad {\rm where} \qquad 
G(t) = \frac1{2\sqrt{\pi r}}\,e^{- \frac{t^2}{4 r}}  \,.
\end{equation}
As far as our formalism is concerned, we shall take the second option. We shall treat the operator $e^{-\partial^2_t}$ through the convolution since, as studied in detail in \cite{Heredia2023_1,Carlsson2016}, the functional space of the convolution operator is broader than that of infinite derivatives. Therefore, the nonlocal action integral (\ref{chap31-SR}) for the above Lagrangian can be written as
\begin{equation}  \label{pS2}
S(\tilde{q},R) = \int_{|t|\leq R} \D t\,\int_\R\D t^\prime  \left[-\frac12\,G(t^\prime)\,\tilde{q}(t)\,\tilde{q}(t-t^\prime) + \frac1{p+1}\,\tilde{q}^{p+1}(t)\,\delta(t-t^\prime)  \right]\,,
\end{equation}
where the nonlocal Lagrangian is 
\begin{align}
L\left(T_t \tilde{q}\right) &= -\frac12\, \tilde{q}(t)\,(G\ast \tilde{q})_{(t)} + \frac1{p+1}\,\tilde{q}^{p+1}(t)\nonumber\\ 
&= -\frac12\, T_t\tilde{q}(0)\,\int_\R \D t^\prime G(t^\prime) T_t\tilde{q}(-t^\prime) + \frac1{p+1}\,T_t\tilde{q}^{p+1}(0)\,. 
\end{align}
\indent Furthermore, the functional derivative (\ref{chap31-Lambda}) is 
\begin{equation}  \label{pS3}
\lambda(\tilde{q},t,\sigma) = \delta(t-\sigma)\,\left[-\frac12\,(G\ast \tilde{q})_{(t)} +\tilde{q}^p(t) \right] - \frac12\,\tilde{q}(t)\,G(t-\sigma)\,,
\end{equation}
and consequently, the nonlocal Euler-Lagrange equations (\ref{chap31-EOMS}) easily follow and have the form of a convolution equation:
\begin{equation}  \label{pS4}
(G \ast \tilde{q})_{(t)} = \tilde{q}^p(t)  \,.
\end{equation}
\indent According to Vladimirov and Volovich \cite{Vladimirov2006}, the only solutions in the space of tempered distributions \cite{Vladimirov_GF} are 
\begin{equation}  \label{pS4a}
 \tilde{q}_{\mathrm{o}}(t) = \left\{\begin{array}{ll}
                   \pm 1, \; 0\,, \qquad &\; \mbox{if $p$ odd} \\
									  \hspace*{.7em} 1, \; 0\,, \qquad &\; \mbox{if $p$ even\,.}
										\end{array}   \right.
\end{equation}	
If we loosen the requirement that solutions must be tempered distributions, the convolution equation can have additional solutions. We shall focus on perturbative ones around $q_0(t)$, namely,
\begin{equation} 
 \tilde{q}(t) = \tilde{q}_\mathrm{o}(t) + \kappa\,\tilde{y}(t) \,, 
\end{equation}										
where $\kappa \ll 1$ is the expansion parameter. Substituting the latter in (\ref{pS4}) and using Newton's formula for $\tilde{q}^p(t)$, we obtain
\begin{equation}  \label{pS5}
G \ast \tilde{y} - p\,\tilde{q}_\mathrm{o}^{p-1}\, \tilde{y} = \kappa\, F \,, \qquad {\rm with } \qquad F := \sum_{n=0}^{p-2} {p\choose n+2}\,\tilde{q}_\mathrm{o}^{p-n-2}\,\kappa^n\,\tilde{y}^{n+2} \,.
\end{equation}
For the sake of simplicity, we will focus on the case $p=2$, which implies that $\tilde{q}_\mathrm{o}(t)=1$ and $F = \tilde{y}^2$. Thus, writing $\tilde{y} = \sum_{n=0}^\infty \kappa^n \tilde{y}_n$ and expanding equation (\ref{pS5}) as a power series of $\kappa$ leads to\vspace{0.2cm}
\begin{equation}  \label{pS6}
G \ast \tilde{y}_n - 2\, \tilde{y}_n = \sum_{m+k=n-1}\,\tilde{y}_k \,\tilde{y}_m\,, \qquad \quad n\geq 0\,,
\end{equation}
which provides an iteration sequence that can be solved order by order: at each level, the equation to be solved is linear and contains a non-homogeneous term that depends on the lower order solutions. The general solution is the result of adding a particular solution to the general solution of the homogeneous equation. \\
\indent At the lowest order $n=0$, it reads
\begin{equation}  \label{pS6a}
G \ast \tilde{y}_\mathrm{o} - 2\, \tilde{y}_\mathrm{o} =  0 
\end{equation}
and admits exponential solutions $\tilde{y}_\mathrm{o}(t) = e^{i \alpha t}$, where the exponent factor $\alpha$ is the roots of the spectral equation
\begin{equation}  \label{pS7}
f(\alpha) := e^{-r \alpha^2} - 2 = 0\,, 
\end{equation}
and the parameter $r$ comes from the kernel $G$ defined in (\ref{pS1}). The general solution of (\ref{pS6a}) is then
\begin{equation} 
  \tilde{y}_\mathrm{o}(t) = \sum_{\alpha\in\mathcal{S}} C_\alpha \,e^{i \alpha t}  \,, \qquad {\rm where} \qquad 
	\mathcal{S} = \left\{\alpha\in\mathbb{C} \,\,\, | \,\,\, f(\alpha) = 0\right\}\,,
\end{equation}
is the infinite set of all complex solutions of $f(\alpha) = 0$. Substituting the latter in the following order ($n=1$) of equation (\ref{pS6}), we obtain that
\begin{equation}  \label{pS6b}
G \ast \tilde{y}_1 - 2\, \tilde{y}_1 =  \sum_{\beta,\gamma\in\mathcal{S}} C_\gamma C_\beta \,e^{i (\gamma +\beta) t}  \,.
\end{equation}
Its Fourier transform is $f(\alpha) \,\tilde{y}_1(\alpha) = \sqrt{2\pi}\,\sum_{\beta,\gamma\in\mathcal{S}} C_\gamma C_\beta\,\delta(\alpha-\beta -\gamma)$; hence, we arrive at
\begin{equation} 
\tilde{y}_1(t) = \sum_{\beta,\gamma\in\mathcal{S}} C_\gamma C_\beta\,\frac{1}{f(\beta+\gamma)}\,e^{i (\beta +\gamma)t} 
\end{equation}
by inverting the Fourier transform. It is worth remarking that the right-hand side is finite because, as it can be easily proved, if $\beta,\gamma \in \mathcal{S}\,$, then $\beta + \gamma \notin \mathcal{S}\,$. Furthermore, it can be easily inferred that $y_n(t)$ is a polynomial of degree $n+1$ in the variables $C_\alpha\,$.\\
\indent Since the nonlocal Lagrangian is not explicitly time-dependent, the nonlocal Euler-Lagrange equations in moving and static coordinates are equivalent; therefore, the structure of the solution in moving coordinates shall be the same. Thus, bearing in mind $q(\sigma) = \tilde q(\sigma+t)$, we find that the perturbative solution in moving coordinates is
\begin{equation}  \label{pS4b}
q(\sigma) = q_{\mathrm{o}}(\sigma) + \kappa\,y_{\mathrm{o}}(\sigma) + \kappa^2\,y_1 (\sigma) + \mathcal{O}(\kappa^3)\,,
\end{equation}
where 
\begin{equation}\label{pS7a}
y_\mathrm{o}(\sigma) = \sum_{\alpha\in\mathcal{S}} Y_\alpha \,e^{i \alpha \sigma}\,, \qquad {\rm with} \qquad 
	\mathcal{S} = \left\{\alpha\in\mathbb{C} \,\,\, | \,\,\, f(\alpha) = 0\right\}\,
\end{equation}
and
\begin{equation} \label{pS7b}
y_1(\sigma) = \sum_{\beta,\gamma\in\mathcal{S}} Y_\gamma Y_\beta\,\frac{1}{f(\beta+\gamma)}\,e^{i (\beta +\gamma)\sigma}\,,\qquad {\rm with} \qquad Y_\alpha = C_\alpha e^{i\alpha t}\,.
\end{equation}

\subsubsection{The symplectic form}

We now compute (\ref{L13}) to obtain the contact form
\begin{equation}  \label{pS8}
\omega =  \frac12\,\int_{\R} \D s\,G(s) \,\int_0^s \D\sigma\,\delta q(\sigma) \wedge \delta q(\sigma-s)  \,,
\end{equation}
where we have used $s:=\sigma-\zeta$. Thus, by combining (\ref{pS4b}), (\ref{pS7a}), and (\ref{pS7b}), we get that 
\begin{equation}
\delta q(\sigma) = \kappa\,\sum_{\alpha} \delta Y_\alpha\,\left(e^{i \alpha \sigma} + 2  \kappa \,\sum_{\beta} Y_\beta\, \frac{e^{i(\alpha+\beta) \sigma}}{f(\alpha+\beta)} \right) + O(\kappa^3)\,.
\end{equation}
Using this expression, the integral on the right-hand side of (\ref{pS8}) yields
\begin{align}\label{pS9} 
&\int_0^s \D\sigma\,\delta q(\sigma) \wedge \delta q(\sigma-s) = \kappa^2\,\left[\sum_{\alpha}s\,e^{i\alpha s}\,\delta Y_\alpha \wedge \delta Y_{-\alpha} - \sum_{\alpha+\gamma\neq 0} \frac{i \left(e^{i\alpha s} - e^{-i\gamma s}\right)}{\alpha+\gamma}\,\delta Y_\alpha \wedge \delta Y_{\gamma}  \right.
\nonumber\\
& \,\, \left.- 2 \kappa\,\sum_{\alpha,\gamma,\beta} \left(\frac{i \left(e^{i(\alpha+\beta) s} - e^{-i\gamma s}\right)}{f(\alpha+\beta)} +\frac{i \left(e^{i\alpha s} - e^{-i(\gamma+\beta) s}\right)}{f(\beta+\gamma)}\right)\,\frac{Y_\beta\,\delta Y_\alpha \wedge \delta Y_{\gamma}}{\alpha+\beta+\gamma}  \right] + O(\kappa^4)\,,
\end{align}
where we have included the fact that
\begin{equation}  \label{pS9a}
\int_0^s \D\sigma\,e^{i \nu \sigma - i\mu s} = 
\left\{  \begin{array}{ll}
    \displaystyle{\frac{- i}{\nu}\, \left(e^{i(\nu-\mu) s} - e^{-i\mu s}\right)   } & \quad \nu\neq 0 \\[2ex]
		  s \,e^{-i \mu s} & \quad \nu=0
			\end{array}  \right.  
\end{equation}			
and, if $\alpha,\beta,\gamma \in \mathcal{S}\,$, then $\alpha +\beta \notin \mathcal{S}\,$, and $\alpha+\beta+\gamma \neq 0$. Then, substituting (\ref{pS9}) into (\ref{pS8}), we can write
\begin{align}
\omega = \kappa^2\sum_{\alpha} \omega_\alpha\,\delta Y_\alpha \wedge \delta Y_{-\alpha} &+ \kappa^2\sum_{\alpha+\gamma\neq 0} \omega_{[\alpha\gamma]}\,\delta Y_\alpha \wedge \delta Y_{\gamma}+ \kappa^3\sum_{\alpha,\gamma,\beta} \omega_{[\alpha\gamma]\beta}\,  Y_\beta\,\delta Y_\alpha \wedge \delta Y_{\gamma} +O(\kappa^4) 
\end{align}
with
\begin{align}
 \omega_\alpha &= \frac12\,\int_\R \D s \,G(s)\,s\,e^{i\alpha s} = 2\,i\,\alpha \,r \nonumber\\
\omega_{\alpha\gamma} &= \frac{-i}{2(\alpha+\gamma)}\,\int_\R \D s \,G(s)\,\left(e^{i\alpha s} - e^{-i\gamma s}\right) = \frac{-i (e^{-r \alpha^2} - e^{-r \gamma^2})}{2(\alpha+\gamma)} = 0 \nonumber\\
\omega_{\alpha\gamma\beta}&= \frac{-i}{\alpha+\gamma+\beta}\,\int_\R \D s \,G(s)\,\left(\frac{e^{i(\alpha+\beta)s} - e^{-i\gamma s}}{f(\alpha+\beta)} + \frac{e^{i\alpha s} - e^{-i(\gamma+\beta) s}}{f(\gamma+\beta)} \right) \nonumber\\
 & = \frac{-i}{\alpha+\gamma+\beta}\,\left(\frac{e^{-r (\alpha+\beta)^2} - e^{-r \gamma^2}}{f(\alpha+\beta)} + \frac{e^{-r \alpha^2} - e^{-r(\gamma+\beta)^2}}{f(\gamma+\beta)} \right) = 0 \,,
\end{align}
where the fact that $f(\alpha)=f(\gamma)=f(\beta)=0$ has been included. Finally, we arrive at
\begin{equation}  \label{pS10}
\omega = 4\, i \,r\,\kappa^2 \,\sum_{\alpha\in \mathcal{S}^+} \alpha\,\delta Y_\alpha \wedge \delta Y_{-\alpha} + O(\kappa^4)\,,
\end{equation}
where $\mathcal{S}^+ = \{\alpha\in \mathbb{C}\,|\;f(\alpha)=0,\; \mathbb{R}\mathrm{e}(\alpha)> 0,\; {\rm or}\; \mathbb{R}\mathrm{e}(\alpha)= 0,\; {\rm and}\;\mathbb{I}\mathrm{m}(\alpha)>0\}$. It is apparent that $\omega$ is non-degenerate, hence symplectic. Furthermore, $Y_\alpha, Y_{-\alpha}$ are multiples of a pair of canonical conjugated coordinates so the elementary non-vanishing Poisson bracket is  
\begin{equation}  \label{pS10a}
 \left\{ Y_{\alpha}, Y_{-\alpha} \right\} = \frac{i}{4 \kappa^2 r \alpha} \,.
\end{equation} 

\subsubsection{The Hamiltonian}

In order to obtain the Hamiltonian, we proceed similarly. By substituting  (\ref{pS3}) into (\ref{L12}), we get
\begin{equation}  \label{pS11} 
h(q) = h^1(q) - L(q) \,, \qquad {\rm with} \qquad 
h^1(q) = \frac12\,\int_{\R} \D s\,G(s) \,\int_0^s \D\sigma\, q(\sigma) \dot{q}(\sigma-s)\,, 
\end{equation}
and
\begin{equation}  \label{pS11a} 
L(q) = -\frac12\,q(0)\,(G\ast {q})_{(0)} + \frac13\,q^3(0) \,.
\end{equation}
Including equations (\ref{pS4b}), (\ref{pS7a}), and (\ref{pS7b}), the second integral on the right-hand side becomes
\begin{align}
 &\int_0^s \D\sigma\, q(\sigma) \dot{q}(\sigma-s) = \int_0^s \D\sigma\,\left[1 + \kappa \sum_{\alpha} Y_\alpha e^{i\alpha \sigma}  \right]\,\kappa\nonumber\\
 &\qquad \qquad \times\left[ \sum_{\gamma} Y_\gamma \,i \,\gamma \,e^{i\gamma(\sigma- s)}  + \kappa \sum_{\gamma,\mu} Y_\gamma Y_\mu \,\frac{e^{i(\gamma+\mu) (\sigma-s)}}{f(\gamma+\mu)} \,i(\gamma+\mu) \right] + O(\kappa^3) \,,
\end{align}
which can be integrated using (\ref{pS9a}), and yields
\begin{align} 
&\int_0^s \D\sigma\, q(\sigma) \dot{q}(\sigma-s) = \kappa \sum_{\alpha} Y_\alpha \left(1 -e^{-i\alpha s} \right) + \kappa^2 \,\left[\sum_{\alpha} Y_\alpha Y_{-\alpha}\,\left(-i \alpha\,s\, e^{i\alpha s} \right) + \right. \nonumber\\
 & \qquad \qquad +\,\left. \sum_{\alpha+\gamma\neq 0} Y_\alpha Y_\gamma \,\left(\frac{\gamma \,e^{i\alpha s}}{\alpha +\gamma} + \frac1{f(\alpha+\gamma)}\right) \,\left(1- e^{-i (\alpha+\gamma) s}\right) \right] +  O(\kappa^3)  \,.
\end{align}
Substituting the latter into (\ref{pS9}), we can write
\vspace{0.1cm}
\begin{equation}
h^1 = \kappa\,\sum_{\alpha} h_\alpha\,Y_\alpha + \kappa^2\,\left[\sum_{\alpha+\gamma\neq 0} h_{(\alpha\gamma)}\,Y_\alpha Y_{\gamma} + \sum_{\alpha} k_{\alpha}\,Y_\alpha Y_{-\alpha} \right]  +O(\kappa^3)
\end{equation}
with
\begin{align}
 h_\alpha &= \frac12\,\int_\R \D s\,G(s)\,\left(1-e^{-i\alpha s}\right) = \frac12\,\left(1-e^{-r \alpha^2}\right)= -\frac12 \nonumber\\
k_\alpha &= - \frac12\,\int_\R \D s\,G(s)\,i\,\alpha\,s\,e^{i\alpha s}= 2\,r\,\alpha^2 \nonumber\\
h_{\alpha\gamma}&= \int_\R \D s \,G(s)\,\left( \frac{\gamma \,e^{i\alpha s} }{2 (\alpha+\gamma)} + \frac1{2\,f(\alpha+\gamma)} \right)\cdot
\left(1 - e^{-i(\alpha+\gamma) s}\right) \nonumber\\
 &= \frac{\gamma \left(e^{-r \alpha^2} - e^{-r \gamma^2}\right)}{2 (\alpha+\gamma)} +  \frac1{2\,f(\alpha+\gamma)}\, \left(1 - e^{-r(\alpha+\gamma)^2}\right) = -\frac{1 + f(\alpha+\gamma)}{2\,f(\alpha+\gamma)} \,,
\end{align}
where the fact that $f(\alpha)=f(\gamma)=0$ has been included. We finally obtain 
\begin{equation}  \label{pS12} 
h^1 = -\frac{\kappa}2\,\sum_{\alpha} Y_\alpha + \kappa^2\,\left[\sum_{\alpha} 2\,r\,\alpha^2\,Y_\alpha Y_{-\alpha} - \sum_{\alpha+\gamma\neq 0} \frac{1 + f(\alpha+\gamma)}{2\,f(\alpha+\gamma)} \,Y_\alpha Y_{\gamma} \right] + O(\kappa^3)  \,.
\end{equation}
 
Proceeding similarly with equation (\ref{pS11a}), we arrive at
\begin{equation}  \label{pS12a} 
L(q)= -\frac16 -\frac{\kappa}2\,\sum_{\alpha} Y_\alpha - \kappa^2\,\sum_{\alpha,\gamma} Y_\alpha Y_{\gamma} \,\frac{1 + f(\alpha+\gamma)}{2 f(\alpha+\gamma)}+O(\kappa^3)\,,
\end{equation}
and, therefore, the reduced Hamiltonian is
\begin{equation}  \label{pS13} 
h = \frac16 + 4\,r\,\kappa^2\,\sum_{\alpha\in \mathcal{S}^+} \alpha^2\,Y_\alpha Y_{-\alpha} +O(\kappa^3)  \,,
\end{equation}
which agrees with \cite{Gomis2004}. The Hamilton equation following this Hamiltonian with the Poisson bracket (\ref{pS10a}) is
\begin{equation}
\mathbf{X}_h Y_\alpha = \{Y_\alpha, h\} = \frac{\partial h}{\partial Y_{-\alpha}}\,\{Y_\alpha, Y_{-\alpha}\} = i\,\alpha \,Y_\alpha \,.
\end{equation}

\section{Nonlocal Lagrangian fields}\label{chap43}
\subsection{$p$-adic open string field}\label{chap432}
We consider the Lagrangian density for the $p$-adic open string 
\begin{equation}  
\label{padic:Lnt}
\mathcal{L} = - \frac{1}{2} \psi \:e^{-r\Box} \psi + \frac{1}{p+1} \psi^{p+1}\,, \qquad \mathrm{with} \qquad r = \frac{1}{2} \ln(p)\,,
\end{equation}
where $\Box$ is the d'Alembert operator $\Box = \eta^{\alpha\beta}\partial_\alpha\partial_\beta$, $\eta^{\alpha\beta}$ is the inverse Minkowski metric with the signature $(+,+,+,-)$, and $p$ is a prime number. 

As for the kinematic space, $\psi(x)$ is a smooth function $\mathcal{C}^\infty(\mathbb{R}^{3+1})$ such that the operation 
\begin{equation}
e^{-r\Box} \psi(x) := \sum_{n=0}^\infty \frac{(-r)^n}{n!}\,\Box^n\psi(x)
\end{equation}
is ``well-defined''. Since it is a part of the Lagrangian density (\ref{padic:Lnt}), we need the series in $e^{-r\Box} \psi(x)$ to converge in some appropriate functional space. The consequences of this requirement have been thoroughly analyzed in \cite{heredia2022_2} and led to the conclusion that:

\textbf{a)} Each function $\psi(x)$ in the kinematic space is the result of a convolution
\begin{equation}  \label{p2}
\psi(x) := (\E \ast\:\phi)_{(x)} \,, \qquad {\rm where} \qquad \E(\mathbf{x},t):= \G_3(\mathbf{x})\,\delta(t)\,,
\end{equation}
for some smooth function $\phi(\mathbf{x},t)$ that grows slowly at $|\mathbf{x}|\rightarrow\infty\,$, that is, 
$\forall \alpha = (\alpha_1, \ldots, \alpha_n)\,,\,\alpha_j = 1,\ldots, 4\,$, they exist 
\begin{equation}
C_\alpha(t)> 0\,,\quad m_\alpha(t) \in \mathbb{Z}^+ \quad \mid \quad \left|\partial_{\alpha_1 \ldots \alpha_n} \phi(\mathbf{x},t)  \right|\leq C_\alpha(t) \,\left(1+|\mathbf{x}|^2 \right)^{m_\alpha(t)} \,.
\end{equation}
No restriction on the behavior of $\phi(\mathbf{x},t)$ at large $|t|$ is imposed. 

\textbf{b)} The operator $e^{-r\Box}$ acts as 
\begin{equation}  \label{p3}  
e^{-r\Box}\psi(x) := \left(\T \ast\:\phi \right)_{(x)} \,, \qquad {\rm where} \qquad \T(\mathbf{x},t):= \delta(\mathbf{x})\,\G_1(t)\,.
\end{equation}
The functions $\G_1(x)$ and $\G_3(x)$ are defined by\footnote{They are the heat kernels in one-dimension and three-dimension space where $r$ plays the role of the evolution parameter of the heat equation. See, for instance, \cite{Kolar2022}. }
\begin{equation}  \label{p5}
\G_n(x) = \frac{1}{(2\sqrt{\pi r})^n} e^{-\frac{|x|^2}{4r}}, \qquad x \in \R^n \,, \qquad n= 1\;{\rm or}\; 3\,.
\end{equation}

Including all this analysis, in terms of the new kinematic variables $\phi(x)\,$, the nonlocal Lagrangian (\ref{padic:Lnt}) becomes in static coordinates
\begin{equation}
\label{padic:Lntc}
\mathcal{L}(T_x \tilde{\phi}) = - \frac{1}{2}(\E \ast\:\tilde{\phi})_{(x)} \: (\T \ast\:\tilde{\phi})_{(x)} + \frac{1}{p+1}\left[(\E \ast\:\tilde{\phi})_{(x)}\right]^{p+1}\;.
\end{equation}
For the nonlocal Lagrangian density (\ref{padic:Lntc}), the functional derivative (\ref{L2o}) is
\begin{equation}
\label{padic:Lambda}
\begin{split}
\lambda(\tilde{\phi},x,z) =& \E(x-z) \left\{ -\frac{1}{2}(\T \ast\:\tilde{\phi})_{(x)} +\left[(\E \ast\:\tilde{\phi})_{(x)}\right]^p\right\}- \frac{1}{2}  \T(x-z)(\E \ast\:\tilde{\phi})_{(x)}
\end{split}
\end{equation}
where $z^a:= (\mathbf{z},\xi)$. Consequently, the nonlocal Euler-Lagrange equations (\ref{L2o}) quickly follow and have the following convolution equation form
\begin{equation}   \label{p6a}
\E \ast\left[\T \ast\:\tilde{\phi} - (\E \ast\:\tilde{\phi})^p\right] = 0 \,,
\end{equation}
which amounts to --see Appendix A.1 for details--\footnote{Considering the spatially homogeneous case, $\tilde{\phi}(x) := \tilde{\phi}(t)$, the field equations are the same as for the $p$-adic particle case.} 
\begin{equation}   \label{p6}
(\T \ast\:\tilde{\phi})_{(x)} - (\E \ast\:\tilde{\phi})^p_{(x)} = 0 \,.
\end{equation}
A possible solution of (\ref{p6}) is\vspace{0.2cm}
\begin{equation}
\label{padic:phisolution}
\tilde{\phi}_{\mathrm{o}}(x) =
\left\{
	\begin{array}{ll}
		\pm 1,0  & \quad \mbox{if } p \text{ is odd} \\
		1,0 & \quad \mbox{if } p \text{ is even} \:.
	\end{array}
\right.
\end{equation}
Equation (\ref{p6}) admits other solutions \cite{Vladimirov2012}; however, we shall focus on the perturbative ones around $\phi_{\mathrm{o}}(x)\neq 0 $, namely,
\begin{equation}
\tilde{\phi}(x) = \tilde{\phi}_{\mathrm{o}}(x) + \kappa\:\tilde\Phi(x)\;,
\end{equation}
where $\kappa\ll1$ is the expansion parameter. For the sake of simplicity, we choose $p=2$, which is even, and, therefore, $\tilde{\phi}_{\mathrm{o}}(x) = 1$. Thus, by substituting $\tilde{\Phi}(x) = \sum^\infty_{n=0} \kappa^n \tilde{\Phi}_n (x)$ in equation (\ref{p6}), we get
\begin{equation}
\label{padic:relsol}
\T \ast\:\tilde{\Phi}_n  - 2\,\E \ast\:\tilde{\Phi}_n  = \sum_{l+m = n-1} (\E \ast\:\tilde{\Phi}_l)(\E \ast\:\tilde{\Phi}_m)\:.
\end{equation}
\indent At the lowest order, $n=0$, it reads
\begin{equation}
\label{padic:relsol0}
\T \ast\:\tilde{\Phi}_0  - 2\,\E \ast\:\tilde{\Phi}_0  = 0\:.
\end{equation}
The latter is an integral equation that might be solved using the Fourier transform, but this would restrict the search to summable functions that vanish at infinity, both spatial and temporal. From a physical point of view, this makes sense for spatial dependence; however, this restriction does not seem appropriate as far as time dependence is concerned. For this reason, we propose that the general solution is a superposition of ``monochromatic'' solutions such as $\tilde{\Phi}_0(z) = C(\mathbf{z})\,e^{i \alpha\xi} $, where $C(\mathbf{z})$ is a summable function. Therefore, by using that
\begin{equation}
\label{padic:relation}
\G_{1}(\xi)\ast\:e^{i \alpha \xi} = e^{-r \alpha^2 + i \alpha \xi}
\end{equation}
and plugging $\tilde{\Phi}_{\mathrm{o}}(z)$ into $(\ref{padic:relsol0})$, we get
\begin{equation}
\label{padic:FT}
e^{i\alpha\xi}\,\left(e^{-r\alpha^2}\,C(\mathbf{z}) - 2 \,(\G_3 \ast\:C)_{(\mathbf{z})} \right) = 0  \:,
\end{equation}
whose spatial Fourier transform yields
\begin{equation} 
\label{padic:cond}
C(\mathbf{k})\,e^{i\alpha\xi}\,\left(e^{-r \alpha^2 } - 2\:e^{-r|\mathbf{k}|^2}\right) = 0\:.
\end{equation}
Whereas $C(\mathbf{k}) = 0$ leads to the trivial solution, non-trivial solutions are connected with the spectral equation
\begin{equation} \label{spect1}
e^{-r\alpha^2 } - 2\:e^{-r|\mathbf{k}|^2}= 0 \, ,
\end{equation}
whose solution is the infinite set of complex numbers
\begin{equation} \label{spect2}
\mathcal{N}= \left\{\alpha_\nu(\mathbf{k}) = s\, \sqrt{|\mathbf{k}|^2 - 2\left(1 + \frac{i\pi l}{r}\right)} \,, \quad \mathbf{k}\in \R^3 \,, \quad \nu=(s,l)\,, \;\, s =\pm\,,\;\, l \in \mathbb{Z} \right\} \,.
\end{equation}
We shall write
\begin{equation}
\nu^\prime =(-s,l)  \,, \qquad \quad \tilde\nu=(s,-l) \,, \qquad \quad -\nu=(-s,-l)\;,
\end{equation}
then
\begin{equation}
\label{padic:alphas}
\alpha_{\tilde\nu}(\mathbf{k}) = \overline\alpha_\nu(\mathbf{k}) \,, \qquad \alpha_{\nu^\prime}(\mathbf{k}) = -\alpha_\nu(\mathbf{k}) \,,
\end{equation}
where $\bar{\alpha}$ denotes the complex conjugate of $\alpha$. Thus, the general solution of (\ref{padic:relsol0}) is
\begin{equation}  \label{p10}
\tilde{\Phi}_0(z) = \frac{1}{(2\pi)^3} \int_{\R^3} \D\mathbf{k} \sum_{\nu} C_\nu(\mathbf{k}) e^{i [\alpha_\nu(\mathbf{k})\xi + \mathbf{k} \cdot \mathbf{z}]}
\end{equation}
and, as $\tilde{\Phi}_0(z)$ has to be real,
\begin{equation}
\label{padic:As}
 C_{-\nu}(-\mathbf{k}) = \overline{C}_{\nu}(\mathbf{k})\,.
\end{equation}
\indent Notice that, as $\alpha(\mathbf{k})$ is complex, the integral might diverge at $|\mathbf{k}| \rightarrow \infty$; however, this is not the case, as shown in Appendix A.2\,. At the next perturbative order, $n=1$, equation (\ref{padic:relsol}) yields
\begin{equation}
\label{padic:relsol1}
\T \ast\:\tilde{\Phi}_1 - 2\, \E\ast\:\tilde{\Phi}_1 =  (\E \ast\:\tilde{\Phi}_0)^2\,.
\end{equation}
Using (\ref{padic:relation}), the right-hand side of this equation can be written as
\begin{equation}
(\E \ast\:\tilde{\Phi}_0)^2_{(z)} = \int_{\mathbb{R}^6} \frac{\D\mathbf{k} \D\mathbf{p}}{(2\pi)^6}\,\sum_{\nu,\mu} C_\nu(\mathbf{k}) C_\mu(\mathbf{p})\,
 e^{-r \left(|\mathbf{k}|^2 + |\mathbf{p}|^2 \right)+ i (\mathbf{k} + \mathbf{p})\cdot\mathbf{z}}\, 
e^{i \left[\alpha_\nu(\mathbf{k})+\alpha_\mu (\mathbf{p})\right]\xi} \,.
\end{equation}
\indent Again, a particular solution (\ref{padic:relsol1}) can be obtained as a superposition of ``monochromatic'' solutions like $\tilde\Phi_1(\mathbf{k},\mathbf{p})\,e^{i (\mathbf{k}+\mathbf{p})\cdot \mathbf{z}} \, e^{i \left[\alpha_\nu(\mathbf{k})+\alpha_\mu (\mathbf{p})\right]\xi}$. Following the same steps as above, we arrive at
\begin{equation}
\tilde{\Phi}_1(z) = \int_{\mathbb{R}^6} \frac{\D\mathbf{k} \D\mathbf{p}}{(2\pi)^6} \, \sum_{\nu,\mu} \frac{C_\nu(\mathbf{k}) C_\mu(\mathbf{p})}{f_{\nu\mu}(\mathbf{k},\mathbf{p})} e^{i\,[\alpha_\nu(\mathbf{k})+\alpha_\mu(\mathbf{p})]\xi + i\,(\mathbf{k} +\mathbf{p})\cdot\mathbf{z}}  \,,
\end{equation}
where
\begin{equation}
f_{\nu\mu}(\mathbf{k},\mathbf{p}):= 2 \left[2e^{-2r \alpha_\nu(\mathbf{k})\alpha_\mu(\mathbf{p})}- e^{-2r\mathbf{k}\cdot\mathbf{p}}\right]\,,
\end{equation}
and we have used that $\alpha_\nu(\mathbf{k})$ is a solution of the spectral equation (\ref{spect1}). Therefore, the general perturbative solution up to the second order in static coordinates is 
\begin{align}  \label{pertsol}
\tilde{\phi}(z) = 1 + &\kappa\:\int_{\R^3} \frac{\D \mathbf{k}}{(2\pi)^3} \sum_{\nu} C_\nu(\mathbf{k}) e^{i [\alpha_\nu(\mathbf{k})\xi + \mathbf{k} \cdot \mathbf{z}]}\left\{1 + \kappa \int_{\R^3} \frac{\D \mathbf{p}}{(2\pi)^3}\sum_{\mu} \frac{C_\mu(\mathbf{p})}{f_{\nu\mu}(\mathbf{k},\mathbf{p})} e^{i\,[\alpha_\mu(\mathbf{p})\xi + \mathbf{p}\cdot\mathbf{z}]} \right\} \,.
\end{align}
\indent As in the $p$-adic particle case, since the nonlocal Lagrangian density does not explicitly depend on the spacetime coordinates $x^a$, the nonlocal Euler-Lagrange equations in moving coordinates are equivalent to the static ones, so the solution structure is equivalent. Therefore, if we take into account that $\phi(z) = \tilde{\phi}(x+z)$ , we can arrive at
\begin{align}  \label{pertsol-mc}
\phi(z) = 1 + &\kappa\:\int_{\R^3} \frac{\D \mathbf{k}}{(2\pi)^3} \sum_{\nu} A_{\nu}(\mathbf{k}) e^{i [\alpha_\nu(\mathbf{k})\xi + \mathbf{k} \cdot \mathbf{z}]}\left\{1 + \kappa \int_{\R^3} \frac{\D \mathbf{p}}{(2\pi)^3}\sum_{\mu} \frac{A_{\mu}(\mathbf{p})}{f_{\nu\mu}(\mathbf{k},\mathbf{p})} e^{i\,[\alpha_\mu(\mathbf{p})\xi + \mathbf{p}\cdot\mathbf{z}]} \right\} \,,
\end{align}
where $A_{\mu}(\mathbf{k}):= C_\mu(\mathbf{k})\,e^{i [\alpha_\nu(\mathbf{k})t + \mathbf{k} \cdot \mathbf{x}]}$ is the relation that keeps the transformation of the solution between moving and static coordinates. 

\subsubsection{The symplectic form \label{SS5.1}}
We find that the momenta (\ref{H7}) are
\begin{equation}
\label{padic:momenta}
P(\phi,y) = -\frac{1}{2} \int_{\R} \D\xi \left[\theta(\tau) - \theta(\xi)\right] \G_1(\xi-\tau) \left(\E \ast\:\phi \right)_{(\mathbf{y},\xi)}\,,
\end{equation}
and, therefore, the contact form (\ref{H13}) becomes
\begin{equation}
\label{padic:simpletic0}
\omega = - \frac{1}{2} \int_{\mathbb{R}^4} \D\mathbf{y}\:\D s \G_1(s) \int^s_0 \D\tau \left(\E \ast\:\delta\phi\right)_{(\mathbf{y},\tau-s)} \wedge \delta\phi(\mathbf{y},\tau)\:,
\end{equation}
where we have introduced the change $s :=\xi-\tau$. Taking now equation (\ref{pertsol-mc}),  we obtain that, after a bit of algebra\footnote{As the calculations are very similar to the $p$-adic particle, we shall not make them explicit. If the reader is interested in the detail of the derivation, see Appendix A.3 of \cite{Heredia2022}.}, the contact form is 
\begin{equation}
\label{padic:simpletic}
\omega = i\,\int_{\R^3} \D\mathbf{k}\,\sum_{l\in\mathbb{Z}} \delta B_l(\mathbf{k})\wedge \delta B^\dagger_l(\mathbf{k}) + \mathcal{O}(\kappa^4)\,, 
\end{equation}
where the new variables 
\begin{equation}  \label{f3p}
B_{l}(\mathbf{k}):= \frac{2\,\kappa\,e^{-r|\mathbf{k}|^2}}{(2\pi)^{3/2}} \,\sqrt{r\: \alpha_l(\mathbf{k})} \, A_{(+,l)}(\mathbf{k}) \qquad {\rm and} \qquad B^\dagger_{l}(\mathbf{k}):= \overline{B}_{-l}(\mathbf{k}) 
\end{equation}
have been introduced. It is apparent that: \textbf{(a)} $\omega$ is non-degenerate, hence symplectic, and \textbf{(b)} the modes $B_l(\mathbf{k})$ and $ B^\dagger_{j}(\mathbf{k})$ are a system of canonical coordinates whose elementary Poisson brackets are
\begin{align}  \label{f4p}
 \{B_l(\mathbf{k}),B^\dagger_{j}(\mathbf{k}^\prime)\} &= i\,\delta_{lj}\,\delta(\mathbf{k}-\mathbf{k}^\prime) + \mathcal{O}(\kappa^4)\,, \quad
\{B_l(\mathbf{k}), B_{j}(\mathbf{k}^\prime)\} = \{B^\dagger_l(\mathbf{k}), B^\dagger_{j}(\mathbf{k}^\prime)\} =  \mathcal{O}(\kappa^4) \,.
\end{align}

\subsubsection{The Hamiltonian \label{SS5.2}}
Substituting  the momenta (\ref{padic:momenta}) in equation (\ref{H12}), we obtain that the Hamiltonian is
\begin{equation}   \label{ham1}
h(\phi) = - L(\phi) + \frac12\int_{\R^3} \D\mathbf{y} \int_{\R} \D s \G_1(s) \int^s_0 \D\xi\:\dot\phi(\mathbf{y},\xi-s)\left(\E \ast\:\phi\right)_{(\mathbf{y},\xi)}
\end{equation}
where we have again defined $s:=\xi-\tau$ and $L(\phi) := \int_{\R^3} \D \mathbf{x}\, \mathcal{L}(T_\mathbf{x} \phi,\mathbf{x},0)$ with $\mathcal{L}$ given by (\ref{padic:Lntc}).  

As above, taking the perturbative expansion of $\phi(z)$ up to $\kappa^2$ terms, we obtain that the Hamiltonian is\footnote{As above, see Appendix A.4 of \cite{Heredia2022} for details.}
\begin{equation}  \label{ham1a}
h = \frac{V_x}{6} + \int_{\R^3}  \D\mathbf{k}\,\sum_{l\in\mathbb{Z}} \alpha_l(\mathbf{k})\:B_l(\mathbf{k}) B^\dagger_l(\mathbf{k}) + \mathcal{O}(\kappa^3) \,,
\end{equation}
where $V_x :=\int_{\R^3} \D\mathbf{x}$ is an infinite contribution to the Hamiltonian, which is associated with the (divergent) vacuum energy. This problem may be highly complex to treat when gravity is present \cite{Martin2012}. However, as gravity is absent, we can drop it \cite{Peskin1995}. Moreover, the Hamiltonian is treated as the generator of the dynamics of our system. Therefore,  this term will not affect Hamilton's equations because it is merely a constant. Hence, the Hamilton equations for this Hamiltonian with the Poisson brackets (\ref{f4p}) are 
\begin{equation}
\mathbf{X}_h B_j(\mathbf{k}) = i\:\alpha_j(\mathbf{k}) \,B_j(\mathbf{k})\qquad {\rm and} \qquad \mathbf{X}_h B^\dagger_j(\mathbf{k}) = - i\:\alpha_j(\mathbf{k}) \,B^\dagger_j(\mathbf{k}) \:.
\end{equation}

\subsubsection{The energy-momentum tensor \label{SS5.3}}
In \cite{Moeller2002}, an expression of the energy-momentum tensor for the homogeneous infinite-order $p$-adic Lagrangian is obtained in a non-closed form (i.e., expressed as an infinite series).  Since our formalism allows us to calculate both the canonical energy-momentum tensor $\mathcal{T}^{a b}$ and the Belinfante-Rosenfeld energy-momentum tensor $\Theta^{a b}$ in a closed form (i.e. with the infinite series summed), we shall now particularise these expressions for the perturbative $p$-adic open string case. Before calculating these tensors, we must take into account these two observations:

\textbf{a)} The nonlocal Lagrangian density (\ref{padic:Lnt}) is Poincar\'e invariant if the $\tilde\psi$  field transforms as a scalar, i.e., $\tilde{\psi}^\prime(x^\prime) = \tilde\psi(x)$. However, note that the $\phi$ field cannot be a Poincar\'e scalar because its definition (\ref{p2}) is not. Therefore, we will require that the $\tilde\psi$ field transforms as a scalar to obtain the  transformation rule of $\tilde\phi$ that leaves the nonlocal Lagrangian density (\ref{padic:Lntc}) Poincar\'e invariant (i.e., with $W^b=0$). Indeed, this transformation is --see Appendix A.5 for details--
\begin{equation}
\label{padic:t1_2}
 \delta \tilde\phi(x) = -\left(\varepsilon^c+ \omega^{cb} x_b\right)\partial_c\tilde\phi(x) + \omega^{ab}\left[2r\,\delta^4_{[a} \delta^i_{b]}\, \partial_i\dot{\tilde{\phi}}(x)\right]\,.
\end{equation}
As one can observe, the last term causes the $\tilde{\phi}$ field not to be a Poincar\'e scalar. Indeed, this term shall contribute to the spin part. Fortunately, due to this additional term's structure, we do not need to recalculate Section \ref{chap521-TS} since the structure of (\ref{padic:t1_2}) is equivalent to (\ref{chap331-P3}). We only need to change the $\omega^{ab} M^A_{B[ab]}\phi^b$ term by the last term of (\ref{padic:t1_2}).  

\textbf{b)} Both the canonical and the Belinfante-Rosenfeld energy-momentum tensor are conserved at any solution of (\ref{p6}); therefore, we have to consider the $p$-adic Lagrangian density and $\lambda(\tilde\phi,x,z)$ with the nonlocal Euler-Lagrange (EL) equations applied to obtain them, that is, 
\begin{equation}
\label{Los}
 \Xi(T_x\tilde{\phi}):=\mathcal{L}_{(EL)}(T_x \tilde{\phi}) = -\frac{1}{6} \left[(\E \ast\:\tilde{\phi})_{(x)}\right]^3
\end{equation}
and 
\begin{equation}
\label{Laos}
\Upsilon(x,z):= \lambda_{(EL)}(\tilde{\phi},x,z) =\frac{1}{2}\left[\E(x-z)(\T \ast\:\tilde{\phi})_{(x)} - \T(x-z)(\E \ast\:\tilde{\phi})_{(x)}\right]\,.
\end{equation}

Thus, bearing in mind the second observation and using (\ref{Los}) and (\ref{Laos}), the canonical energy-momentum tensor in closed form is
\begin{equation}
\label{padic:tc}
\mathcal{T}_a^{\:\:b}(T_x\tilde\phi) = - \Xi(T_x\tilde\phi)\:\delta^b_a + \int^1_0 \D s \int_{\R^4} \D z \:\Upsilon(x+[s-1]z, x+sz)\:z^b\:\tilde\phi_{|a}(x+sz)\:.
\end{equation}
Now, using (\ref{Laos}) and bearing in mind both the first and second observations, the spin current (\ref{P6a}) is
\begin{align}
\label{padic:Sc}
\mathcal{S}^{\;\,\;b}_{ac}(T_x\tilde\phi) =  2 \int_{\R^4} \D z\,z^b\,\int_0^1 \D s\,&\Upsilon(x+[s-1]z, x+sz)\nonumber\\
&\times \left[s\,z_{[c}\tilde\phi_{|a]}(x+sz) + 2\,r\,\delta^4_{[c}\delta^i_{a]}\,\dot{\tilde{\phi}}_{|i}(x+sz)\right]\,.
\end{align}
With the last expression (\ref{padic:Sc}), we find that $\mathcal{W}^{cba}(T_x\tilde\phi)$ is
\begin{align}
\label{padic:W}
\mathcal{W}^{cba}=& \int_{\R^4} \D z\int_0^1 \D s\,\Upsilon(x+[s-1]z, x+sz)\left[s\,(z^a z^b\,\delta^c_g -z^az^c\,\delta^b_g)\,\tilde{\phi}^{|g}(x+sz)\right.\nonumber\\
&\left. \qquad\quad+2\,r\left(z^{(a}\eta^{b)4}\eta^{cf} - z^{(a}\eta^{b)f}\eta^{4c} - z^c\eta^{4[a}\eta^{b]f}\right) \dot{\tilde{\phi}}_{|f}(x+sz)\right]\,,
\end{align}
where we have used the fact that $\eta^{4[c}\eta^{a]i}\,\dot{\tilde{\phi}}_{|i}(x+sz) = \eta^{4[c}\eta^{a]f}\, \dot{\tilde{\phi}}_{|f}(x+sz)$ because of the antisymmetry.  Finally, deriving concerning $x^c$ equation (\ref{padic:W}) and using the property $z^c\, A_{|_c}(x+s z) = \frac{\D}{\D s}A(x+sz)$ and integration by parts, we obtain
\begin{align}
\label{padic:cW}
&\partial_cW^{cba} =- \int_{\R^4} \D z\left\{ z^a\,\Upsilon(x, x+z)\tilde{\phi}^{|b}(x+z) +2\,r\,\eta^{4[a}\left( \Upsilon(x, x+z) \dot{\tilde{\phi}}^{|b]}(x+z) - \Upsilon(x-z, x) \dot{\tilde{\phi}}^{|b]}(x) \right) \right.\nonumber\\
&\left. - \int_0^1 \D s\left[ s\,z^a z^b\partial_c[\Upsilon(x+[s-1]z, x+sz)\tilde{\phi}^{|c}(x+sz)] +z^a\Upsilon(x+[s-1]z, x+sz)\tilde{\phi}^{|b}(x+sz)\right.\right.\\
&\left.\left. + 2\,r\,z^{(a}\left(\eta^{b)4}\,\partial^f[\Upsilon(x+[s-1]z, x+sz)\dot{\tilde{\phi}}_{|f}(x+sz)] -\partial^4[\Upsilon(x+[s-1]z, x+sz)\dot{\tilde{\phi}}^{|b)}(x+sz)] \right)\right] \right\}\,. \nonumber
\end{align}
Therefore, putting (\ref{padic:tc}) and (\ref{padic:cW}) together, we obtain the Belinfante-Rosenfeld tensor in closed form
\begin{align}
&\Theta^{ab}(T_x\tilde{\phi}) = - \Xi(T_x\tilde\phi) \delta^{ab} - \int_{\R^4} \D z\left\{\Upsilon(x, x+z) z^a \tilde{\phi}^{|b}(x+z) + 2\,r\,\eta^{4[a}\left(\Upsilon(x, x+z)\,\dot{\tilde{\phi}}^{|b]}(x+z)\right.\right.\nonumber\\
&\left.\left.  - \Upsilon(x-z, x)\, \dot \phi^{|b]}(y) \right) -  \int_0^1 \D s\left[ s\,z^a z^b\,\partial_c[\Upsilon(x+[s-1]z, x+sz)\tilde{\phi}^{|c}(x+sz)] \right.\right.\nonumber\\
&\left.\left. +2\,\Upsilon(x+[s-1]z, x+sz)\,z^{(a}\tilde{\phi}^{|b)}(x+sz)+ 2\,r\,z^{(a}\left(\eta^{b)4}\,\partial^f[\Upsilon(x+[s-1]z, x+sz)\dot{\tilde{\phi}}_{|f}(x+sz)]\right.\right.\right.\nonumber\\
&\left.\left.\left. - \partial^4[\Upsilon(x+[s-1]z, x+sz) \dot{\tilde{\phi}}^{|b)}(x+sz)] \right)\right]\right\}\,.
\end{align}
It is worth mentioning that the Belinfante-Rosenfeld tensor depends on the solution that the theory might present. For this reason, we use the perturbative solution (\ref{pertsol}) to obtain its components explicitly.

The first element to be calculated is the $(4,4)$-component, which indicates the energy density of the system associated with the perturbative solution. Therefore, the energy density is
\begin{align}
&\Theta^{\,\,\,4}_4(T_x\tilde{\phi}) = - \Xi(T_x\tilde{\phi}) + \int_0^1\D s\int_{\R^4} \D z\,z^4\left\{\Upsilon(x+[s-1]z, x+sz)\dot{\tilde{\phi}}(x+sz)\right.\nonumber\\
&\left. - s\,z^i\,\partial_i\left[\Upsilon(x+[s-1]z, x+sz)\dot{\tilde{\phi}}(x+sz)\right] - \partial_i\left[s\,z^4\,\Upsilon(x+[s-1]z, x+sz)\dot{\tilde{\phi}}^{|i}(x+sz)\right.\right.\nonumber\\
&\left.\left. - 2\,r\, \Upsilon(x+[s-1]z, x+sz) \dot{\tilde{\phi}}^{|i}(x+sz)\right] \right\}\,,
\end{align}
and, including
\begin{equation}
\label{padic:Ups}
\Upsilon(x+[s-1]z, x+sz) = \frac{1}{2}\left[\delta(\xi)\mathcal{G}_3(\mathbf{z})(\mathcal{T}\ast\tilde{\phi})_{(x+[s-1]z)} - \delta(\mathbf{z})\mathcal{G}_1(\xi)(\mathcal{E}\ast\tilde{\phi})_{(x+[s-1]z)}\right]\,,
\end{equation}
 it simplifies as
\begin{align}
\Theta^{\,\,\,4}_4(T_x\tilde{\phi}) =& - \Xi(T_x\tilde{\phi}) - \frac{1}{2} \int_0^1\D s \int_\R \D\xi\,\xi\,\mathcal{G}_1(\xi)\left\{(\mathcal{E}\ast\tilde{\phi})_{(\mathbf{x},t+[s-1]\,\xi)}\,\dot{\tilde{\phi}}(\mathbf{x},t+s\,\xi)\right.\nonumber\\
&\quad \left. -\partial_i\left[(\mathcal{E}\ast\tilde{\phi})_{(\mathbf{x},t+[s-1]\xi)}\left(s\,\xi\,\tilde{\phi}^{|i}(\mathbf{x},t+s\,\xi) -2\,r\,\dot{\tilde{\phi}}^{|i}(\mathbf{x},t+s\,\xi)\right)\right] \right\}\,.
\end{align}
Taking (\ref{pertsol}) and computing the integrals, it becomes --after a tedious computation--
\begin{equation}
\Theta_{4}^{\:\:4}(x) = \frac{1}{6}+ \kappa\:\overset{1}{\Theta}\,_{4}^{\:\:4}(x) + \kappa^2\:\overset{2}{\Theta}\,_{4}^{\:\:4}(x) + \mathcal{O}(\kappa^3)\,,
\end{equation}  
where 
\begin{align}
\overset{1}{\Theta}\,_{4}^{\:\:4}(x):= \frac{1}{2} \int_{\R^3}\frac{\D \mathbf{k}}{(2\pi)^3}& \sum_\nu C_\nu(\mathbf{k}) \:e^{i(\alpha_\nu(\mathbf{k})t + \mathbf{k}\cdot\mathbf{x})}\nonumber\\
&\times \left[1-e^{-r|\mathbf{k}|^2} - \frac{|\mathbf{k}|^2}{2\,\alpha_\nu(\mathbf{k})^2}\left(e^{-r|\mathbf{k}|^2}+r\,\alpha_\nu(\mathbf{k})^2 - \frac{1}{2}\right)\right]
\end{align}
and
\begin{align}
\overset{2}{\Theta}\,_{4}^{\:\:4}(x)&:= \frac{1}{2} \int_{\R^6}\frac{\D \mathbf{k}\:\D\mathbf{p}}{(2\pi)^6} \sum_{\nu,\mu}C_\nu(\mathbf{k})C_\mu(\mathbf{p})\,e^{i([\alpha_\nu(\mathbf{k})+\alpha_\mu(\mathbf{p})]t + [\mathbf{k}+\mathbf{p}]\cdot\mathbf{x})} \left[\frac{1-e^{-r(\mathbf{k}+\mathbf{p})^2}}{f_{\nu\mu}(\mathbf{k},\mathbf{p})} \right.\nonumber\\
&\quad \left. - \frac{(\mathbf{k}+\mathbf{p})^2}{f_{\nu\mu}(\mathbf{k},\mathbf{p})[\alpha_{\nu}(\mathbf{k})+\alpha_{\mu}(\mathbf{p})]^2}\left(e^{-r[\alpha_\nu(\mathbf{k}) +\alpha_\mu(\mathbf{p})]^2} + 2r\,[\alpha_{\nu}(\mathbf{k})+\alpha_{\mu}(\mathbf{p})]^2 -1\right)\right.\nonumber\\
&\quad \left. + \frac{\mathbf{k}\cdot(\mathbf{k}+\mathbf{p})\,e^{-2r|\mathbf{p}|^2}}{2\,[\alpha_{\nu}(\mathbf{k})+\alpha_{\mu}(\mathbf{p})]^2}\left(e^{-r(|\mathbf{k}|^2-|\mathbf{p}|^2)} +2r\,\alpha_\nu(\mathbf{k})[\alpha_\nu(\mathbf{k})+\alpha_\mu(\mathbf{p})] -1 \right)\right.\nonumber\\
&\quad \left. + \frac{\alpha_\nu(\mathbf{k})\,e^{-r|\mathbf{p}|^2}}{[\alpha_{\nu}(\mathbf{k})+\alpha_{\mu}(\mathbf{p})]}\left(e^{-r\alpha_{\mu}(\mathbf{p})^2} - e^{-r\alpha_\nu(\mathbf{k})^2}\right) \right]\:.
\end{align}
From the last two expressions, it is easy to prove that if we calculate the total energy of the system at $t=0$, it coincides with the Hamiltonian (\ref{ham1a}) since\vspace{0.2cm}
\begin{equation}
\int_{\R^3} \D \mathbf{x}\:\overset{1}{\Theta}\,_{4}^{\:\:4}(\mathbf{x},0) = 0,
\end{equation}
and 
\begin{equation}
E - \frac{V_x}{6} = h - \frac{V_x}{6}  = \int_{\R^3} \D \mathbf{x}\:\overset{2}{\Theta}\,_{4}^{\:\:4}(\mathbf{x},0) = \int_{\R^3}  \D\mathbf{k}\,\sum_{l\in\mathbb{Z}} \alpha_l(\mathbf{k})\:B_l(\mathbf{k}) B^\dagger_l(\mathbf{k})\:.
\end{equation}
where $V_x =\int_{\R^3}\D\mathbf{x}$ is the vacuum energy. It is necessary to highlight this result. Note that the system's total energy (or the Hamiltonian for this case) is not affected by choice of tensor. As we have just shown, either through the canonical or the Belinfante-Rosenfeld energy-momentum tensor, the result remains the same, as might be expected, since the system's total energy is not modified. 

The second element is the $(i,j)$-component, which indicates the system's pressure. Thus, the pressure is
\begin{align}
\Theta^{ij}&(T_x\tilde\phi) = - \Xi(T_x\tilde\phi) \delta^{ij} - \int_{\R^4} \D z\left\{\Upsilon(x, x+z) z^i \tilde\phi^{|j}(y+z) -  \int_0^1 \D s\left[ s\,z^i z^j\,\right. \right.\nonumber\\
&\left.\left.\times \partial_c[\Upsilon(x+[s-1]z, x+sz)\tilde\phi^{|c}(x+sz)]  +2\,\Upsilon(x+[s-1]z, x+sz)\,z^{(i}\tilde\phi^{|j)}(x+sz) \right.\right.\nonumber\\
&\left.\left.+ 2\,r\,z^{(i} \partial_4[\Upsilon(x+[s-1]z, x+sz) \dot{\tilde{\phi}}^{|j)}(x+sz)]\right]\right\}
\end{align}
that, including (\ref{padic:Ups}), simplifies as
\begin{align}
\Theta^{ij}&(T_x\tilde\phi) = - \Xi(T_x\tilde\phi) \delta^{ij} - \frac{1}{2} (\mathcal{T}\ast\tilde\phi)_{(x)} \int_{\R^3}\D\mathbf{z}\,\mathcal{G}_3(\mathbf{z})\,z^i \tilde\phi^{|j}(\mathbf{x}+\mathbf{z},t)\nonumber\\
& + \frac{1}{2} \int^1_0 \D s\int_{\R^3}\D\mathbf{z} \,\mathcal{G}_3(\mathbf{z})\left\{ s\,z^i\,z^j \partial_c\left[(\mathcal{T}\ast\tilde\phi)_{(\mathbf{x}+(s-1)\,\mathbf{z}, t)}\,\tilde\phi^{|c}(\mathbf{x}+s\,\mathbf{z}, t)\right]\right.\\
&\left.+  2\,(\mathcal{T}\ast\tilde\phi)_{(\mathbf{x}+(s-1)\,\mathbf{z},t)}\,z^{(i}\tilde\phi^{|j)}(\mathbf{x}+s\,\mathbf{z},t) +2\,r\,z^{(i}\partial_4\left[(\mathcal{T}\ast\tilde\phi)_{(\mathbf{x}+(s-1)\,\mathbf{z},t)}\dot{\tilde{\phi}}^{j)}(\mathbf{x}+s\,\mathbf{z},t)\right]\right\}\;.\nonumber
\end{align}
By taking (\ref{pertsol}), the last expression becomes
\begin{equation}
 \Theta^{ij}(x) = \frac{\delta^{ij}}{6} +  \kappa\,\overset{1}{\Theta}\,^{ij}(x) + \mathcal{O}(\kappa^2)\,,
\end{equation} 
where
\begin{align}
&\overset{1}{\Theta}\,^{ij}(x) = \frac{1}{2}\int_{\R^3}\frac{\D\mathbf{k}}{(2\pi)^3}\sum_\nu C_\nu(\mathbf{k})\,e^{i(\alpha_\nu(\mathbf{k})t + \mathbf{k}\cdot\mathbf{x})}\left[ \left(\delta^{ij}+2\,r\,k^i\,k^j\right)\,e^{-r|\mathbf{k}|^2}\right.\nonumber\\
&\qquad \left. + 2\frac{k^i\,k^j}{|\mathbf{k}|^2}\left\{\left(e^{-r|\mathbf{k}|^2}-1\right)\left(1- r\,\alpha_\nu(\mathbf{k})^2\right) + \frac{|\mathbf{k}|^2-\alpha_\nu(\mathbf{k})^2}{|\mathbf{k}|^2}\left(1-\left[1+r\,|\mathbf{k}|^2\right]e^{-r|\mathbf{k}|^2} \right)\right\} \right]\,.
\end{align}
Again, it is important to highlight this result. Note that the tensor is symmetric in $(i,j)$-indices due to the Belinfante-Ronsenfeld symmetrization technique, as expected. Therefore, this fact ensures that we can use this tensor in theories that need to be symmetric, for instance, General Relativity.  Likewise, if we calculate the pressure exerted on a spherical surface $A$ of radius $R$ at $t=0$, we get  
\begin{equation}
\Sigma^i:= \int_A \Theta^{ij}(\mathbf{x},0)\,\D^2A_j =  \kappa\:\overset{1}{\Sigma}\,^{i} + \mathcal{O}(\kappa^2)\,,
\end{equation}
where 
\begin{align}
\overset{1}{\Sigma}\,^{i} :=&\,4\,\pi\,i\,\int_{\R^3}\frac{\D\mathbf{k}}{(2\pi)^3}\sum_\nu C_\nu(\mathbf{k})\,\frac{\hat{k}^i}{|\mathbf{k}|^2}\,\left[\sin(R\,|\mathbf{k}|)-|\mathbf{k}|\,R\,\cos(R\,|\mathbf{k}|)\right] \left[ \frac{1}{2}\left(1+2\,r\,|\mathbf{k}|^2\right)e^{-r|\mathbf{k}|^2}\right.\nonumber\\
&\left. + \left(e^{-r|\mathbf{k}|^2}-1\right)\left(1- r\,\alpha_\nu(\mathbf{k})^2\right) +  \frac{|\mathbf{k}|^2-\alpha_\nu(\mathbf{k})^2}{|\mathbf{k}|^2}\left(1-\left[1+r\,|\mathbf{k}|^2\right]e^{-r|\mathbf{k}|^2} \right)  \right]\,,
\end{align}
and $\hat{k}^i$ is the unitary vector of $k^i$. One might think that the pressure is imaginary because of the $i$ factor in front; however, it can be proved that using equations (\ref{padic:alphas}) and (\ref{padic:As}), it is indeed a real value, as expected. \\
\indent By analogy with the electromagnetic case, the last two remaining elements are the elements of the Poynting vector 
\begin{align}
&\Theta^{i4}(T_x\tilde\phi) = \int_{\R^4} \D z\left\{\Upsilon(x, x+z)\,z^i\,\dot{\tilde{\phi}}(x+z) + \int^1_0 \D s\left[\,s\,z^4\,z^i\partial_c\left[\Upsilon(x+[s-1]z, x+sz)\tilde\phi^{|c}(x+sz)\right]\right.\right.\nonumber\\
&\left.\left. + 2\,\Upsilon(x+[s-1]z, x+sz)z^{(4}\,\tilde\phi^{|i)}(x+sz)- 2\,r\,z^{(i}\partial_j\left[\Upsilon(x+[s-1]z, x+sz) \dot{\tilde{\phi}}^{|j)}(x+sz)\right] \right] \right\}
\end{align}
and
\begin{align}
&\Theta^{4i}(T_x\tilde\phi) = \int_{\R^4} \D z\left\{ -\Upsilon(x,x+z)\,z^4\,\tilde\phi^{|i}(x+z) +2\,r\,\left[\Upsilon(x, x+z)\dot\phi^{|i}(x+z)-\Upsilon(x-z, x)\dot{\tilde{\phi}}^{|i}(x)\right]\right.\nonumber\\
&\left. + \int^1_0 \D s\left[\,s\,z^4\,z^i\partial_c\left[\Upsilon(x+[s-1]z, x+sz)\tilde\phi^{|c}(x+sz)\right]+ 2\,\Upsilon(x+[s-1]z, x+sz)z^{(4}\,\tilde\phi^{|i)}(x+sz)\right.\right.\nonumber\\
&\left.\left. - 2\,r\,z^{(i}\partial_j\left[\Upsilon(x+[s-1]z, x+sz) \dot{\tilde{\phi}}^{|j)}(x+sz)\right] \right] \right\}\,,
\end{align}
respectively. Including (\ref{padic:Ups}) and 
\begin{align}
\Upsilon(x, x+z) &= \frac{1}{2}\left[\delta(\xi)\,\mathcal{G}_3(\mathbf{z})\,(\mathcal{T}\ast\tilde\phi)_{(x)} - \delta(\mathbf{z})\,\mathcal{G}_1(\xi)\,(\mathcal{E}\ast\tilde\phi)_{(x)}\right]\nonumber\\
\Upsilon(x-z, x) &=\frac{1}{2}\left[\delta(\xi)\,\mathcal{G}_3(\mathbf{z})\,(\mathcal{T}\ast\tilde\phi)_{(x-z)} - \delta(\mathbf{z})\,\mathcal{G}_1(\xi)\,(\mathcal{E}\ast\tilde\phi)_{(x-z)}\right]\,,
\end{align}
they become
\begin{align}
\Theta^{i4}(T_x\tilde\phi) =& \frac{1}{2}(\mathcal{T}\ast\tilde\phi)_{(x)}\int_{\R^3}\D\mathbf{z}\,z^i\,\mathcal{G}_3(\mathbf{z})\, \dot{\tilde{\phi}}(\mathbf{x}+\mathbf{z},t)\nonumber\\
&-\frac{1}{2}\int^1_0\D s\left[\int_\R \D\xi\,\xi\,\mathcal{G}_1(\xi)(\mathcal{E}\ast\tilde\phi)_{(\mathbf{x},t+(s-1)\xi)}\,\tilde\phi^{|i}(\mathbf{x},\tau+s\,\xi)\right.\nonumber\\
&\left. + \int_{\R^3} \D\mathbf{z}\,z^i\,\mathcal{G}_3(\mathbf{z})(\mathcal{T}\ast\tilde\phi)_{(\mathbf{x}+(s-1)\,\mathbf{x},t)}\,\dot{\tilde{\phi}}(\mathbf{x}+s\,\mathbf{z},\tau)\right.\nonumber\\
&\left.  + 2\,r\, \int_{\R^3}\D\mathbf{z}\,\mathcal{G}_3(\mathbf{z})\,z^{(i}\,\partial_j\left[(\mathcal{T}\ast\tilde\phi)_{(\mathbf{x}+(s-1)\,\mathbf{z},t)}\,\dot{\tilde{\phi}}^{|j)}(\mathbf{x}+s\,\mathbf{x},\tau)\right]  \right]
\end{align}
and\vspace{0.2cm}
\begin{align}
\Theta^{4i}(T_x\tilde\phi) =& (\mathcal{E}\ast\phi)_{(x)}\int_\R\D\xi\,\mathcal{G}_1(\xi)\left[\frac{1}{2}\,\xi\, \tilde\phi^{|i}(\mathbf{x},t+\xi) -r\,\dot{\tilde{\phi}}^{|i}(\mathbf{x},t+\xi)\right]\nonumber\\
&+ r\,\dot{\tilde{\phi}}^{|i}(y) \left[\int_\R\D\xi\,\mathcal{G}_1(\xi)(\mathcal{E}\ast\tilde\phi)_{(\mathbf{x},t-\xi)} - \int_{\R^3} \D\mathbf{z}\,\mathcal{G}_3(\mathbf{z})(\mathcal{T}\ast\tilde\phi)_{(\mathbf{x}-\mathbf{z},t)}\right]\nonumber\\
& + r\,(\mathcal{T}\ast\tilde\phi)_{(x)} \int_{\R^3}\D\mathbf{z} \,\mathcal{G}_3(\mathbf{z})\,\dot{\tilde{\phi}}^{|i}(\mathbf{x}+\mathbf{z},t)\nonumber\\
&-\frac{1}{2}\int^1_0\D s\left[\int_\R \D\xi\,\xi\,\mathcal{G}_1(\xi)(\mathcal{E}\ast\tilde\phi)_{(\mathbf{x},t+(s-1)\xi)}\,\tilde\phi^{|i}(\mathbf{x},t+s\,\xi)\right.\nonumber\\
&\left. + \int_{\R^3} \D\mathbf{z}\,z^i\,\mathcal{G}_3(\mathbf{z})(\mathcal{T}\ast\tilde\phi)_{(\mathbf{x}+(s-1)\,\mathbf{z},t)}\,\dot{\tilde{\phi}}(\mathbf{x}+s\,\mathbf{z},t)\right.\nonumber\\
&\left.  + 2\,r\int_{\R^3}\D\mathbf{z}\,\mathcal{G}_3(\mathbf{z})\,z^{(i}\,\partial_j\left[(\mathcal{T}\ast\tilde\phi)_{(\mathbf{x}+(s-1)\,\mathbf{z},t)}\,\dot{\tilde{\phi}}^{|j)}(\mathbf{x}+s\,\mathbf{x},t)\right] \right]\,.
\end{align}
As above, taking (\ref{pertsol}), we finally obtain --after a tedious computation--
\begin{equation}
 \Theta^{4i}(x) = \Theta^{i4}(x) =  \kappa\:\overset{1}{\Theta}\,^{i4}(x) + \kappa^2\:\overset{2}{\Theta}\,^{i4}(x) + \mathcal{O}(\kappa^3)\,,
\end{equation}  
where 
\begin{align}
\overset{1}{\Theta}\,^{i4}(x) = \frac{\kappa}{2}\int_{\R^3}\frac{\D \mathbf{k}}{(2\pi)^3}& \sum_\nu C_\nu(\mathbf{k}) \:e^{i(\alpha_\nu(\mathbf{k})t + \mathbf{k}\cdot\mathbf{x})}\,k^i\nonumber\\
&\times\left[\frac{\alpha_\nu(\mathbf{k})}{|\mathbf{k}|^2}\left(1-e^{-r|\mathbf{k}|^2}\right) + \frac{1}{\alpha_\nu(\mathbf{k})}\left(1-2\,e^{-r|\mathbf{k}|^2}\right) -2\,r\,\alpha_\nu(\mathbf{k})\right]
\end{align}
and
\begin{align}
&\overset{2}{\Theta}\,^{i4}(x)=\frac{1}{2} \int_{\R^6}\frac{\D \mathbf{k}\:\D\mathbf{p}}{(2\pi)^6} \sum_{\nu,\mu}C_\nu(\mathbf{k})C_\mu(\mathbf{p})\,e^{i([\alpha_\nu(\mathbf{k})+\alpha_\mu(\mathbf{p})]t + [\mathbf{k}+\mathbf{p}]\cdot\mathbf{x})}\nonumber\\
&\times\left[k^i\left(\frac{e^{-r|\mathbf{p}|^2}}{\alpha_\nu(\mathbf{k})+ \alpha_\mu(\mathbf{k})}\left[e^{-r\alpha_\mu(\mathbf{k})^2} - e^{-r\alpha_\nu(\mathbf{k})^2}\right] - 4\,r\, \alpha_\nu(\mathbf{k})\,e^{-r(|\mathbf{k}|^2+|\mathbf{p}|^2)}\right)\right.\\
&\left. + \frac{(\mathbf{k}+\mathbf{p})^i(\alpha_\nu(\mathbf{k}) + \alpha_\mu(\mathbf{p}))}{f_{\nu\mu}(\mathbf{k},\mathbf{p})}\left(2\,r\,\left[1-2\,e^{-r|\mathbf{k}+\mathbf{p}|^2}\right] + \frac{1-e^{-r(\alpha_\nu(\mathbf{k})+\alpha_\mu(\mathbf{p}))^2}}{(\alpha_\nu(\mathbf{k})+\alpha_\mu(\mathbf{p}))^2} + \frac{e^{-r|\mathbf{k}+\mathbf{p}|^2}-1}{|\mathbf{k}+\mathbf{p}|^2}\right) \right]\,. \nonumber
\end{align}
Now, if we integrate the whole volume at $t = 0$, we obtain the $i$-components of linear momentum $P^i$, 
\begin{align}
P^i &:= \int_{\R^3}\D\mathbf{x}\,\Theta^{i4}(\mathbf{x},0) = -2\,r\,\kappa^2 \int_{\R^3}\frac{\D\mathbf{k}}{(2\pi)^3}\sum_{\mu\neq\nu^\prime,\nu} C_\nu(\mathbf{k}) C_{\mu}(-\mathbf{k}) \,k^i\,\alpha_\nu(\mathbf{k})\,e^{-2\,r\,|\mathbf{k}|^2} + \mathcal{O}(\kappa^3)\,.
\end{align}
With equations (\ref{padic:alphas}) and (\ref{padic:As}), it can be proved that the last expression is real, as expected. Therefore, everything holds.

\subsection{The energy-momentum tensor for a dispersive medium}\label{chap431}

Let us postulate that the following nonlocal action integral describes a homogeneous isotropic dispersive medium\footnote{For the reader who is more interested in the motivation of postulating this action integral to describe a dispersive media, we suggest reading \cite{Heredia1}.}
\begin{equation}   \label{e19}
 S(\tilde{A},R) = \frac1{4}\int_{|x|\leq R} \D x\, \tilde{F}_{ab}(x) \left(M^{abcd}\ast\tilde{F}_{cd}\right)_{(x)} \,, 
\end{equation}
where $\tilde{F}_{ab}(x) = \tilde{A}_{b|a}(x) - \tilde{A}_{a|b}(x)\,$, and $M^{abcd}(x)$ is known as the dielectric tensor and has the following structure:
 \begin{equation}\label{e18}
 M^{abcd}(x) = (2\pi)^{-2}\,\left[m(x) \,\hat\eta^{a[c}\hat\eta^{d]b} + 2 \varepsilon(x)\, u^{[a}\hat\eta^{b][c} u^{d]} \right]\,.  
 \end{equation}
The functions $m(x)$ and $\varepsilon(x)$ are the Fourier transforms of $\mu^{-1}(\omega,\mathbf{k})$ and $\varepsilon(\omega,\mathbf{k})\,$, namely, the dielectric and magnetic functions; moreover, $\hat\eta^{ab} := \eta^{ab}+ u^a u^b$ is the projector onto the hyperplane orthogonal to $u^b$, where this $u^b$ is the proper velocity for the laboratory frame; namely, a four-vector $u^b=(\gamma\,\mathbf{v},\gamma)$ with $\gamma=(1-|\mathbf{v}|^2)^{-1/2}$ and the standard three-velocity $\mathbf{v}$. 

The four-vector $u^a$ is a timelike unit vector, $u^a u_a = u^a u^b \eta_{ab} = -1$, which we shall take the particular case $u^a=(0,0,0,1)$ corresponding to the laboratory reference frame. Given any skewsymmetric tensor as $\tilde{F}_{ab}$, they exist $\tilde{E}_b$ and $\tilde{B}^d$ such that \cite{Choquet1982}
\begin{equation}   \label{s632-NLE}
    \tilde{F}_{ab} = 2\,u_{[a} \tilde{E}_{b]} + \hat{\tilde{F}}_{ab}  \,, \qquad \qquad    \hat{\tilde{F}}_{ab}= \varepsilon_{abcd} u^c \tilde{B}^d \,, \qquad \qquad  u^a \tilde{E}_a = u_d \tilde{B}^d = 0\,, 
\end{equation}
where $\varepsilon_{abcd}$ is the totally skewsymmetric Levi-Civita symbol in four dimensions:
$$ \varepsilon_{abcd} = \left\{ \begin{array}{cl}
                              -1 & \mbox{if $abcd$ is an even permutation of 1234} \\
                              1 & \mbox{if $abcd$ is an odd permutation of 1234} \\
                              0 & \mbox{if there is some repeated index} \,.
															\end{array}  \right.   $$
It can be easily checked that 
\begin{equation}   \label{e13a}
   \tilde{E}_a = \tilde{F}_{ab} u^b  \qquad {\rm and} \qquad \tilde{B}^d = \frac12\, \varepsilon^{cdab} u_c \tilde{F}_{ab}
\end{equation}
where we have used that $u^a$ is a unit vector and, as well as 
\begin{equation}
\varepsilon^{abcd}\varepsilon_{mned} = -\delta^{abc}_{mne} = \sum_{\sigma} {\rm sign}(\sigma) \,\delta_m^{\sigma_a}\,\delta_n^{\sigma_b}\,\delta_e^{\sigma_c} \,,
\end{equation}
with $\sigma$ running over the permutation group $S_3$. Notice that the particular case $u^a= (0,0,0,1)$ yields $\tilde{E}_a=(\tilde{E}_1,\tilde{E}_2,\tilde{E}_3,0)$ and $\tilde{B}^a=(\tilde{B}^1,\tilde{B}^2,\tilde{B}^3,0)$, namely, the electric field and the magnetic function, respectively. 

Note that $M^{abcd}(x)$ is a skewsymmetric tensor in both pairs of indices
\begin{equation}   \label{e19a}
M^{abcd}(x) = -M^{bacd}(x)= -M^{abdc}(x) \qquad \mathrm{and} \qquad M^{abcd}(-x) = M^{cdab}(x)\,,
\end{equation}
which, particularized to the specific form (\ref{e18}), amounts to require that $m$ and $\varepsilon$ are even functions,
\begin{equation}   \label{e19aa}
m(-x) = m(x) \qquad\quad {\rm and} \qquad\quad \varepsilon(-x) =\varepsilon(x) \,.
\end{equation}
See that the nonlocal action (\ref{e19}) also includes the local (namely, the non-dispersive) case since $M^{abcd}(x-y)=M_{\mathrm{o}}^{abcd}\,\delta^4(x-y)$, where $M_{\mathrm{o}}^{abcd}$ is a constant tensor.

The nonlocal Lagrangian density for dispersive media is then
\begin{equation}\label{chapEM-L}
\mathcal{L}(T_x\tilde A) = \frac{1}{4} \tilde{F}_{ab}(x) \left(M^{abcd}\ast\tilde{F}_{cd}\right)_{(x)}\,, 
\end{equation}
which is nonlocal because $\mathcal{L}$ depends on the field derivatives $\tilde{A}_{b|a}$ and, due to the convolution, it also depends on the values $\tilde{A}_{b|a}$ at any other point. The functional derivative (\ref{L2o}) is 
\begin{equation}  \label{A10_EM}
 \lambda^b(\tilde{A},x,z) :=  \frac12\,\delta_{|a}(x-z) \,\tilde{H}^{ab}(x) +  \tilde{A}_{e|a}(x)\, M^{aecb}_{\quad\,\,\,|c}(x-z) \,,
\end{equation}
where 
\begin{equation}\label{chapEM-H}
\tilde{H}^{ab}(x):= 2\left(M^{abcd}\ast \tilde{A}_{d|c}\right)_{(x)} = \left(M^{abcd}\ast \tilde{F}_{cd}\right)_{(x)}
\end{equation}
has been included and is known as the displacement tensor. Consequently, the nonlocal Euler-Lagrange equations (\ref{L2o}) become
\begin{equation}  \label{A11}
\partial_a \tilde{H}^{ab}(x) = 0\,.
\end{equation}

\subsubsection{The canonical energy-momentum tensor and the spin current}

Let us find the canonical energy-momentum tensor and the spin current for the case we are considering. Using (\ref{A10_EM}) and the symmetry condition (\ref{e19a}), the integrand of (\ref{P5}) becomes
\begin{align} 
&\lambda^b(\tilde{A},x+[s-1]z,x+sz)\,\tilde{A}_{b|a}(x+sz) = \nonumber\\
 & \qquad -\frac12\,\left\{\delta_{|a}(z)\,\tilde{H}^{ab}(x+[s-1]z) + \tilde{F}_{ae}(x+[s-1]z)\,M^{cbae}_{\quad\;\,|c}(z)\right\}\,\tilde{A}_{b|a}(x+sz)\,. 
\end{align}
Therefore, the canonical energy-momentum tensor (\ref{P5}) is
\begin{align}\label{D1}  
&\mathcal{T}^{ab} = -\mathcal{L}(T_x\tilde{A})\,\eta^{ab} + \frac12\,\tilde{H}^{bc}(x)\,\tilde{A}_c^{\,\,\,|a}(x)\nonumber\\
&\qquad \qquad \qquad- \frac12\,\int_{\mathbb{R}^4} \D z\,z^b \int_0^1\D s\,
\tilde{F}_{fe}(x+[s-1]z)\,M^{dnfe}_{\quad\,\,\,\,\,|d}(z)\,\tilde{A}_n^{\,\,\,|a} (x+s z)\,. 
\end{align}
\indent As the field $\tilde{A}_a$ transforms as a covariant vector $\tilde{A}^\prime_a(x^\prime) = \tilde A_a(x) - \omega^b_{\,\,a} \tilde A_b(x)$, it implies that equation (\ref{chap331-P3}) is $\delta \tilde{A}_a(x) = - \omega^b_{\,\,a} \tilde A_b(x) - \tilde{A}_{a|c} \delta x^c$, which we can identify the term $M^A_{B[ab]}$ of equation (\ref{chap331-P3}) as
\begin{equation}\label{chapEM_M}
M^A_{B[ab]} = \delta^A_{[a}\eta_{b] B}\,.
\end{equation}
Thus, plugging (\ref{chapEM_M}) into (\ref{P6a}), we find that the spin current is\vspace{0.2cm}
\begin{align}\label{N7}
\mathcal{S}^{\; \,\;b}_{ac} &= \tilde{H}^b_{\;[c}(x)\tilde{A}_{a]}(x) + \int_{\mathbb{R}^4} \D z\,z^b\,\int_0^1\D s\,\tilde{F}_{fe}(x+[s-1]z)\,M^{dnfe}_{\quad\,\,\,|d}(z) \nonumber  \\  
&\hspace*{15em} \times \left\{\,\eta_{n[a} \tilde{A}_{c]}(x+s z) + s z_{[a} \tilde{A}_{n|c]}(x+sz)\right\}\,.
\end{align}
\indent It is easy to check that for the local case (i.e., for the non-dispersive case) $M^{dnfe}(x)\propto\delta(x)$, equations (\ref{D1}) and (\ref{N7}) become
\begin{equation}  \label{N7z}
\mathcal{T}^{\; b}_a = -\mathcal{L}\,\delta^b_a + \tilde{H}^{bc}\,\tilde{A}_{c|a}  \qquad {\rm and} \qquad \mathcal{S}^{\; \,\;b}_{ac} = 2 \tilde{H}^b_{\;[c}\tilde{A}_{a]} \,,
\end{equation}
as expected.

\subsubsection{The Belinfante-Rosenfeld energy-momentum tensor}

In the case of our nonlocal Lagrangian density (\ref{chapEM-L}), the canonical energy-momentum tensor  (\ref{D1}) is not symmetric; however, let us use the  Belinfante-Rosenfeld symmetrization technique to construct the Belinfante-Rosenfeld energy-momentum tensor $ \Theta^{ba}$.\\
\indent First, let us substitute (\ref{N7}) in (\ref{chap331-W}); it leads to
\begin{align}\label{N11}
\mathcal{W}^{cab} &= \frac12\,\tilde{H}^{ca}(x) \tilde{A}^b(x) + \frac12\,\int_{\mathbb{R}^4} \D z\,\int_0^1\D s\,\tilde{F}_{fe}(x+[s-1]z)\,M^{dnfe}_{\quad\,\,\,|d}(z)\nonumber\\
&\times \left\{2 s\,z^bz^{[c} \tilde{A}_{\,\,\,|n}^{a]}(x+sz) + \left(\delta_n^{[c} \eta^{a]m} z^b + \delta_n^{[c} \eta^{b]m} z^a  - \delta_n^{[a} \eta^{b]m} z^c \right)\tilde{A}_m(x+sz) \right\}\,.
\end{align}
When calculating $\partial_c \mathcal{W}^{cab}$, the combination $z^c\partial_c$ shall occur in several instances as
\begin{equation}
z^c\partial_c P(x+sz,z) = \frac{\partial}{\partial s}\,P(x+sz,z) \,,
\end{equation}
where $P(x+sz,z)$ is a product of $\tilde{F}_{fe}(x+[s-1]z)$ times either $\tilde{A}_m(x+sz)$ or its derivative. This fact allows us to compute some integrals on $s$ as
\begin{align}
\int_0^1\D s\,z^c\partial_c P(x+sz,z) &= P(x,x+z) - P(x-z,x) \nonumber\\
\int_0^1\D s\,s \,z^c\partial_c P(x+sz,z) &= P(x,x+z) - \int_0^1\D s \,P(x+sz,z)\,.  
\end{align}
Hence, from (\ref{N11}) and after a bit of algebra, it follows that
\begin{align} \label{N12} 
&\partial_c\mathcal{W}^{cab}  =  \frac12\tilde{H}^{ca}(x) \tilde{A}^b_{\;|c}(x) + \frac12 \tilde{F}_{fe}(x)\int_{\mathbb{R}^4} \D z\,M^{dnfe}_{\quad\,\,\,|d}(z) \left[z^b\,\tilde{A}_n^{\;|a}(x+z) + \delta_n^{[b} \tilde{A}^{a]}(x+z)\right] \nonumber\\ 
 &\quad + \frac12 \int_{\mathbb{R}^4} \D z\,\int_0^1\D s\,M^{dnfe}_{\quad\;\,\,\,|d}(z) \left\{ \delta_n^{[c}\left(\eta^{a]m} z^b + \eta^{b]m} z^a \right) \partial_c\left[\tilde{F}_{fe}(x+[s-1]z)\,\tilde{A}_m(x+sz) \right] \right. \nonumber\\
 &\quad \left.- z^b \tilde{F}_{fe}(x+[s-1]z)\,\tilde{A}_n^{\;\,\,|a}(x+sz) - s z^b z^a \partial_c\left[\tilde{F}_{fe}(x+[s-1]z)\,\tilde{A}_n^{\;\,\,\,|c}(x+sz) \right] \right\} \,.
\end{align}
Substituting in (\ref{chap331-BT}) leads to\vspace{0.2cm}
\begin{align} \label{D4A}
&\Theta^{ba}  = \frac12 \tilde{H}^{ca}(x) \tilde{F}_c^{\;\,b}(x) -\frac14\eta^{ab} \tilde{F}_{ed}(x) \tilde{H}^{ed}(x) \nonumber\\
&\quad +\frac12\tilde{F}_{fe} \left[\left(M^{fed[b}\ast\tilde{F}^{a]}_{\;\,d}\right)_{(x)} +  \left(M^{fed(b|a)} \ast \tilde{A}_d\right)_{(x)} + \frac12 \left(\left[y^b M^{fedn}\right]\ast \tilde{F}_{dn}^{\,\,\,|a}\right)_{(x)} \right] \nonumber \\
 &\quad- \frac12 \int_{\R^4} \D z M^{ndfe}(z)\frac{\partial}{\partial z^d}\int_0^1 \D s\,z^{(a}\left\{\tilde{F}_{fe}(x+[s-1]z) \left[\tilde{A}_n^{\,\,|b)}(x+sz) + \tilde{F}^{b)}_{\,\,\,n}(x+sz)\right] \right. \nonumber\\
 &\quad - \tilde{F}_{fe|n}(x+[s-1]z) \tilde{A}^{b)}(x+sz)  + \left. \delta^{b)}_n \left[\tilde{F}_{fe}(x+[s-1]z) \tilde{A}^c(x+sz) \right]_{|c} \right.\nonumber\\
 &\left. \hspace*{6cm}+ s z^{b)} \left[\tilde{F}_{fe}(x+[s-1]z) \tilde{A}_n^{\,\,\,|c}(x+sz)  \right]_{|c} \right\}\,.
\end{align}
To check that the result is consistent, for the local case (non-dispersive media), we obtain that $\Theta^{ba}$ is
\begin{equation}  \label{e17d}
\Theta^{ba} = \tilde{H}^{ac} \tilde{F}^b_{\,\;c} - \frac14 \,\eta^{ba}\,\tilde{H}^{mn} \tilde{F}_{mn}  \,,  
\end{equation}
which is indeed the Minkowski energy-momentum tensor \cite{Ramos2015}. \\
\indent Of course, equation (\ref{D4A}) is not symmetric and does not have to be. Recall that the Belinfante-Rosenfeld tensor $\Theta^{ba}$ is symmetric if the angular momentum current $\mathcal{J}^{\;\;\;b}_{ca}$ is conserved \cite{dixon1978}, which would follow from the Noether theorem and the Lorentz invariance of the Lagrangian. However, notice that the Lagrangian (\ref{chapEM-L}) is not Lorentz invariant because the dielectric tensor $M^{abcd}$ privileges the time vector $u^a$, which breaks boost invariance. Therefore, the nonlocal Lagrangian density (\ref{chapEM-L}) is only invariant under the Lorentz subgroup that preserves $u^a$. For instance, it can be checked that the part of (\ref{e17d}) that is orthogonal to $u^b$ is indeed symmetric.

\subsubsection{Gauge dependence and conservation}

Since both expressions depend linearly on the electromagnetic potential, they are gauge-dependent. The potential $\tilde{A}_b$ can be eliminated by employing the inverse of the definition $\tilde{F}_{ed} = \tilde{A}_{d|e} - \tilde{A}_{e|d}$; by the Poincar\'e lemma \cite{Spivak1965,Choquet1982}, the inverse is indeed
\begin{equation}   \label{D1a}  
 \tilde{A}_b(x) = \int_0^1 \D\tau\, \tau\,x^c\, \tilde{F}_{cb}(\tau x) + \partial_b f(x)\,, 
\end{equation}
where $f(x)$ is an arbitrary function that is related to gauge transformations. Thus, due to the linear dependence, both energy-momentum tensors split into one part that only depends on $\tilde{F}_{cb}$ and is gauge-independent and another one that depends linearly on $\partial_b f$, i.e., a gauge-dependent contribution.  It can be proved that the gauge parts are conserved; hence, we can take the gauge-independent parts as the definitions of the energy-momentum tensors.  \\
\indent Let us now check whether these tensors are locally conserved or not. To begin with, we have that, by construction, $\partial_b\Theta^{ab} = \partial_b\mathcal{T}^{ab}$. Therefore, proving the conservation of the canonical energy-momentum tensor is sufficient. Thus, from (\ref{D1}), we have that
\begin{align}
\partial_b\mathcal{T}^{ab} &= - \frac14 \tilde{F}_{cd}(x) \tilde{H}^{cd|a}(x) \nonumber\\ 
&\quad -\frac12\int_{\mathbb{R}^4} \D z\,M^{dnfe}_{\quad\,\,\,|d}(z)\, \int_0^1\D s\, z^b \,\partial_b\left[\tilde{F}_{fe}(x+[s-1]z) \tilde{A}_n^{\,\,\,|a} (x+sz) \right] \,,
\end{align}
where we have used the nonlocal Euler-Lagrange equations (\ref{A11}). Including now the identity
\begin{equation}
z^b \,\partial_b\left[\tilde{F}_{fe}(x+[s-1]z) \tilde{A}_n^{\,\,\,|a} (x+sz) \right] = \frac{\partial}{\partial s}\,\left[\tilde{F}_{fe}(x+[s-1]z) \tilde{A}_n^{\,\,\,\,|a} (x+sz) \right] \,,
 \end{equation}
we can perform the integral and arrive at
\begin{align}
 &\partial_b\mathcal{T}^{ab} = - \frac14 \tilde{F}_{cd}(x) \tilde{H}^{cd|a}(x) - \frac12\int_{\mathbb{R}^4} \D z\,M^{dnfe}_{\quad\,\,\,|d}(z)\,
\left[\tilde{F}_{fe}(x) \tilde{A}_n^{\,\,\,|a} (x+z) - \tilde{F}_{fe}(x-z) \tilde{A}_n^{\,\,\,|a} (x)\right] \nonumber\\
 &\quad = - \frac14 \tilde{F}_{cd}(x) \tilde{H}^{cd|a}(x) - \frac12\int_{\mathbb{R}^4}\D z\,M^{dnfe}_{\quad\;|d}(z)\,\tilde{F}_{fe}(x) \tilde{A}_n^{\,\,\,|a}(x+z) + \frac12 \tilde{H}^{dn}_{\,\,\,\,|d}(x) \tilde{A}_n^{\,\,\,|a}(x)\,.
\end{align}
The last term vanishes due to the nonlocal Euler-Lagrange equations, and by integrating by parts and using the $dn$-skewsymmetry, we have that
\begin{equation}
\partial_b\mathcal{T}^{ab} = - \frac14 \tilde{F}_{cd}(x) \tilde{H}^{cd|a}(x) + \frac14 \tilde{F}_{fe}(x)\int_{\mathbb{R}^4}\D z\,M^{dnfe}(-z)\,\tilde{F}_{dn}^{\,\,\,\,|a}(x-z) \,,
\end{equation}
that is,
\begin{equation}  \label{D5}
 \partial_b\mathcal{T}^{ab}  = -\frac12 \tilde{F}_{fe}(x) \left(M_-^{fedn} \ast \tilde{F}_{dn}^{\quad|a}\right)_{(x)}\,,
\end{equation}
where $M_-^{cdef}(z) := \frac12[M^{cdef}(z) - M^{efcd}(-z)]$. In case the constraints (\ref{A11}) can be derived from a Lagrangian, then the symmetry relation (\ref{e19a}) implies that $M_-^{fedn}$ vanishes and both energy-momentum tensors, (\ref{D1}) and (\ref{D4A}), are locally conserved.

\subsubsection{Real dispersive media: absorption and causality}
Recall that the symmetry relation (\ref{e19a}) implies that the dielectric and magnetic functions are even. Hence, their Fourier transforms $\varepsilon$ and $\mu$ are real-valued; therefore, is so the refractive index $n$. However, causality implies that the real and imaginary parts of the functions $\varepsilon$ and $\mu$ must fulfill either the Kramers-Kr\"onig relations \cite{Llosa2004} (in the optical approximation, i.e., $\varepsilon$ and $\mu$ only depend on the angular frequency $\omega$) or the Leontovich relations \cite{LEONTOVICH1961,Llosa2019} in the general case.  Consequently, if $\varepsilon$ and $\mu$ were real-valuated, they should be constant and the medium would be non-dispersive.

For a real dispersive medium, $\varepsilon$ and $\mu$ are not constant; therefore, the symmetry relation (\ref{e19a}) is not fulfilled, and the right-hand side of (\ref{D5}) does not vanish. However, including (\ref{e18}), we have
\begin{equation}
M_-^{fedn} = (2\pi)^{-2}\,\left[m_- \,\hat\eta^{f[d}\hat\eta^{n]e} + 2 \varepsilon_-\, u^{[f}\hat\eta^{e][d} u^{n]} \right] \,,
\end{equation}
with $m_-(y) = \frac{\tilde{m}(y) - \tilde{m}(-y)}2$ and similarly for $\varepsilon_-(y)$. By using (\ref{s632-NLE}), we can divide the electric and magnetic parts and write (\ref{D5}) as
\begin{equation}
\partial_b\mathcal{T}^{ab}  = (2\pi)^{-2}\,\left[ \tilde{E}_d(x)\,\left(\varepsilon_-\ast \tilde{E}^{d|a}\right)_{(x)} - \tilde{B}_d(x)\,\left(m_-\ast \tilde{B}^{d|a}\right)_{(x)} \right]\,.
\end{equation}
Now, as the Fourier transforms of $m_-(x)$ and $\varepsilon_-(x)$ are connected with $\imap\,m(k)$ and $\imap\, \varepsilon(k)$, i.e., the absorptive parts of the magnetic and dielectric functions, we have that the failure of local conservation of energy-momentum in a real medium is due to absorption.

\subsubsection{Plane wave solutions \label{S3.4}}
Let us study the implications of Maxwell's equations for plane wave solutions.  Considering these particular solutions 
\begin{equation}
\tilde{F}_{cd}(x) = f_{cd}\,e^{i k_b x^b}\,,\qquad {\rm with} \qquad f_{cd} +f_{dc} = 0
\end{equation} 
and substituting them into the constraints (\ref{A11}), we get
\begin{equation}   \label{OP1}  
M^{abcd} (k)\, f_{cd} k_b = 0\,, 
\end{equation}
where $M^{abcd} (k)$ is the Fourier transform of $M^{abcd}(x)$. Now, as $\tilde{F}_{cd}$ can be derived from an electromagnetic potential $\tilde{A}_b$, it must fulfill the first pair of Maxwell equations, which for plane waves reads $\, k_b f_{cd} +  k_c f_{db} +  k_d f_{bc} = 0 \,$ and whose general solution is   
\begin{equation}   \label{OP2}  
 f_{cd} = f_c k_ d - f_d k_c\,,
\end{equation}
where $f_c$ is the wave polarization vector and is determined apart from the addition of a multiple of $k_c\,$.
Substituting this into equation (\ref{OP1}), we arrive at
\begin{equation}   \label{OP3}  
M^{abcd}(k) \, k_b \, k_d \, f_c = 0\,, 
\end{equation}
which is a linear homogeneous system and admits non-trivial solutions for the polarization vector if, and only if, 
\begin{equation}
 \det\left[ M^{abcd}(k) \, k_b \, k_d \right] = 0\,.  
 \end{equation}

We shall assume that the dielectric tensor $M^{abcd}$ has the form (\ref{e18}); hence, its Fourier transform is
\begin{equation}   \label{OP6}  
 M^{abcd}(k) = m(k) \,\hat\eta^{a[c}\hat\eta^{d]b} + 2  \varepsilon(k)\, u^{[a}\hat\eta^{b][c} u^{d]}\,. 
\end{equation}
If the medium is spatially isotropic, the functions $\varepsilon(k) $ and $m(k)=\mu^{-1}(k)$ depend on the wave vector $k^b$ through the scalars $\omega = - k^b u_b$ and $q^2:= k^b k_b + \omega^2\,$, where we have taken  
\begin{equation}
k^a = \omega u^a + q\,\hat{\mathbf{q}}^a \,,   \qquad {\rm with} \qquad \hat{\mathbf{q}}^a\hat{\mathbf{q}}_a =  1 \,, \qquad \quad \hat{\mathbf{q}}^a u_ a = 0\,. 
\end{equation}
\indent As the polarization is determined up to the addition of a multiple of $k^c$, we can choose it so that
\begin{equation}
f^c = \psi\,\hat{\mathbf{q}}^c + f_\perp^c \qquad {\rm with} \qquad f_\perp^c u_c = f_\perp^c \hat{\mathbf{q}}_c = 0 \,.
\end{equation}
Substituting this in equation (\ref{OP3}), we arrive at\vspace{0.2cm}
\begin{equation}
\left( q^2 - \omega^2  n^2  \right)\,f^a_\perp  = 0 \,, \qquad \quad \psi=0 \,,
\end{equation}
where $n(q,\omega) =\sqrt{\varepsilon \mu}$ is the refractive index. Hence, there are non-trivial solutions if, and only if,
\begin{equation}     \label{OP4}
q^2 - \omega^2  n^2   = 0 \qquad \qquad {\rm and}  \qquad \qquad  f_b k^b = f_b u^b  = 0\,.  
\end{equation}

Maxwell's equations thus imply that: \textbf{a)} the waves are polarized transversely to the plane spanned by $k^b$ and $u^b$, and \textbf{b)} the velocity phase satisfies the dispersion relation $q = \omega\,n(q,\omega)$.

\subsubsection{The general solution of field equations}
The initial value problem for a nonlocal partial differential system such as Maxwell's equations (\ref{A11}) is not as simple as the Cauchy problem for a first-order one. We shall now see that the nature of the initial data problem for (\ref{A11}) depends on whether the number of real roots of the dispersion relations (\ref{OP4}) is finite.

As Maxwell equations (\ref{A11}) are linear and involve a convolution, the Fourier transform is an excellent tool to solve them. We write 
\begin{equation}     \label{GS1}
\tilde{F}_{cd}(x) = (2\pi)^{-2} \,\int_{\R^4} \D k \,f_{cd}(k)\,e^{ik_a x^a} 
\end{equation}
that, substituted in Maxwell equations, yields
\begin{equation}     \label{GS2}
f_{cd} = f_c k_d - f_d k_c \qquad {\rm and} \qquad M^{abcd}(k) \, k_b \, k_d \, f_c = 0\,.
\end{equation}
As discussed above, this result means that $f_c$ vanishes unless the dispersion relation (\ref{OP4}) is fulfilled, i.e.,
\begin{equation}
f_{cd}(k) = f^+_{cd}(k)\,\delta\left[q - \omega \,n(q,\omega)\right] + f^-_{cd}(k)\,\delta\left[q + \omega \,n(q,\omega)\right] 
\end{equation}
or
\begin{equation}     \label{GS3}
 f_{cd}(k) = \sum_\alpha f^{\alpha}_{cd}(\mathbf{q})\,\delta\left[\omega-\omega_\alpha(q)\right] + \sum_\beta f^{\beta}_{cd}(\mathbf{q})\,\delta\left[\omega-\omega_\beta(q)\right] \, , 
\end{equation}
where $\omega_\alpha$ (resp. $\omega_\beta$) are the positive (resp. negative) real roots of the dispersion relation (\ref{OP4}). The arbitrary coefficients $ f^{\alpha}_{cd}(\mathbf{q}) $ and $ f^{\beta}_{cd}(\mathbf{q})$ depend on the initial data. Indeed, the $n^{\rm{th}}$-time derivative of (\ref{GS1}) at $x^a= (\mathbf{x},0)$ yields
\begin{equation}
\partial^n_t \tilde{F}_{cd}(0,\mathbf{x}) = (2\pi)^{-2} \,\int_{\R^3} \D \mathbf{q} \,e^{i\mathbf{q}\cdot\mathbf{x}} \,\left[\sum_\alpha f^{\alpha}_{cd}(\mathbf{q})\,(-i\omega_\alpha)^n + \sum_\beta f^{\beta}_{cd}(\mathbf{q}) \,(-i\omega_\beta)^n \right]\,,
\end{equation}
where (\ref{GS3}) has been used to perform the integration on $\omega$, whence it follows that
\begin{equation}     \label{GS4}
\sum_\alpha f^{\alpha}_{cd}(\mathbf{q})\,\omega^n_\alpha + \sum_\beta f^{\beta}_{cd}(\mathbf{q}) \,\omega^n_\beta = \frac{i^n}{2\pi}\, 
\int_{\R^3} \D \mathbf{x} \,e^{-i\mathbf{q}\cdot\mathbf{x}} \, \partial^n_t \tilde{F}_{cd}(0,\mathbf{x})\,. 
\end{equation}
Since we have as many unknowns $f^{\alpha}_{cd}$ and $f^{\beta}_{cd}$ as the number of real roots of the dispersion relation (\ref{OP4}), the number of initials ($t=0$) time derivatives of $\tilde{F}_{cd}(t,\mathbf{x}) $ that are needed at least to determine a solution is the real root numbers.  If it is $N <\infty\,$, then giving $\tilde{A}_d(\mathbf{x},0) $ and its time derivatives up to the order $N-1$, the solution of the system is determined; otherwise, the initial value problem requires further study.

\subsubsection{The energy-momentum tensor for a wave packet}

Aiming to compare the energy-momentum tensor obtained here with other proposals advanced in literature, for instance \cite{Landau1984, Jackson1999, schwinger1998}, we shall particularize the Belinfante-Rosenfeld tensor expression (\ref{D4A}) to the wave packet 
\begin{equation}     \label{OP5}
\tilde{F}_{cd}(x) =  \realp \left(\underline{\tilde{F}}_{cd}(x) \, e^{ik_a x^a} \right) = |\underline{\tilde{F}}_{cd}(x)|\,\cos\left(k_a x^a + \varphi\right)    \,,  \qquad k^a = \mathbf{q}^a + \omega u^a\,.
\end{equation}
$\underline{\tilde{F}}_{cd}(x)$ denotes a ``slowly'' varying complex amplitude (compared with the rapidly oscillating carrier $e^{ik_c x^c}$), $|\underline{\tilde{F}}_{cd}(x)|$ is the modulus of each component, and $\varphi$ is the phase.
Furthermore, we shall restrict to the optical approximation, that is, $\varepsilon$ and $\mu$ only depend on the frequency $\omega = - u_a k^a$; therefore, we shall take
\begin{equation}     \label{OP7a}
 M^{abcd}(x) = (2\pi)^{-1/2}  \delta^3(\mathbf{x})\, m^{abcd}(t) \,,
\end{equation} 
where $m^{abcd}(t) = m(t) \,\hat\eta^{a[c}\hat\eta^{d]b} + 2 \varepsilon(t)\,u^{[a}\hat\eta^{b][c} u^{d]}$,  $t= -x^a u_a\,$, and $\mathbf{x}^a = x^a - t\,u^a$ is the spatial part of $x^a\,$. Using this, the displacement tensor (\ref{chapEM-H}) yields
\begin{equation}     \label{OP7}
 \tilde{H}^{ca}(x) =  \realp \left(\underline{\tilde{H}}^{ca}(x) \, e^{ik_b x^b } \right)
  \qquad {\rm with} \qquad  \underline{\tilde{H}}^{ca}(x) \approx m^{caed}(\omega) \underline{\tilde{F}}_{ed}(x)\,,
\end{equation}
where $\approx$ means that the ``slow variation'' approximation\footnote{In other words, $\partial_a \underline{\tilde{F}}_{ed}(x) \approx 0$. } has been included to evaluate the convolution, and $m^{caed}(\omega)$ is the Fourier transform of $m^{caed}(t)\,$, that is,\vspace{0.2cm}
\begin{equation}
(2\pi)^{-1/2}  \int_\R \D t\, m^{caed}(t) \underline{\tilde{F}}_{ed}(x^b-t u^b)\,e^{ik_b x^b + i \omega t } \approx m^{caed}(\omega) \underline{\tilde{F}}_{ed}(x)\,e^{ik_b x^b} \,.
\end{equation}
\indent By employing (\ref{OP7}), the Maxwell equations become
\begin{equation} 
D_c \underline{\tilde{H}}^{ca} \approx 0 \qquad {\rm and} \qquad D_b\underline{\tilde{F}}_{cd}+ D_c\underline{\tilde{F}}_{db}+ D_d\underline{\tilde{F}}_{bc} \approx 0\,,
\end{equation}
where $D_b := \partial_b + i k_b\,$. For slowly varying amplitudes, they reduce to the Maxwell equations for a plane wave, and we can write (see Section \ref{S3.4})
\begin{equation}     \label{OP8}
\underline{\tilde{F}}_{cd} \approx -\frac1\omega\,\left( \tilde{\underline{E}}_c k_d - \tilde{\underline{E}}_d k_c \right) \qquad{\rm and}\qquad m^{caed}(\omega) k_c \tilde{\underline{E}}_e k_d \approx 0\,,
\end{equation}
where $\tilde{\underline{E}}_c = \underline{\tilde{F}}_{cd} u^d\,$ is the electric field, and, furthermore, we have taken into account that $f_c u^c =0$. 
\vspace{0.2cm}
Similarly, as in Section \ref{S3.4}, the second equation implies that
\begin{equation}     \label{OP8a}
\tilde{\underline{E}}_c k^c = 0   \qquad \qquad{\rm and}\qquad \qquad q^2 = \omega^2 \varepsilon(\omega) \mu(\omega)\,.
\end{equation}
Moreover, from (\ref{OP5}) and (\ref{OP8}), it follows that the electromagnetic potential is 
\begin{equation}     \label{OP9}
 \tilde{A}_b(x) =  \realp \left(\underline{\tilde{A}}_b(x) \, e^{ik_cx^c} \right)
  \qquad {\rm with} \qquad  \underline{\tilde{A}}_c \approx - \frac{i}\omega \,\underline{\tilde{E}}_c + \underline\alpha\,k_c \,,
\end{equation}
where $\underline\alpha(x)$ is an arbitrary gauge function. If we now substitute the wave packet (\ref{OP5}) in the Belinfante-Rosenfeld tensor expression (\ref{D4A}), we find that:\\
\indent \textbf{a)} The evaluation of the convolution products in the first line yields
\begin{align}
M^{edh[b}\ast \tilde{F}^{a]}_{\;\,h} &\approx \realp\left( m^{edh[b}(\omega) \underline{\tilde{F}}^{a]}_{\;\,h} e^{i k_c x^c}   \right)\,,\nonumber\\
M^{edh(b;a)}\ast \tilde{A}_h &\approx \realp\left( m^{edh(b}(\omega) D^{a)}\underline{\tilde{A}}_h e^{i k_c x^c}   \right)\,, \quad \mathrm{and} \nonumber\\
\left( y^b M^{edhf}\right)\ast \tilde{F}_{hf}^{\,\,\,\,|a} &\approx \realp\left(-i\,D^a \underline{\tilde{F}}_{hf} \partial_\omega m^{edhf}(\omega) \right)\,. 
\end{align}
Thus, every term in the first line of (\ref{D4A}) has the form\vspace{0.2cm}
\begin{equation}  \label{OP10}
\Phi\,\Psi = \realp\left(\underline{\Phi} e^{i k_c x^c}\right)\,\realp\left( \underline{\Psi} e^{i k_c x^c}   \right) = \frac12\, \realp\left(\underline{\Phi}\,\underline{\Psi} e^{2 i k_c x^c} + \underline{\Phi}\,\underline{\Psi}^\ast \right) 
\end{equation}
and consists of a slowly varying part plus a rapidly oscillating one. Taking the average for the carrier period, we obtain
\begin{equation}
\langle \Phi\,\Psi \rangle = \frac12\, \realp\left(\underline{\Phi}\,\underline{\Psi}^\ast \right) \,,
\end{equation}
where we have taken into account that the number of oscillations are very high compared to the average interval $T$. Due to the periodicity of the wave function, this causes that only the contribution that is constant remains.\\
\indent \textbf{b)} To evaluate the integral in the second and third lines in (\ref{D4A}), we realize that each term contains a group of the kind of (\ref{OP10}) and a Fourier integral of $m^{hfed}(t)\,$. We shall use the slow variation approximation and take the mean over a carrier period.\\
\indent A tedious calculation leads to 
\begin{align}
\langle \Theta^{ba} \rangle &\approx \frac14\,\realp \left[2 \underline{\tilde{H}}^{ca} \underline{\tilde{F}}_c^{\ast \,b} - \underline{\tilde{H}}^{c[a} \underline{\tilde{F}}_c^{\ast \,b]} - \frac12\,\underline{\tilde{H}}^{cd} \underline{\tilde{F}}_{cd}^\ast \,\eta^{ab} + \underline{\tilde{F}}_{ed}^\ast  m^{edh[b} \underline{\tilde{F}}^{a]}_{\;\;h}   \right. \nonumber\\
&\qquad \qquad \left. + i\,m^{edh(a} k^{b)} \left( \underline{\tilde{F}}_{ed}^\ast \underline{\tilde{A}}_h - \underline{\tilde{F}}_{ed}  \underline{\tilde{A}}_h^\ast \right)- \frac12\,u^a k^b \underline{\tilde{F}}_{ed}\, \partial_\omega m^{edhf} \underline{\tilde{F}}_{hf}^\ast\right]\,,
\end{align}
which, using the relations (\ref{OP8}-\ref{OP9}), can be simplified to
\begin{equation}  \label{OP11}
\langle \Theta^{ba} \rangle \approx \frac12\,\realp \left[\underline{\tilde{H}}^{ca} \underline{\tilde{F}}_c^{\ast \,b}  - \frac14\,\underline{\tilde{H}}^{cd} \underline{\tilde{F}}_{cd}^\ast \,\eta^{ab} - \frac14\,u^a k^b \underline{\tilde{F}}_{ed}\, \partial_\omega m^{edhf} \underline{\tilde{F}}_{hf}^\ast \right]\,.
\end{equation}
From the latter, we easily obtain that the energy density in the medium rest frame ($u^a =\delta_4^a$) is\vspace{0.2cm}
\begin{equation}
\mathcal{U}  = \langle \Theta^{44} \rangle \approx \frac14\,\realp \left[(\varepsilon + \omega\partial_\omega\varepsilon) \underline{\tilde{\mathbf{E}}} \cdot \underline{\tilde{\mathbf{E}}}^\ast + (m - \omega \partial_\omega m) \underline{\tilde{\mathbf{B}}} \cdot \underline{\tilde{\mathbf{B}}}^\ast \right] \,, 
\end{equation}
where $\underline{\tilde{\mathbf{E}}} \cdot \underline{\tilde{\mathbf{E}}}^\ast = \tilde{E}_a \tilde{E}^{\ast\,a}\,$, or
\begin{equation}  \label{OP12}
\mathcal{U} \approx \frac14\,\realp \left[\frac{\D (\varepsilon \omega)}{\D\omega}\, \underline{\tilde{\mathbf{E}}} \cdot \underline{\tilde{\mathbf{E}}}^\ast + \frac{\mu^\ast}{\mu} \,\frac{\D (\mu \omega)}{\D\omega}\, \underline{\tilde{\mathbf{H}}} \cdot \underline{\tilde{\mathbf{H}}}^\ast \right] \,, 
\end{equation}
where $\tilde{\mathbf{H}}$ is the magnetic field that is related via the magnetic induction $\tilde{\mathbf{H}}= (2\pi)^{-2}(m \ast \tilde{\mathbf{B}})$. This procedure reproduces previous results in the literature --see \cite{Landau1984, Jackson1999, schwinger1998}.

The momentum density in the medium rest frame is $G^i := \langle \Theta^{i4} \rangle$, and, in an obvious vector notation, we obtain from (\ref{OP11}) that
\begin{equation}
\mathbf{G} \approx \frac12\,\realp \left[\varepsilon\, \underline{\tilde{\mathbf{E}}} \times \underline{\tilde{\mathbf{B}}}^\ast + \frac12\,\left(\partial_\omega\varepsilon\,\underline{\tilde{\mathbf{E}}} \cdot \underline{\tilde{\mathbf{E}}}^\ast - \partial_\omega m\,\underline{\tilde{\mathbf{B}}} \cdot \underline{\tilde{\mathbf{B}}}^\ast \right)\, \mathbf{q} \right]\,.
\end{equation}
Now, from Maxwell's equations (\ref{OP8a}) and the second equation of (\ref{s632-NLE}), it follows that
\begin{equation}
\underline{\tilde{\mathbf{E}}} \times \underline{\tilde{\mathbf{B}}}^\ast = \frac{\underline{\tilde{\mathbf{E}}} \cdot \underline{\tilde{\mathbf{E}}}^\ast}\omega\,\mathbf{q}
\qquad {\rm and} \qquad \underline{\tilde{\mathbf{B}}} \cdot \underline{\tilde{\mathbf{B}}}^\ast = \varepsilon\mu\,\underline{\tilde{\mathbf{E}}} \cdot \underline{\tilde{\mathbf{E}}}^\ast\,,
\end{equation}
which substituted above yields
\begin{equation}  \label{OP13}
\mathbf{G} \approx \frac14\,\realp \left[\frac1{\omega\mu}\,\frac{\D (\varepsilon\mu \omega^2)}{\D\omega}\,\underline{\tilde{\mathbf{E}}} \times \underline{\tilde{\mathbf{B}}}^\ast \right]\,.
\end{equation}

The Poynting vector is $S^i = \langle \Theta^{4i} \rangle$; therefore, we get that
\begin{equation}
\mathbf{S} \approx \frac12\,\realp \left[\underline{\tilde{\mathbf{E}}}^\ast\times\underline{\tilde{\mathbf{H}}}^\ast \right]\,,  
\end{equation}
and the Maxwell stress tensor is
\begin{equation}
T^{ij} = - \langle \Theta^{ij} \rangle \approx  \frac12\,\realp \left[\underline{\tilde{E}}^{\ast\,i}\underline{\tilde{D}}^j + \underline{\tilde{H}}^i \underline{\tilde{B}}^{\ast\,j} -\frac12\,\left(\underline{\tilde{\mathbf{D}}} \cdot \underline{\tilde{\mathbf{E}}}^\ast + \underline{\tilde{\mathbf{H}}} \cdot \underline{\tilde{\mathbf{B}}}^\ast \right)\,\delta^{ij}\right]\,,
\end{equation}
where $\tilde{D^j}$ or $\tilde{\mathbf{D}}$ is the electric displacement that is related to the electric field via $\tilde{D}=(2\pi)^{-2} (\varepsilon\ast\tilde E)$.

\chapter{Conclusions}\label{concl}

We have studied dynamical systems and field theories governed by a nonlocal Lagrangian. Their treatment differs from the local Lagrangian approach usually considered in mechanics textbooks, especially concerning time evolution and the initial data space.

 In the local (first-order) mechanics case, the Euler-Lagrange equations form an ordinary differential system and, due to the existence and uniqueness theorems: (a) the initial data space has a finite number of dimensions --twice as many as degrees of freedom (plus one if they depend explicitly on time)--, (b) the instantaneous coordinates and velocities $(q^i_0,\,\dot q^i_0,t_0)$ give the system's state, and (c) it evolves according to the solution of the Euler-Lagrange equations for these initial data. For the local field case, the field equations form a partial differential system with a well-posed initial value problem, and the Cauchy-Kowalevski theorem provides a solution determined by the field itself and its first-time derivative on a non-characteristic hypersurface.

In contrast, both for mechanics and fields, the Euler-Lagrange equations in the nonlocal case are integrodifferential, in the simplest case in convolutional form. No general existence and uniqueness theorems exist concerning the solution for such a system. Consequently, the picture of a ``state of the system'' that evolves in time according to the Euler-Lagrange equations breaks down. Each system requires specific treatment to determine a set of parameters characterizing each dynamic solution, and this number of parameters may be infinite.

In our approach, we have opted for: (1) taking the Euler-Lagrange equations as constraints that select the dynamic trajectories, $\mathcal{D}^\prime$, as a subclass among all kinematic trajectories, $\mathcal{K}^\prime$, and (2) the time evolution as the trivial correspondence $q(t)\rightarrow q(t+\sigma)$, i.e., the trajectory evolves in time by advancing its initial point an amount $\sigma$ towards the future. For the field case, it follows the same idea.  The spacetime translation is considered the trivial correspondence $\phi(x) \rightarrow \phi(x+y)\,$, i.e., moving the ``initial'' spacetime point an amount $y^a$.

We then posed the principle of least action and derived the nonlocal Euler-Lagrange equations. It has the peculiarity that, because the nonlocality makes all the values of the trajectory or field intervene in the action integral $\mathcal{S}$, the support of the integral action has to be overall $\R$ or $\R^4$, respectively. This fact could lead to the integral action $\mathcal{S}$ being infinite. For this reason, we introduced the one-parameter family of finite nonlocal action integrals $\mathcal{S}(R)$. Moreover, the trajectory or field variations are required to be of bounded support to avoid this issue. 

In addition, we have studied the Noether symmetry for nonlocal Lagrangians. As is well known, when a local Lagrangian can be written as a total derivative, the equations of motion vanish identically. However, when we have a nonlocal Lagrangian, although we can always write it as a total derivative of a nonlocal function, it is observed that the nonlocal Euler-Lagrange equations do not vanish identically. For this reason, we have studied what sufficient condition the total derivative must meet for the Noether symmetry to be fulfilled.  We have concluded that a sufficient condition is (\ref{chap32-ACW}) for mechanics and (\ref{TD3}) for fields.

Both for mechanics and fields, we have proved an extension of Noether's theorem for nonlocal Lagrangians. In particular, for the mechanics case, once we get the closed form of the boundary term, we make an educated guess of the canonical momenta definition, which is used to set up the Hamiltonian formalism for such Lagrangians. This fact could not be based on a Legendre transformation in the usual manner because the latter consists of replacing half of the coordinates in the initial data space, namely the velocities, with the same number of conjugated momenta. In contrast, in the nonlocal case, the initial data space is infinite-dimensional, and half of infinity is still infinity. Therefore, infinite dimensions cannot be handled with the same tools as finite ones.  

Hence, the method we have proposed to build a Hamiltonian formalism is as follows: we start by considering an almost trivial Hamiltonian formalism on the kinematic phase space $\Gamma^\prime$. We show that the Hamiltonian flow preserves a submanifold that is diffeomorphic to the extended dynamic space $\mathcal{D}^\prime$. This fact enables us to pull the Hamiltonian formalism in the larger space $\Gamma^\prime$ back onto $\mathcal{D}^\prime$. We opt for the symplectic formalism instead of Dirac's method for constrained Hamiltonian systems since it is better suited  for pullback techniques. This way, we derive the formulae for the Hamiltonian and the (pre)symplectic form, provided we can find an appropriate coordinatization of the dynamic space. Furthermore, we try to see that the (pre)symplectic form is non-degenerate when we introduce the constraints of $\calD^\prime$, i.e., when we specify the trajectory in terms of the ``coordinates" of $\calD^\prime$. This procedure must be explicitly done for each case. To extend the Hamiltonian formalism from mechanics to fields, we have used the same procedure as the local case; we have relied on the transformations we have reflected in the table (\ref{chap34-TableMF}). 

Whereas in the local case, the Euler-Lagrange equations are an ordinary differential system, the theorems of existence and uniqueness allow to coordinate the dynamic space without needing the general solution in an explicit form. For nonlocal systems, we do not have such theorems as a rule, and to apply our method thoroughly (discerning whether the presymplectic form is indeed non-degenerate), one needs to know the explicit general solution of the classical problem. Therefore, the method proposed here is not suitable for solving the nonlocal Euler-Lagrange equations, in contrast to what happens in the local case. Nevertheless, it provides a way to set up a Hamiltonian formalism for nonlocal Lagrangians, which, up to now, is a necessary step towards quantization.

We have then applied our results to a regular first-order and $r^{\rm th}$-order Lagrangian, the nonlocal harmonic oscillator, $p$-adic particle, $p$-adic open string, and electrodynamics of dispersive media. 

For the nonlocal harmonic oscillator example, we have solved its nonlocal Euler-Lagrange equations. They have a peculiarity; they can be transformed into a local one, allowing us to obtain the general solution. Furthermore, we have studied for which conditions this solution is stable, and we concluded that stability only exists if all characteristic roots are simple and imaginary, namely, whenever the parameters $(g,\omega^2)$ lie in regions 1 or 4 in Figure \ref{F1}. In addition, we have calculated the symplectic form and the Hamiltonian. In particular, the Hamiltonian has no definite sign, so it is not bounded from below. However, our system is still stable, contradicting the widespread belief that nonlocal and local higher-order Lagrangians also suffer from Ostrogradsky instability. In our view, this conception results from taking a sufficient condition of stability, namely ``energy is bounded from below," as a necessary condition.

As for the $p$-adic particle and open string, we have focused on the perturbative solutions allowed by these models to obtain both the Hamiltonian and the symplectic form. Furthermore, for the $p$-adic open string, the canonical momentum energy tensor and the Belinfante-Rosenfeld tensor were calculated in closed form, and the components of the Belinfante-Rosenfeld tensor were explicitly computed for a perturbative solution.

Other methods previously studied the $p$-adic examples. These methods transform the nonlocal Lagrangian into an infinite-order Lagrangian by replacing the whole trajectories in the nonlocal Lagrangian with a formal Taylor series (that includes all the derivatives of the coordinates) and then deal with it as a $r^{\rm th}$-order Lagrangian with $r\rightarrow\infty$. The value of those methods can only be heuristic unless the convergence of the series is proved or the ``convergence'' for $r\rightarrow\infty$ is adequately defined. Furthermore, these methods are cumbersome in that they often imply handling infinite series with many subindices, square rank $\infty$ matrices, formal inverses, regularizations, etc. In contrast, our approach is based on functional methods and, as it involves integrals instead of series, is much easier to handle, and the results are the same.

Regarding the electromagnetic field in dispersive media, we have obtained an explicit expression for the canonical energy-momentum tensor, which depends quadratically and nonlocally on the Faraday tensor and its first-order derivatives. In the non-dispersive (local case) limit, this tensor does not coincide with the Minkowski energy-momentum tensor; the difference is the four-divergence of an antisymmetric tensor. Therefore, we have derived this correction by applying the Belinfante-Rosenfeld technique and obtained the Belifante-Rosenfeld energy-momentum tensor, which in the non-dispersive limit does reduce to Minkowski tensor. The Belinfante-Rosenfeld is generally not symmetric --nor is the Minkowski tensor-- because the angular momentum current is not conserved. After all, the nonlocal Lagrangian is not Lorentz invariant, as expected, because the proper reference system of the medium is privileged.

It must be said that our nonlocal Lagrangian model has the disadvantage that its scope is restricted to non-absorptive media. Indeed, the nonlocal action (\ref{e19}) implies the symmetry conditions (\ref{e19a}) and (\ref{e19aa}), so it follows that $\varepsilon(\omega,k)$ is real for real $\omega$ and $k$. It must be recalled that the absorptive behavior of a medium is connected with the imaginary part of its dielectric function $\varepsilon\,$. Moreover, if this imaginary part vanishes, it follows from Kramers-Kr\"onig relations that $\varepsilon$ and $\mu$ must be constant. Therefore, if the nonlocal Lagrangian model does not violate causality, it must be non-dispersive, i.e., local.

Giving up the nonlocal Lagrangian model and basing the description of the causal non-dispersive medium on Maxwell's equations, we have been still able to propose (\ref{D1}) and (\ref{D4A}) as two possible definitions for energy-momentum currents, respectively, the canonical and the Belinfante-Rosenfeld tensors. Evaluating their four-divergences provided that the nonlocal Euler-Lagrange equations (\ref{A11}) hold, we have then found that they are not locally conserved, and this fact is due to the absorptive components of the dielectric and magnetic functions, i.e., $\imap(\varepsilon)$ and $\imap(\mu)$.

We have then particularised our Belinfante-Rosenfeld energy-momentum tensor to the electromagnetic field of slowly varying amplitude over a rapidly oscillating carrier wave. Furthermore, we have considered the medium in the optical approximation, that is, $\varepsilon$ and $\mu$ only depend on the frequency $\omega$. Taking the average over one period of the carrier and using the slow motion approximation, we have evaluated the energy and momentum densities, the Poynting vector, and the Maxwell stress tensor in the rest reference frame. Energy density is the only of these quantities that are given in some textbooks, and our result agrees with them \cite{Landau1984, Jackson1999, schwinger1998}.

\section{Outlook}

Although the results are promising, there are still some points that are worth studying:

The first point is clear. Note that all the examples we have applied our formalism are based on nonlocal Lagrangians that do not explicitly depend on time. Therefore, it would be exciting to study the case of nonlocal Lagrangians that depend explicitly on time \cite{Barnaby2011} and see what results (or information) can be extracted from them. Recall that this formalism contemplates it, so the great difficulty will come from coordinating the extended dynamical space for such cases. 

The second point to study is also evident. Note that we have focused on applications of Noether's first theorem. In none of the examples have we applied the second one. It would be interesting to study a model for which it can be applied, for example, the non-commutative spacetime theory U(1) \cite{Gomis2001_2}. On this point, we can advance that the results we have obtained in Section \ref{chap522-GS}  by applying Noether's second theorem \cite{Heredia2023} agree with the results obtained in \cite{Gomis2001_2}. 

The third point to explore further would be the possible covariance of the proposed Hamiltonian formalism. Note that the extension of Noether's theorem is completely covariant; it is at the point of defining the momenta, where the covariance is broken. It would be interesting to follow the line of \cite{Leon1985} and propose an extension of the De Donder-Weyl theory for nonlocal Lagrangians. 

Finally, the last interesting avenue for further research would be to investigate an extension of this formalism that does not specify coordinates and allows for more complex temporal evolutions beyond those proposed in this thesis. Such an approach could be particularly valuable in the context of nonlocal gravity theories.


%

\appendix

\chapter{Appendix: p-adic open string field}\label{App2}

In the following, we shall detail some of the steps of the example ``p-adic open string field." We shall assume that the reader is familiar with the theory of generalized functions (distributions); if not, we suggest \cite{Vladimirov_GF, Vladimirov_MF, Vladimirov}.

\section{Equation (\ref{p6}) derived from (\ref{p6a})}

Equation (\ref{p6a}) amounts to
\begin{equation}
 \E \ast\:X = 0\,, \qquad {\rm where} \qquad X:=\T \ast\:\phi -\left(\E \ast\:\phi\right)^p \,.
 \end{equation}

For any fixed value of $t$, $\phi(\mathbf{x},t)$ belongs to the class of slowly growing smooth functions $\theta_M(\R^3)$, i.e., it grows more slowly than any power of $|\mathbf{x}| $ at spatial infinity. This class is a subset of the space of tempered distributions $\mathcal{S}^\prime(\mathbb{R}^3)$; therefore,  $\left(\mathcal{T}\ast \phi\right)_{(\mathbf{x},t)} \,, \left(\mathcal{E}\ast \phi\right)_{(\mathbf{x},t)}$ and $X(\mathbf{x},t)$ also belong to the same class for any fixed value of $t$ \cite{Vladimirov_GF}.

We shall now prove that $\mathcal{E}\ast X =0$ implies $X=0$. Indeed, for any fixed $t$, $X_{(t)}(\mathbf{x}):=X(\mathbf{x},t)$ is a tempered distribution. Furthermore, as $\mathcal{G}_3 \in \mathcal{S}(\mathbb{R}^3)$ -- Schwartz space --, the convolution theorem holds \cite{Vladimirov_GF} and 
\begin{equation}
\mathcal{G}_3 \ast X_{(t)} \in \theta_M(\mathbb{R}^3) \qquad {\rm and} \qquad 
\mathcal{F}\left(\mathcal{G}_3 \ast X\right) = \mathcal{F}\left(\mathcal{G}_3\right) \,\mathcal{F}\left(X_{(t)}\right) \in \mathcal{S}^\prime(\mathbb{R}^3)\,, 
\end{equation}
where $\mathcal{F}$ means the Fourier transform in $\mathcal{S}^\prime(\mathbb{R}^3)\,$.  Therefore, $\mathcal{E}\ast\: X = 0$ implies $\mathcal{G}_3\ast\:X_{(t)} = 0\,$, whose Fourier transform yields
\begin{equation}
e^{-r \mathbf{k}^2} \,\mathcal{F}\left(X_{(t)}\right)  = 0 \,,
\end{equation}
that is, $\forall \varphi(\mathbf{k})\in \mathcal{S}(\mathbb{R}^3)\,,\,(\mathcal{F}\left(X_{(t)}\right), e^{-r \mathbf{k}^2}\varphi(\mathbf{k})) = 0 \,$. This fact does not yet imply that $\mathcal{F}\left(X_{(t)}\right) = 0$ because not all $\psi\in \mathcal{S}(\mathbb{R}^3)$ can be written as $e^{-r \mathbf{k}^2}\varphi(\mathbf{k})\,$. However, since $\mathcal{S}^\prime \subset \mathcal{D}^\prime$ --tempered distributions are distributions-- and as, for any $\rho(\mathbf{k})\in \mathcal{D}(\mathbb{R}^3)$, then $e^{r\mathbf{k}^2}\rho(\mathbf{k})$ also has compact support, we have that \vspace{0.2cm}
\begin{equation}
\left(\mathcal{F}\left(X_{(t)}\right),\rho(\mathbf{k})\right) = \left(e^{-r \mathbf{k}^2} \,\mathcal{F}\left(X_{(t)}\right), e^{r \mathbf{k}^2} \,\rho(\mathbf{k})\right) = 0\;.
\end{equation}
Now, the space of test functions $\mathcal{D}$ is dense in the Schwartz space $\mathcal{S}$ \cite{Vladimirov_GF}, i.e., any $\psi\in \mathcal{S}$ is the limit of a sequence $\{\rho_n \in \mathcal{D}\,, \; n\in\mathbb{N}\}$ and, therefore,
\begin{equation}
\left(\mathcal{F}\left(X_{(t)}\right),\psi(\mathbf{k})\right) = \lim_{n\rightarrow\infty}\left(\mathcal{F}\left(X_{(t)}\right),\rho_n(\mathbf{k})\right) = 0 \,.
\end{equation}
Namely, $\mathcal{F}\left(X_{(t)}\right)=0\:\:\:\forall\psi(\mathbf{k})\in\calS(\R^3)$, or equivalently, $X_{(t)}(\mathbf{x}) = 0\:\:\:\forall\psi(\mathbf{x})\in\calS(\R^3)$.

\section{The convergence of solution (\ref{p10})}
The space integral in (\ref{p10}) might diverge because the imaginary part of $\alpha_\nu(\mathbf{k})$ makes that $e^{i\,t\, \alpha_\nu(\mathbf{k})}$ grows exponentially for large $\mathbf{k}$ and $t< 0$. We shall see that, despite this fact, the integral does converge. Indeed, 
\begin{equation}
\label{padic:Desig}
|\tilde{\Phi}_0(\mathbf{x},t)|=\frac{1}{(2\pi)^3} \left| \int_{\R^3} \D\mathbf{k} \sum_{\nu} C_\nu(\mathbf{k}) e^{i[\alpha_\nu(\mathbf{k}) t + \mathbf{k}\cdot\mathbf{x}]}\right| \leq \int_{\R^3} \D\mathbf{k} \sum_{\nu}  \left| C_\nu(\mathbf{k})\right| e^{- t\,s\,\alpha_{\nu\mathbb{I}}(\mathbf{k}) }\:,
\end{equation}
where we have written $\alpha_\nu (\mathbf{k})= s\,\left[\alpha_{\nu\mathbb{R}} (\mathbf{k}) + i\,\alpha_{\nu\mathbb{I}}(\mathbf{k}) \right]\,$, with $\nu=(s,n)\,,\,\,n\in\mathbb{Z}$, and 
\begin{align}   
\alpha_{\nu\mathbb{R}} (\mathbf{k}) &:=  \sqrt{\frac{|\mathbf{k}|^2 -2 +\sqrt{\left(|\mathbf{k}|^2-2\right)^2+\left(\frac{2n\pi}{r}\right)^2}}{2}}  \qquad {\rm and} \nonumber\\
\alpha_{\nu\mathbb{I}}(\mathbf{k}) &:=   \sign(n)\,\sqrt{\frac{-|\mathbf{k}|^2 + 2 +\sqrt{\left(|\mathbf{k}|^2-2\right)^2+\left(\frac{2n\pi}{r}\right)^2}}{2}} \,.
\end{align}

If $t\,s\,\sign(n) <0\,$, the exponent on the right-hand side of (\ref{padic:Desig}) is $|t\,\alpha_{\nu\mathbb{I}}(\mathbf{k})| > 0\,$, which is positive. However, it can be easily checked that $\lim_{|\mathbf{k}|\rightarrow\infty} \alpha_{\nu \mathbb{I}}(\mathbf{k}) = 0 \,$. Hence,
\begin{equation}
\forall \varepsilon >0\,,\,\,\exists K > 0\,\quad \mid \quad |\mathbf{k}| > K \,\,\Rightarrow\,\,|t\,\alpha_{\nu\mathbb{I}}(\mathbf{k})| <\varepsilon\,.
\end{equation}
Namely, if  $|\mathbf{k}|$ is large enough, the exponential is bounded by the constant $e^\varepsilon$. Therefore, it follows that $|\tilde{\Phi}_0(\mathbf{x},t)| < \infty$, provided that $\sum_{\nu} \left| C_\nu(\mathbf{k})\right|$ is summable.

\section{The $\tilde{\phi}$ transformation rule for Poincar\'e invariance}
Let us find the  transformation rule of $\tilde{\phi}$ to leave the nonlocal Lagrangian density (\ref{padic:Lntc}) Poincar\'e invariant. Since $\tilde{\psi}$ is a scalar, by equation (\ref{chap331-P3}), we get 
\begin{equation}   \label{A5.1}
 \delta \tilde{\psi}(x) = - \left(\varepsilon^c + \omega^c_{\;b} x^b\right) \,\partial_c\tilde{\psi}(x) \,.
\end{equation}
In order to correctly define the $e^{-r\Box}$ operator, our dynamic variables are the $\tilde{\phi}$ fields. \vspace{0.2cm}
\noindent These fields are related to the $\tilde{\psi}$ fields through equation (\ref{p2}), which its Fourier transform is
\begin{equation}   \label{A5.2}
\mathcal{F}_{\mathbf{x}}[\tilde{\psi}](\mathbf{k},t)=\mathcal{G}(\mathbf{k}) \cdot \tilde\phi(k) \,,  \quad\mathrm{with} \quad \tilde\phi(\mathbf{k},t)=\mathcal{F}_{\mathbf{x}}[\tilde{\phi}](\mathbf{k},t)   \quad \mathrm{and} \quad \mathcal{G}(\mathbf{k}) = e^{-r \mathbf{k}^2}\,.
\end{equation}
On the other hand, the Fourier transform of (\ref{A5.1}) is
\begin{align}   
\label{A5.3}
\mathcal{F}_{\mathbf{x}}[\delta \tilde{\psi}](\mathbf{k},t) & = -\left\{\varepsilon^0\,\partial_0\,\mathcal{F}_{\mathbf{x}}[\tilde{\psi}](\mathbf{k},t) +  \omega^{0i}\,\partial_0\,\mathcal{F}_{\mathbf{x}}[x_i\:\tilde{\psi}](\mathbf{k},t) + (\varepsilon^j +\omega^{j0} x_0) \mathcal{F}_{\mathbf{x}}[\partial_j\tilde{\psi}](\mathbf{k},t) \right.\nonumber\\
 &\qquad \left. +\omega^{ji}\mathcal{F}_{\mathbf{x}}[x_i\,\partial_j \tilde{\psi}](\mathbf{k},t)  \right\}\,.
 \end{align}
Plugging equation (\ref{A5.2}) into (\ref{A5.3}), we find 
\begin{align}   
\label{A5.6}
 \mathcal{G}(\mathbf{k})\:\delta\tilde\phi(\mathbf{k},t)&= -\mathcal{G}(\mathbf{k})\,\left\{\varepsilon^0\:\partial_0\tilde\phi(\mathbf{k},t) + i\,\omega^{0i}\left[ D_i\partial_0\tilde\phi(\mathbf{k},t)-2\,r\,k_i \,\partial_0\tilde\phi(\mathbf{k},t) \right]\right.\nonumber\\
 &\quad \left.+ i\,(\varepsilon^j +\omega^{j0} x_0)\mathbf{k}_j \:\tilde\phi(\mathbf{k},t) -\omega^{ji}\left[\delta_{ij} \:\tilde\phi(\mathbf{k},t) + k_j \,D_i\tilde\phi(\mathbf{k},t)\right.\right.\nonumber\\
 &\quad \left.\left. -2\,r\,k_jk_i \,\tilde\phi(\mathbf{k},t) \right] \right\}\,,
\end{align}
where $D_j$ denotes $\partial/\partial k^j$ to distinguish it from $\partial/\partial x^j$. Bearing in mind that $\omega^{ab}=-\omega^{ba}$, we can simplify it as
\begin{align}   
\label{A5.7}
 \delta\tilde\phi(\mathbf{k},t)=& - i\,(\varepsilon^j +\omega^{j0} x_0)\,k_j\,\tilde\phi(\mathbf{k},t) - \left(\varepsilon^0 - 2\,i\,r\,\omega^{0i}\,k_i\right)\partial_0\tilde\phi(\mathbf{k},t)\nonumber\\
 & + \omega^{ji}k_j\,D_i\tilde\phi(\mathbf{k},t) - i\,\omega^{0i}D_i\partial_0\tilde\phi(\mathbf{k},t)
\end{align} 
that, undoing the Fourier transform, we obtain how the $\tilde{\phi}$ field transforms,
\begin{equation}
\label{padic:t1}
 \delta \tilde{\phi}(x) = -\left(\varepsilon^c+ \omega^{cb} x_b\right)\partial_c\tilde{\phi}(x) + \omega^{ab}\left[2r\,\delta^4_{[a} \delta^i_{b]}\, \partial_i\dot{\tilde{\phi}}(x)\right]\,.
\end{equation}

\clearpage
\phantomsection
\bibliographystyle{utphys}
{\bibliography{References}}
\addcontentsline{toc}{chapter}{Bibliography}

\newpage
\printindex

\backmatter

\end{document}